\documentclass[amsmath,amssymb,showpacs,showkeys,onecolumn]{revtex4}
\usepackage[dvips]{graphicx,color}
\usepackage{times}
\usepackage{mathrsfs}
\usepackage{amsmath}
\usepackage{amssymb}
\usepackage{graphicx}
\usepackage{epsfig}
\usepackage{dcolumn}
\usepackage{bm}
\topmargin {-2.0cm}

\definecolor{r}{rgb}{1.0,0,0}

\begin{document}
\title{A simulation study on few parameters of Cherenkov photons in extensive air showers of different primaries incident at various zenith angles over a high altitude observation level}

\author{G. S. Das\footnote{gsdas@dibru.ac.in}, P. Hazarika\footnote{poppyhazarika1@gmail.com} and U. D. Goswami\footnote{umananda@dibru.ac.in}}
\affiliation{Department of Physics, Dibrugarh University,
Dibrugarh 786 004, Assam, India}

\begin{abstract}
We have studied the distribution patterns of lateral density, arrival time and 
angular position of Cherenkov photons generated in Extensive Air Showers (EASs) initiated by $\gamma$-ray, proton and iron primaries incident with various 
energies and at various zenith angles. This study is the extension of our 
earlier work \cite{Hazarika} to cover a wide energy range of ground 
based $\gamma$-ray astronomy with a wide range of zenith angles 
($\le 40^\circ$) of primary particles, as well as the extension to study the 
angular distribution patterns of Cherenkov photons in EASs. This type of study 
is important for distinguishing the $\gamma$-ray initiated showers from the 
hadronic showers in the ground based $\gamma$-ray astronomy, where Atmospheric 
Cherenkov Technique (ACT) is  being used. Importantly, such study 
gives an insight on the nature of $\gamma$-ray and hadronic showers in general. In this work, the CORSIKA 6.990 simulation code is used for generation 
of EASs. Similarly to the case of Ref.\cite{Hazarika}, this study also revealed 
that, the lateral density and arrival time distributions of Cherenkov photons 
vary almost in accordance with the functions: $\rho_{ch}(r) = 
\rho_{0}\;e^{-\beta r}$ and $t_{ch}(r) = t_{0}e^{\Gamma/r^{\lambda}}$ 
respectively by taking different values of the parameters of functions for the 
type, energy and zenith angle of the primary particle. The 
distribution of Cherenkov photon's angular positions with respect to shower
axis shows distinctive features depending on the primary type, its energy 
and the zenith angle. As a whole this distribution pattern for the iron primary
is noticeably different from those for $\gamma$-ray and proton primaries. The 
value of the angular position at which the maximum number of Cherenkov photons 
are concentrated, 
increases with increase in energy of vertically incident primary, but for 
inclined primary it lies within a small value ($\le 1^\circ$) for almost all 
energies and primary types. 
No significant difference in the 
results obtained by using the high energy hadronic interaction models, viz., 
QGSJETII and EPOS has been observed.   
\end{abstract}

\maketitle

\section{Introduction}
The primary objective of the $\gamma$-ray astronomy is to detect $\gamma$-rays 
from celestial sources. For this purpose, in the ground based $\gamma$-ray 
astronomy, the Atmospheric Cherenkov Technique (ACT) 
\cite{Ong, Hoffman, Weekes, Lorenz, Holder0, Funk, Degrange, Goumard, Bhat,
Acharya, Holder} is being used most widely 
within its operational energy range from around hundred GeV to tens of TeV. 
This 
technique is based on detection of Cherenkov photons emitted in the Extensive 
Air Showers (EASs) that are generated due to the interaction between the 
primary $\gamma$-rays and air nuclei. The 
$\gamma$-ray sources also emit Cosmic Rays (CRs), which are deflected by the 
intragalactic magnetic fields because of their charge 
and hence they loose their directional property. Whereas $\gamma$-rays, being 
neutral, they retain their direction of origin. Thus, by the detection of 
$\gamma$-rays one can make an estimate of the positions of those astrophysical 
objects.

Because of the indirect nature of experiments in ACT as well as due to the 
presence of huge CR background, a complete Monte Carlo simulation study on 
atmospheric Cherenkov photons needs to be carried out for the detection of 
$\gamma$-rays with proper estimation of their energy from the observational 
data of such experiments. It is to be noted that, although both $\gamma$-ray 
and CR can generate EAS, the nature of two are different. EAS generated by 
$\gamma$-ray is purely electromagnetic (EM) in nature, whereas it is an 
admixture of EM and hadronic cascades in the case of CR.
Although many studies have already been done, specially on the lateral density 
and arrival time distributions of Cherenkov photons in EASs using available 
simulation techniques \cite{Bardan, Chitnis, Hillas, Lafebre, Nerling}, still 
it would be a worthy task to have detailed studies on angular distributions as 
well as on lateral density and arrival time distributions of Cherenkov photons, 
initiated by $\gamma$-ray and hadronic particles, incident at various zenith 
angles with a wide range of energy, particularly at high altitude 
observation levels. Keeping this point in mind, in this work we have studied 
the lateral density, arrival time and angular distributions of Cherenkov 
photons in EASs at different energies and zenith angles over a high altitude 
observation level, using two different high energy hadronic interaction models, viz., QGSJETII and EPOS with FLUKA low energy hadronic interaction model 
available in the CORSIKA simulation package \cite{Heck}. This is the extension 
of our earlier work \cite{Hazarika} to cover a wide energy range of 
ground based $\gamma$-ray astronomy with a wide range of zenith 
angles ($\le 40^\circ$) of primary particles, and to study the angular 
distribution patterns of Cherenkov photons in EASs.

CORSIKA is a four dimensional detailed Monte Carlo simulation code developed
to study the evolution and various properties of EASs in the 
atmosphere. It can be used to simulate interactions and decays of nuclei, 
hadrons, 
muons, electrons and photons in the atmosphere up to energies of the order of 
10$^{20}$eV. For the simulation of hadronic interactions, presently CORSIKA 
has the option of seven high energy hadronic interaction models and three low 
energy hadronic interaction models \cite{Heck}. It uses the EGS4 code 
\cite{Nelson} for the simulation of EM component of the air shower.

The rest of the paper is organized as follows. In the next section, we discuss 
about the simulation process involved in this work. The analysis of the 
simulation work and consequent results are discussed in the Section III. 
We summarized our work in the Section IV.    

\section{Simulation of Cherenkov photons in extensive air showers}
As mentioned in the previous section, we have used the CORSIKA 6.990 simulation
package by selecting two high energy hadronic interaction models, viz., 
QGSJETII.3 and EPOS 1.99 with the low energy hadronic interaction model FLUKA 
to simulate Cherenkov photons in EAS. In fact, we have generated EASs for the 
monoenergetic $\gamma$-ray, proton and iron primaries incident vertically as 
well as inclined at zenith angles 10$^{\circ}$, 20$^{\circ}$, 30$^{\circ}$ and 
40$^{\circ}$ using the QGSJETII-FLUKA and EPOS-FLUKA hadronic interaction model 
combinations. The intention to use two different high energy hadronic 
interaction models in our simulation is to test robustness of conclusions of 
our work. Moreover, this will also provide an opportunity to compare the 
performance of the models concerned. QGSJETII is the improved version of the 
model QGSJET01, which is based on the Gribov-Regge theory \cite{Heck, Goswami}.
QGSJET is one of the most extensively used high energy hadronic interaction 
models in the simulation works of CR and $\gamma$-ray experiments.
On the other hand, EPOS is based on quantum mechanical multiple scattering 
approach based on partons and strings. The performance of EPOS is better in 
comparison to RHIC data \cite{Pierog}. In our earlier work 
\cite{Hazarika}, we have studied extensively the QGSJET01C, VENUS 4.12 and
QGSJETII.3 along with all hadronic interaction models at low energy presently 
available in CORSIKA. Using the above cited model combinations,
 the following numbers of showers are generated at different energies and 
zenith angles for the $\gamma$-ray, proton and iron primaries as given in the 
Table \ref{tab1}.

\vspace{-0.4cm}
\begin{center}
\begin{table}[ht]
\caption{\label{tab1} Number of showers generated at different energies for the
$\gamma$-ray, proton and iron primaries incident at 0$^{\circ}$, 10$^{\circ}$, 
20$^{\circ}$, 30$^{\circ}$ and 40$^{\circ}$ zenith angles.}
\begin{tabular}{ccc}\\[-5.0pt]
\hline
Primary particle & ~~~~Energy & ~~~~Number of Showers \\\hline\\[-8.0pt] 
$\gamma$-ray  &  ~~~100 GeV  & 10,000 \\
              &  ~~~250 GeV  &  ~~7000 \\
              &  ~~~500 GeV  &  ~~5000 \\
              &  ~~~~~~~1 TeV    &  ~~2000 \\
              &  ~~~~~~~2 TeV    &  ~~1000 \\
              &  ~~~~~~~5 Tev    &  ~~~400\\
  && \\
 Proton       &  ~~~250 GeV  & 10,000 \\
              &  ~~~500 GeV  &  ~~8000 \\
              &  ~~~~~~~1 TeV    &  ~~5000 \\
              &  ~~~~~~~2 TeV    &  ~~2000 \\
              
              &  ~~~~~~~5 TeV    &  ~~~800 \\
 && \\
 Iron         &  ~~~~~~~1 TeV  & 50,000 \\
              &  ~~~~~~~5 TeV  & ~4000 \\
              &  ~~~~~10 TeV  &  ~~2000 \\
              &  ~~~~~50 TeV    &  ~~1000 \\
              &  ~~~100 TeV    & ~~~600 \\
            
\hline
\end{tabular}
\end{table}
\end{center}
\vspace{-0.5cm}
The energy range of the primaries selected here lies within the typical range 
of ACT energy for different primaries on the basis of their equivalent number 
of Cherenkov photon yields. The generation of these showers is done by taking 
the altitude of HAGAR experiment at Hanle (longitude: 
78$^{\circ}$ 57$^\prime$ 51$^{\prime\prime}$ E, latitude: 32$^{\circ}$ 
46$^\prime$ 46$^{\prime\prime}$ N, altitude: 4270 m) as the observational 
level. However, few other showers are generated at altitudes of 1000 m and 
2000 m with same longitude and latitude as Hanle to study the effect of 
altitude on nature of EASs. The cores of the EASs are considered to 
coincide with  
the centre of the detector array. The geometry of the detector system is taken 
as a flat horizontal detector array, having 25 non-imaging telescopes in the 
E--W direction 
with a separation of 25 m in between two consecutive detectors and also 25 such 
telescopes in the N--S direction with a separation of 20 m. Each telescope is 
considered to have 9 m$^2$ mirror area. To detect a TeV EAS with a large 
zenith angle, a very wide array of detectors is required for the case of 
non-imaging detection. 
Considering the 
range of energies and zenith angles of our simulated showers, the detector 
array taken here is quite sufficient for the purpose of this simulation 
work. 
In case of the longitudinal Cherenkov photon distribution, photons are counted 
in the step where they are emitted and for simplicity of calculation it is 
chosen that their emission angle is wavelength independent. The wavelength 
window for the Cherenkov photon production is selected as 200 -- 650 nm as 
the wavelength sensitivity of usual photomultiplier tube (PMT) used in ACT lies 
well within this range. Hence only photons produced by the secondary charged 
particles within this specified wavelength range are allowed to propagate to 
the observation level. The absorption and scattering of Cherenkov photons in 
the atmosphere 
are not taken into account \cite{Heck}, although there are possibilities that 
Cherenkov photons may undergo Compton scattering, pair production as well as
photoelectric/photonuclear reactions in the  atmosphere. The variable bunch 
size option of 
Cherenkov photon is set to "5". This slightly high value of the variable bunch 
size is used to reduce the size of the simulated data. The parameter STEPFC in 
EGS4 code \cite{Nelson} decides the multiple scattering length for e$^-$ and 
e$^+$, which is set to 0.1. The low energy cutoffs of kinetic energy for 
hadrons, muons, electrons and photons are chosen as 3.0 GeV, 3.0 GeV, 0.003 GeV
and 0.003 GeV respectively. The last two options will help to reduce the
shower production time without affecting the final expected outcome. The 
Linsley's parametrized US standard atmosphere \cite{US} has been used
out of the different atmospheric models available in CORSIKA.  

\section{Analysis of simulated data and results}

To calculate the lateral density of Cherenkov photons, we have counted the 
number of photons incident on each detector per shower. The arrival time of 
a Cherenkov photon over a detector is obtained by calculating the time 
taken by the photon to reach the detector with respect to the first photon of 
the 
shower hitting the array. There are several photons hitting each detector per
shower. So, average of their arrival times is calculated for each detector.
Moreover, there are fluctuations in Cherenkov photon density and arrival time
over each detector from shower to shower. Hence, the variation of Cherenkov 
photon density and arrival time with respect to core distance is found by 
calculating their average values for the specified number of showers. These 
fluctuations of photon density and arrival time as a function of 
core distance (or for each detector) are expressed by calculating  the ratio 
of their r.m.s to mean values. Averaged over the azimuth, the number of 
photons produced per angular bin of photon's angular position (with respect to 
the shower axis) is counted to get angular distribution of Cherenkov photons. 
Furthermore, to investigate the model dependent variation of any parameter, 
as in the case of our earlier work \cite{Hazarika} we have calculated the 
percentage relative deviation between the two models for a particular 
parameter by using the following formula:
\begin{equation}
\Delta_{\chi} = \frac{\chi_{mp} - \chi_{rp}}{\chi_{rp}}\times 100\%.
\label{eq1}
\end{equation}                 
Here $\Delta_{\chi}$, $\chi_{rp}$ and $\chi_{mp}$ represent the relative 
deviation in percentage of a parameter, the reference model parameter
and the given model parameter respectively. 

For such analysis of the simulated data, C++ programs have been developed in 
the platform of the ROOT software \cite{Root}. The ROOT software is being 
developed in CERN for the purpose of the effective handling of huge data of 
high energy physics experiments. In the following subsections, various 
aspects of Cherenkov photons' distribution in terms of their lateral density, 
arrival time and angular position as well as their model dependent features 
are discussed.       

\begin{figure*}[hbt]
\centerline
\centerline{
\includegraphics[scale=0.29]{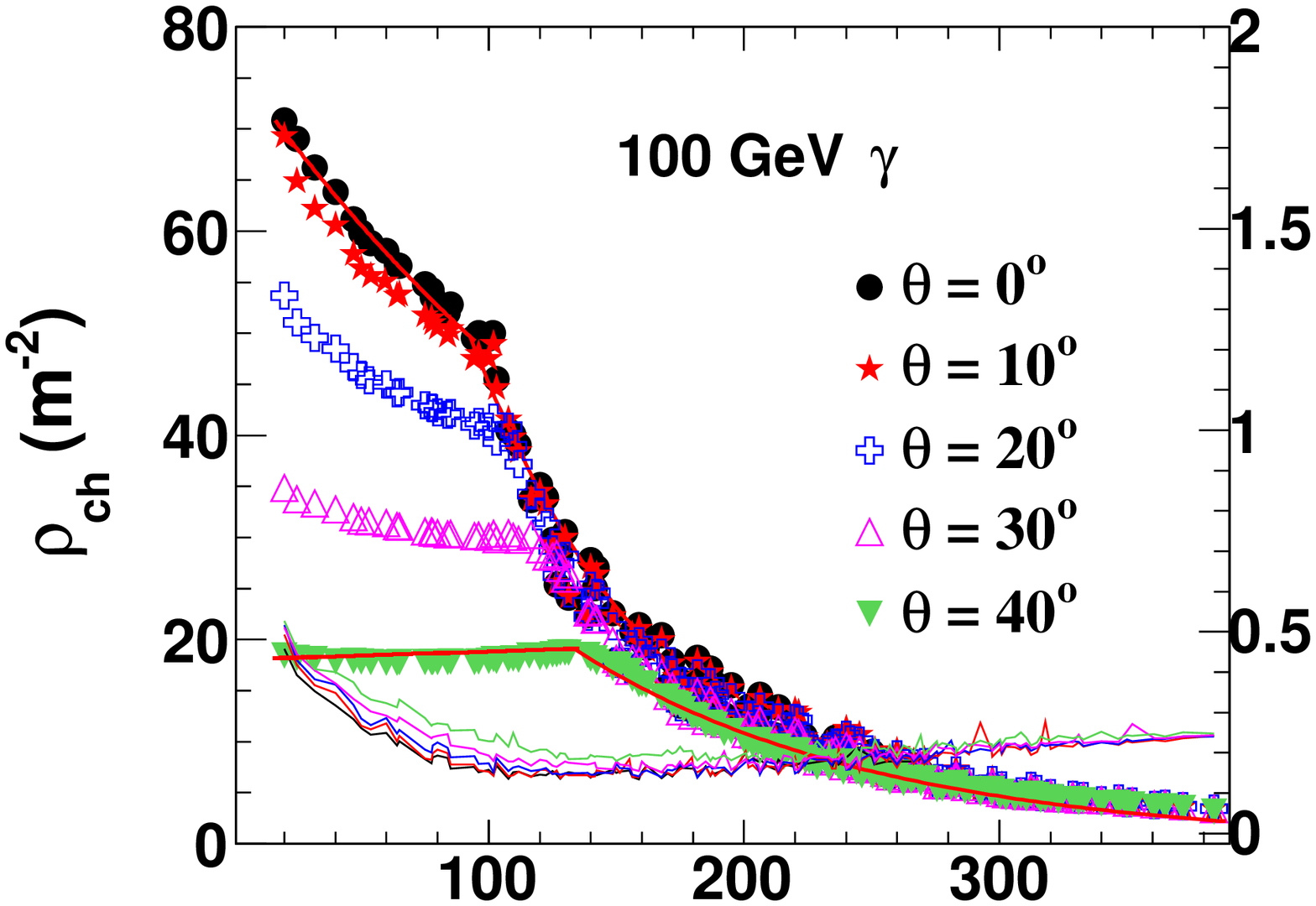}\hspace{-1mm}
\includegraphics[scale=0.29]{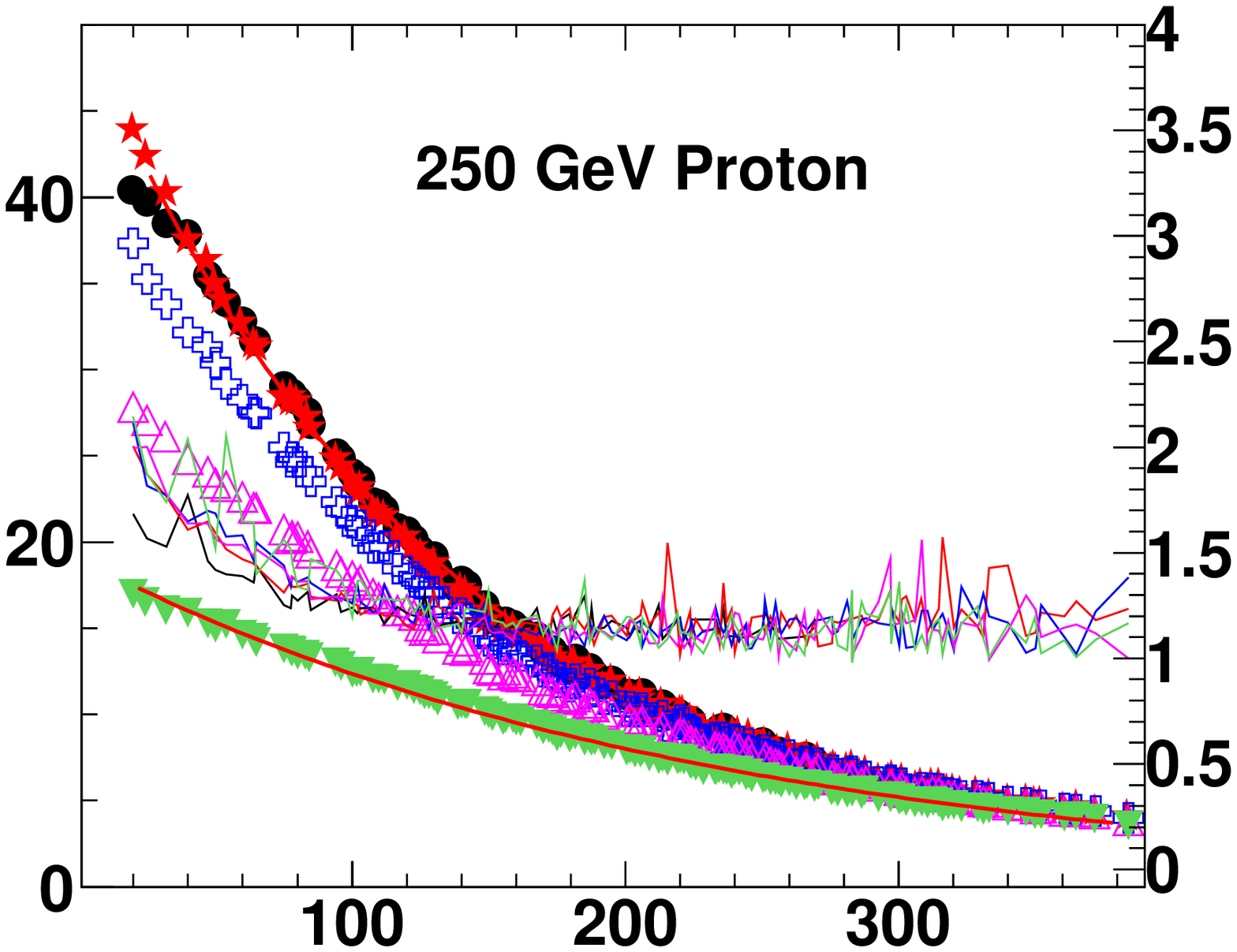}\hspace{-1mm}
\includegraphics[scale=0.31]{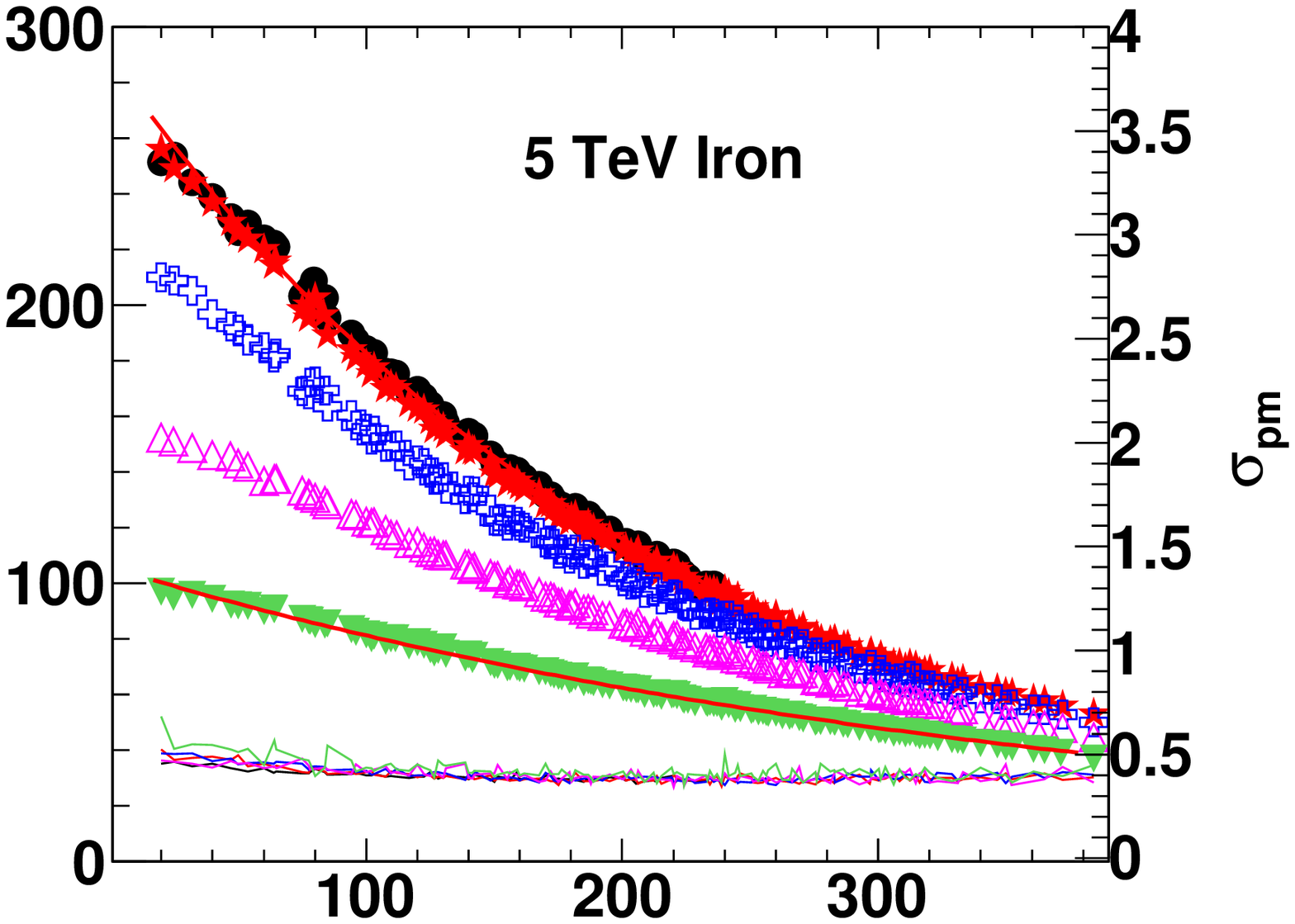}}

\centerline{\hspace{-1mm}
\includegraphics[scale=0.29]{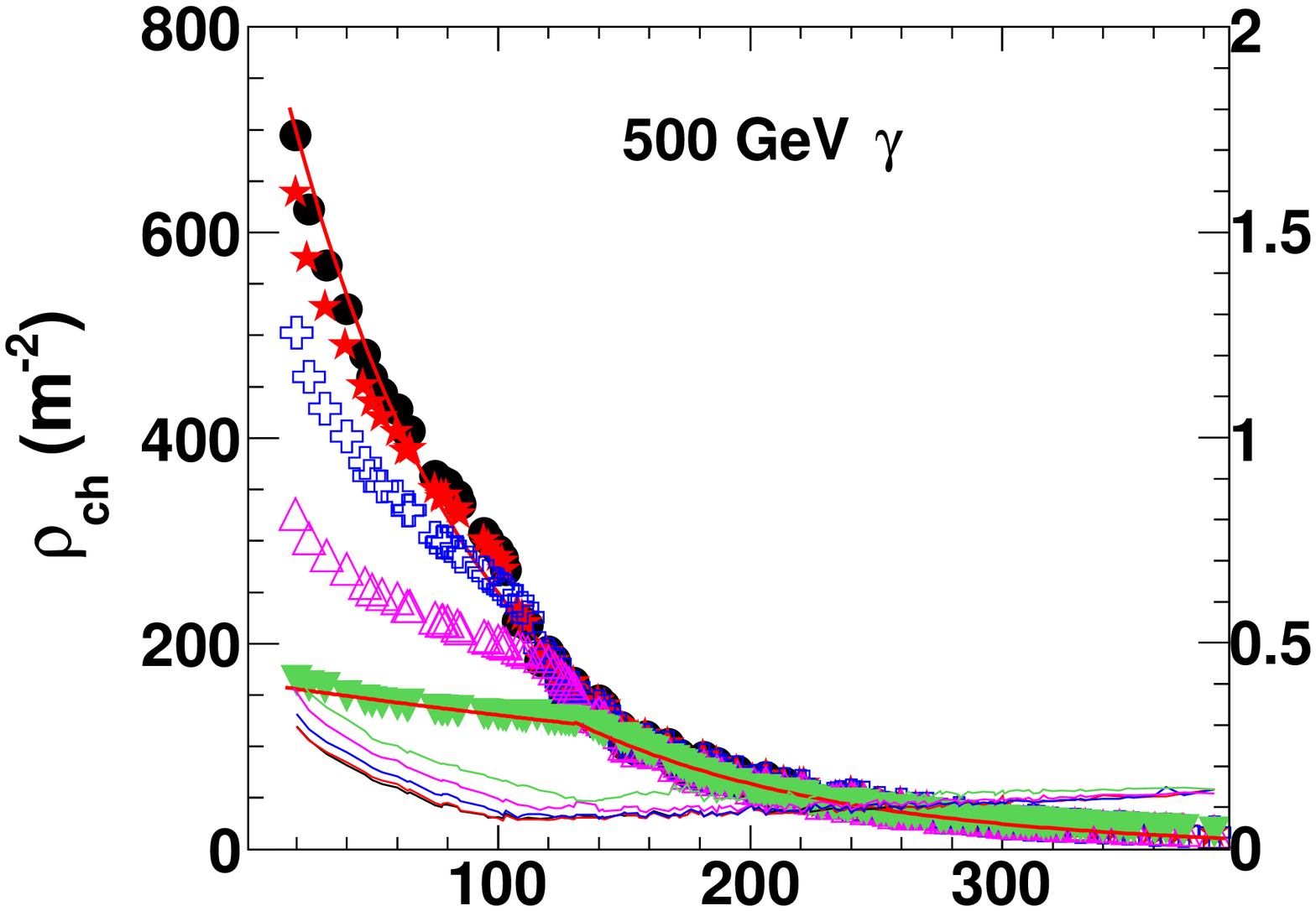} \hspace{0mm}
\includegraphics[scale=0.29]{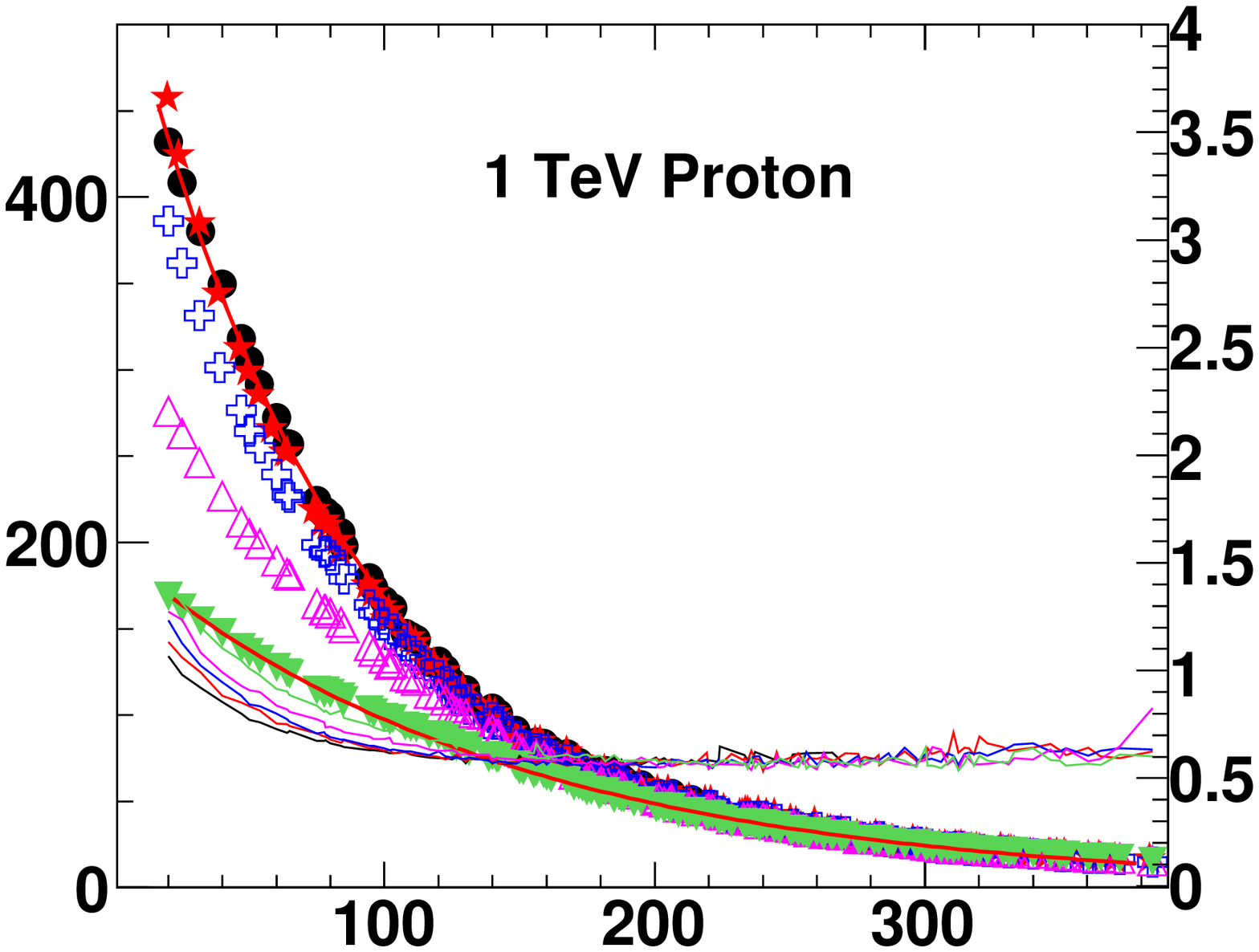} \hspace{-3mm}
\includegraphics[scale=0.31]{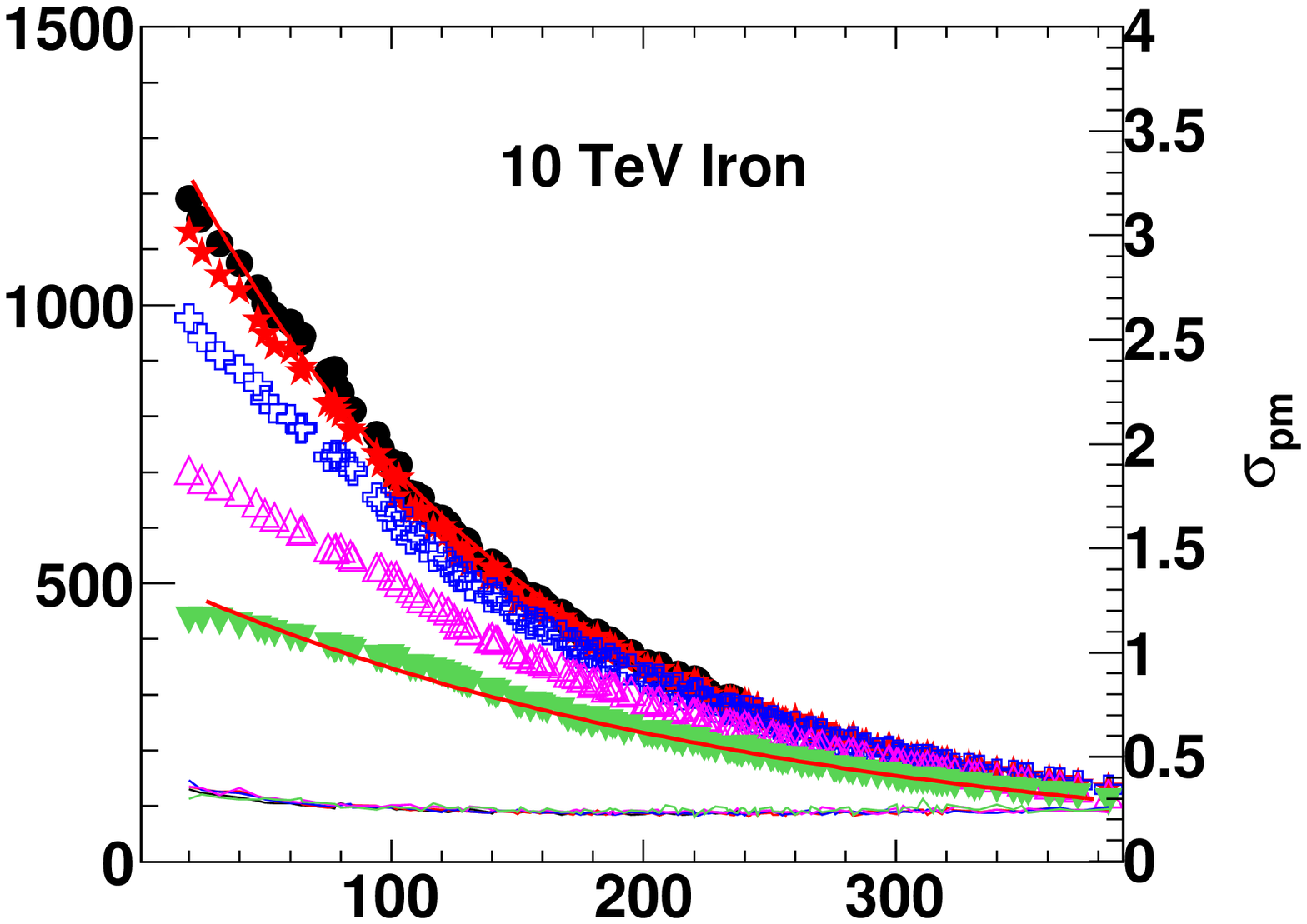}}
\centerline{\hspace{0mm}
\includegraphics[scale=0.30]{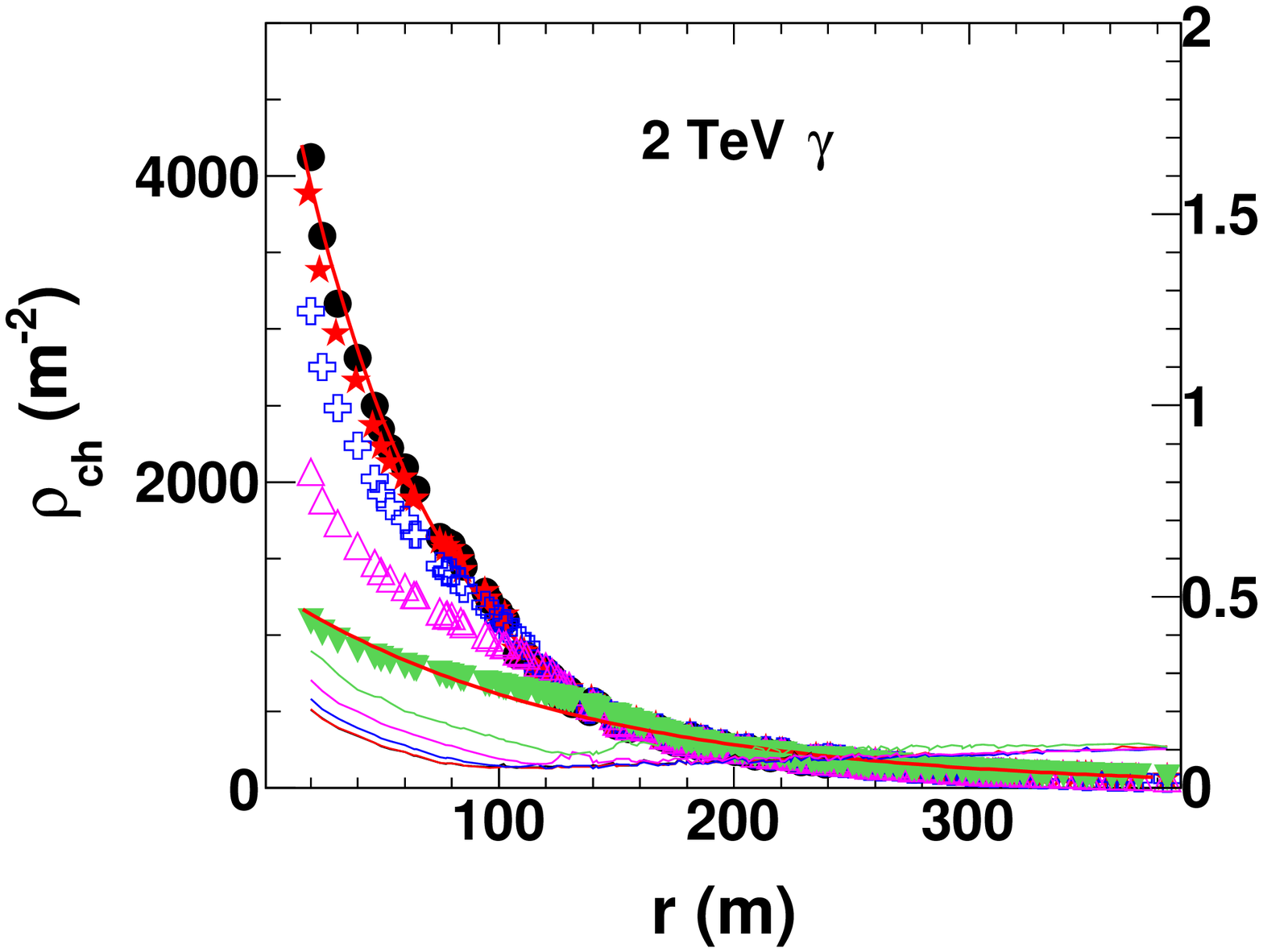} \hspace{0mm}
\includegraphics[scale=0.29]{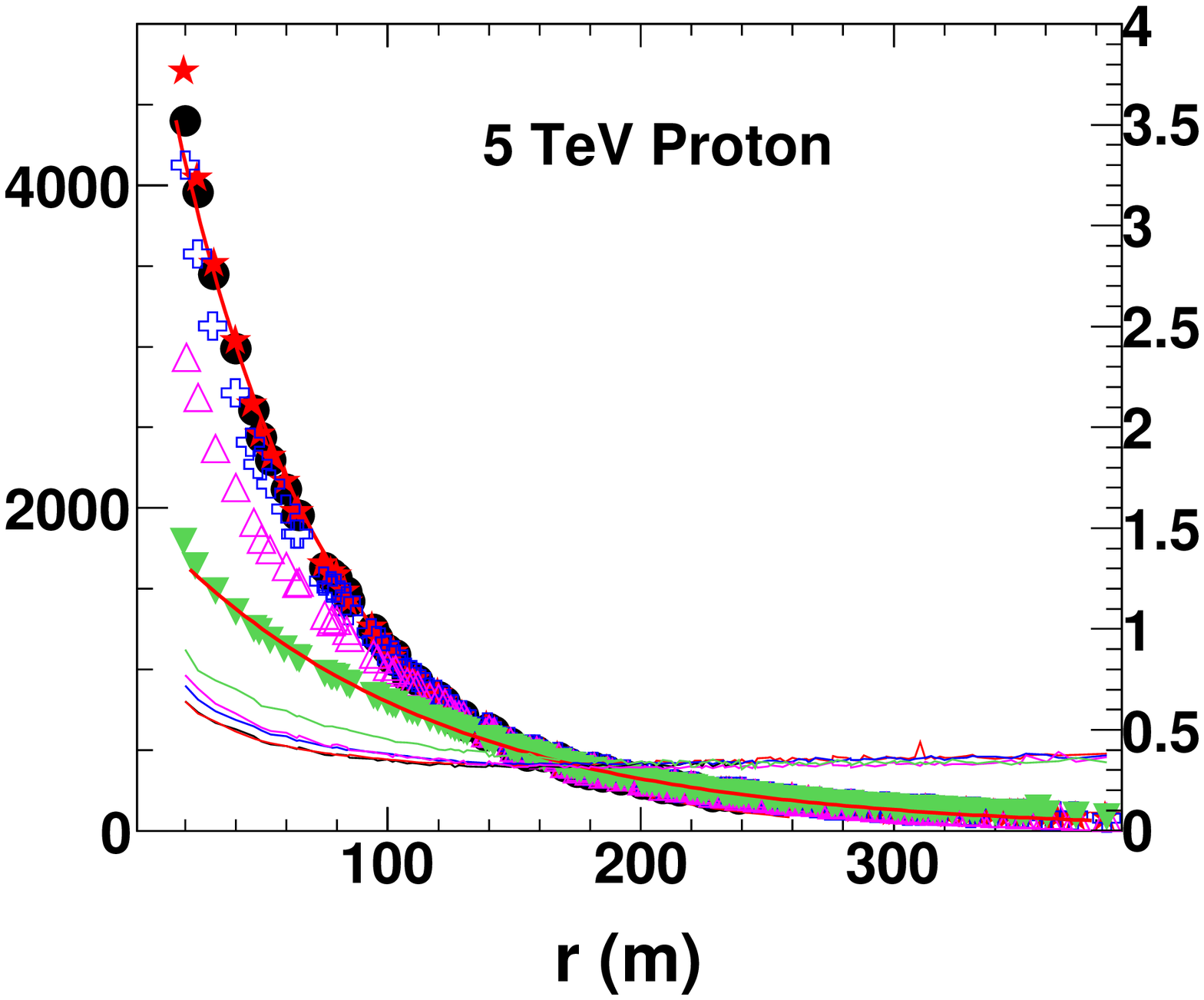} \hspace{0mm}
\includegraphics[scale=0.31]{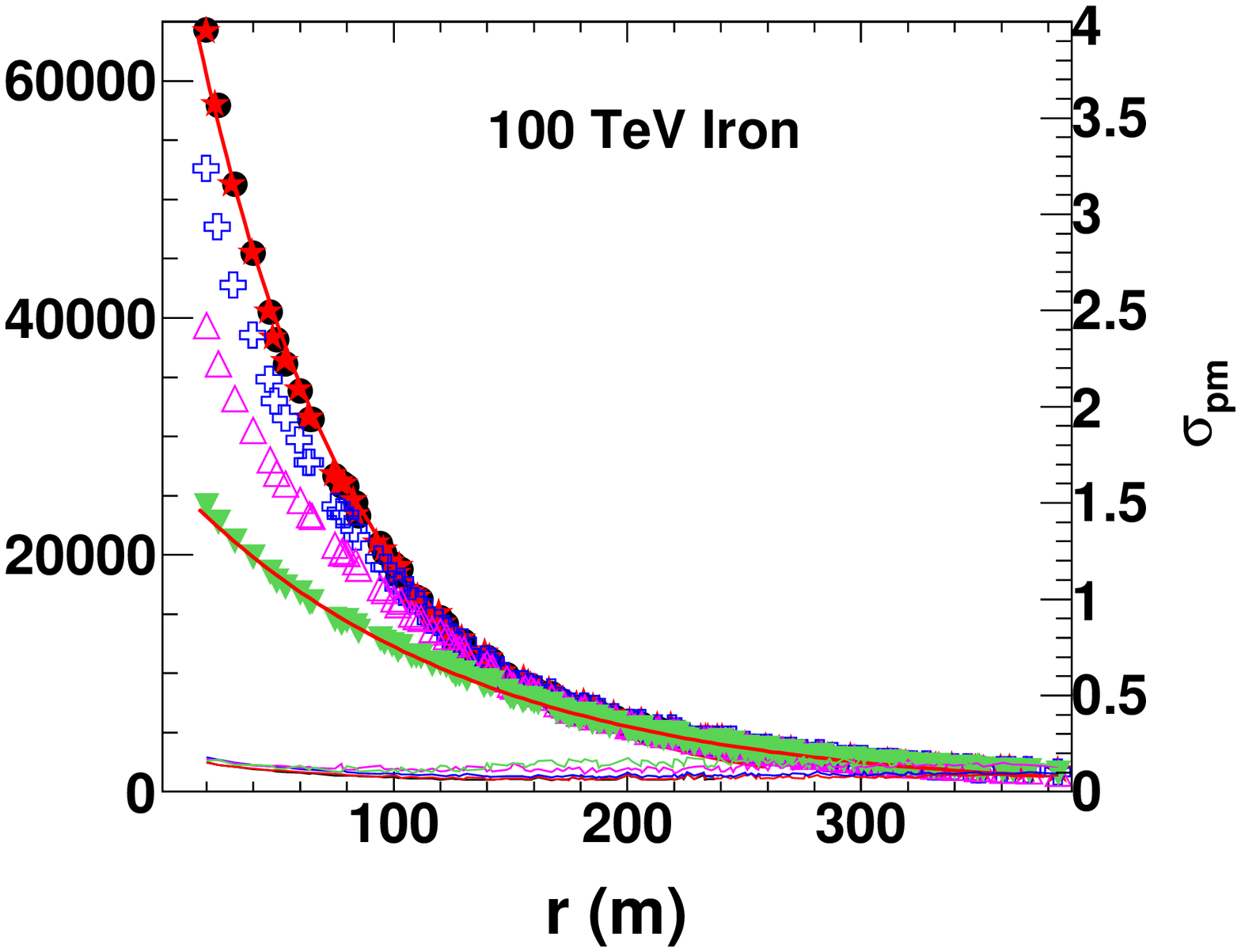}}

\caption{Average density of Cherenkov photons ($\rho_{ch}$) and the ratio of 
r.m.s. to mean of the photon density ($\sigma_{pm}$) are plotted 
with respect to the shower
core distance of $\gamma$-ray, proton and iron primaries for different energy 
and angle of incidence. The scale on the right hand side y-axis (y2-axis) is
used to read the plots of $\sigma_{pm}$. The results of the best fit function 
(\ref{eq2}) are
shown by the solid lines in respective plots. The fit on the plots on both 
sides of the position of hump (wherever is applicable) can be made with the 
same function with different function parameters.}
\label{fig1}
\end{figure*}

\subsection{Lateral density of Cherenkov photon}
\subsubsection{General characteristics }
Fig.\ref{fig1} shows the variation of average density of Cherenkov photons
($\rho_{ch}$) and  the ratio of r.m.s. to mean of the photon density 
($\sigma_{pm}$) as a 
function of the distance from shower core of $\gamma$-ray, proton and iron 
primaries with different energy and at different angle of incidence for the 
EPOS-FLUKA model combination. To save space, in this figure we have shown 
the plots only at three different energies for all three primaries, viz. for 
$\gamma$, proton and iron primaries incident vertically as 
well as inclined at zenith angles 10$^{\circ}$, 20$^{\circ}$, 30$^{\circ}$ and 
40$^{\circ}$ respectively. From the figure it is clear that, the $\rho_{ch}$ 
distribution with core distance falls almost exponentially for all primary 
particles and their energies with gradual reduction in the slope 
with increasing zenith angle. The variation in Cherenkov photon 
density with respect to core distance may be effectively 
represented by the equation \cite{Hazarika}
\begin{equation}
\rho_{ch}(r) = \rho_{0}\;e^{-\beta r},
\label{eq2}
\end{equation}
where $\rho_{ch}(r)$ is the density of Cherenkov photons as a function of 
position, $r$ is the shower core distance, $\rho_{0}$ is the Cherenkov photon's
density at the core of a shower and $\beta$ is the slope. Different primaries 
will have different values of $\rho_{0}$ and $\beta$ depending upon their 
energy and the zenith angle. The best fit negative exponential 
functions, represented by the equation (\ref{eq2}), for the vertically incident 
and also for the most inclined (i.e. incident at 40$^{\circ}$) showers are 
shown by the solid lines in the plots of the Fig.\ref{fig1} as an example. For 
these
fittings we used the $\chi^2$-minimization method available in the ROOT 
software \cite{Root}. It should be noted that, due to the presence of the 
significant characteristic hump, the fit for 100 GeV $\gamma$-ray primary is
made at two segments, one before the position of hump and other
after the position of hump with different function parameters. This technique
is applied to any plot wherever is required. 
For the vertically incident primary, this hump is observed at around 100 m core 
distance. With increasing zenith angle ($\theta$), the distance of the hump 
from the core increases with increasing prominence, even the hump is seen up to 
the energy of 1 TeV for the $\theta$ = 40$^{\circ}$. The geometry of the 
$\rho_{ch}$ distribution is  different for different primary at a 
particular energy and at a particular angle $\theta$ even though the 
distributions follow the same mathematical function almost for all the cases 
with 
different coefficients and slopes. It is observed that for the $\gamma$-ray 
primary at a given energy and $\theta$, the distribution has a larger 
curvature (beyond the position of the hump along the core distance,
wherever applicable) with higher values of the coefficient $\rho_0$ and slope 
$\beta$ of the exponential function (\ref{eq2}) in comparison to proton and 
iron primaries. Moreover, we found that $\beta$ is smaller for iron primary 
than that for proton, i.e. for the primary particle of higher mass composition 
the curvature is smaller. 

\begin{figure*}[hbt]
\centerline
\centerline{
\includegraphics[width=6.0cm, height=4.7cm]{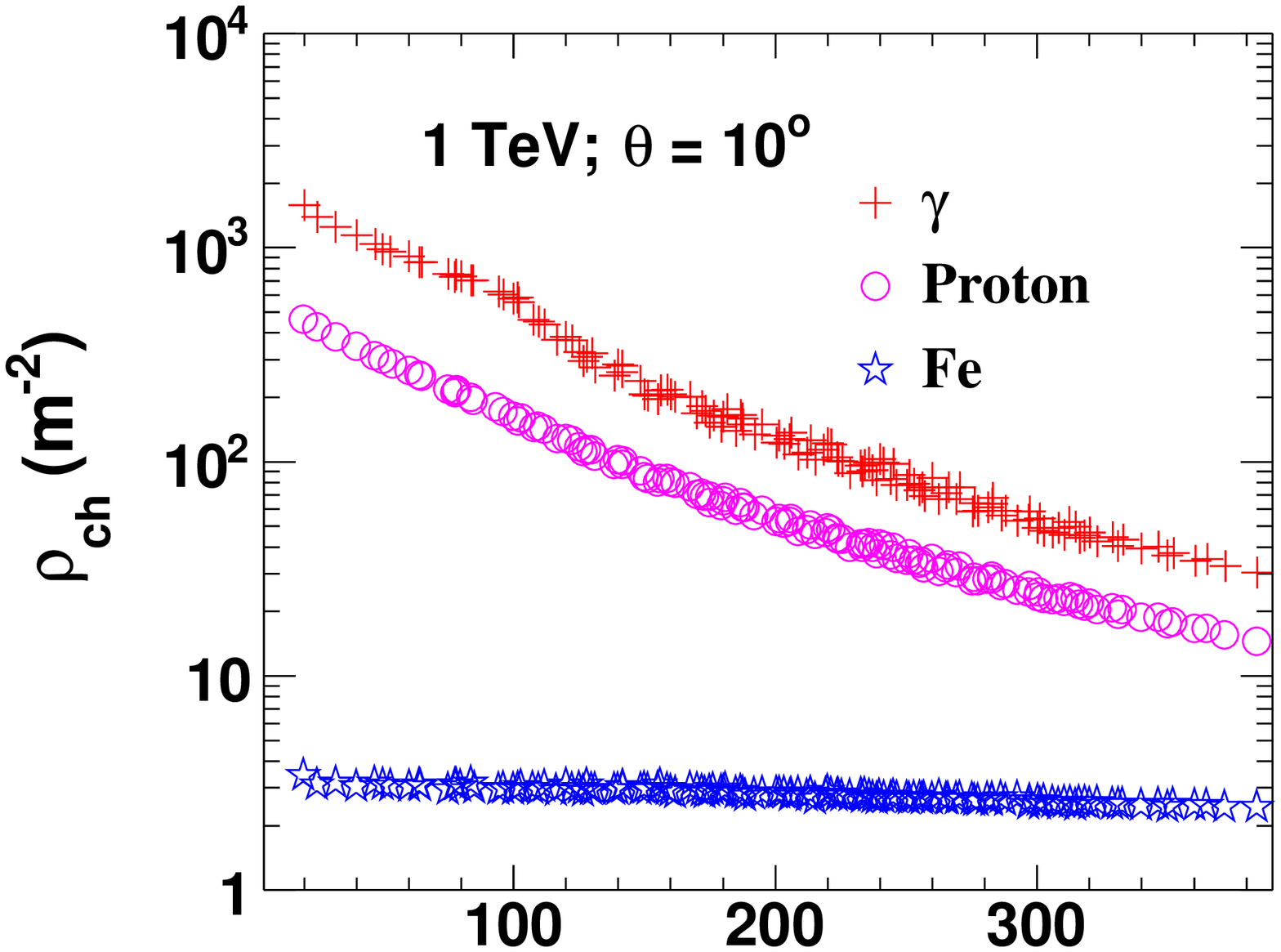}\hspace{-2mm}
\includegraphics[width=6.0cm, height=4.7cm]{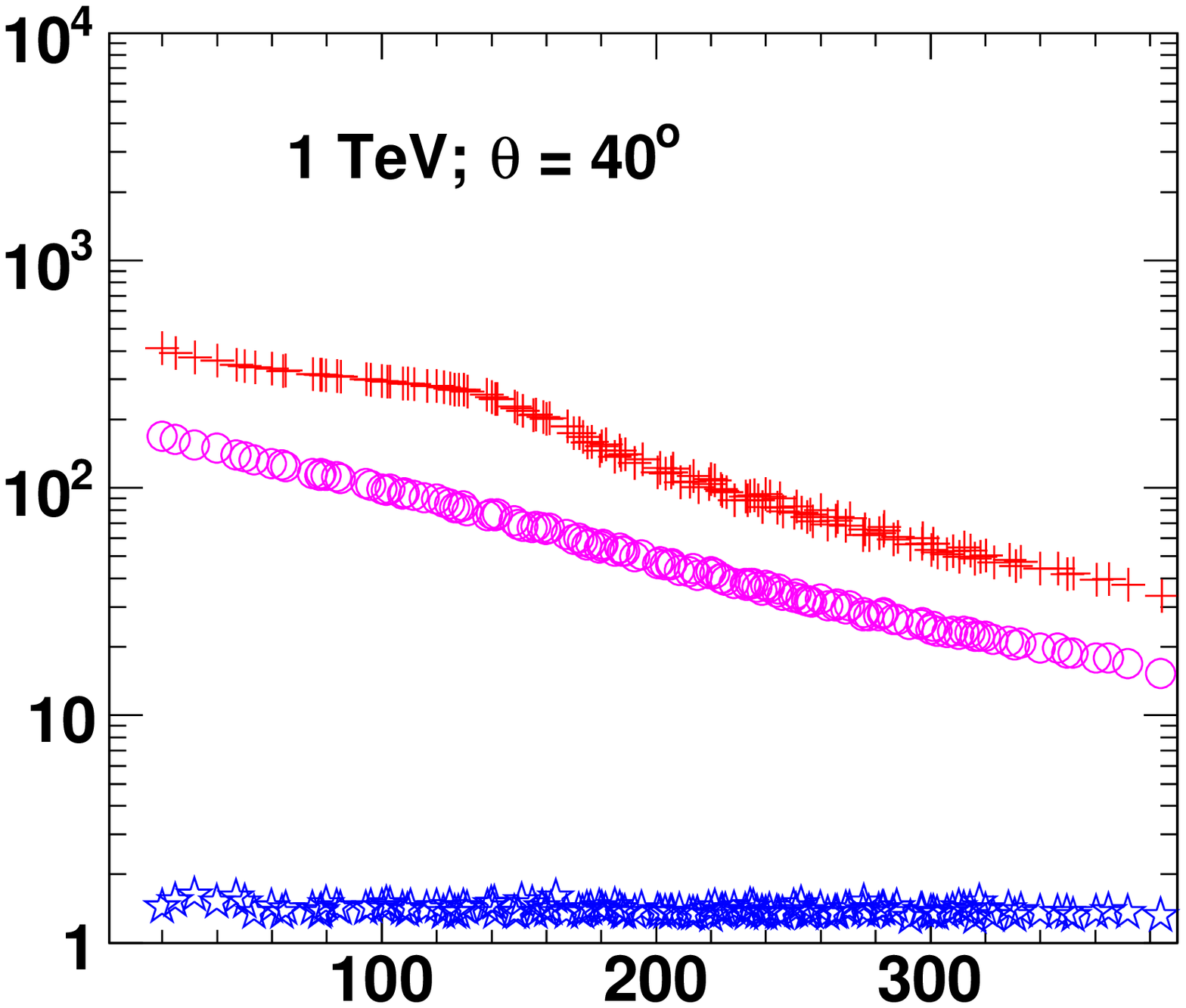}}

\vspace{-2mm}
\centerline{
\includegraphics[width=6.0cm, height=4.7cm]{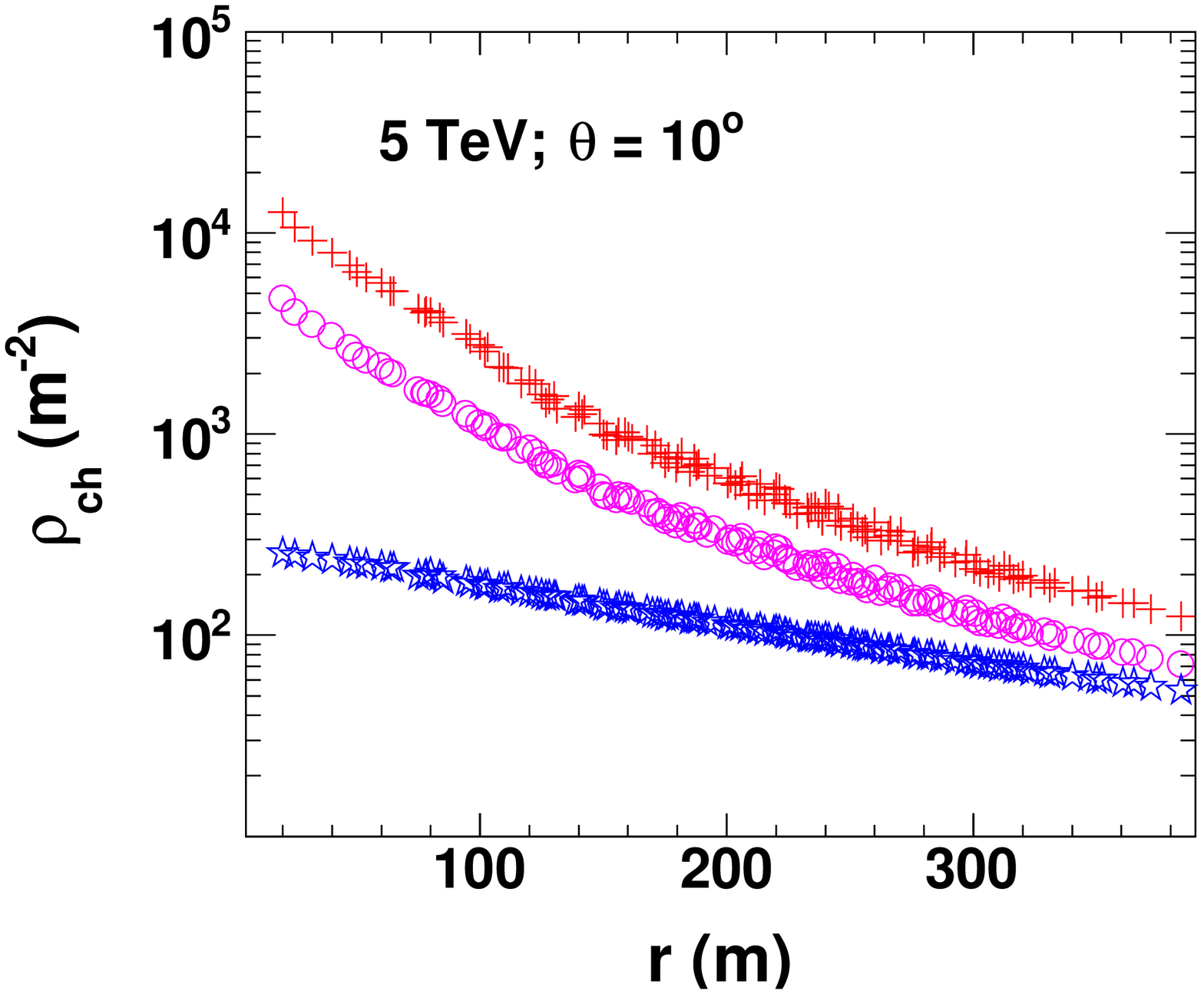}\hspace{-2mm}
\includegraphics[width=6.0cm, height=4.7cm]{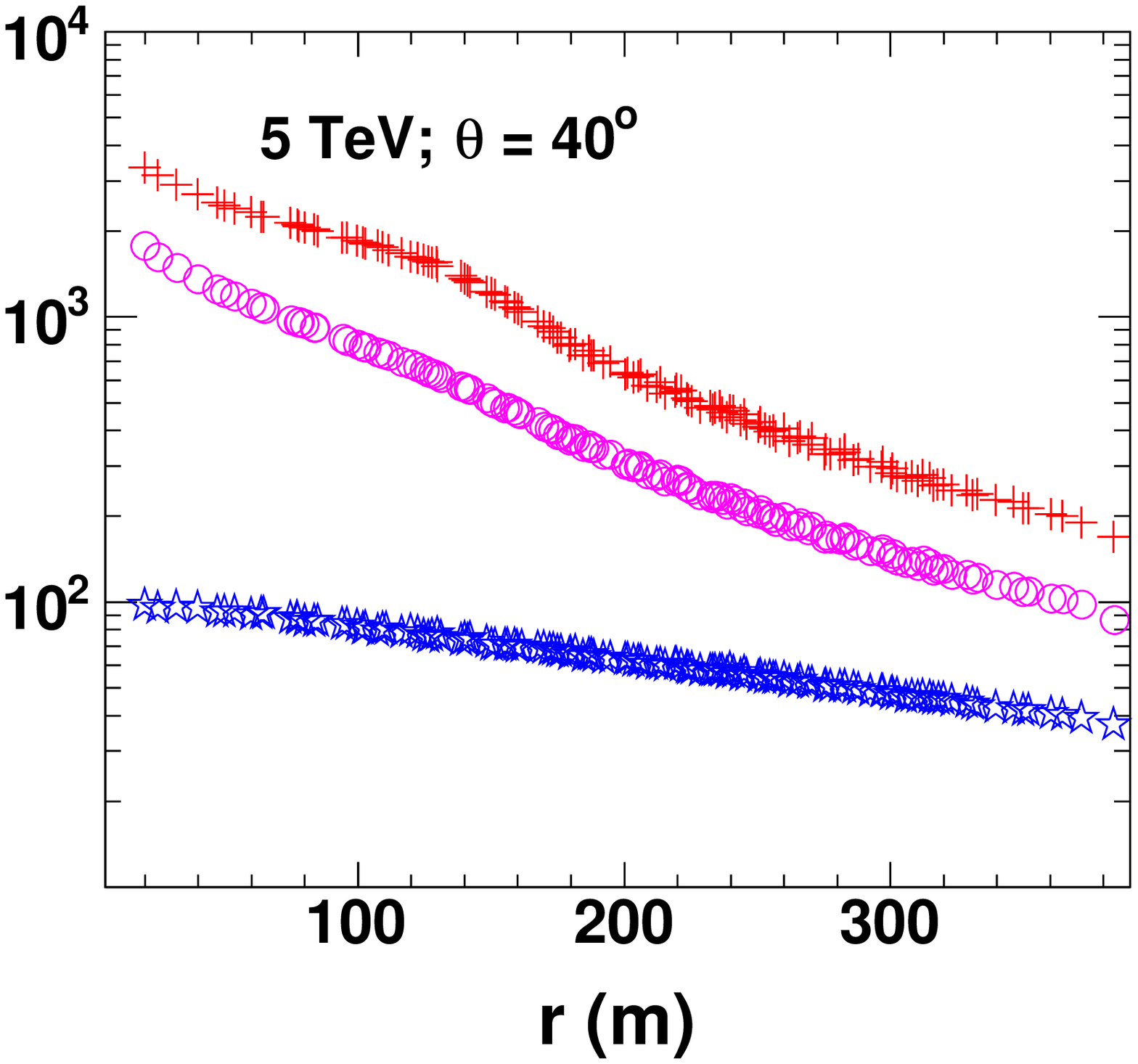}}

\caption{Distributions of $\rho_{ch}$ with respect to core distance of the
showers of $\gamma$, proton and iron primaries at 1 TeV and 5 TeV energies
incident at 10$^{\circ}$ and 40$^{\circ}$ zenith angles.}
\label{fig3}
\end{figure*}

\begin{figure*}[hbt!]
\centerline
\centerline{
\includegraphics[scale=0.27]{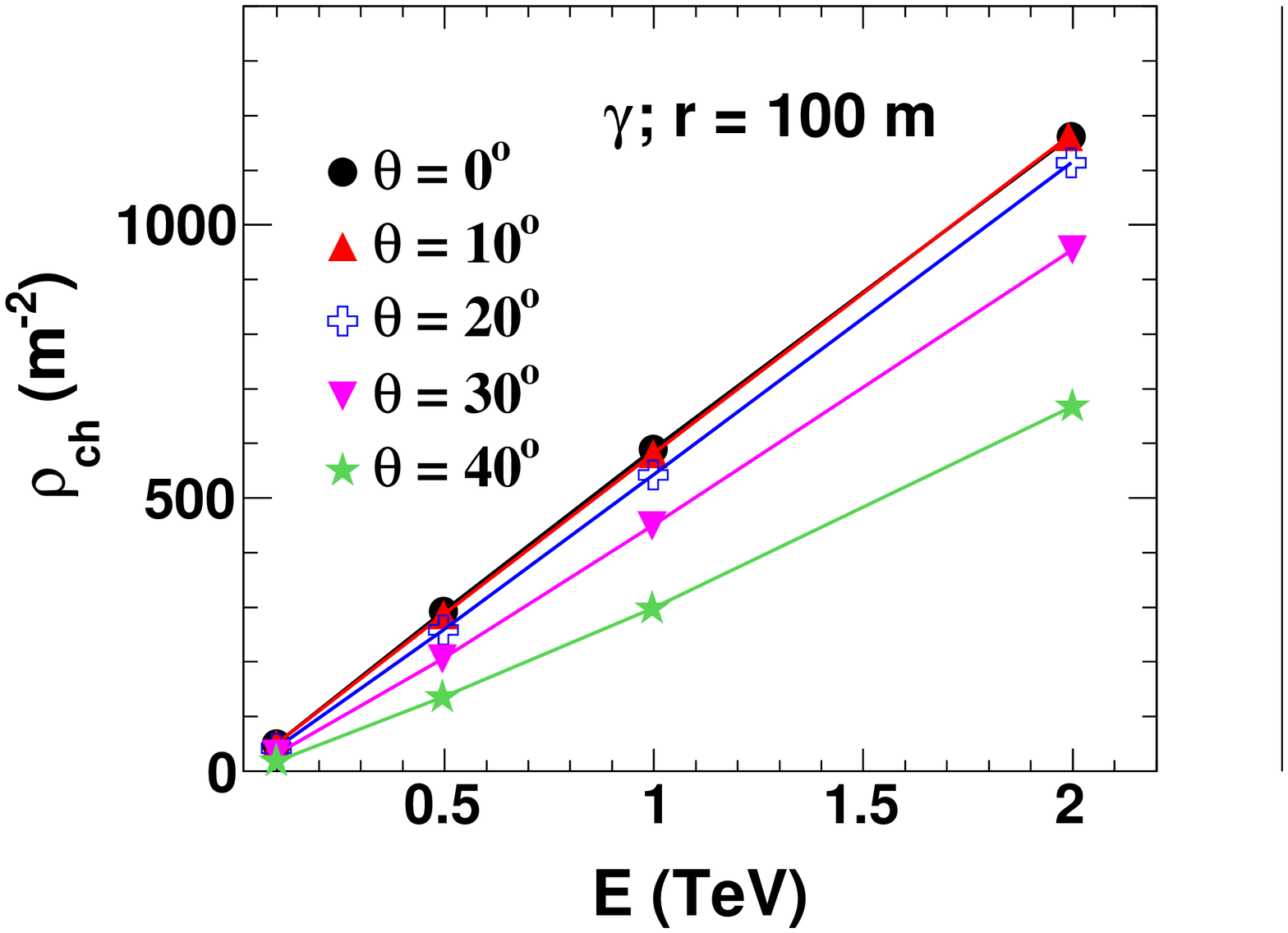} \hspace{1mm}
\includegraphics[scale=0.27]{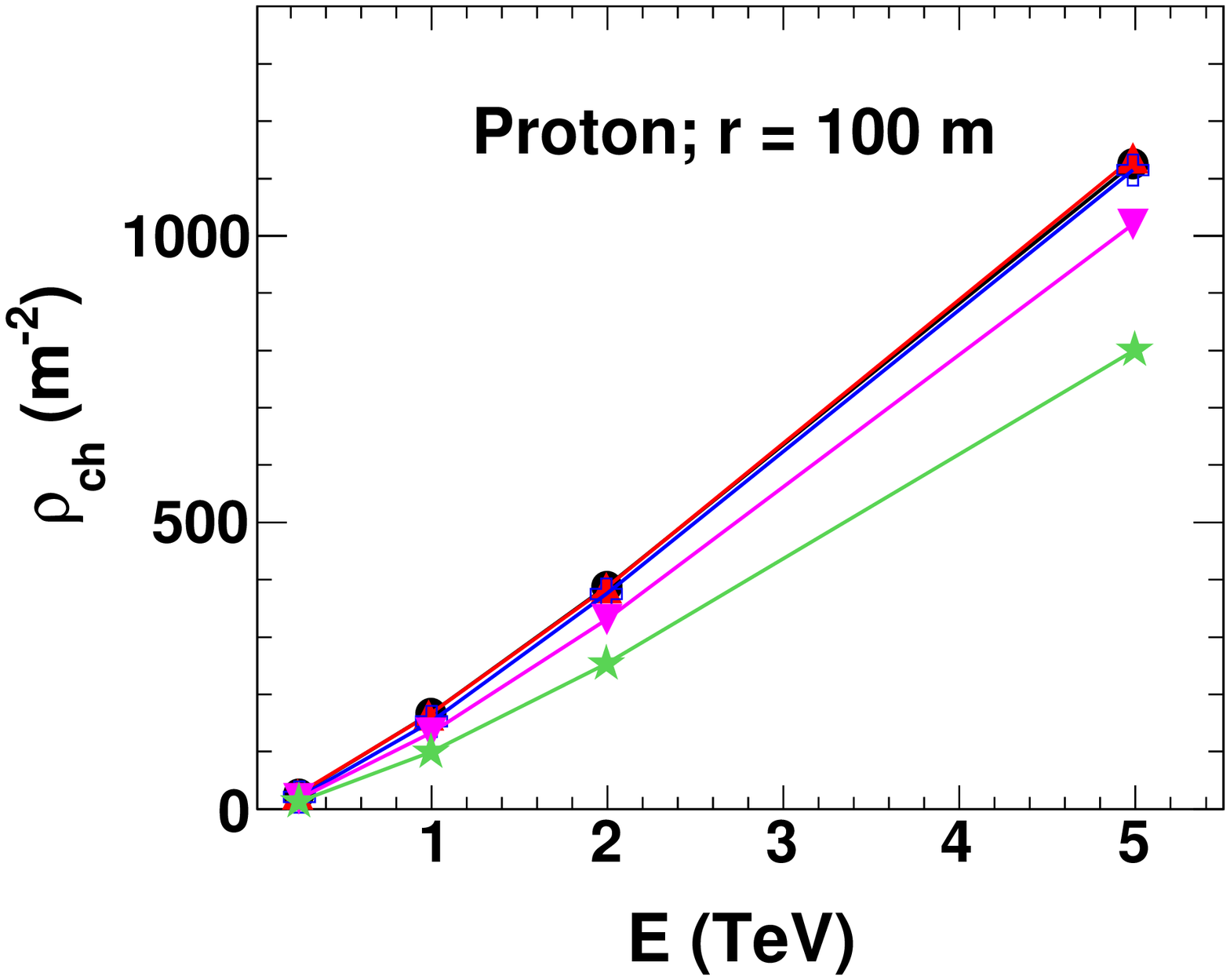} \hspace{1mm}
\includegraphics[scale=0.27]{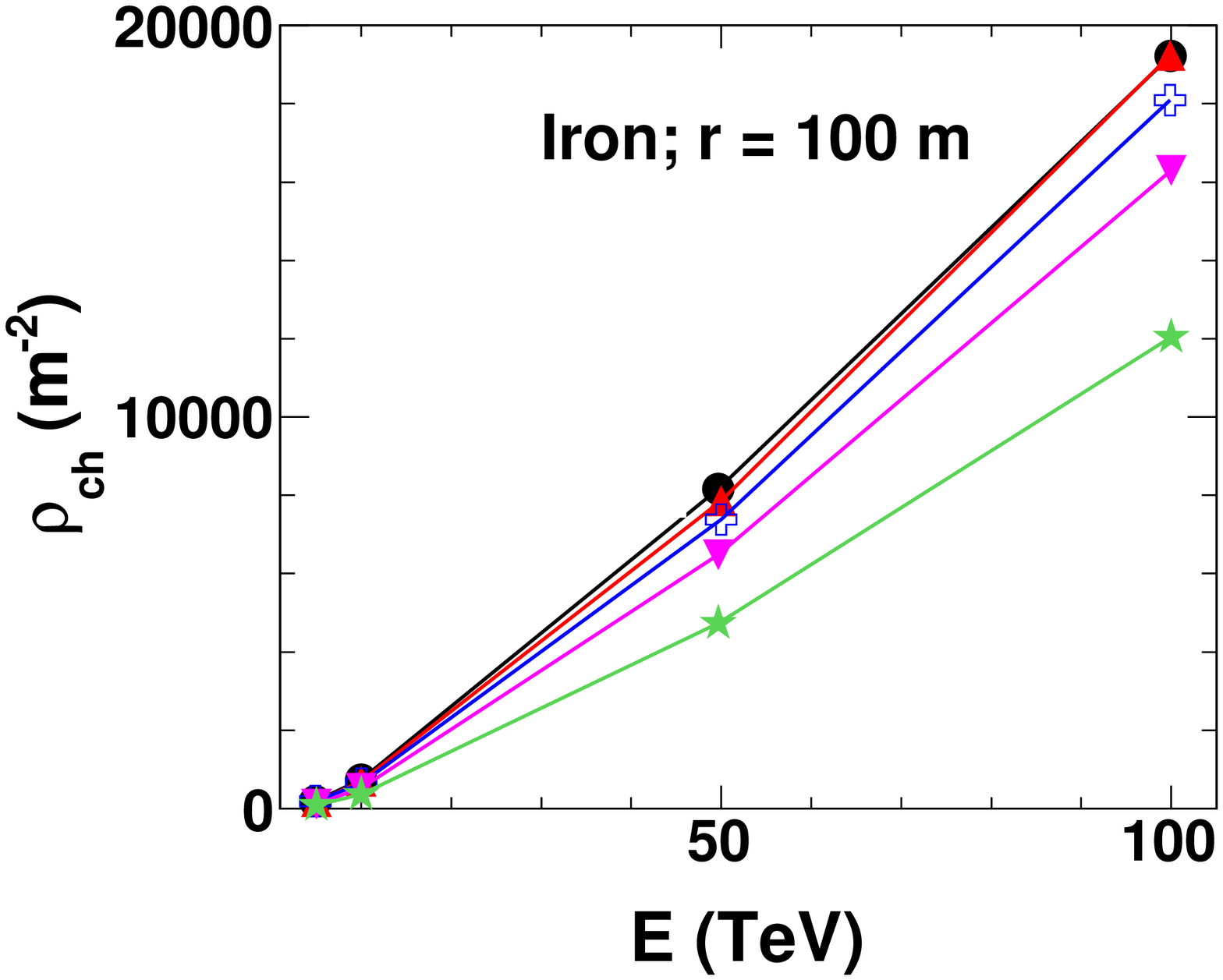}}

\caption{$\rho_{ch}$ as a function of energy of the primary particle at
100 m distance from the core of the showers of $\gamma$, proton and iron
primaries.}
\label{fig4}
\end{figure*}

\subsubsection{Primary particle and energy dependence}
It is quite obvious that the $\gamma$-ray primary generates maximum number of
Cherenkov photons, whereas the Cherenkov photon yield of iron primary is the
lowest at a given energy and zenith angle. This is basically due to the
fact that almost whole energy of the $\gamma$-ray primary are utilized in the
formation of EM shower, whereas in the cases of proton and iron
primaries a portion of their energy is used up in the formation of hadronic
shower along with the EM one. Another contributing factor for such
behaviour of iron primary is the height of shower maximum. At a given energy 
and zenith angle, 
the shower maximum of the iron primary is produced at sufficiently high 
altitude in comparison to that for the $\gamma$-ray and proton primaries. For
the $\gamma$-ray and proton, this height is nearly equal for the same condition 
(see 
the Table \ref{tab2}). So, a considerable fraction of EM charged particles 
(which are sources of Cherenkov photons) of iron initiated shower may be 
absorbed
in the atmosphere before reaching the level of the shower maximum of 
$\gamma$-ray and proton primaries of same energy and zenith angle. In 
Fig.\ref{fig3}, the 
distributions of $\rho_{ch}$ with respect to the distance of the shower core 
produced by
the $\gamma$-ray, proton and iron primaries with 1 TeV and 5 TeV energies and
inclined at 10$^{\circ}$ and 40$^{\circ}$ are shown to visualize this 
observation.

Fig.\ref{fig4} shows the variation of $\rho_{ch}$ at 100 m core
distance with respect to energy for primary particle of all types. It has
been found that, the $\rho_{ch}$ increases at a faster rate, and as well as 
almost linearly
with the energy of the $\gamma$-ray primary. On the other hand, for the cases
of proton and specially of iron primaries, the increasing trend is
comparatively slow and also non-linear with respect to the primary energy. 
These behaviours of primaries with respect to energy can be understood in the 
light of explanation mentioned above. Moreover, this figure also indicates the 
result stated above.

\begin{center}
\begin{table}[ht]
\caption{\label{tab2} Average slant depth corresponding to shower maximum of 
EASs produced by $\gamma$-ray, proton and iron primaries incident with 
different energies and at different zenith angles.}
\begin{tabular}{cccc}\\[-5.0pt]
\hline
Primary particle & ~~~~Energy & ~~~~Zenith angle & ~~~~Shower maximum position (gm/cm$^2$) \\\hline\\[-8.0pt]
$\gamma$-ray  &  ~~~100 GeV   &~ 0$^\circ$ & 255.856 \\
              &  ~~~500 GeV   &~ 0$^\circ$ & 313.383 \\
              &  ~~~~~~~1 TeV &~ 0$^\circ$ & 337.019 \\
              &               & 20$^\circ$ & 314.837 \\
              &               & 40$^\circ$ & 262.001 \\
  &&& \\
 Proton       &  ~~~250 GeV   & ~ 0$^\circ$ & 301.220 \\
              &  ~~~~~~~1 TeV & ~ 0$^\circ$ & 340.016 \\
              &  ~~~~~~~2 TeV & ~ 0$^\circ$ & 360.020 \\
              &               & 20$^\circ$  & 342.191 \\
              &               & 40$^\circ$  & 284.292 \\
 &&& \\
 Iron         &  ~~~~~~~1 TeV & ~ 0$^\circ$ & 200.103 \\
              &  ~~~~~~~5 TeV & ~ 0$^\circ$ & 243.849 \\
              &               &  20$^\circ$ & 231.510 \\
              &               &  40$^\circ$ & 190.410 \\
              &  ~~~~~50 TeV  & ~ 0$^\circ$ & 324.214 \\

\hline
\end{tabular}
\end{table}
\end{center}

\subsubsection{Photon density fluctuations} 

It is clear from the Fig.\ref{fig1} that the value of $\sigma_{pm}$ of 
$\rho_{ch}$ for $\gamma$-ray primaries initially decreases with increasing  
core distance up to a distance of $\sim$ 100 m and then gradually increases. 
This pattern continues for all energies and for vertical as well as inclined 
showers. Moreover, the value of $\sigma_{pm}$ is found to become smaller with 
increase in energy of the primaries. Similar pattern in the variation of 
$\sigma_{pm}$ is also observed for proton primaries. In the case of proton, 
the values are much higher than $\gamma$-ray primaries with a decreasing trend 
with increase in energy of the primary. $\sigma_{pm}$ decrease with increasing 
energy for the iron primaries also, however its value is much smaller than that 
of proton and, unlike proton and $\gamma$-ray primaries it remains almost 
constant for iron primaries with increasing core distance \cite{Hazarika}. 
Furthermore, $\sigma_{pm}$ is 
highest for the most inclined shower and lowest for the vertical shower in 
the cases
of all primaries, energies and almost for all core distances. This is due to
the reason that, as inclination increases the shower has to travel gradually
longer distance to reach the observation level, which creates increasing 
statistical fluctuation with increasing absorption of shower particles within
the atmosphere.

\begin{figure*}[hbt]
\centerline
\centerline{
\includegraphics[scale=0.28]{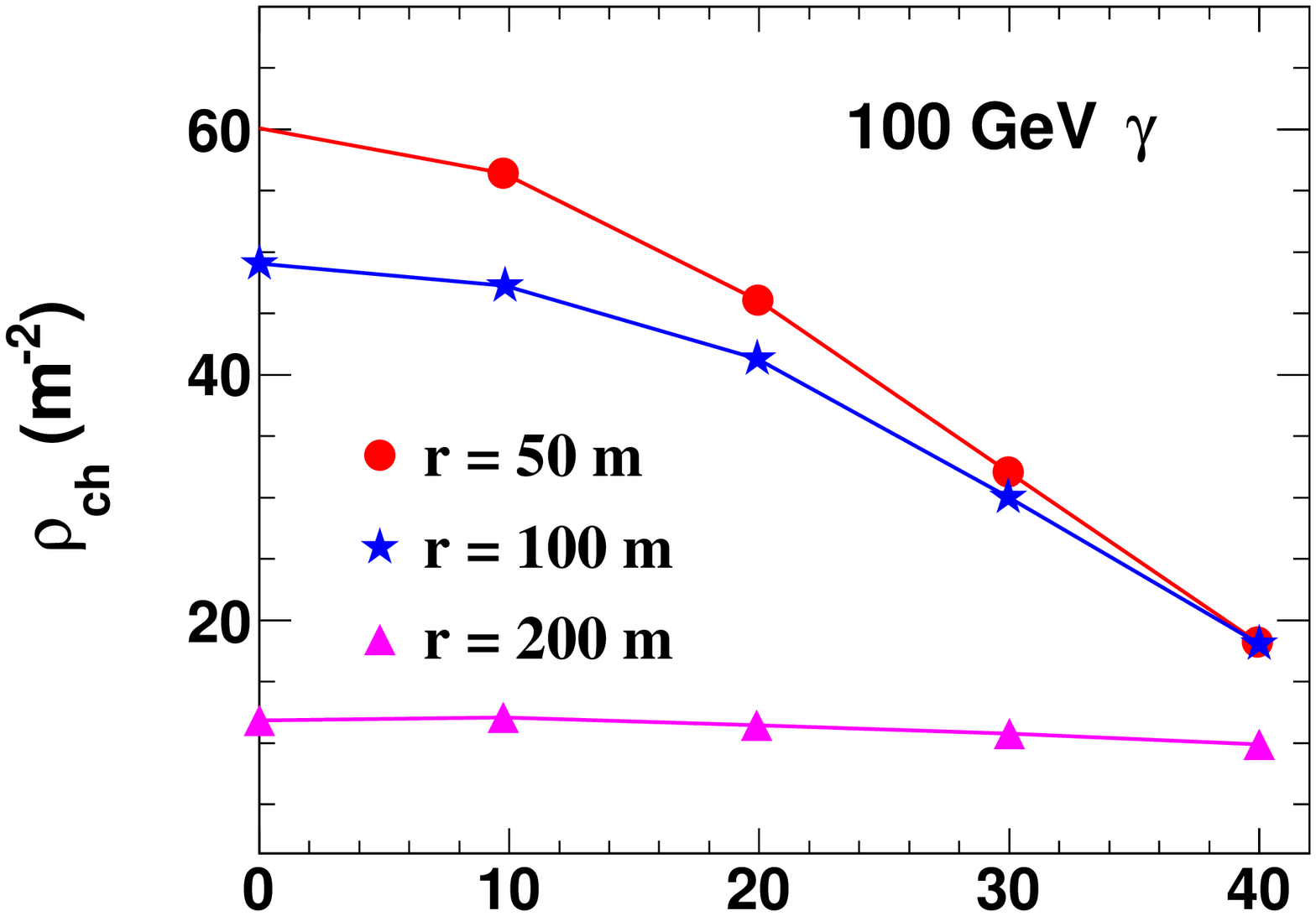} \hspace{-2mm}
\includegraphics[scale=0.28]{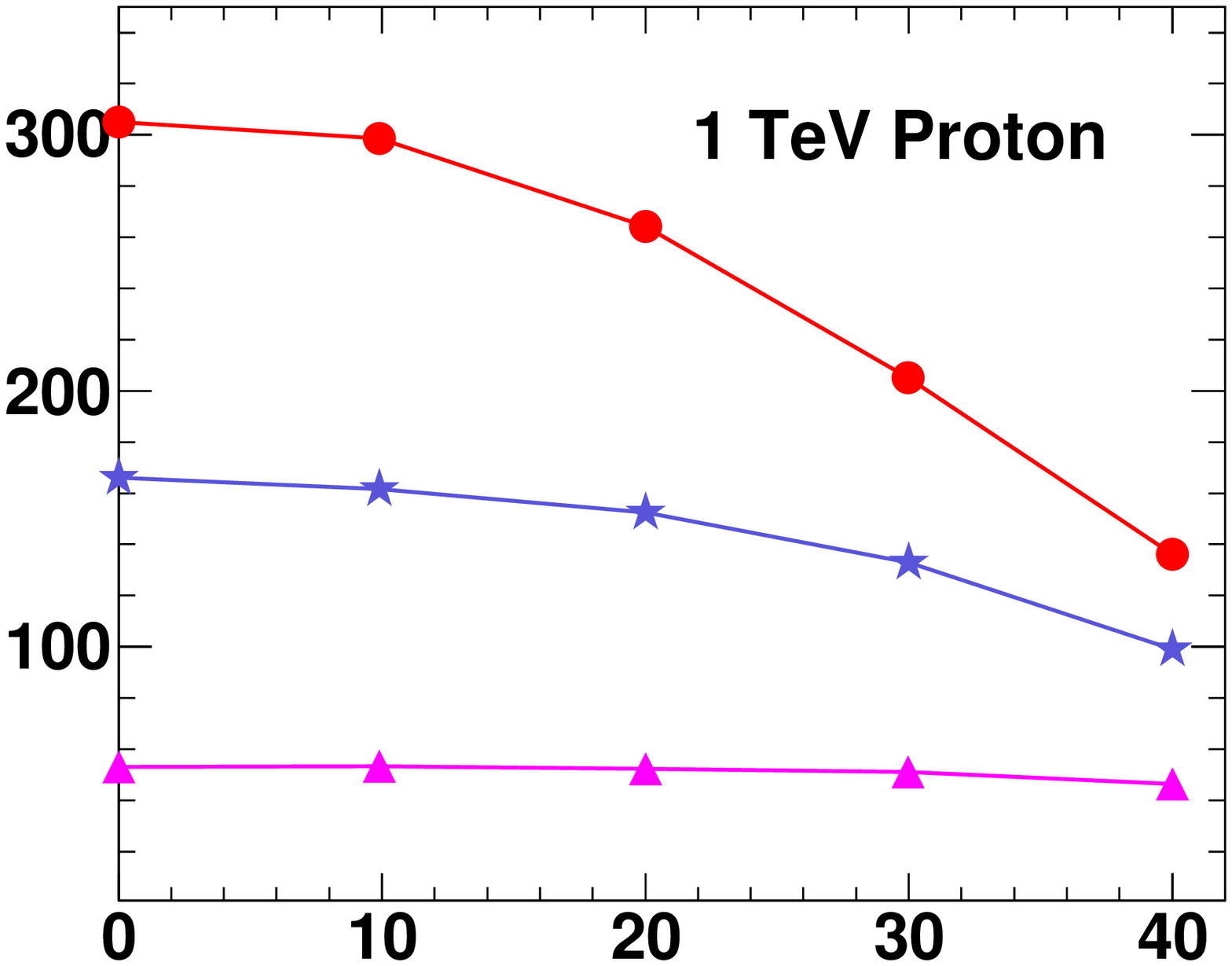} \hspace{-2mm}
\includegraphics[scale=0.28]{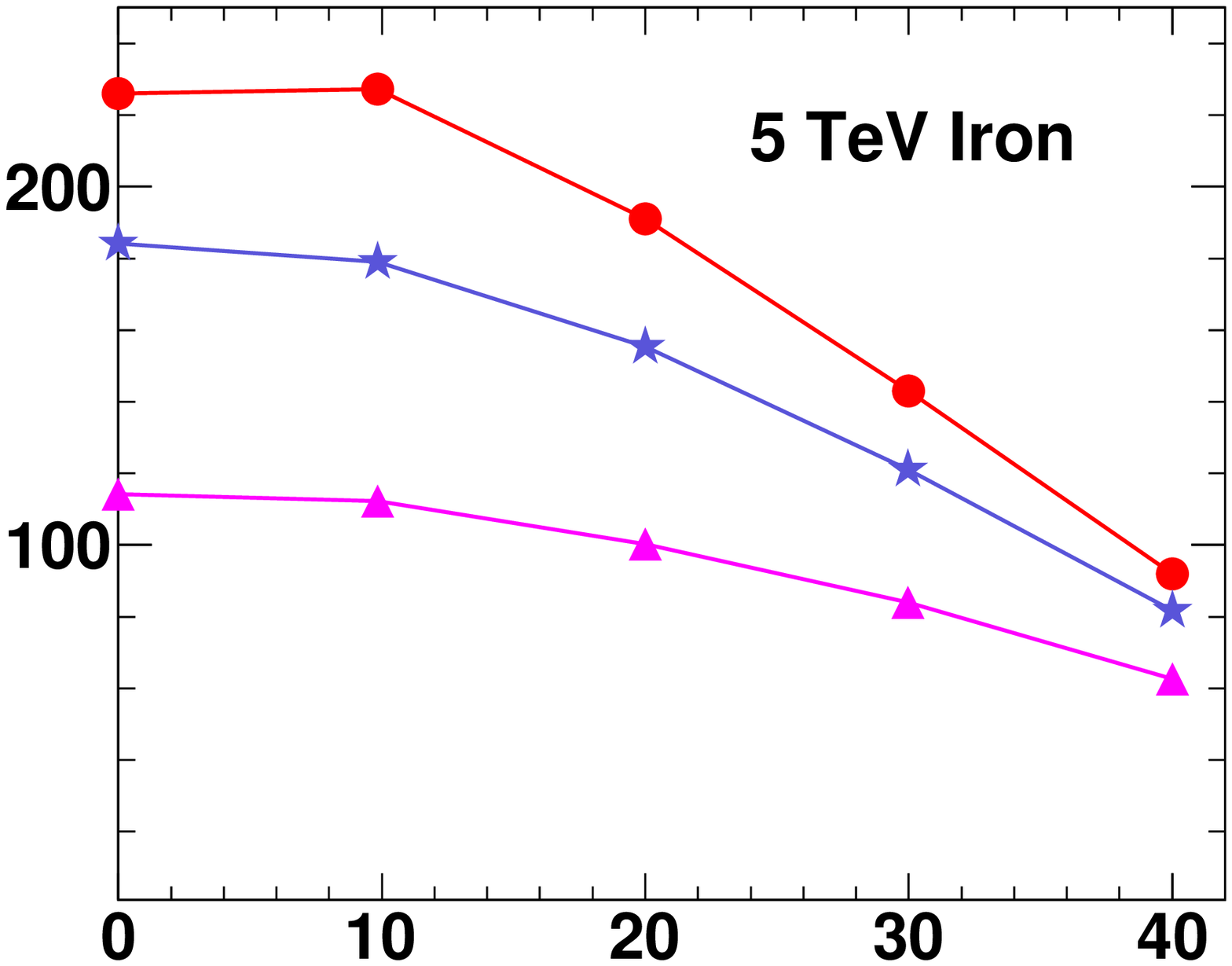}
}
\vspace{-3mm}
\centerline{
\includegraphics[scale=0.28]{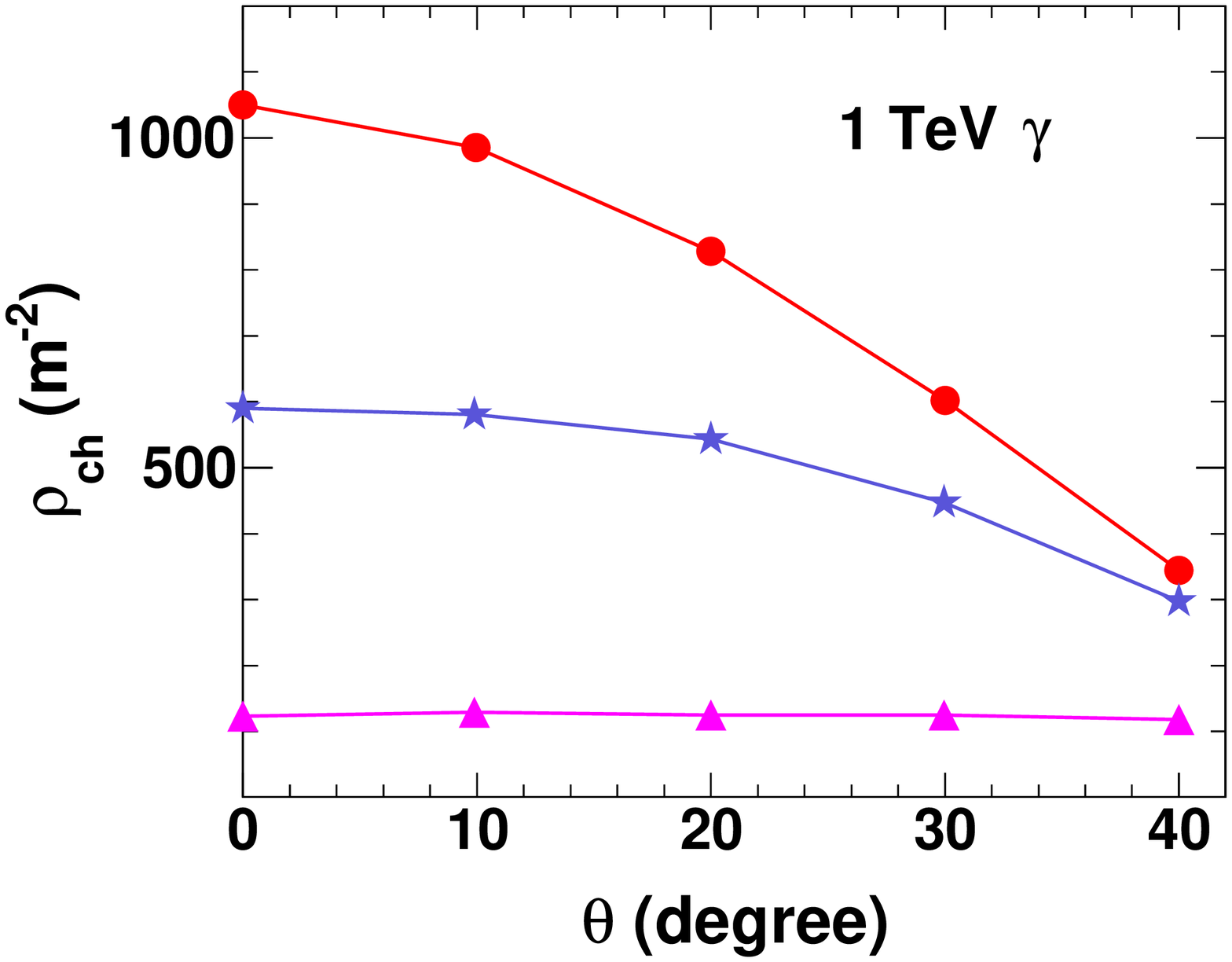} \hspace{-2mm}
\includegraphics[scale=0.28]{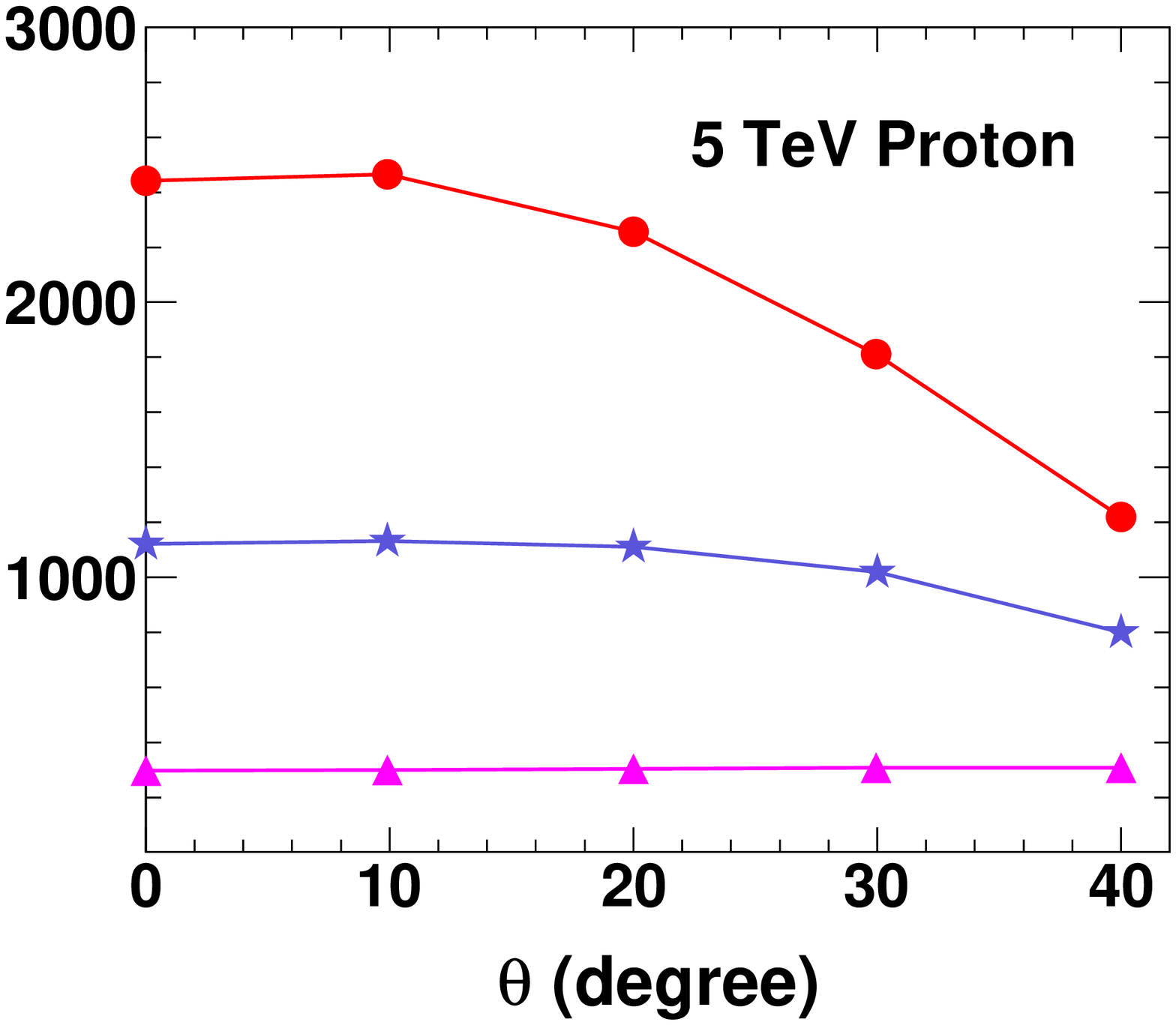} \hspace{-2mm}
\includegraphics[scale=0.278]{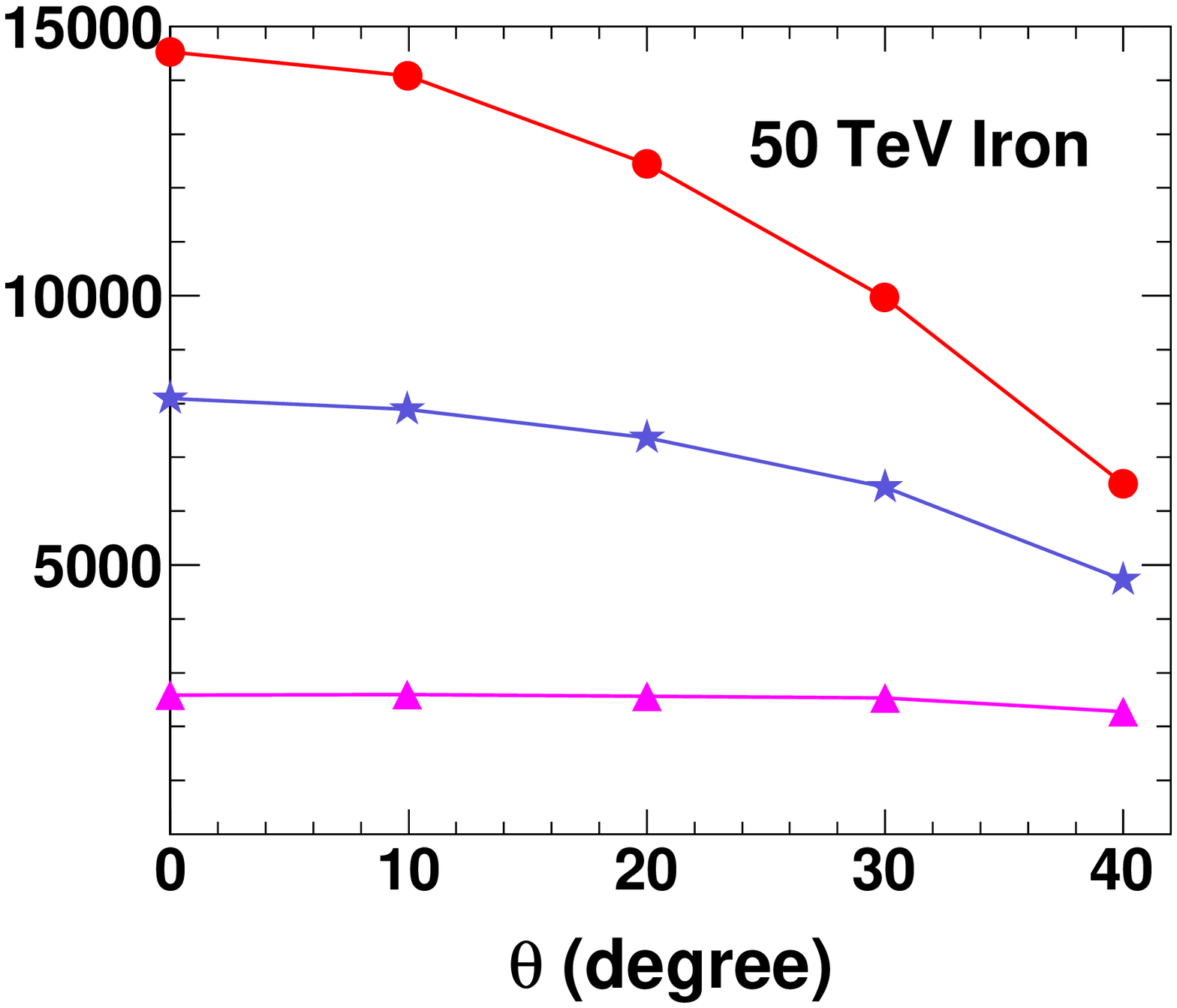}}

\caption{ $\rho_{ch}$s with respect to zenith angle at 50 m,100 m and 
200 m from the core of the showers of $\gamma$, proton and iron primaries.}
\label{fig5}
\end{figure*}

\subsubsection{Zenith angle dependence}
In Fig.\ref{fig5} the variation of density $\rho_{ch}$ with respect to the 
zenith angle at distances from the core equal to 50 m, 100 m and 200 m 
is shown for all the three primaries at two different energies. It is quite 
clear that up to the distance of 50 m there is very small or no difference in 
density between vertically incident shower and  shower inclined at 10$^{o}$. 
Beyond the zenith angle 10$^{\circ}$, the density falls off 
at a faster rate for all combinations of primary particle and energy. At 200 m 
the variation of density with zenith angle is negligible because only 
high energetic charged particles, in fact photons from such particles could 
reach at larger distances from the core over the observation level. 
At 100 m the pattern of 
variation of photon density is
in between the patterns of variation at 50 m and 200 m. The difference in 
densities at 50 m and 100 m decreases with increasing angle of incidence, but
increases with primary energy. This difference is lowest for $\gamma$-ray 
primaries (in fact zero for 100 GeV $\gamma$ at 40$^{\circ}$ zenith angle) 
and highest for the proton primaries. These observed behaviours of 
variation of density with the zenith angle is due to the fact that with 
increasing zenith angle, the (additional) slant depth to be crossed by a 
shower increases (refer to the Table \ref{tab2}) and hence most of the low 
energy charged particles, i.e. low energy photon's sources get absorbed 
depending on energy of the primary as well as the nature of generation of 
shower of the primary itself. 

\subsubsection{Altitude dependence}

\begin{figure*}[hbt]
\centerline
\centerline{
\includegraphics[width=5.8cm, height=4.5cm]{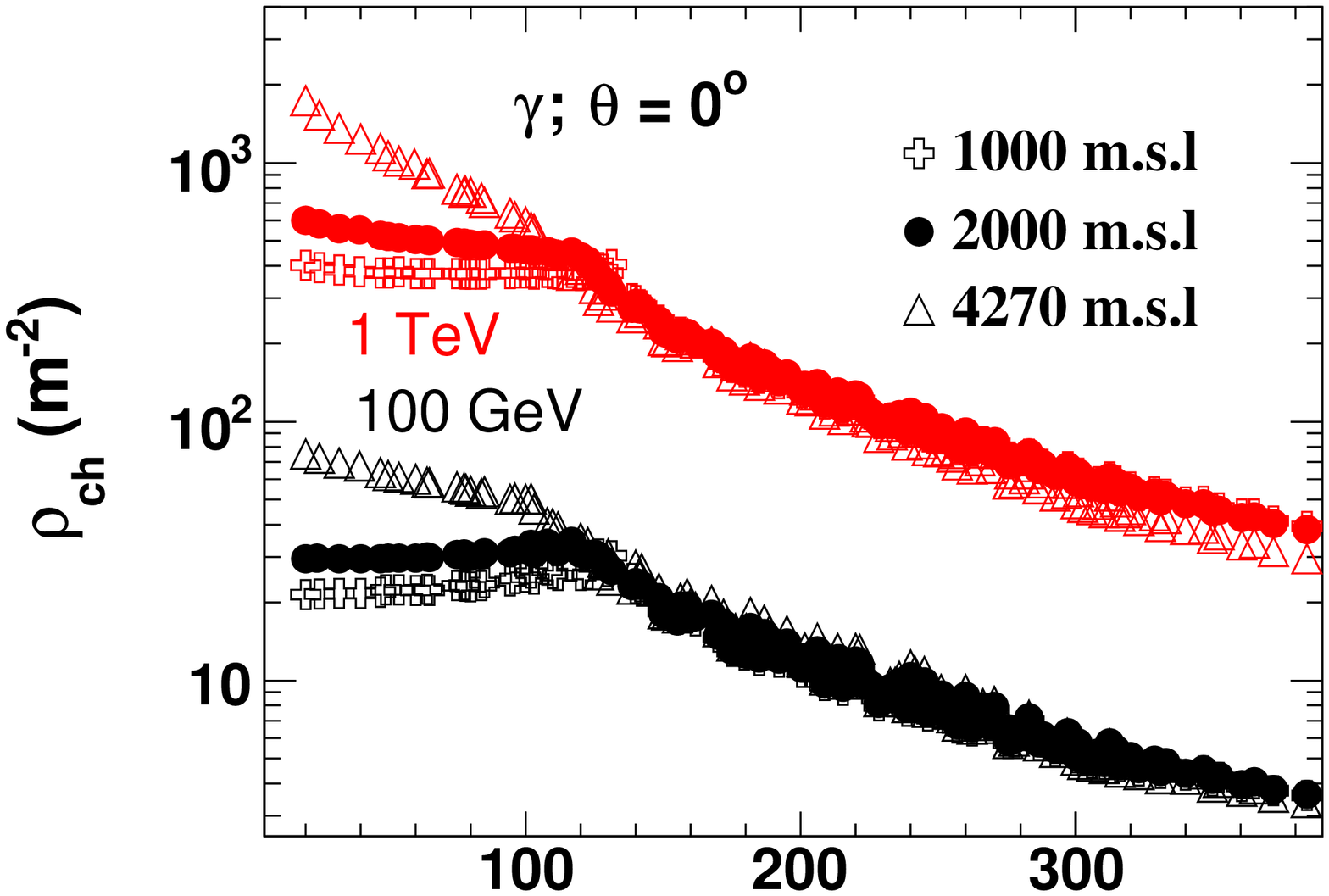} \hspace{-2mm}
\includegraphics[width=5.8cm, height=4.5cm]{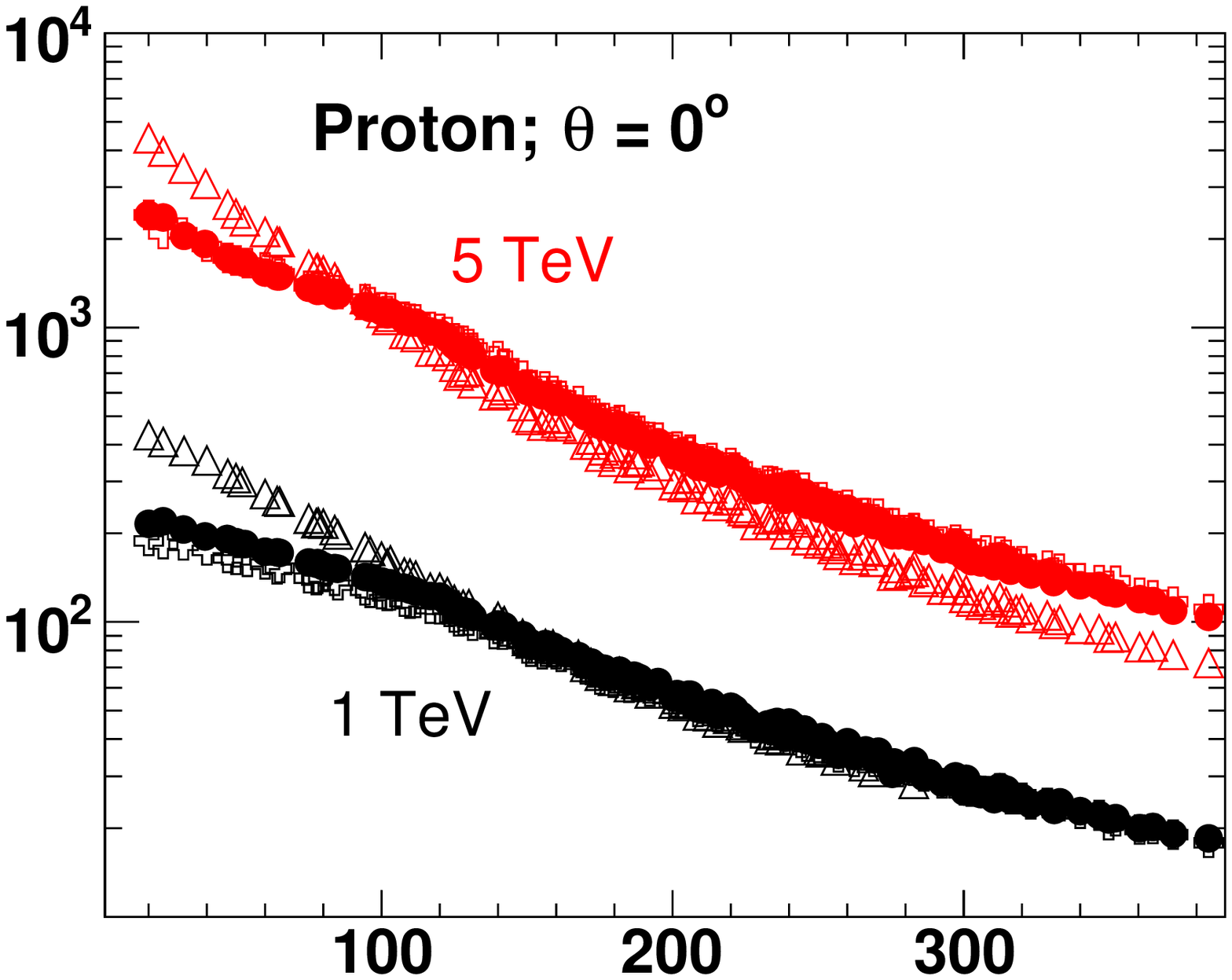} \hspace{-2mm}
\includegraphics[width=5.8cm, height=4.5cm]{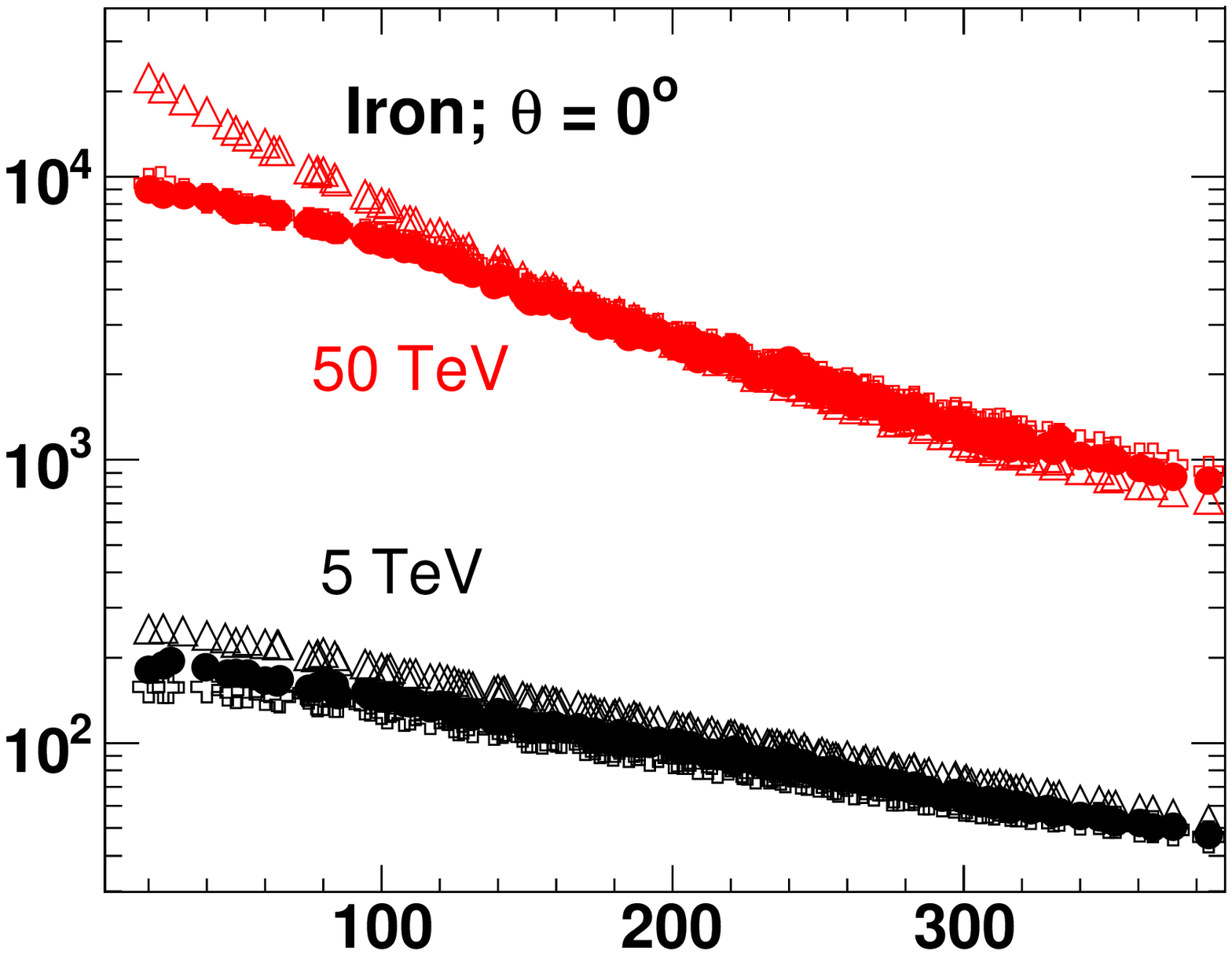}}

\vspace{-2mm}
\centerline{
\includegraphics[width=5.8cm, height=4.5cm]{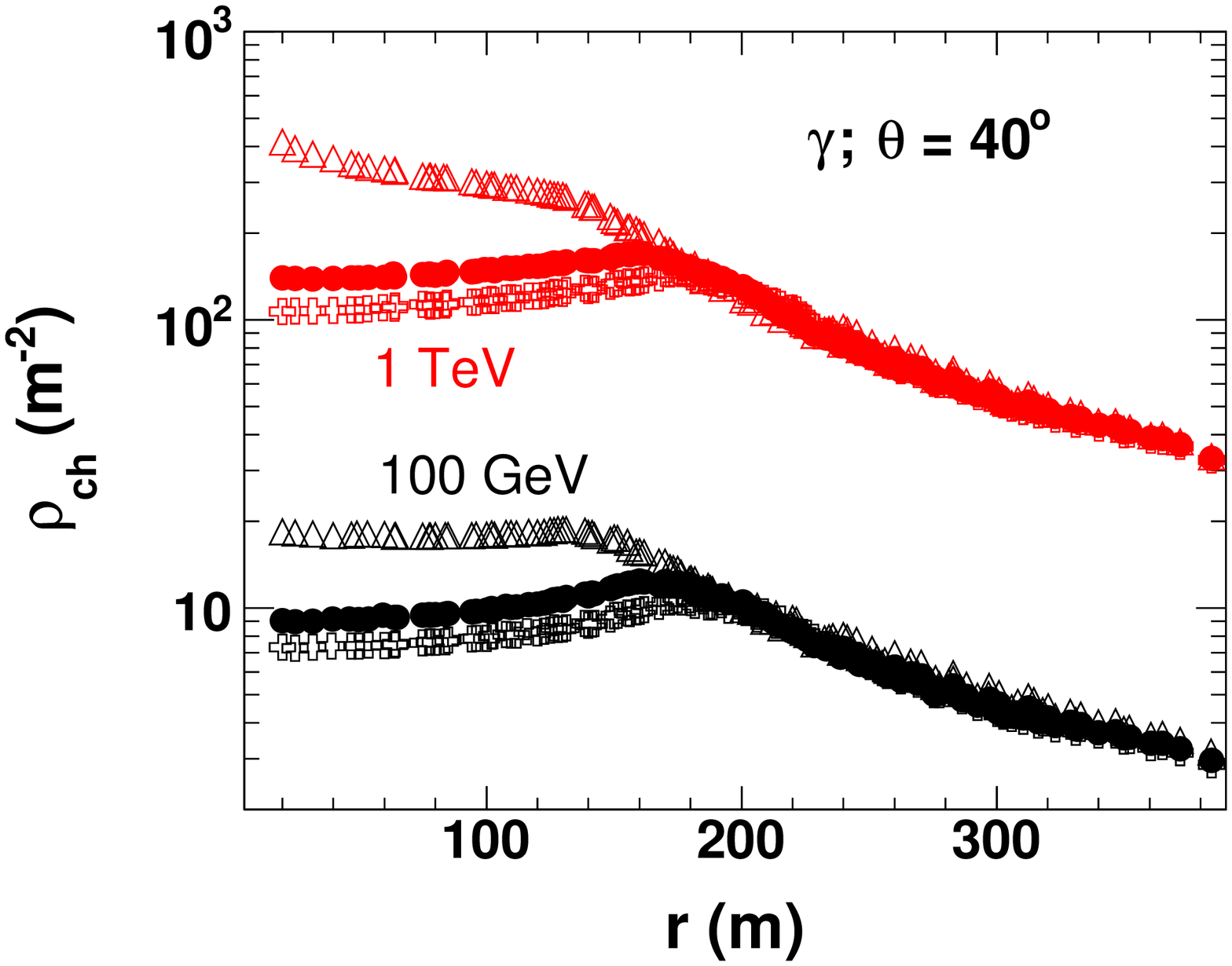} \hspace{-2mm}
\includegraphics[width=5.8cm, height=4.5cm]{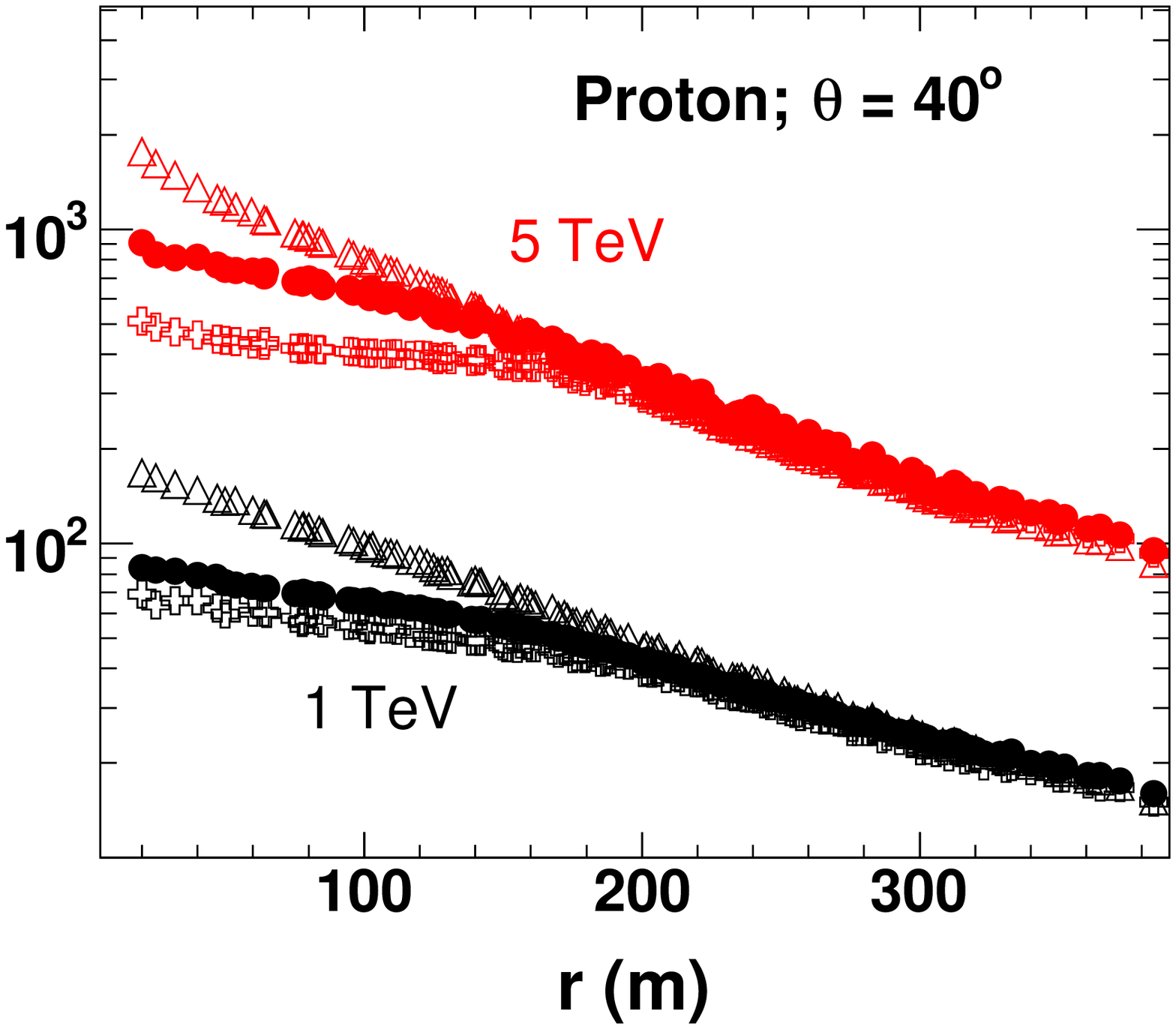} \hspace{-2mm}
\includegraphics[width=5.8cm, height=4.5cm]{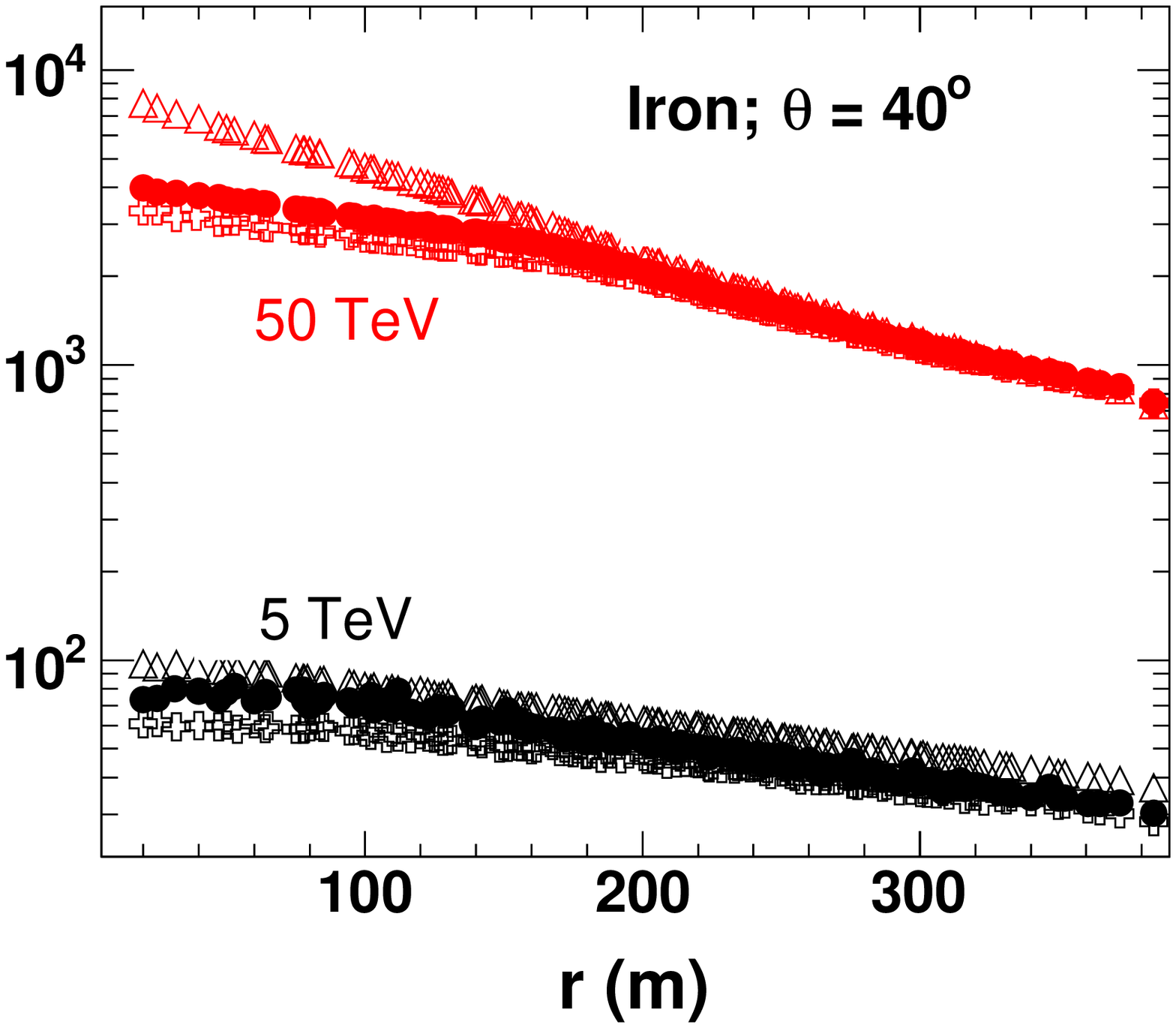}}

\caption{Variation of $\rho_{ch}$ with distance
from shower core at three different altitudes of observation level for 
$\gamma$, proton and iron primary incident at 0$^{\circ}$ and 40$^{\circ}$ 
zenith angles. The black colour is used to represent the low energy plots and
the red colour to represent the high energy plots.}
\label{fig5a}
\end{figure*}

The density of Cherenkov photons increases with the increasing altitude of 
the observational level mostly near the shower core ($< r_{hm}$, $r_{hm}$ 
is the minimum distance from the shower core at which photon density is almost 
same for all altitudes) for all 
primary particles incident at any zenith angle and at any energy as long as the 
observation level remains below the position of the shower maximum of the given 
primary particle (see Fig.\ref{fig5a}). Moreover, the difference in density
of photons between low and high altitudes of observation level becomes larger 
with the increasing value of zenith angle. Also with the increasing value of 
zenith angle, $r_{hm}$ is shifted away from the shower core. These happen 
because, with increasing slant depth more and
more low energetic particles, which are mostly concentrated near the core of 
the shower, get absorbed as the shower moves towards the lower observation 
levels. So at lower observation level and at 40$^\circ$ zenith angle, 
the characteristic hump, which is observed usually for $\gamma$-ray primaries, 
is seen slightly even for the iron primary as well as for the proton primary.

These effects are most observed in the case of $\gamma$-ray primary and least
observed in the case of iron primary, since a $\gamma$-ray primary only 
produce the EM cascade, which is mainly responsible for the 
generation of Cherenkov photons, whereas a proton and an iron primaries 
produce both EM as well hadronic cascade in different proportion,
as mentioned above. Due to subsequent production of secondary charged 
particles by secondary collisions at lower observational level, the Cherenkov
photon's density exceeds that at high altitude observation level at large
distances ($> r_{hm}$) from the shower core. This is specially more noticeable
for the vertical proton shower at high energy. It should be noted that for
the vertical showers of all primary energies, $r_{hm}$ lies in between 100 m to
150 m, whereas for the most inclined shower it lies in between 150 m to 200 m.

\begin{figure*}[hbt]
\centerline
\centerline{
\includegraphics[scale=0.29]{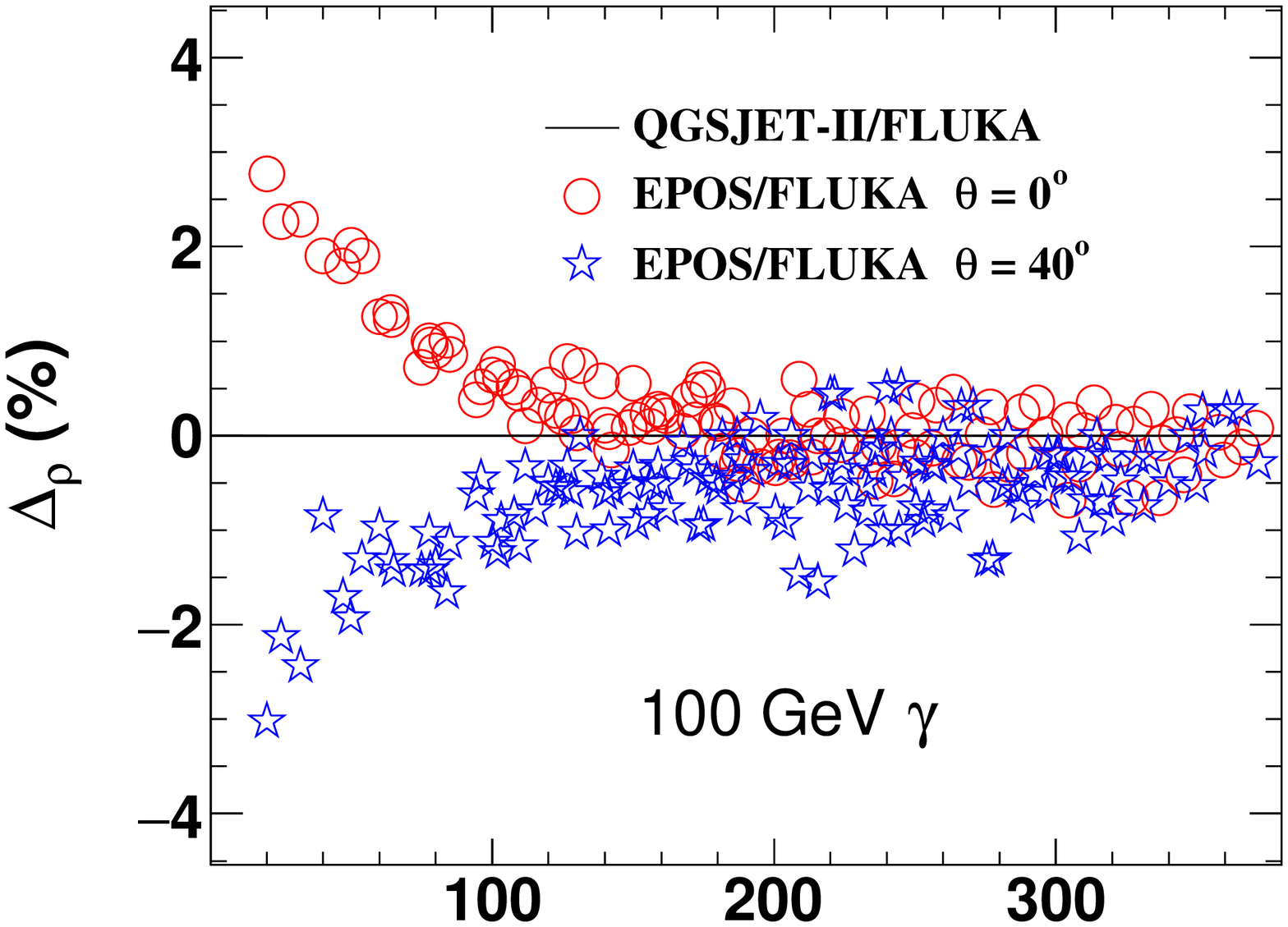} \hspace{-3mm}
\includegraphics[scale=0.29]{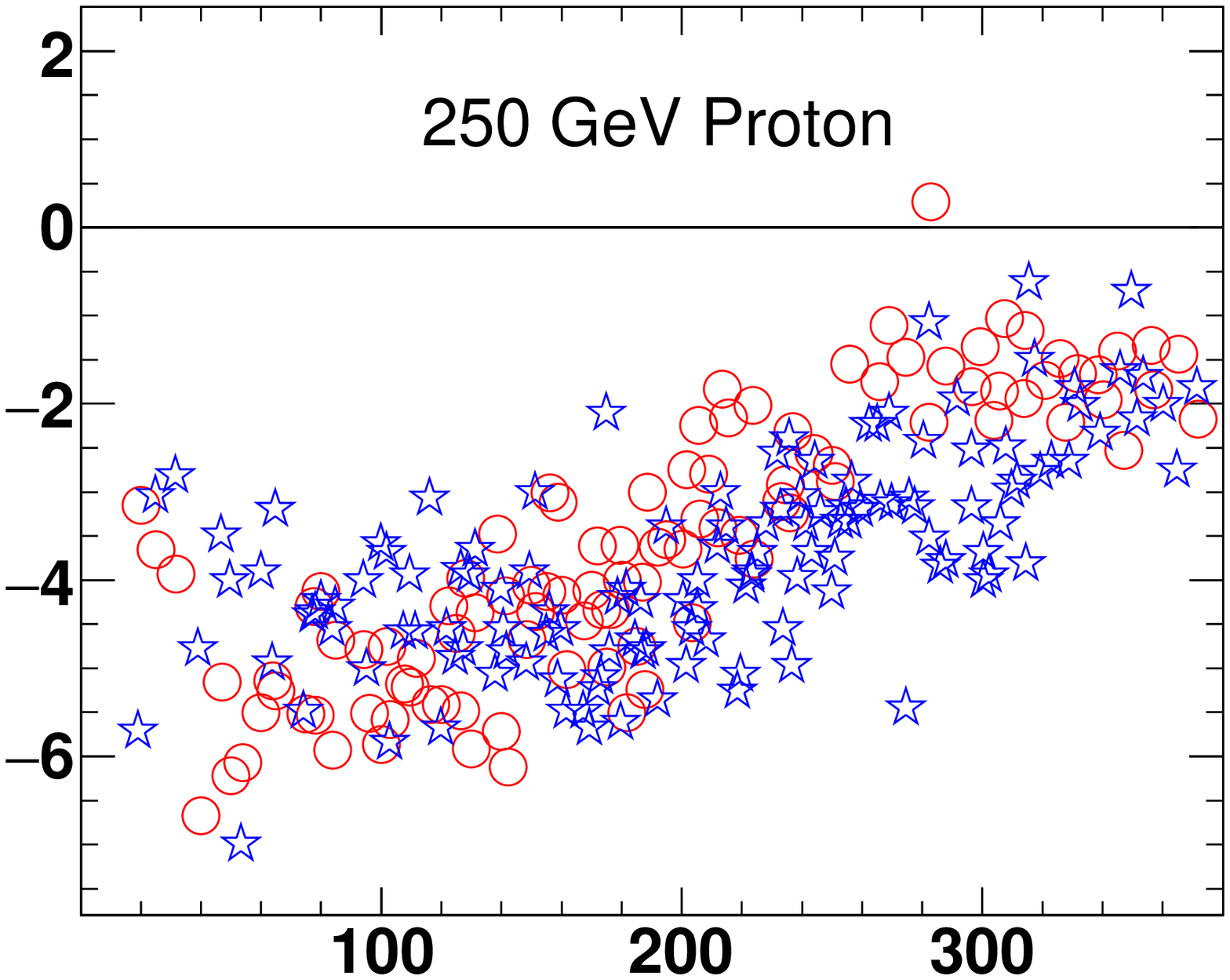} \hspace{-3mm}
\includegraphics[scale=0.29]{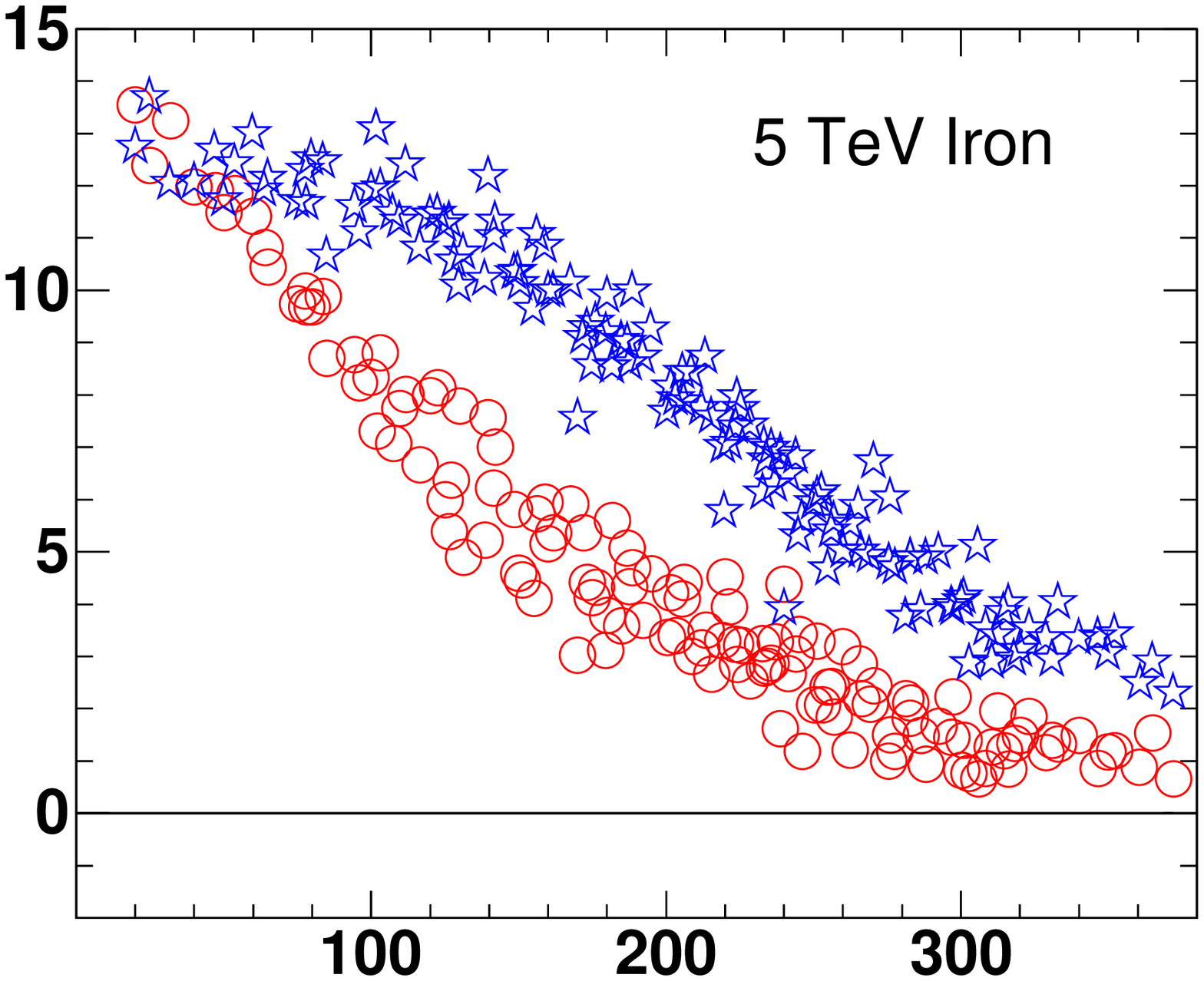}}

\vspace{-3mm}
\centerline{
\includegraphics[scale=0.29]{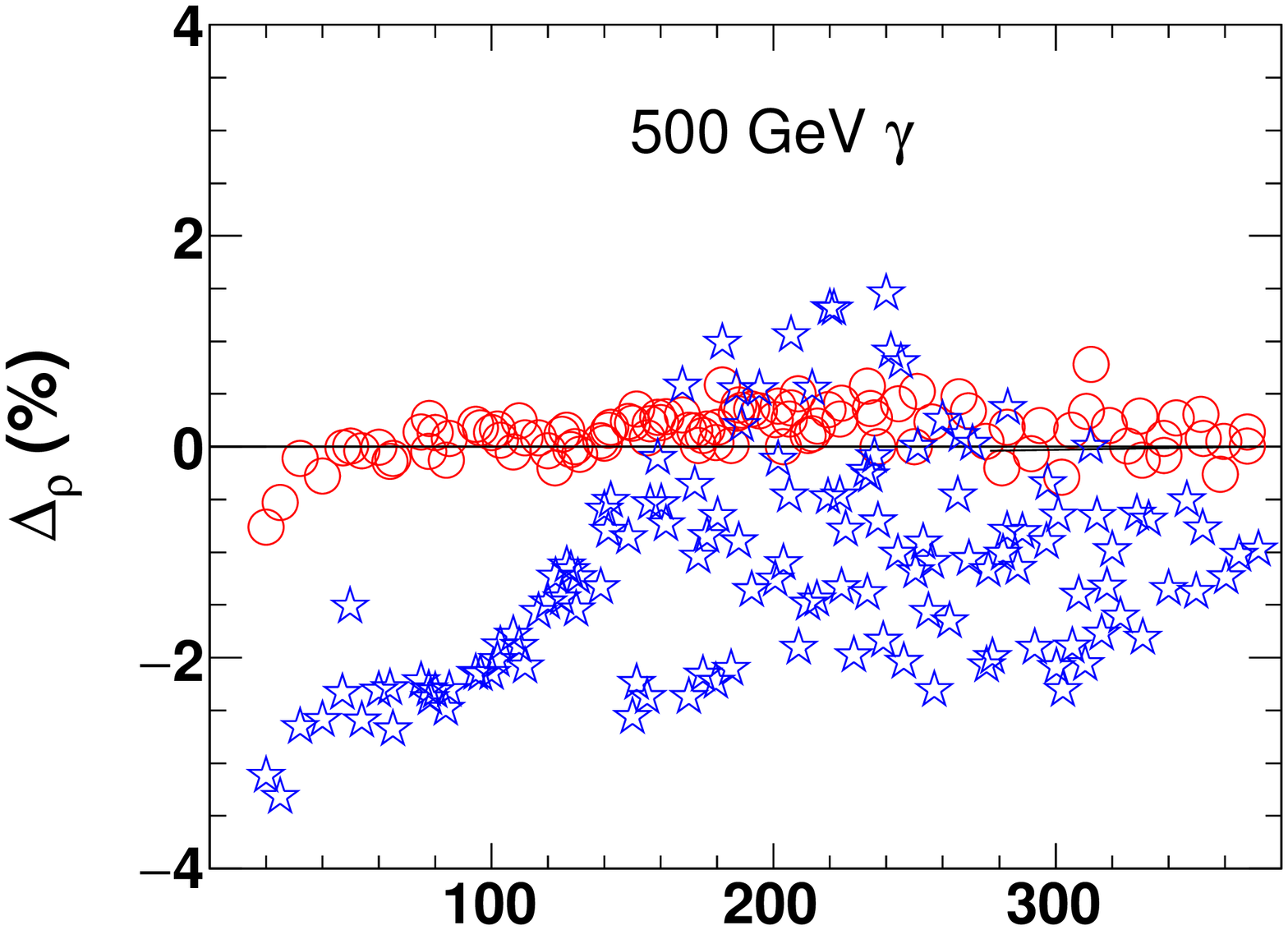} \hspace{-3mm}
\includegraphics[scale=0.29]{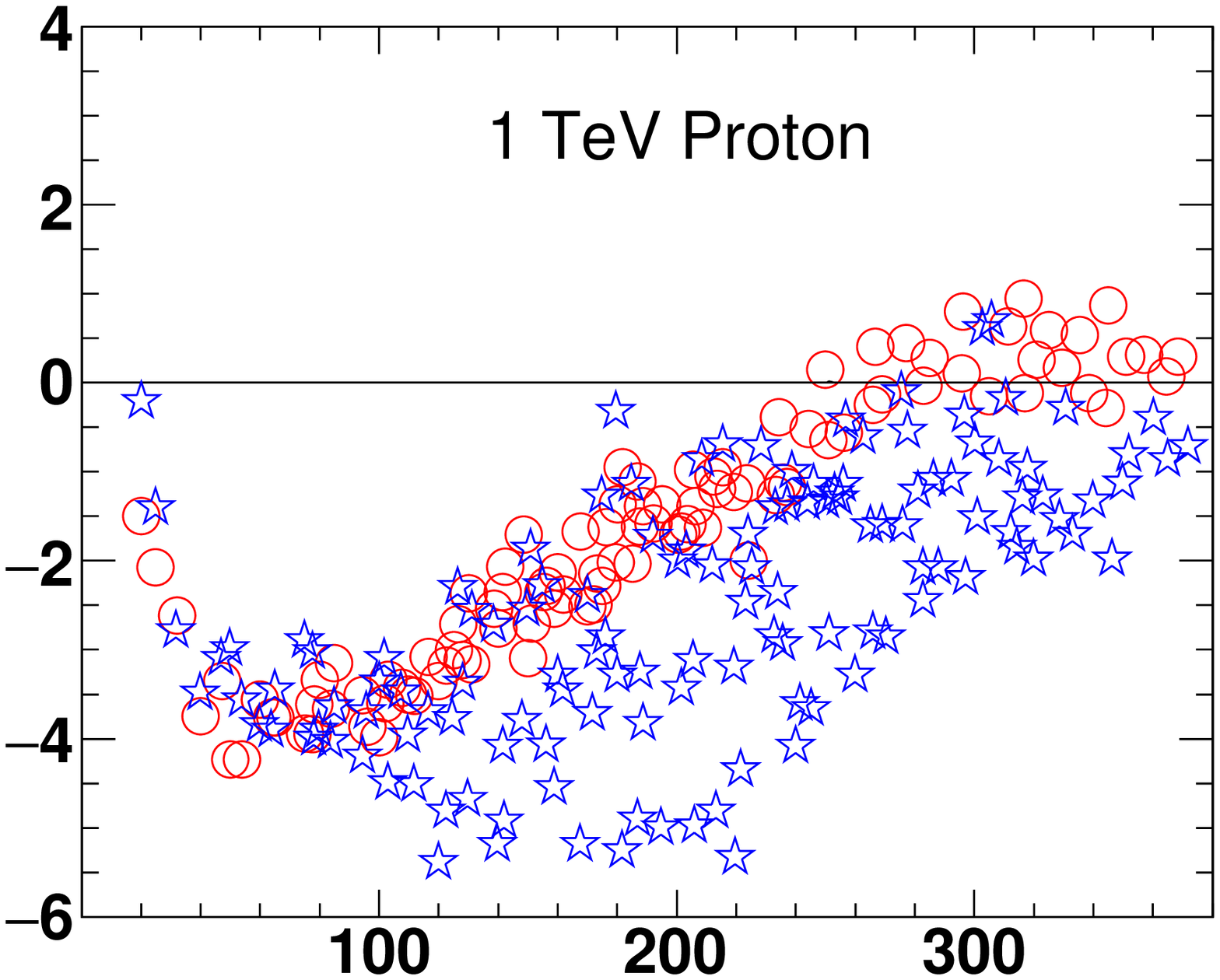} \hspace{-3mm}
\includegraphics[scale=0.29]{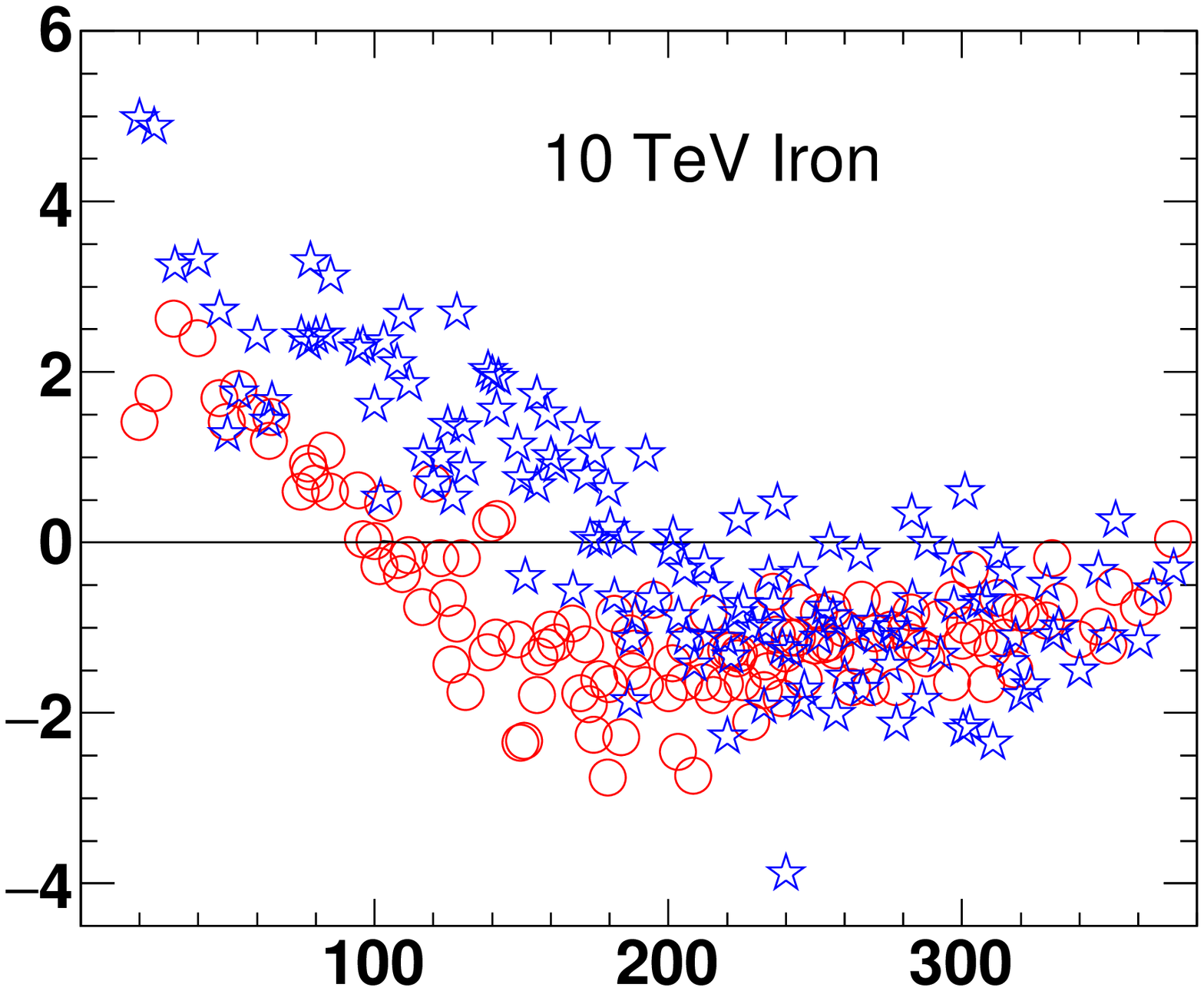}}

\vspace{-3mm}
\centerline{
\includegraphics[scale=0.29]{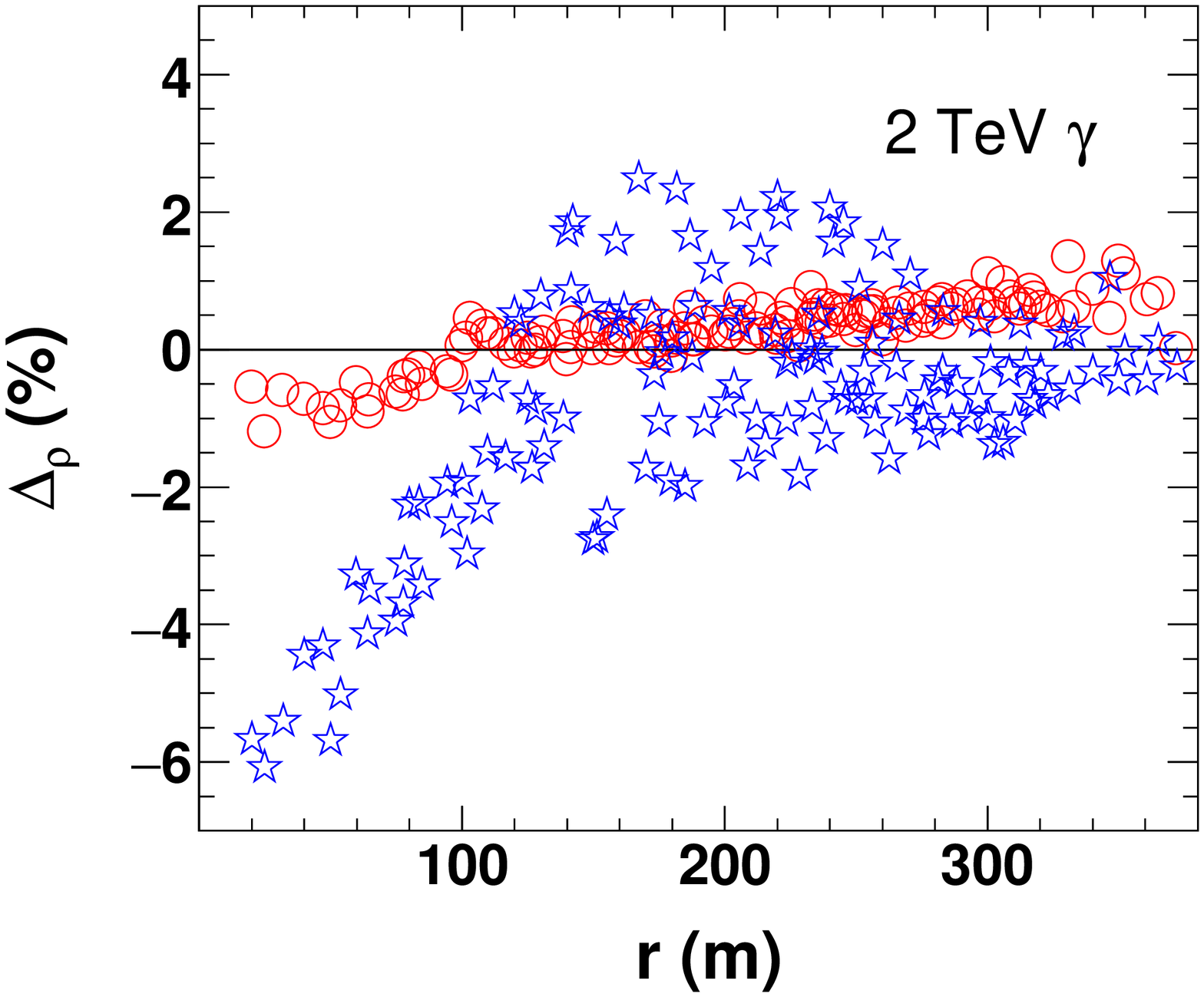} \hspace{-3mm}
\includegraphics[scale=0.29]{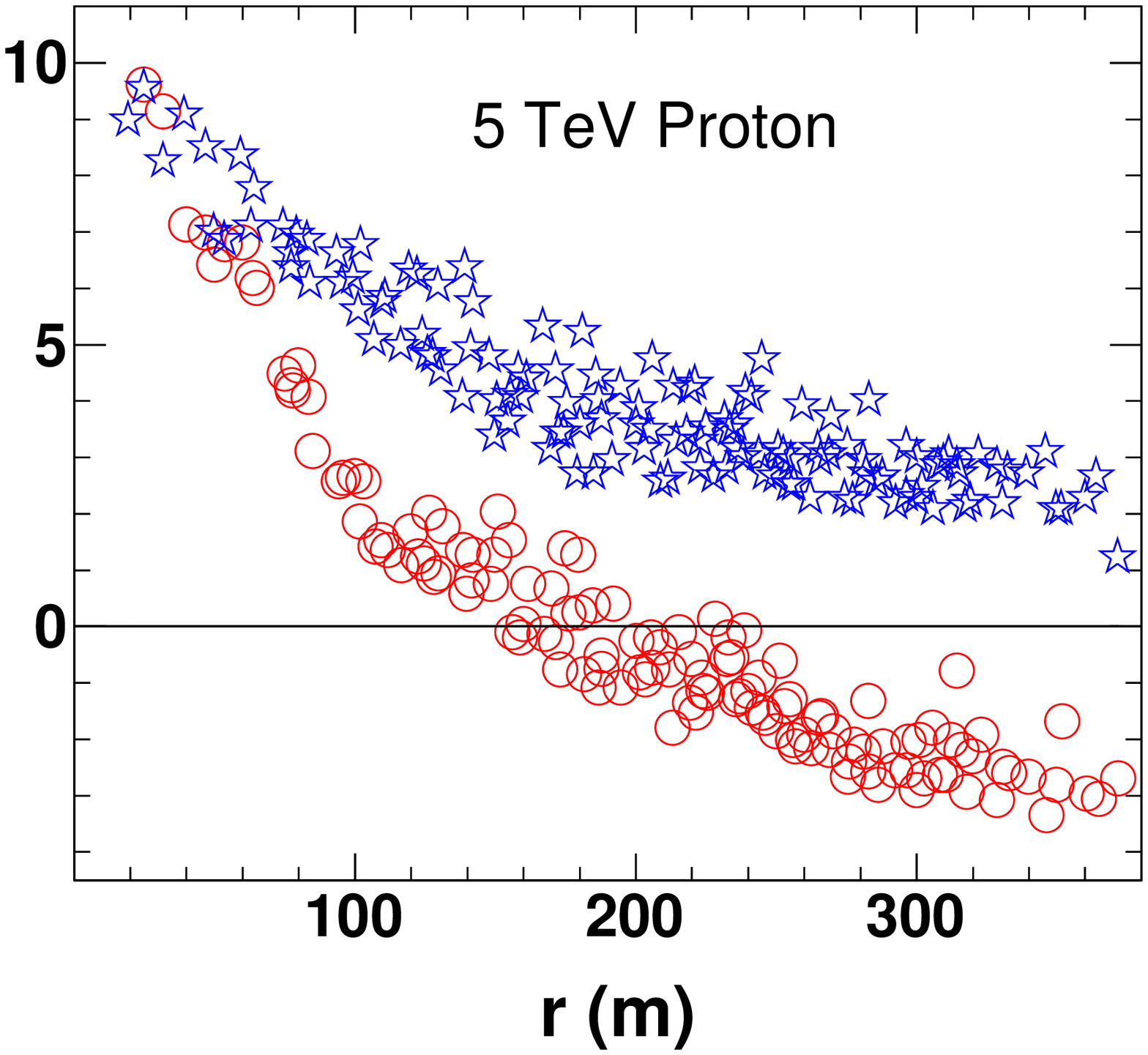} \hspace{-3mm}
\includegraphics[scale=0.29]{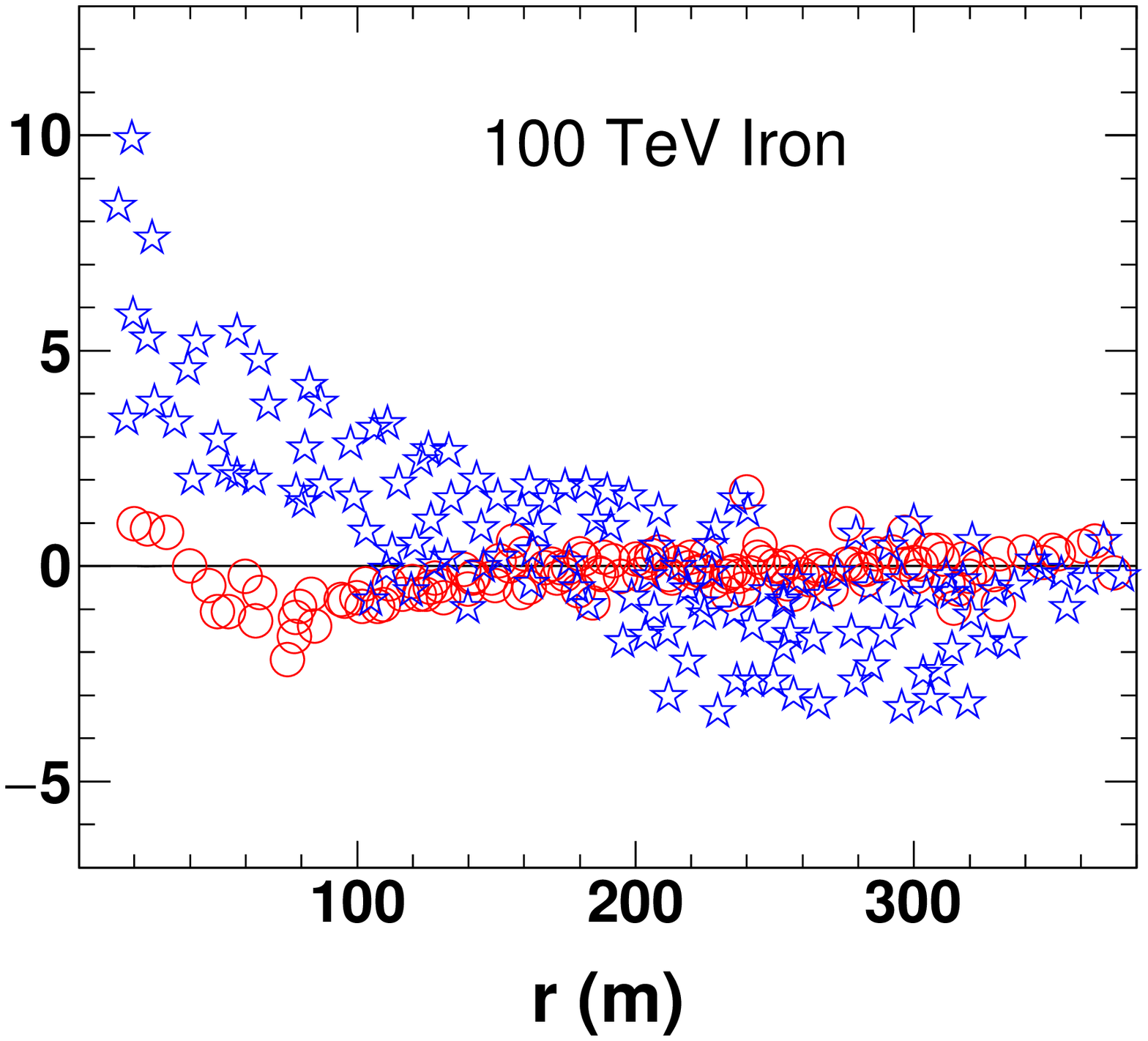}}

\caption{\% Relative deviation of Cherenkov photon densities ($\Delta{\rho}$s)
with respect to core distance of shower of different primaries obtained by
using QGSJETII and EPOS high energy hadronic interaction models at 4270 m 
altitude of the observation level. The
QGSJETII-FLUKA model combination is considered as the reference for the
calculation, which is indicated by a horizontal solid line in all plots.}
\label{fig2}
\end{figure*}         

\subsubsection{Hadronic interaction model sensitivity}
To see the effect of hadronic interaction models (used in this study) on
the lateral density of the Cherenkov photons, the \% relative deviation of
$\rho_{ch}$ ($\Delta_{\rho}$) is calculated for the EPOS-FLUKA model
combination taking QGSJETII-FLUKA as the reference. Results are shown in the
Fig.\ref{fig2} for the vertically incident and 40$^{o}$ inclined showers of
different primaries with different energies only as representation.
It is to be mentioned that, the reference model combination choice is fully
arbitrary, so any one of the two may be used as the reference. For the
$\gamma$-ray primaries, incident vertically at different energies, there is
no considerable differences in densities are seen for these two hadronic 
interaction
model combinations. However, closer to the shower core ($< 100$ m) some
small deviations of $\sim \pm 3 \%$ are observed. Beyond this distance the
deviations are within $\sim \pm 1\%$. However, for the inclined shower the 
density deviation is slightly higher (up to $\sim 6\%$), specially at higher 
energies.

For the proton primary, the deviation between the two models is clearly 
visible for both vertically incident and inclined showers, and is 
significant in
comparison to $\gamma$-ray primary. At lower primary energies these deviations
are mostly negative and limited within $\sim 6\%$, while for higher primary
energies they are mostly positive. In the case of 1 TeV primary, the deviations
are limited to within $\sim - 6\%$ to $1\%$. For the 5 TeV primary with 
vertical incidence, the deviation extends from $\sim -3\%$ to $\sim 10\%$ and 
for the inclined shower it ranges from $\sim 2\%$ to $\sim 10\%$.

The deviation in density for the iron primaries are different from the
primaries of $\gamma$-ray and proton, where the lower primary energy gives
higher deviation. At
5 TeV energy, for both vertical and inclined showers, the deviation is maximum
($\sim 13\%$) near the shower core and gradually decreases with distance from
the core to match with each other at the farthest points. For 10 TeV primary
the deviation is mostly limited within $\sim \pm 3\%$ with the deviation
becoming positive from negative near the core. With increasing energy, the
deviation for the vertical shower decreases substantially in comparison to
inclined shower. For example, at 100 TeV the two models exactly match each
other for the vertical shower while for the inclined shower, the deviation is
mostly distributed between $\sim \pm 5\%$. To understand this behaviour of 
the iron primary, the high energy hadronic interaction models QGSJET-II and 
EPOS shall have to be studied in details carefully.   

It is clear that almost all the cases, the deviation is basically high only 
near the shower core ($< 100$ m). Also seen that for all primary particles and 
energies, the density deviation due to these two models is higher for the 
inclined 
shower than the vertical shower. However, as a whole on average, the range of
deviation is within the acceptable limit ($< \pm 10\%$).

\subsubsection{Influence of atmospheric model}

\begin{figure*}[hbt]
\centerline
\centerline{
\includegraphics[scale=0.29]{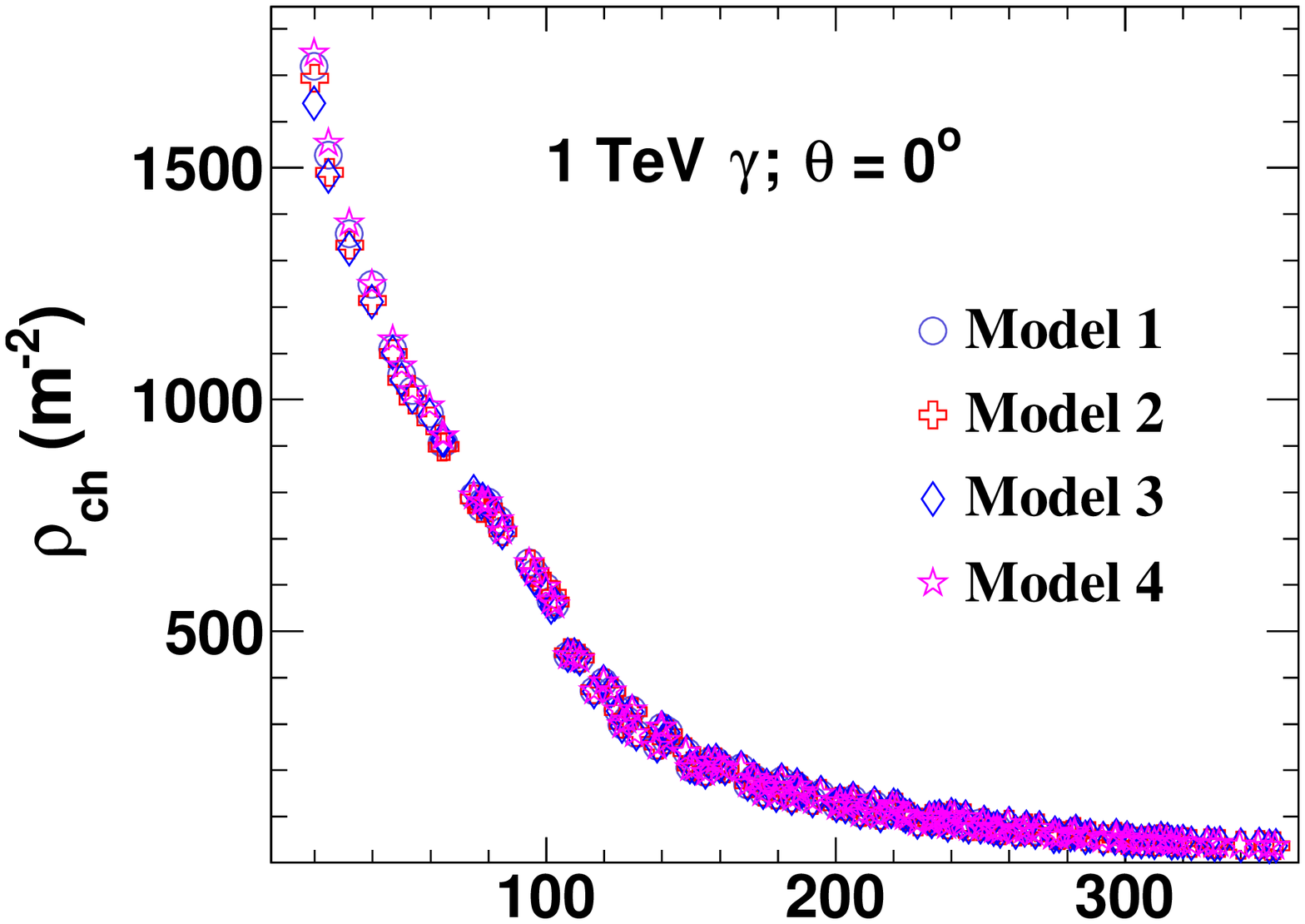}  \hspace{-3mm}
\includegraphics[scale=0.29]{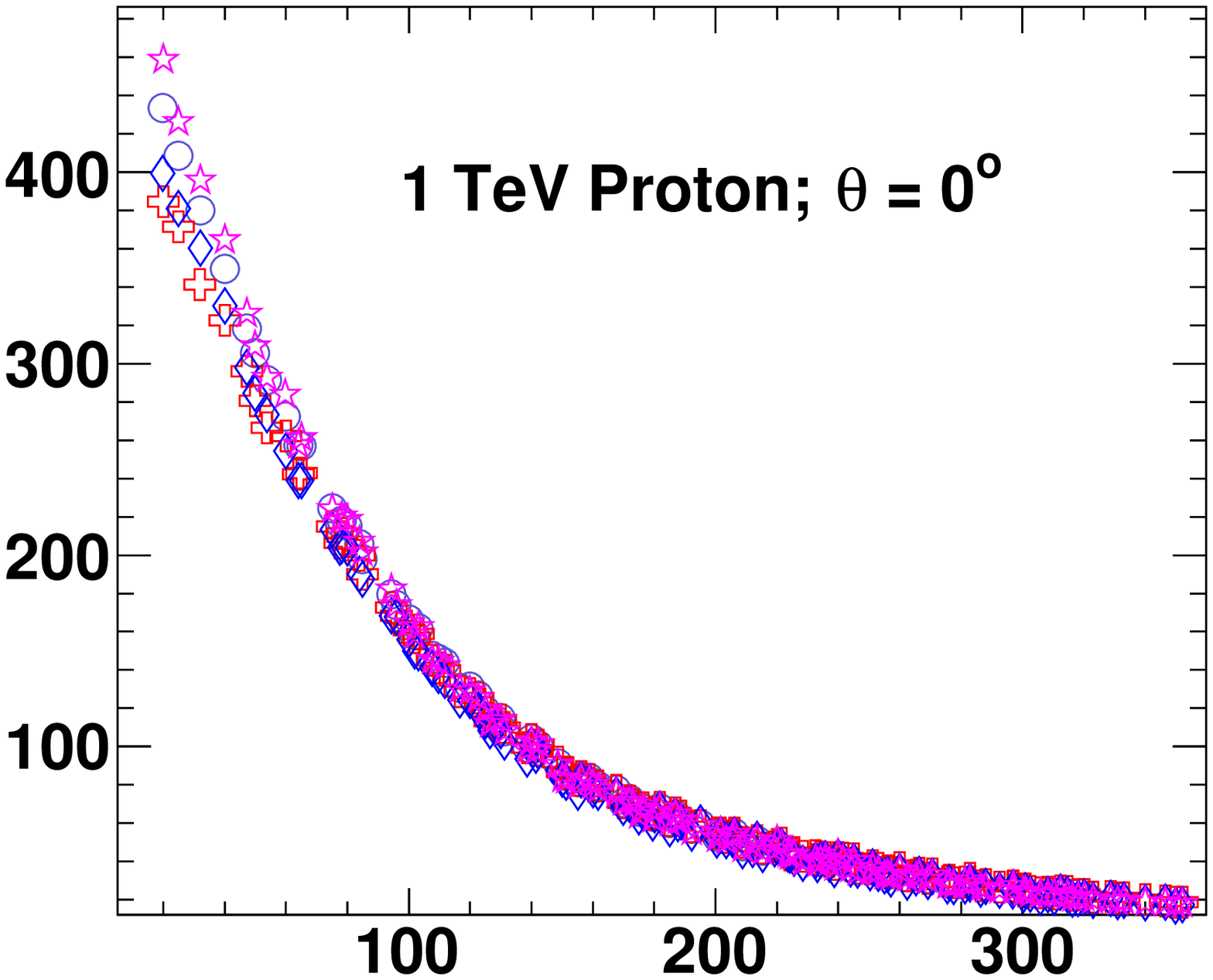} \hspace{-3mm}
\includegraphics[scale=0.29]{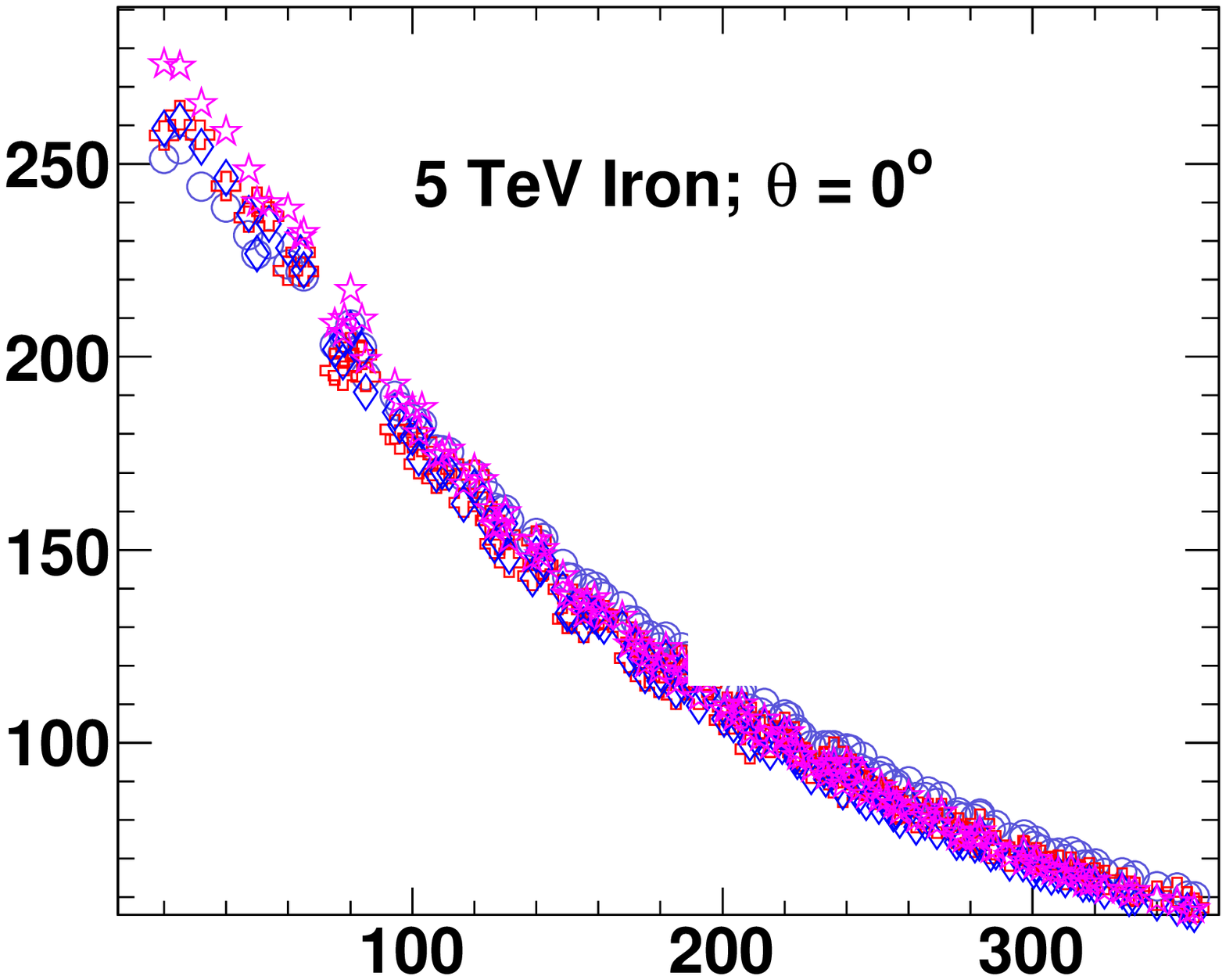}}

\vspace{-3mm}
\centerline{\hspace{2mm}
\includegraphics[scale=0.29]{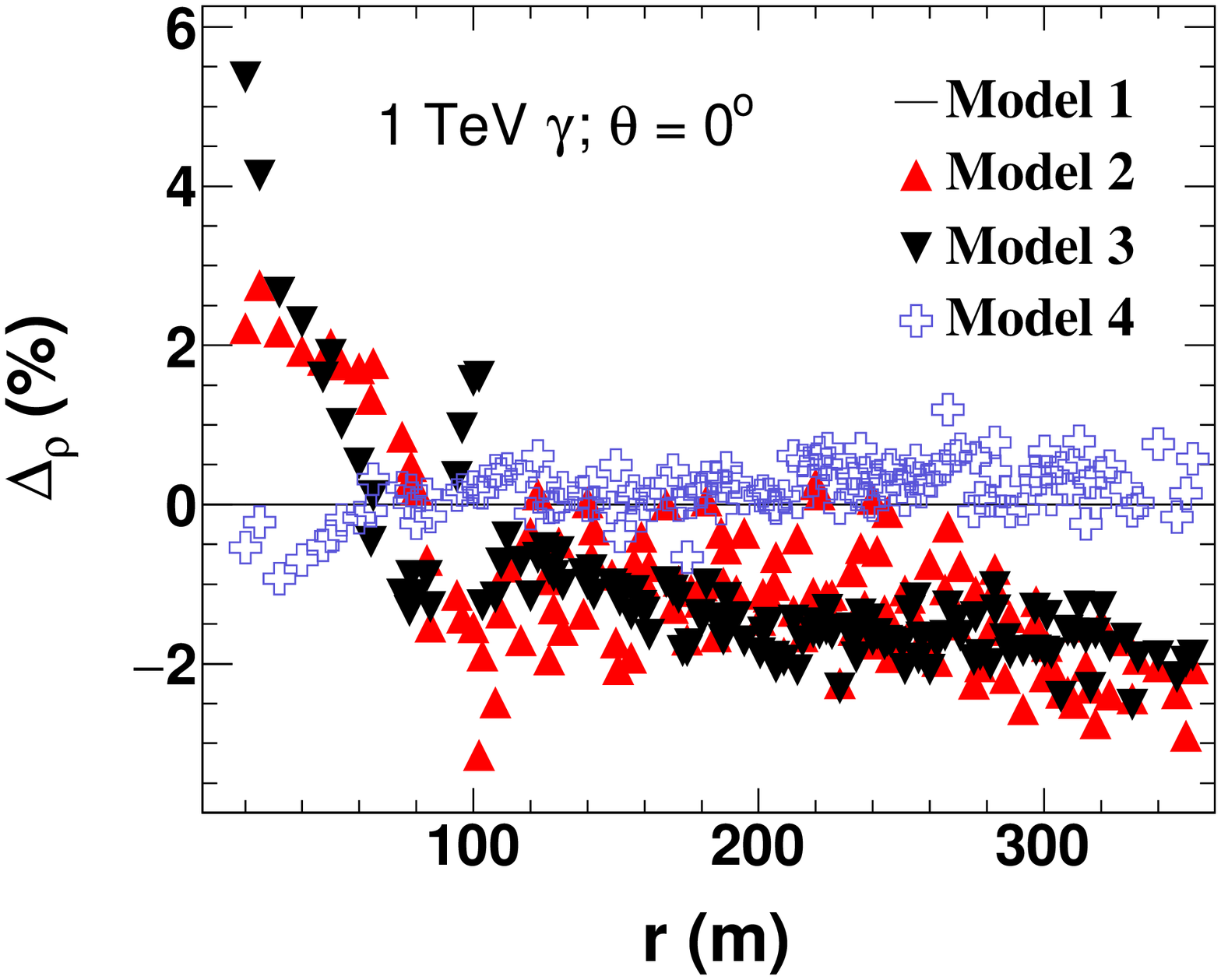}  \hspace{-3mm}
\includegraphics[scale=0.29]{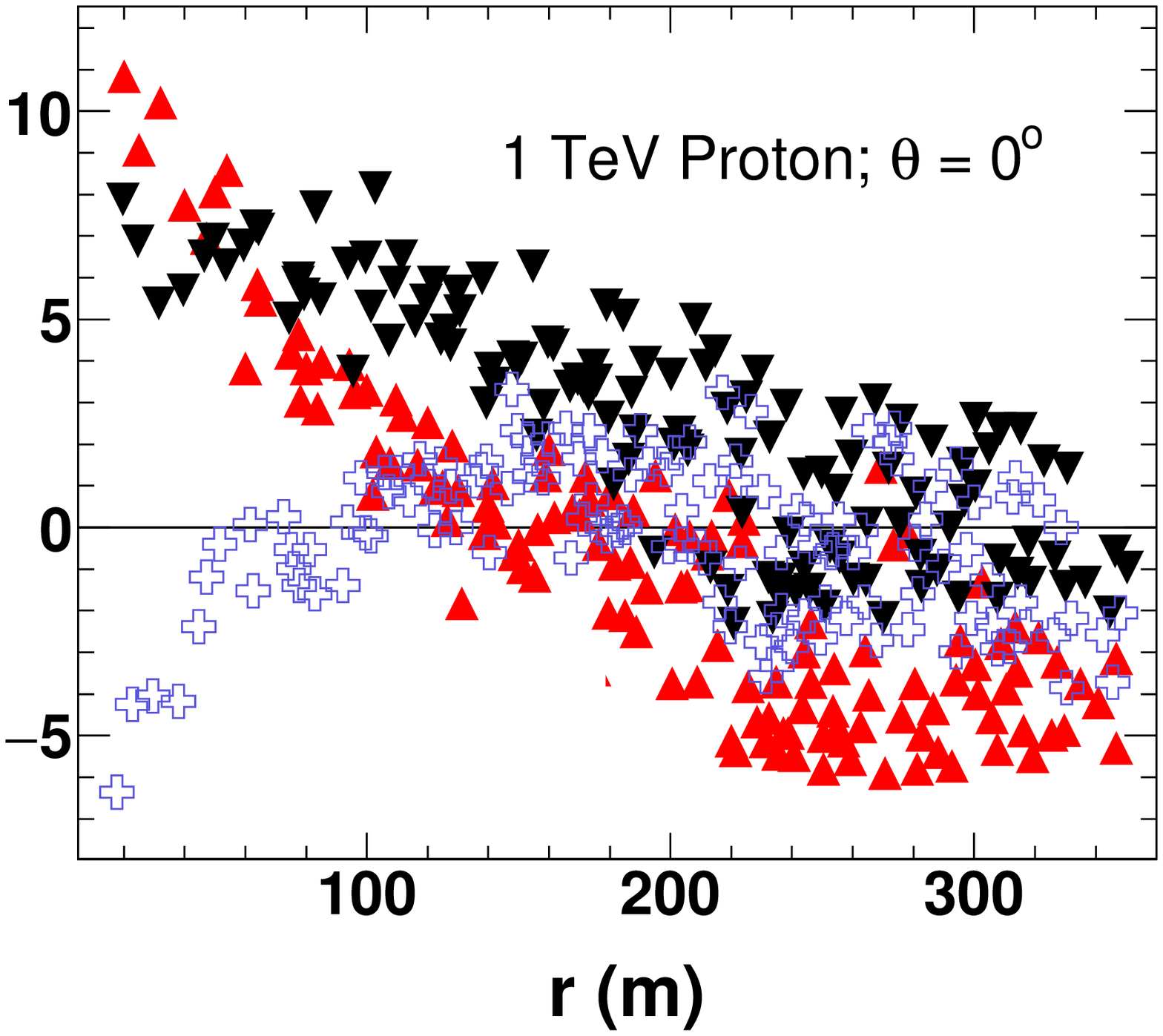} \hspace{-3mm}
\includegraphics[scale=0.29]{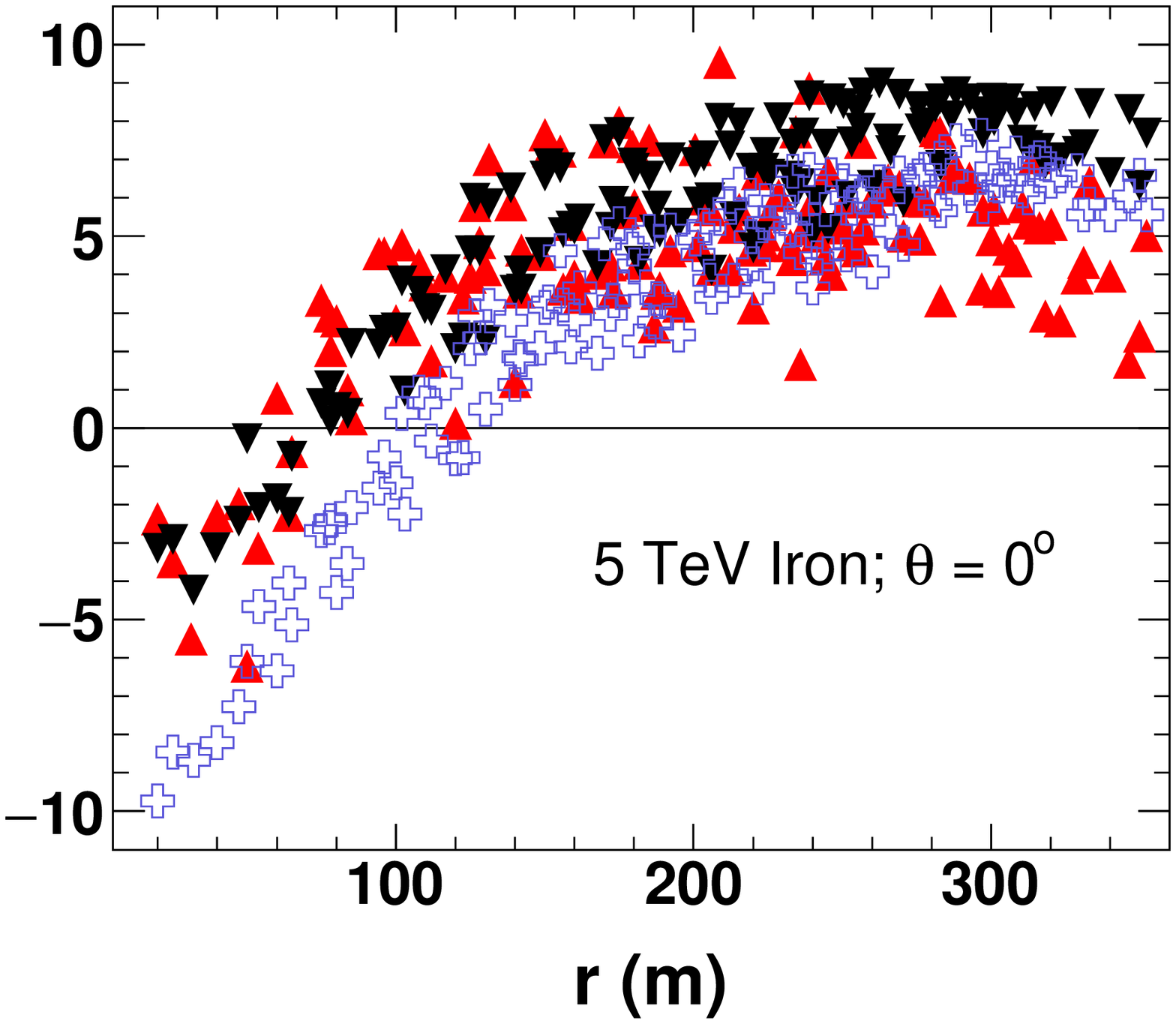}}    

\caption{Variation of $\rho_{ch}$ with distance
from the shower cores  of $\gamma$-ray, proton and iron primaries incident at 
0$^{\circ}$ zenith angle as obtained for four different atmospheric models
and the corresponding \% relative differences ($\Delta_\rho$) between different 
model predictions (bottom panels). 
The Model-1, Model-2, Model-3 and Model-4 represent the U.S. standard
atmosphere as parameterized by Linsley, AT 115 Central European atmosphere for
Jan. 15, 1993, Malarg\"ue winter atmosphere I after Keilhauer and U.S. standard atmosphere as parameterized by Keilhauer \cite{Heck} respectively.}
\label{fig5b}
\end{figure*}
All our simulations of EASs have been done by using the U.S. standard 
atmosphere as parameterized by Linsley as mentioned in the previous section. 
But it is also important to check the 
influence
of different atmospheric models on our results of Cherenkov photon density 
calculations. For this purpose we have used four different atmospheric models
available in CORSIKA (see the caption of the Fig.\ref{fig5b} for detail) by
random selection to 
calculate the Cherenkov photon density for all three primary particles 
incident
vertically at a given energy as shown in the Fig.\ref{fig5b}. It is
found that the difference in densities produced by these atmospheric models are
$<$ 5\% for $\gamma$-ray, $<$ 15\% for proton and $<$ 10\% for the iron 
primaries (see the lower panels of the Fig.\ref{fig5b}). Thus the error 
that may be introduced in the calculation of Cherenkov photon density due to 
the use of a particular atmospheric model
is found to be small for the $\gamma$-ray primary. 

\begin{figure*}[hbt]
\centerline
\centerline{
\includegraphics[scale = 0.28]{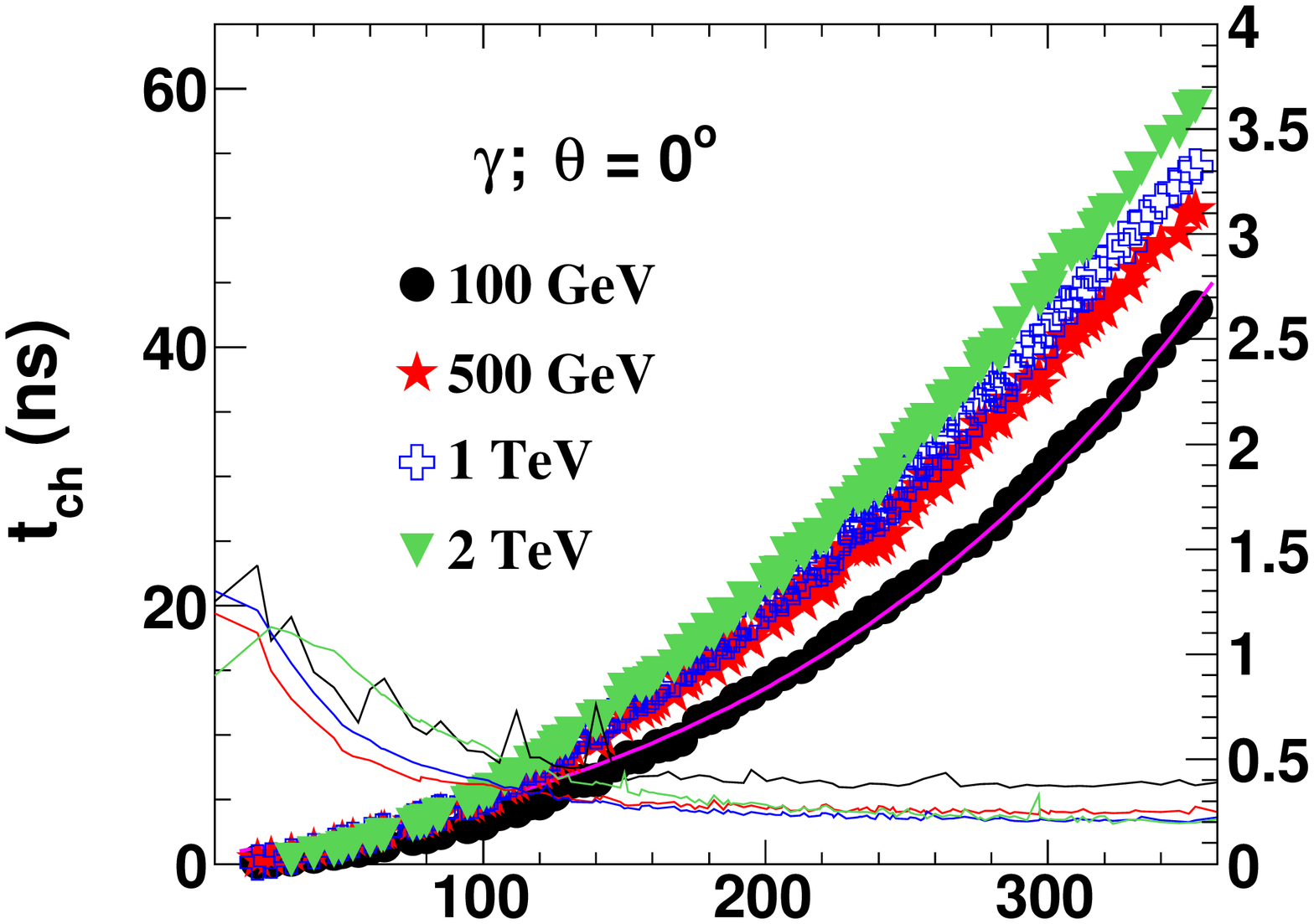}\hspace{-1.5mm}
\includegraphics[scale = 0.28]{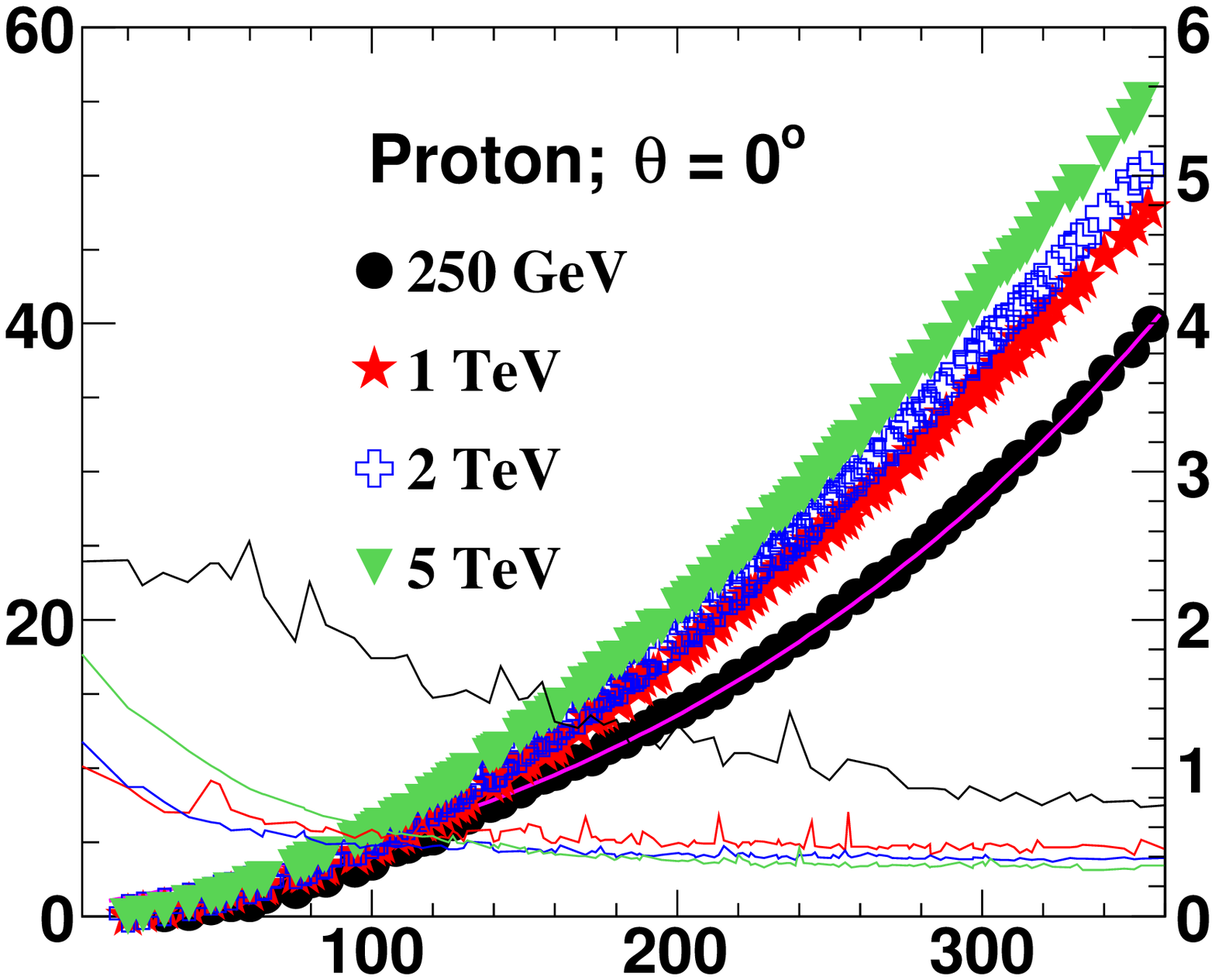}\hspace{-1.5mm}
\includegraphics[scale = 0.29]{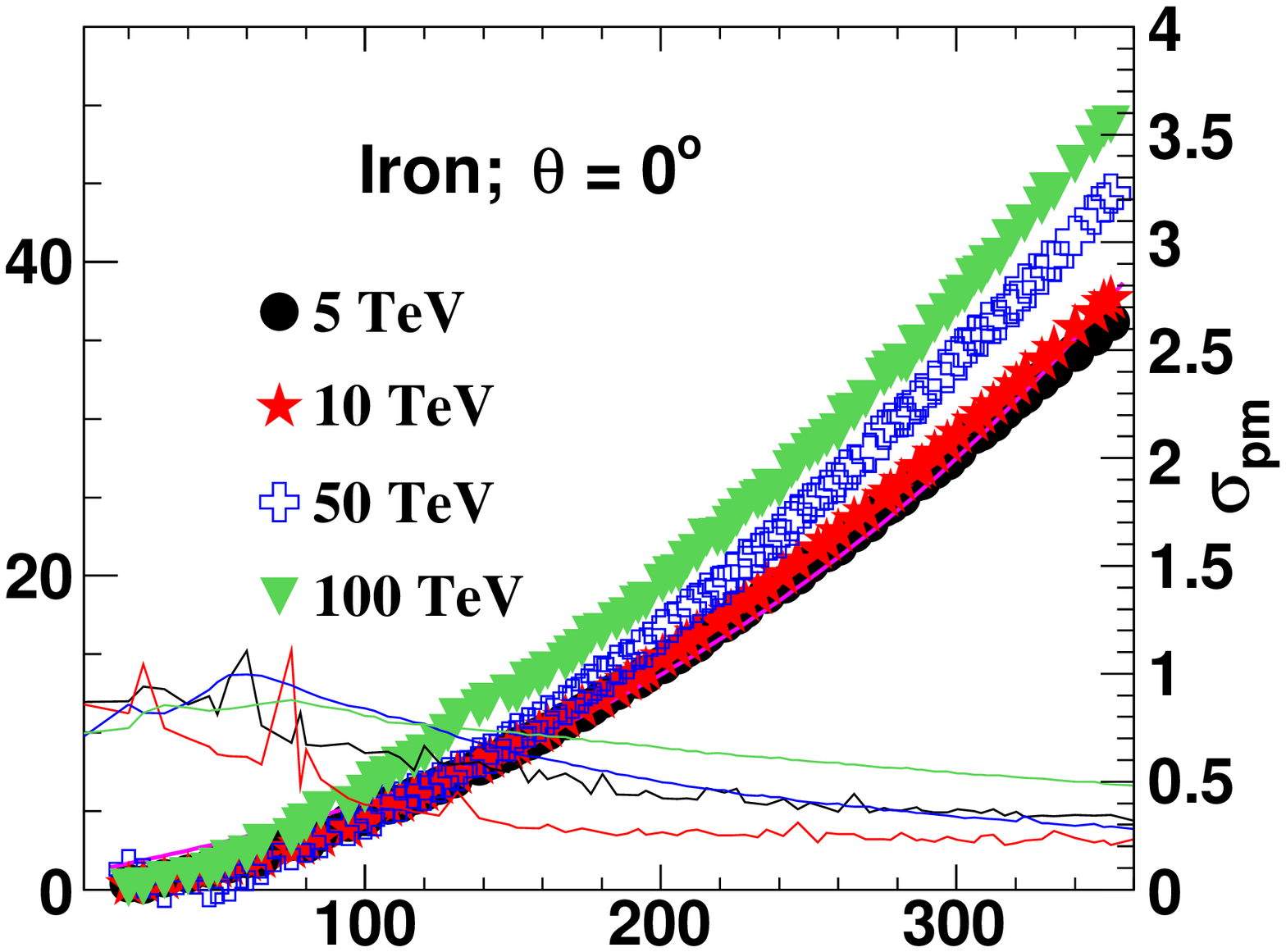}}
\vspace{-2mm}

\centerline{
\includegraphics[scale = 0.28]{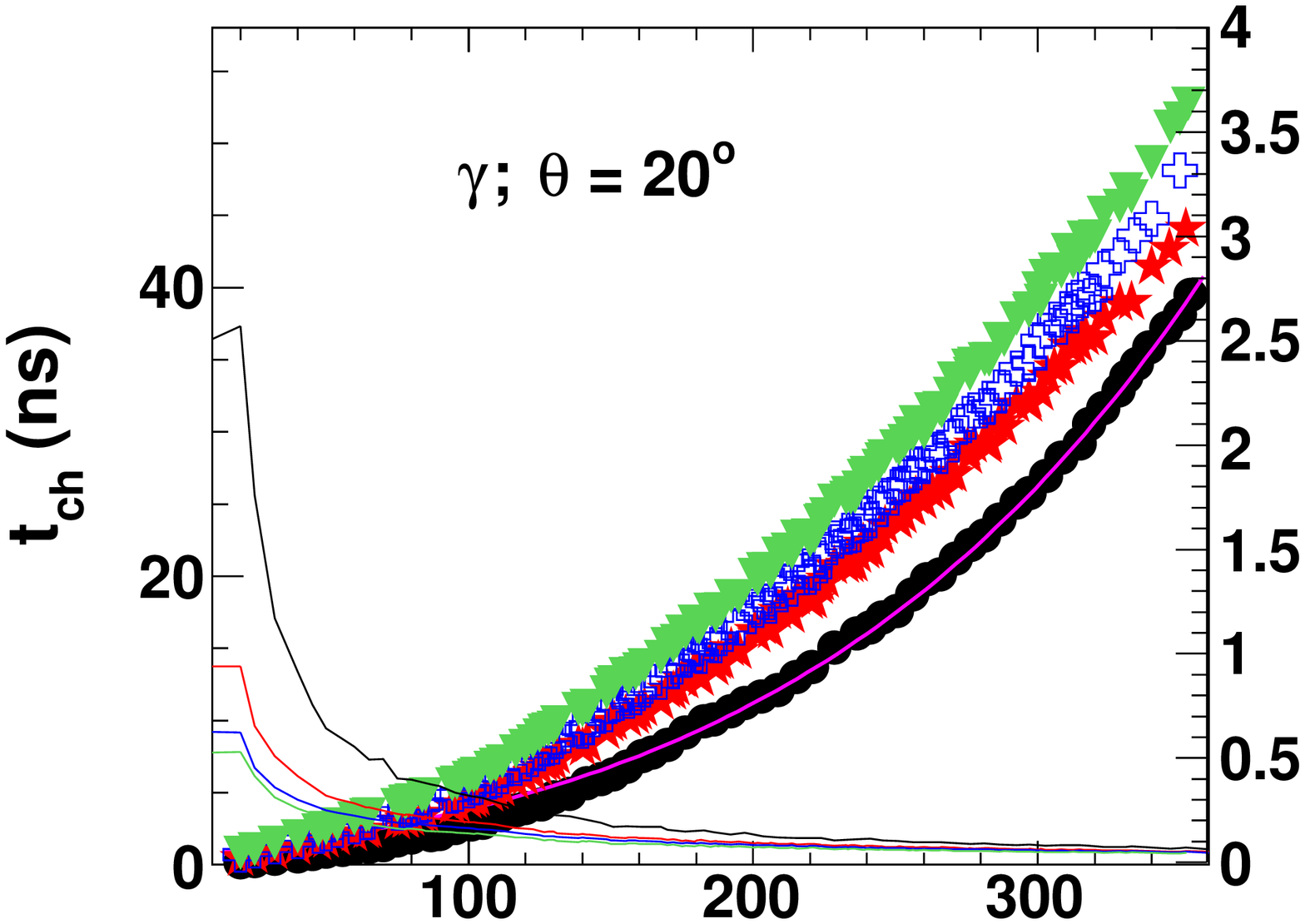} \hspace{-1.5mm}
\includegraphics[scale = 0.28]{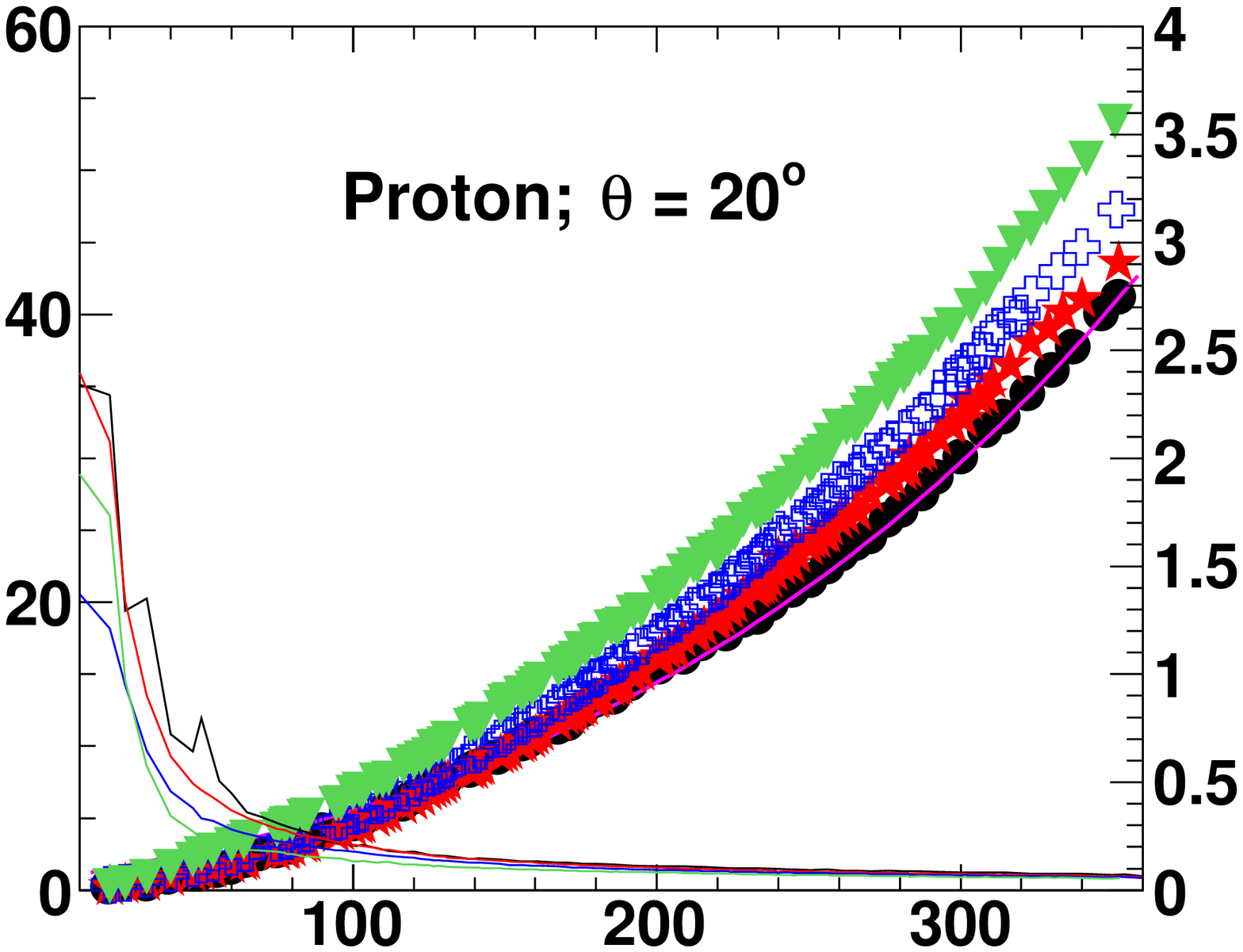} \hspace{-1.5mm}
\includegraphics[scale = 0.29]{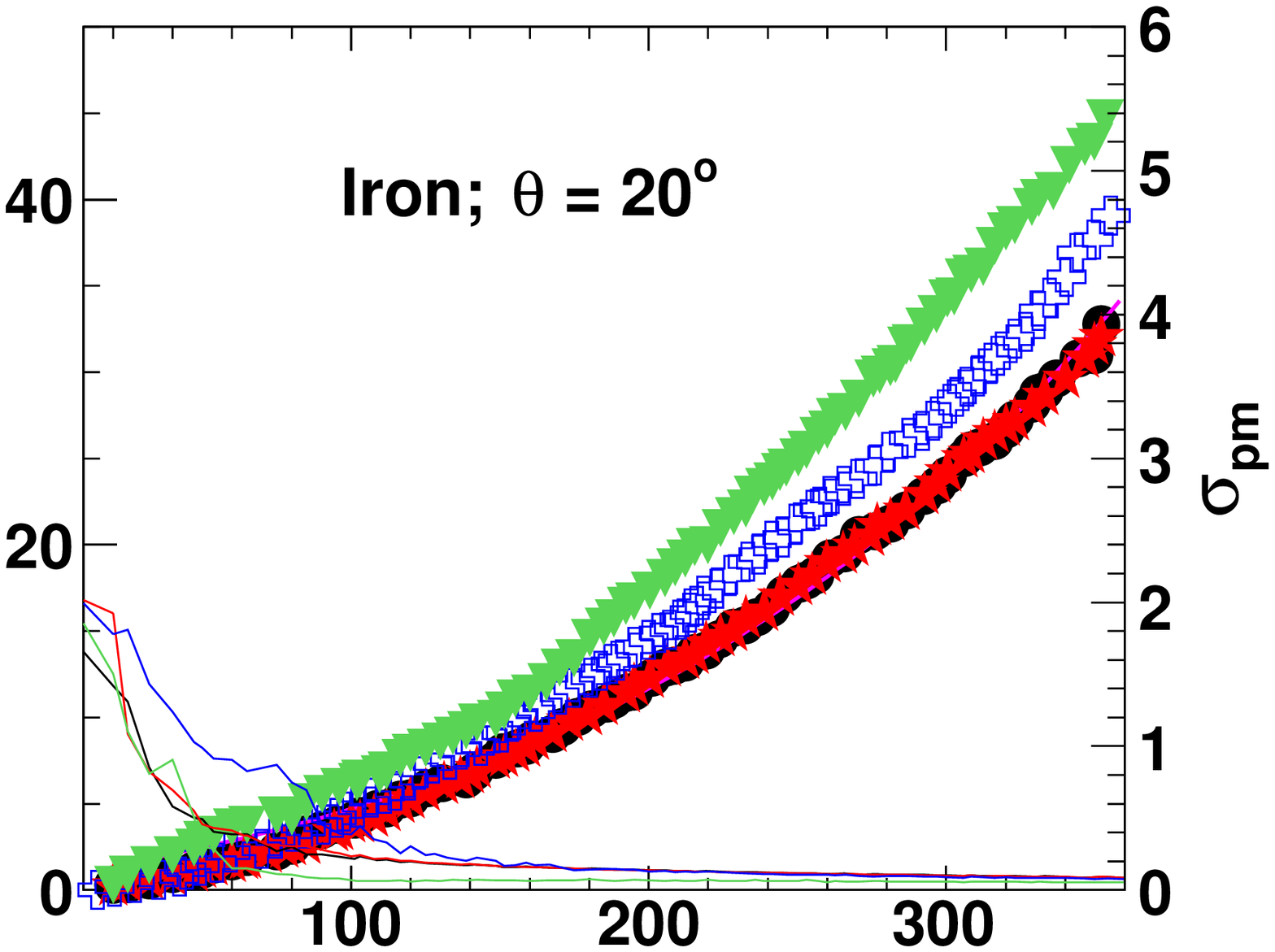}}

\vspace{-2mm}
\centerline{
\includegraphics[scale = 0.28]{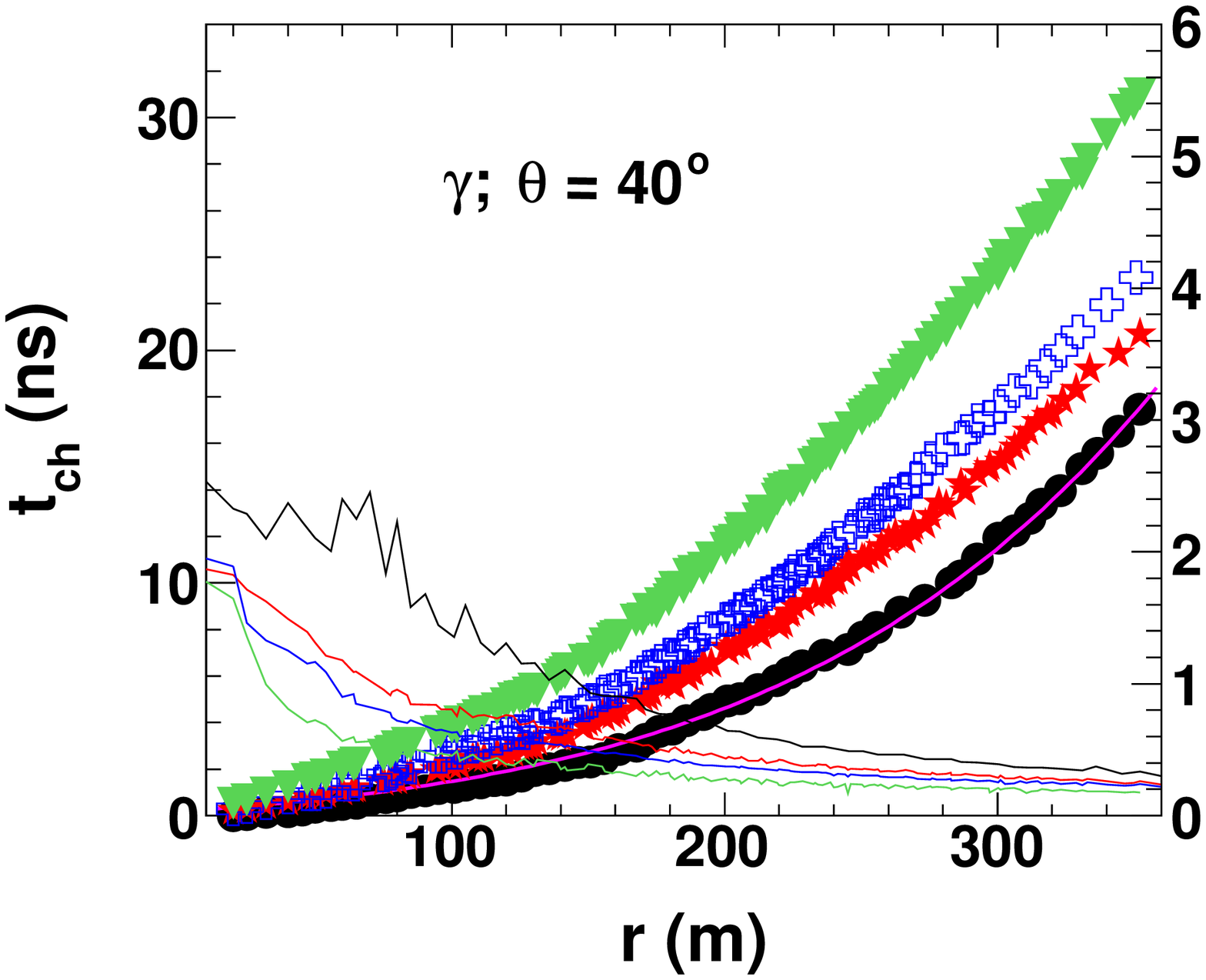} \hspace{-1.5mm}
\includegraphics[scale = 0.28]{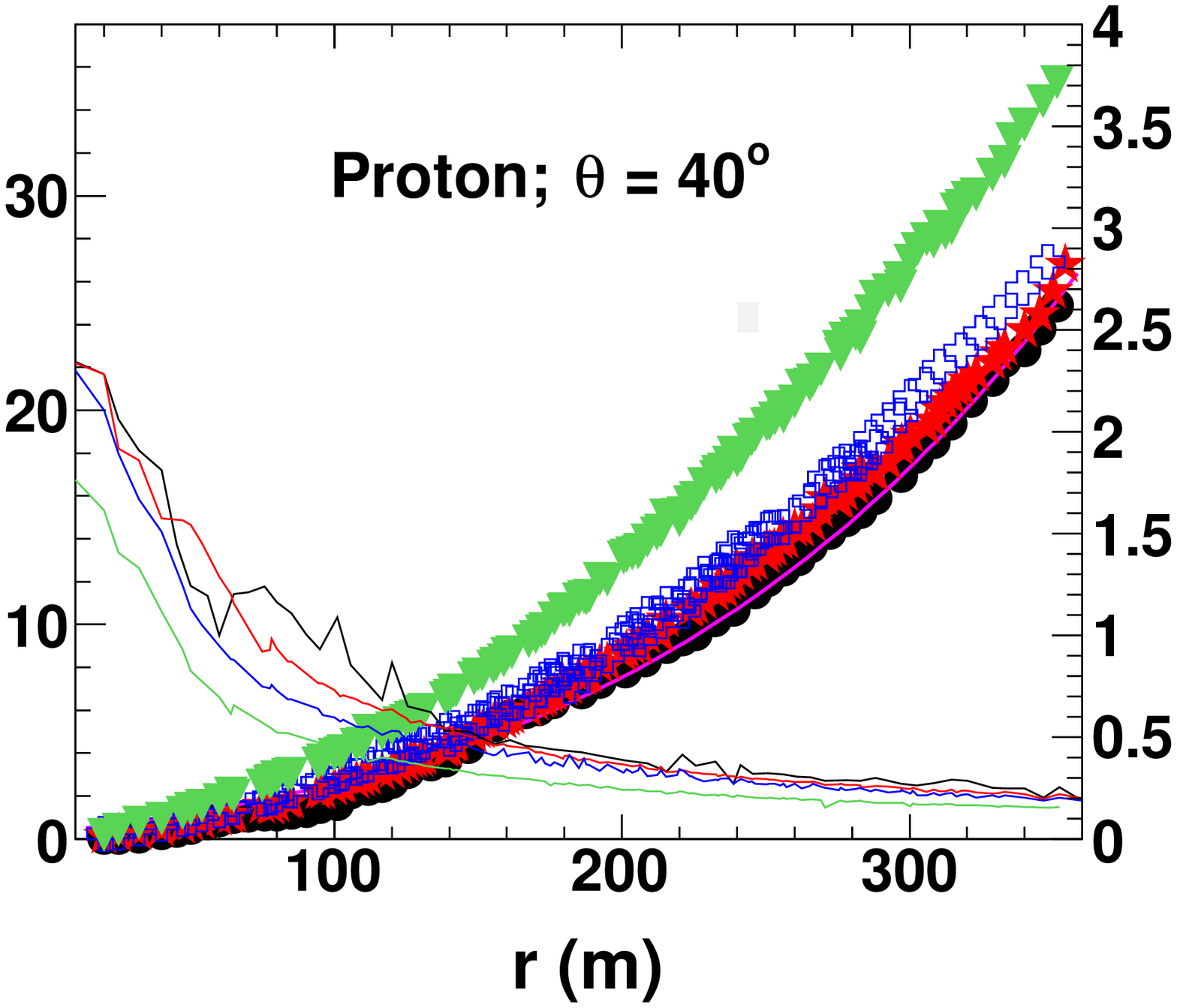} \hspace{-1.5mm}
\includegraphics[scale = 0.29]{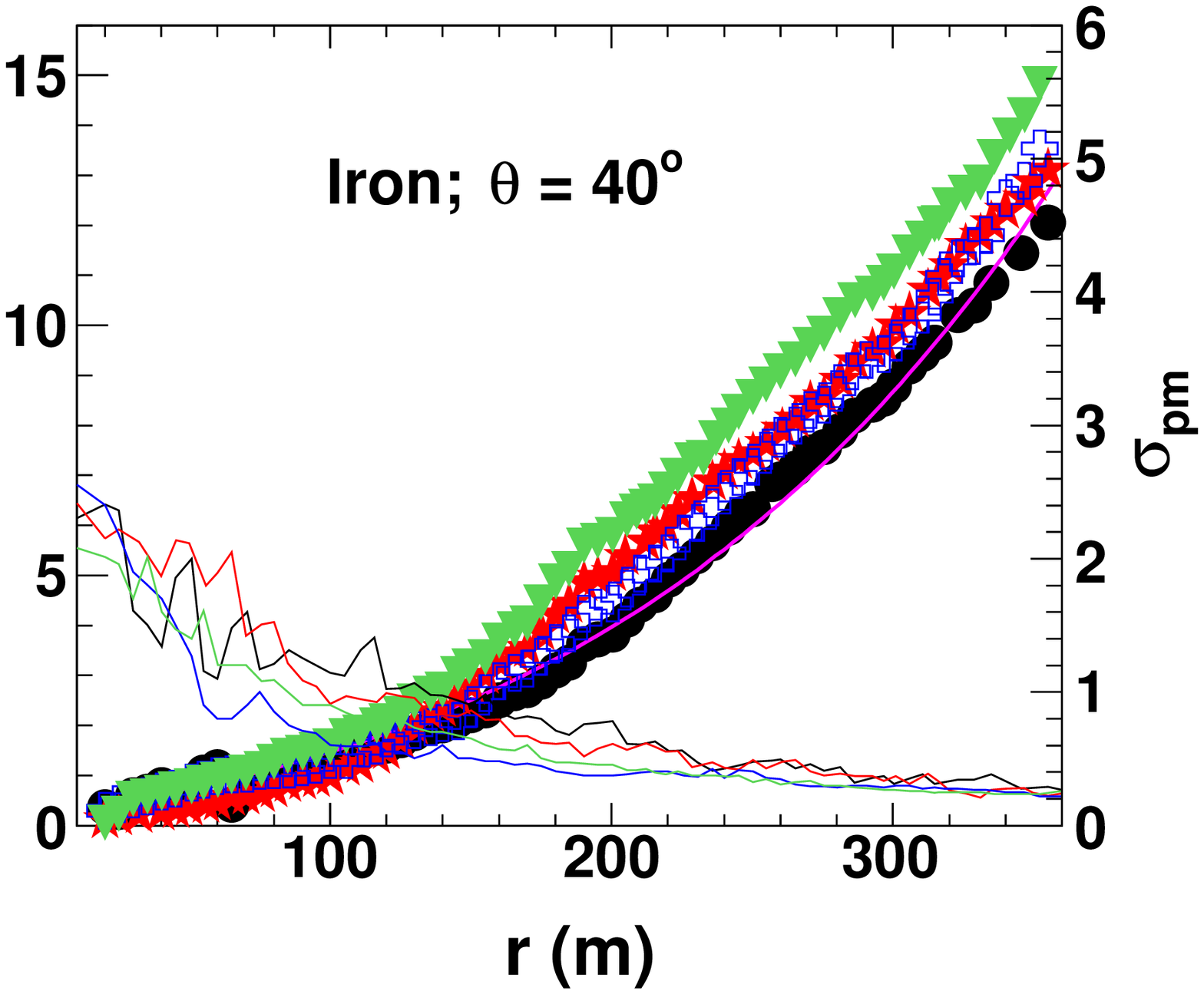}}
\caption{Average arrival time of Cherenkov photon ($t_{ch}$)
and the ratio of r.m.s. to mean of the photon arrival time ($\sigma_{pm}$) as 
a function of the distance from the core of showers of
different primaries with different energies corresponding to some fixed value
of zenith angle. The scale on the y2-axis is used to read the plots
of $\sigma_{pm}$. The solid line in each plot indicates the result of the best 
fit function (\ref{eq3}) into the data for the lowest energy.}
\label{fig6}
\end{figure*}

\subsection{Cherenkov photon's arrival time}
\subsubsection{General characteristics}
\begin{figure*}[hbt]
\centerline
\centerline{
\includegraphics[scale = 0.27]{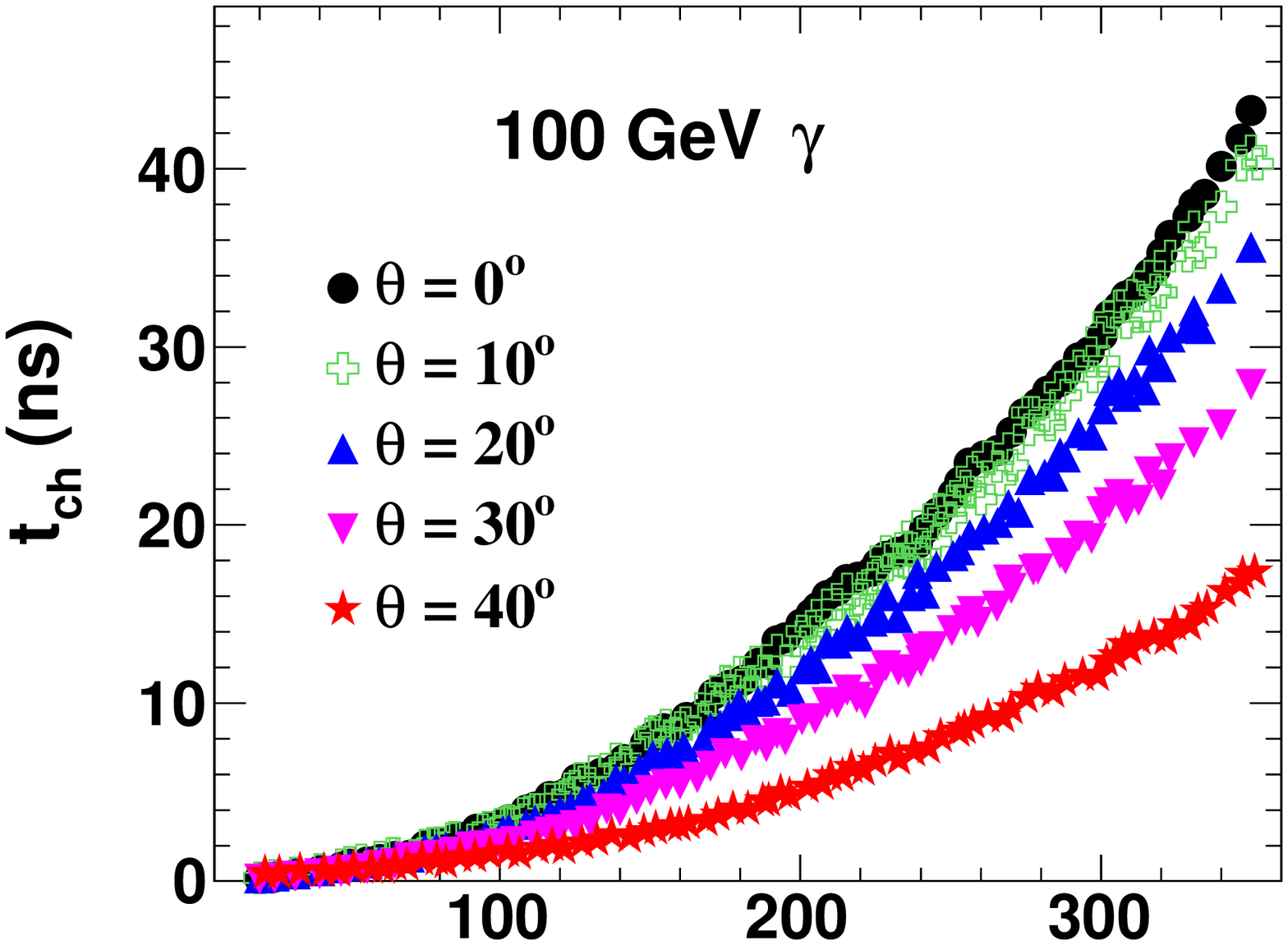} \hspace{-3mm}
\includegraphics[scale = 0.27]{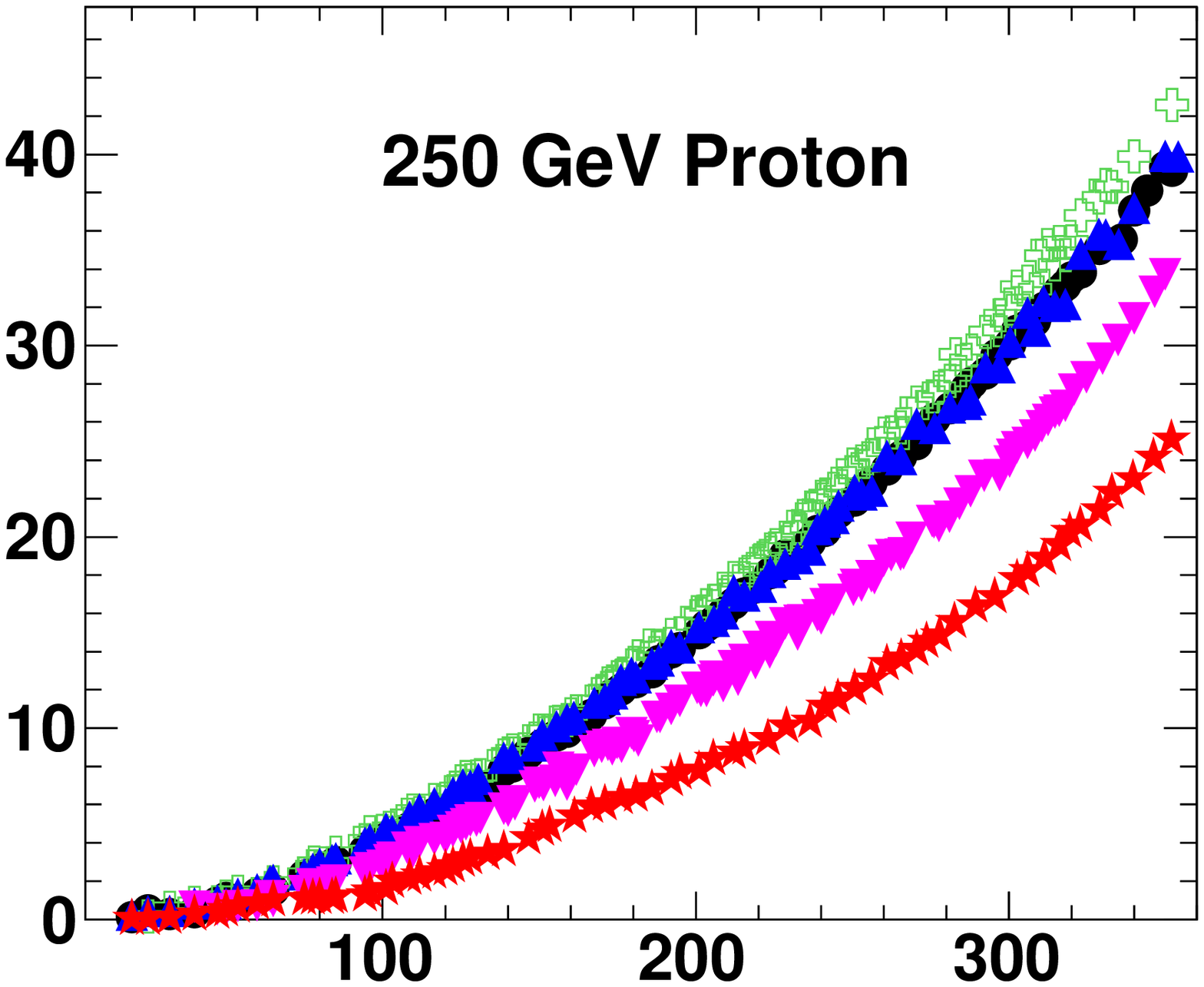} \hspace{-3mm}
\includegraphics[scale = 0.27]{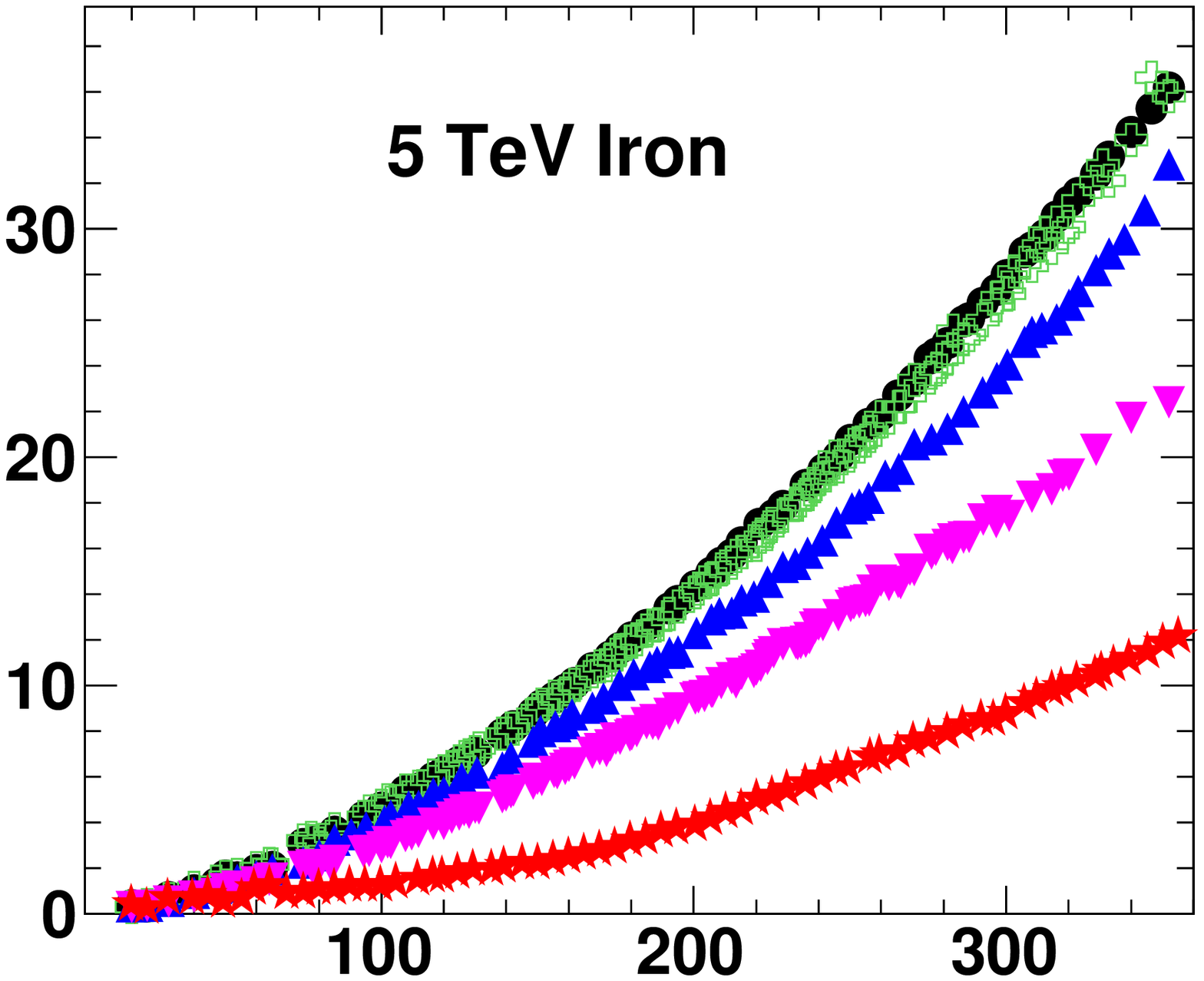}}

\vspace{-3mm}
\centerline{
\includegraphics[scale = 0.27]{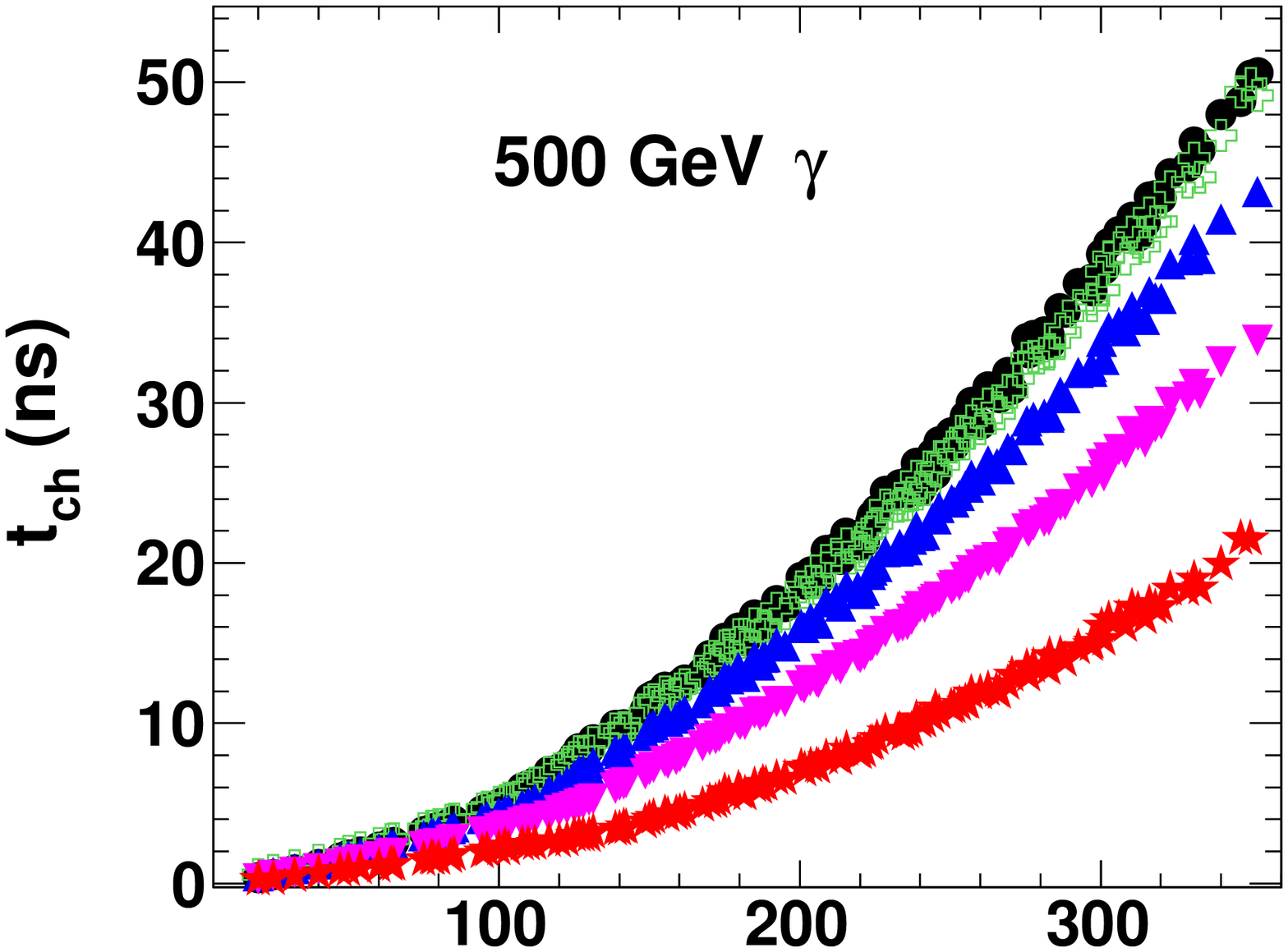} \hspace{-3mm}
\includegraphics[scale = 0.27]{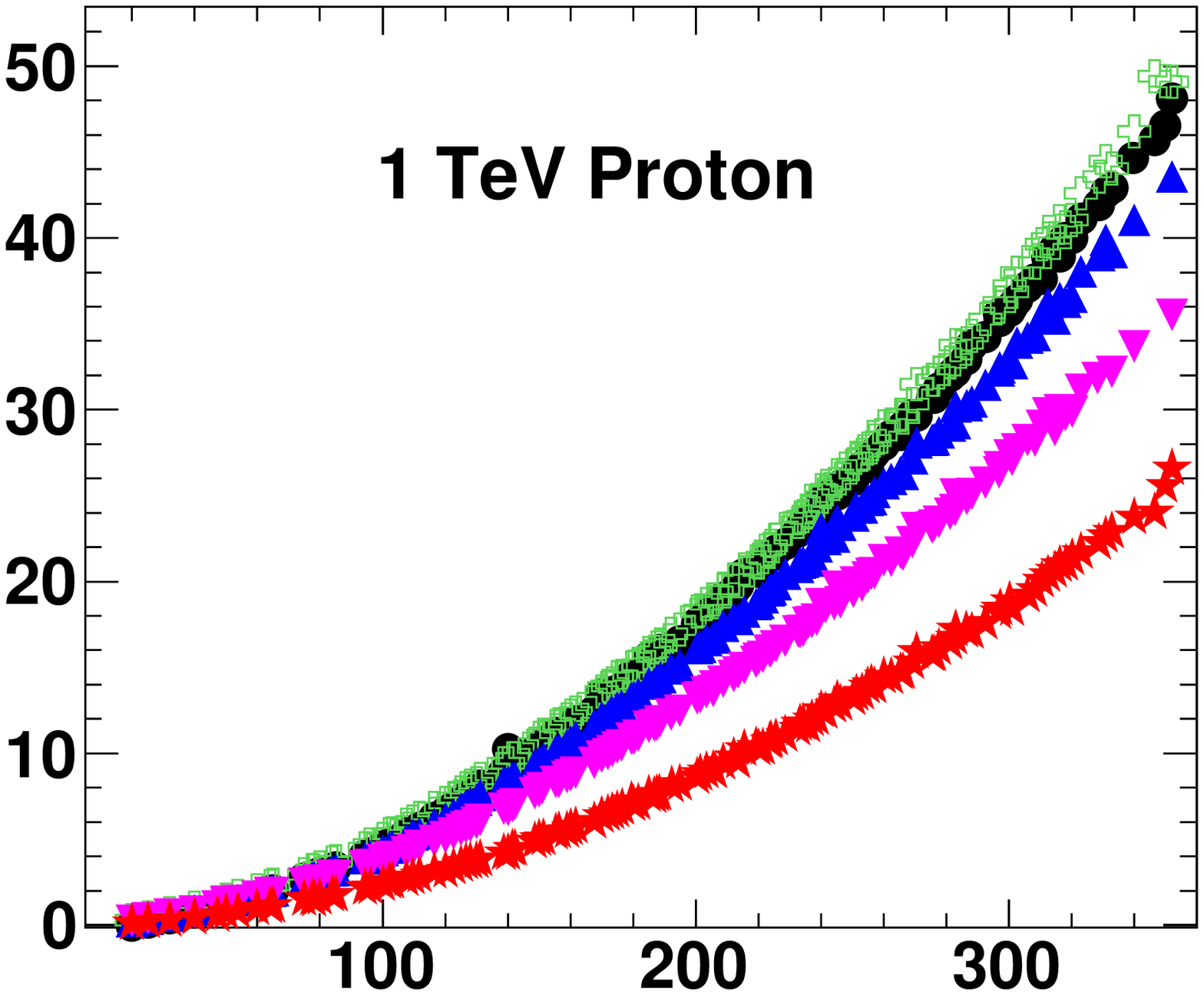} \hspace{-3mm}
\includegraphics[scale = 0.27]{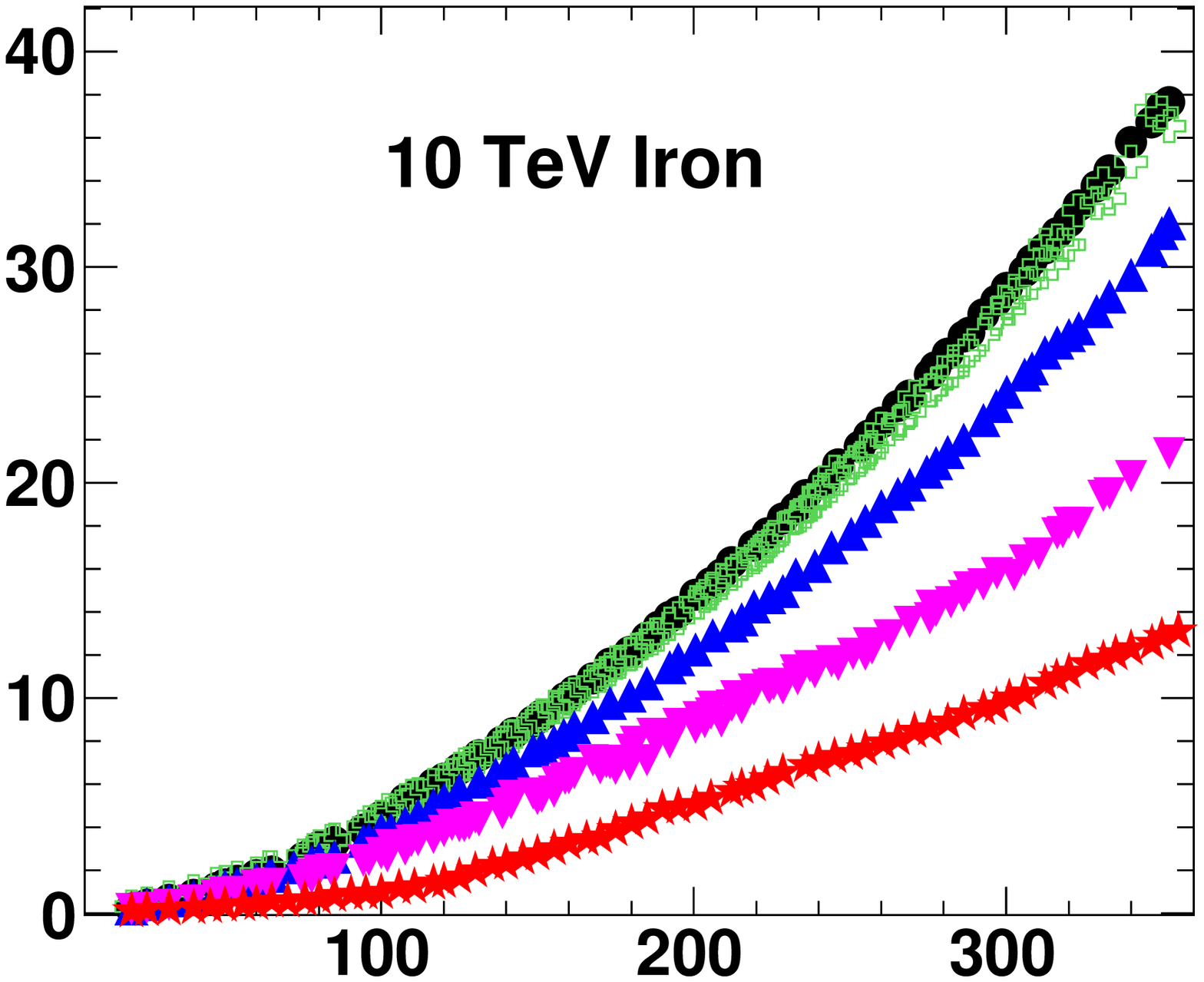}}

\vspace{-3mm}
\centerline{
\includegraphics[scale = 0.27]{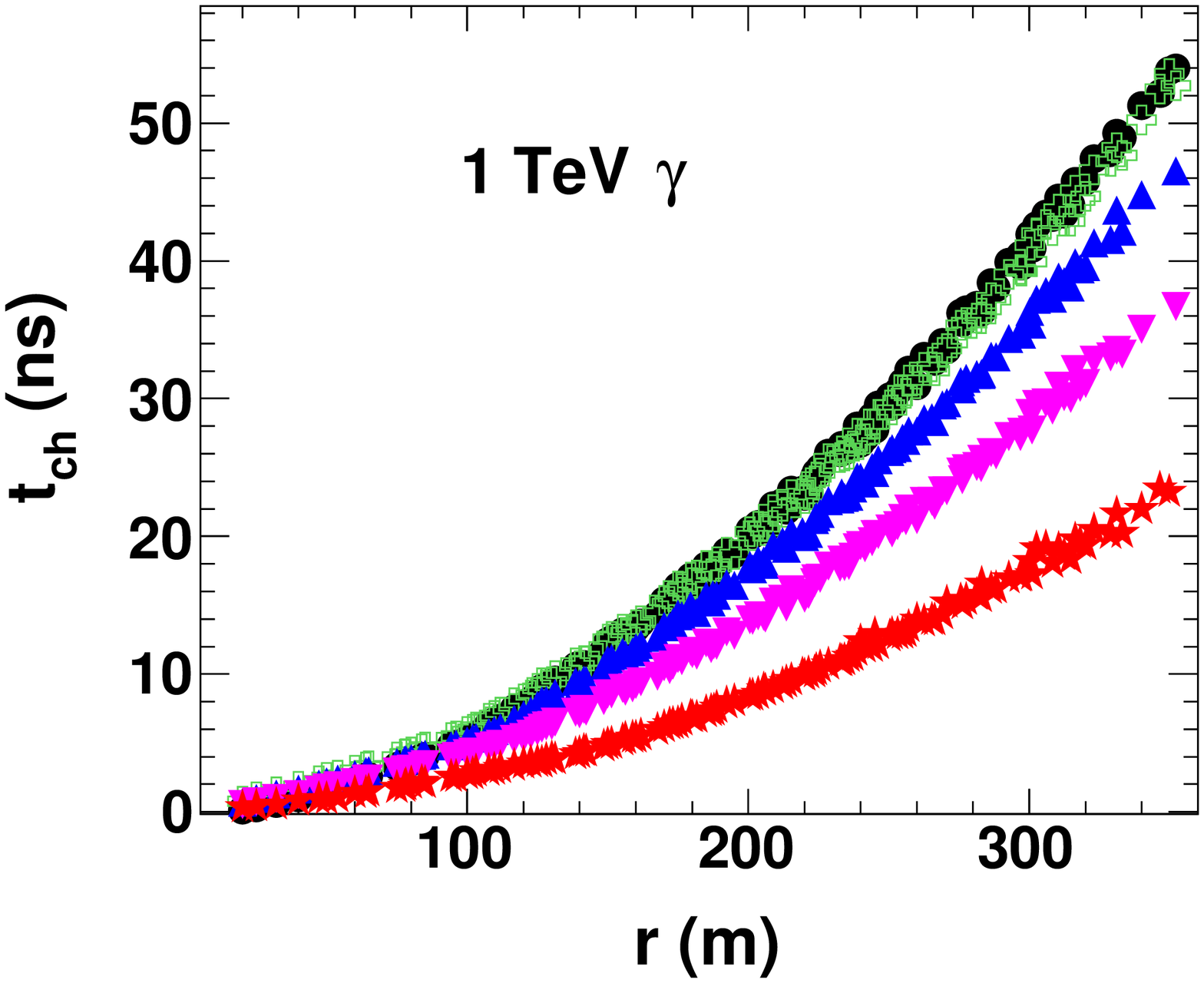} \hspace{-3mm}
\includegraphics[scale = 0.27]{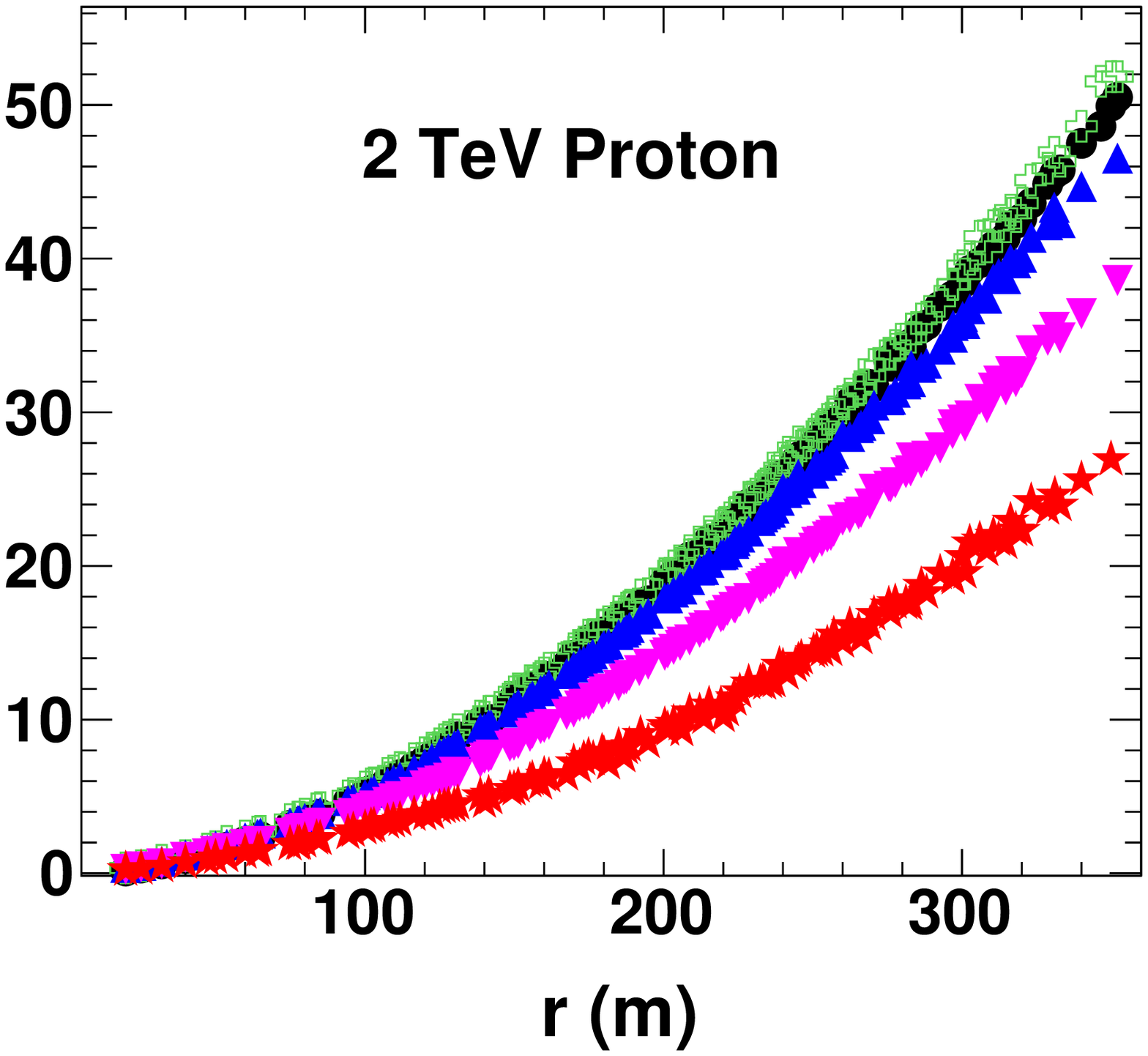} \hspace{-3mm}
\includegraphics[scale = 0.27]{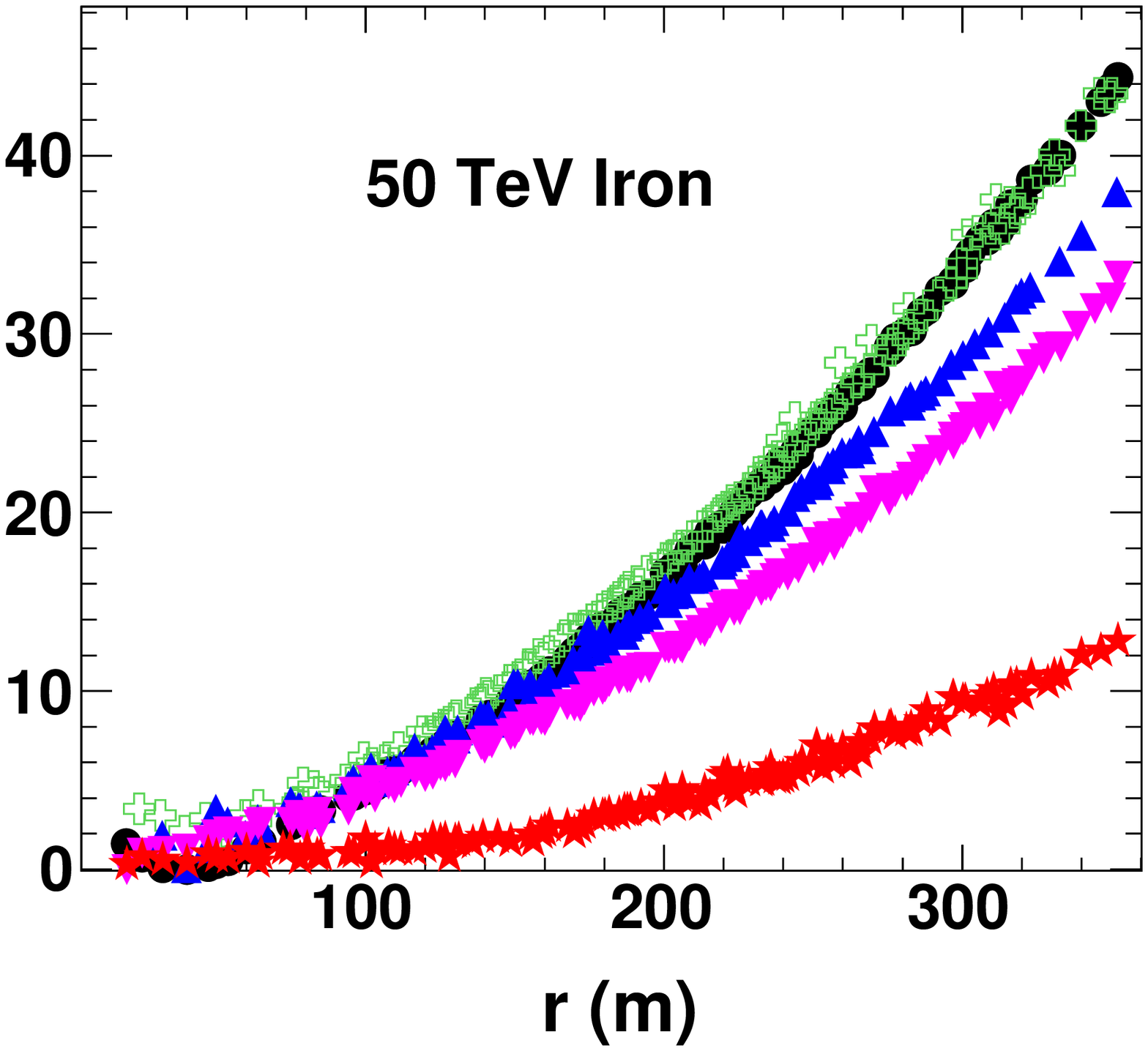}}


\caption{Average arrival time of Cherenkov photon ($t_{ch}$) as a function of
the shower core distance for different primaries at different zenith angle 
corresponding to some fixed value of energy.}
\label{fig7}
\end{figure*}

The average arrival time of Cherenkov photons ($t_{ch}$) as a 
function of core distance has been studied from two perspectives: one with 
a fixed zenith angle and variable energy, while the other with a fixed 
energy and five different zenith angles. Some of the results are 
shown in the Fig.\ref{fig6} and  Fig.\ref{fig7} respectively. For all primary 
particles, energies and zenith angles, the Cherenkov light front is 
found to be nearly spherical in shape. But, near the shower core ($\le$ 100 m), 
$t_{ch}$ increases slowly. The increase of $t_{ch}$ appears to gradually 
faster with a inconsistent rate as we go away from the core. Because of such 
inconsistent nature of rate of increase, 
the distribution deviates slightly from the shape of spherical symmetry. 
It is clear from the Fig.\ref{fig6} that for all values of 
zenith angle, the value of  $t_{ch}$ increases with increasing energy of the 
primary.
Fig.\ref{fig7} shows that with the increasing angle of incidence 
the value of $t_{ch}$ decreases for all combinations of energy and 
primary. The dependence of $t_{ch}$ with the energy and zenith angle of a 
primary particle can be related with the distance of the detector array from 
the shower maximum of the particle. With increasing energy of a primary 
particle this distance decreases (slant depth of the shower maximum increases),
whereas it increases (slant depth of the shower maximum decreases) with 
increasing zenith angle of the particle (see the 
Table \ref{tab2}). Hence above mentioned behaviours of dependence of $t_{ch}$ 
with the energy and zenith angle of a primary particle are due to respective 
increase and decrease of number of charged particles of the shower over a 
detector. Increasing number of charged particles give increasing time spread 
of Cherenkov photons produced by a such particles over the detector 
array and vise verse. Consequently the average arrival time of photons will be 
increased and decreased. Moreover, it is also evident from these figures that for vertically incident as well for inclined showers 
the arrival time follows very similar pattern.
It is found that for all combinations of energy and zenith angle 
irrespective of the primary particle, the variation of $t_{ch}$ can be 
represented by an equation of the form \cite{Hazarika}:
\begin{equation}
t_{ch}(r) = t_{0}e^{\Gamma/r^{\lambda}},
\label{eq3}
\end{equation} 
where $t_{ch}(r)$ represents the mean arrival time of Cherenkov photons as a 
function
of position, $r$ is the shower core distance. $t_{0}$, $\Gamma$ and $\lambda$ 
are constant parameters. For a given zenith angle, the values of these 
parameters depend on the type and energy of the primary particle. In the 
Fig.\ref{fig6} the best fit functions are shown as solid lines for one of the 
energy corresponding to the different zenith angle. The method used for 
the fitting is same as that used for the density distributions. 

\begin{figure*}[hbt]
\centerline
\centerline{
\includegraphics[scale = 0.27]{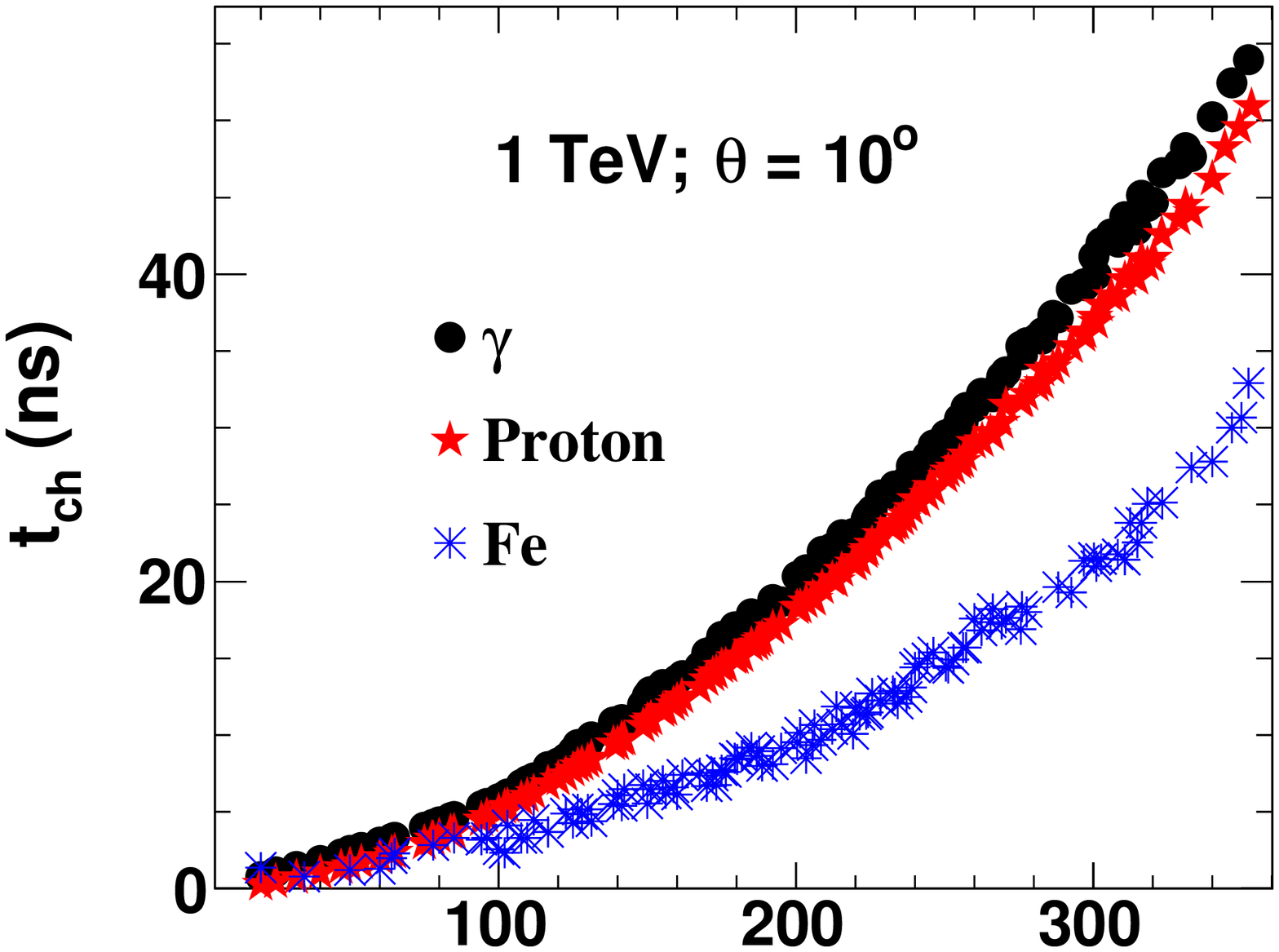} \hspace{-3mm}
\includegraphics[scale = 0.27]{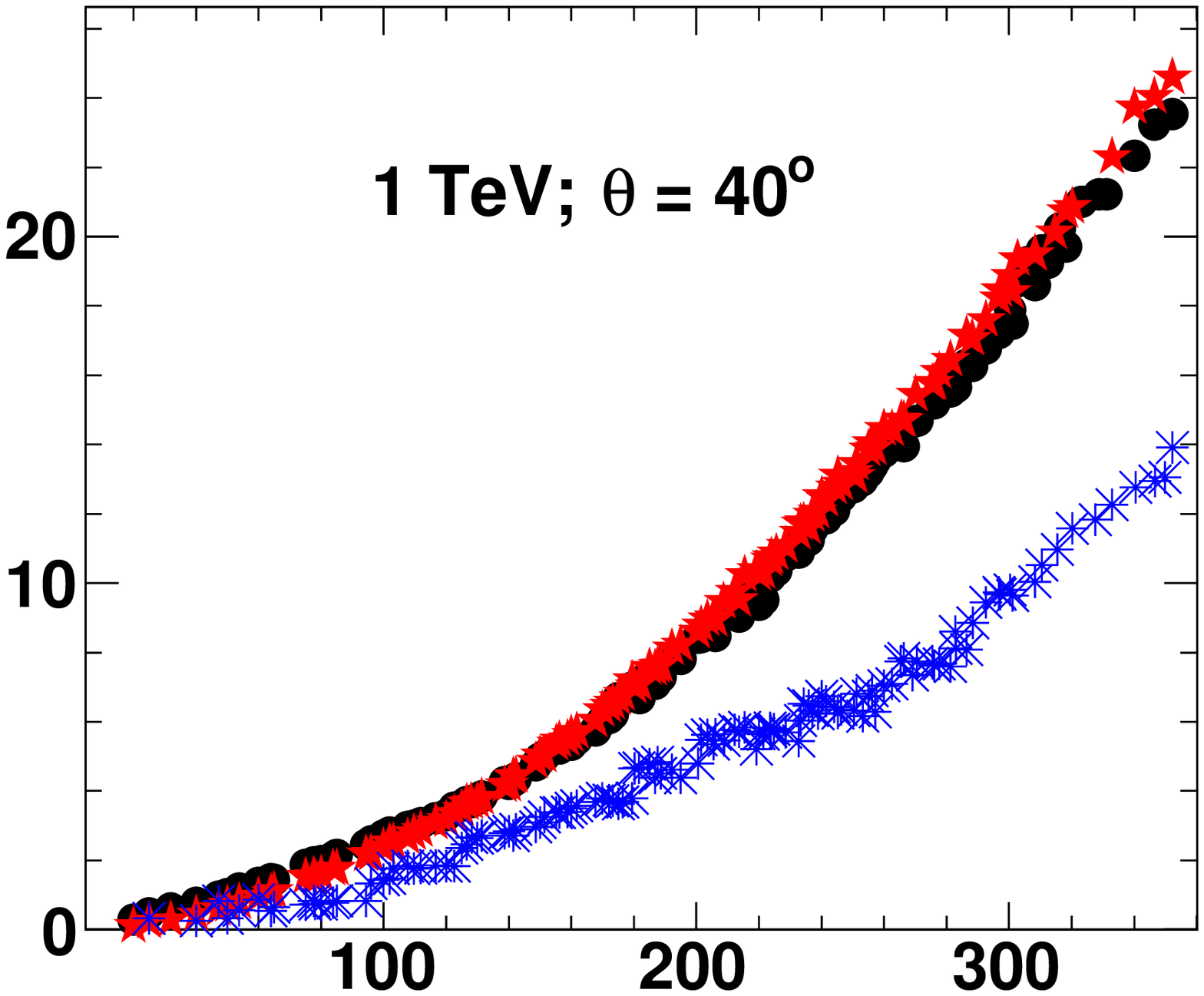}}
\vspace{-3mm}
\centerline{
\includegraphics[scale = 0.27]{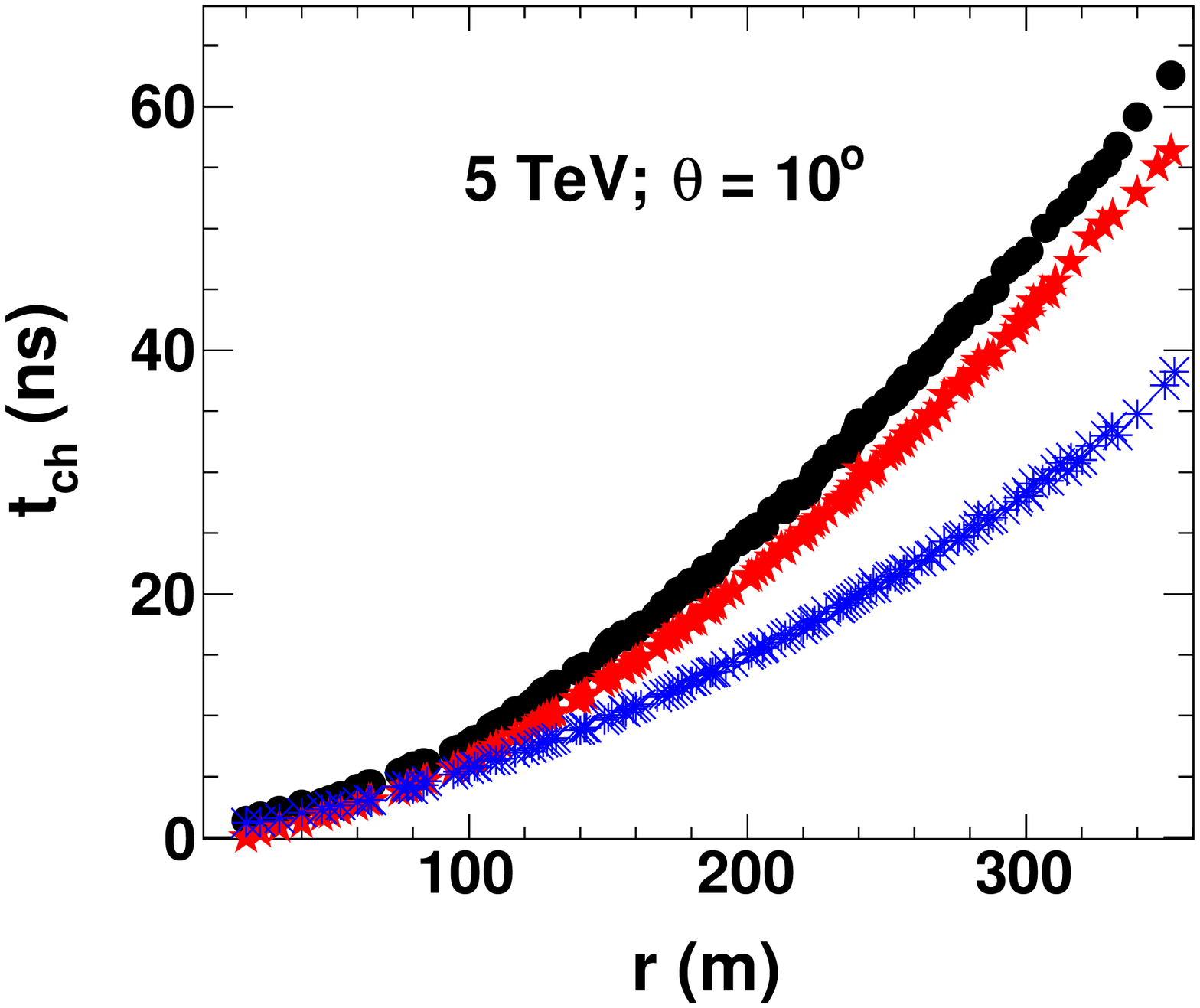} \hspace{-3mm}
\includegraphics[scale = 0.27]{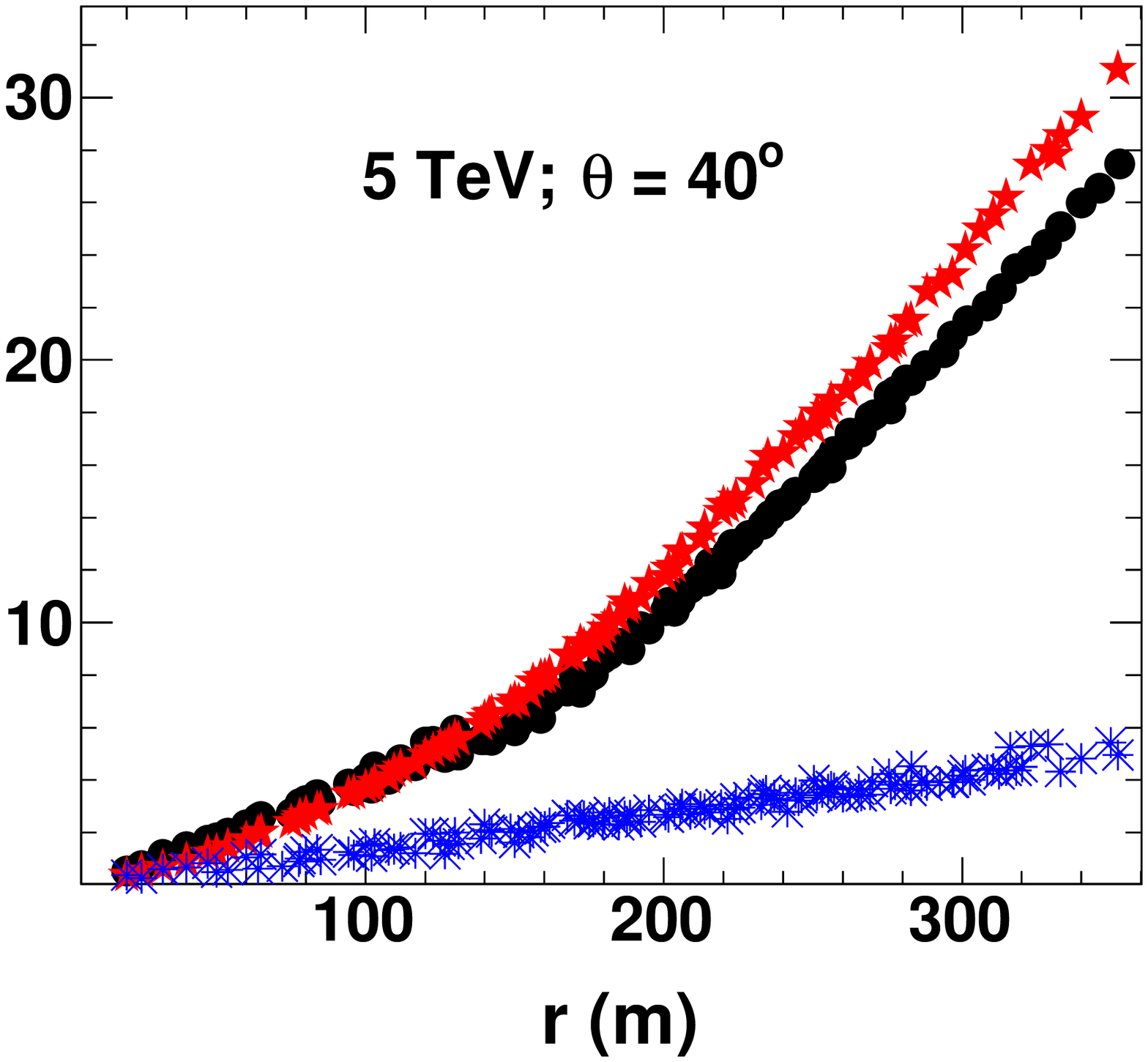}}
\caption{Variation of $t_{ch}$ with respect to the core distance of the showers
of $\gamma$, proton and iron primaries of the same energy.}
\label{fig9}
\end{figure*}

\begin{figure*}[hbt]
\centerline
\centerline{
\includegraphics[scale=0.26]{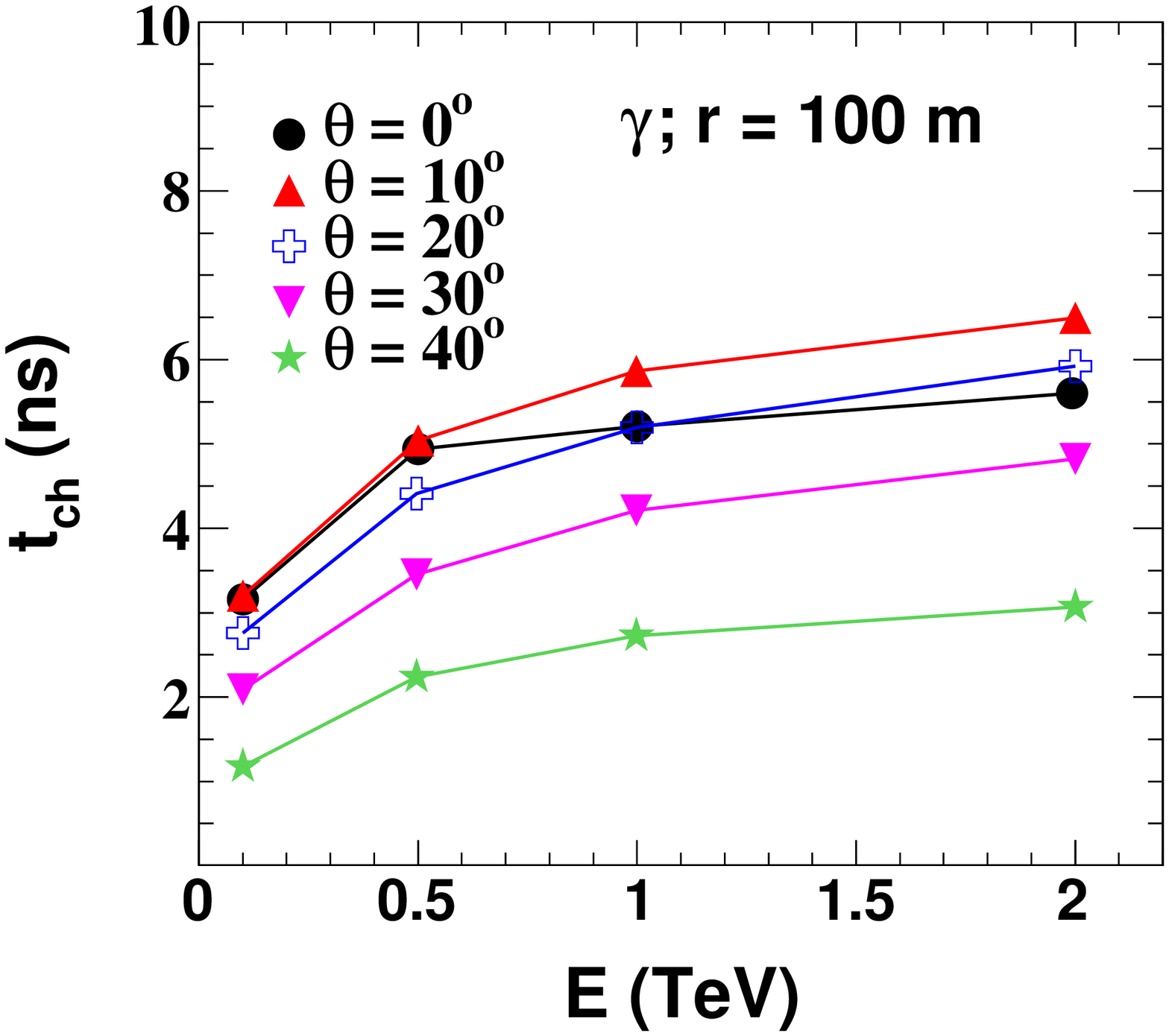}\hspace{5mm}
\includegraphics[scale=0.26]{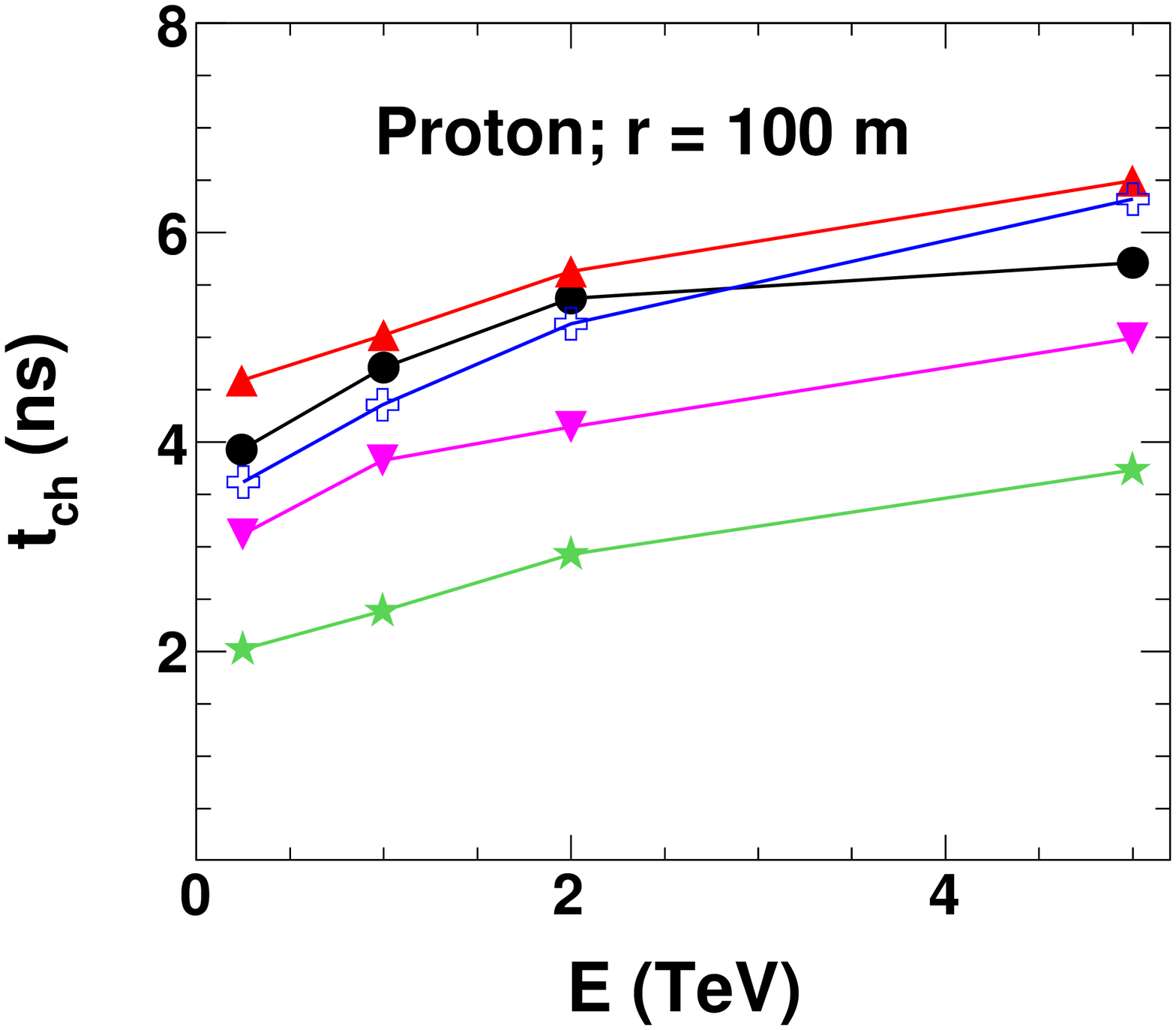}\hspace{5mm}
\includegraphics[scale=0.26]{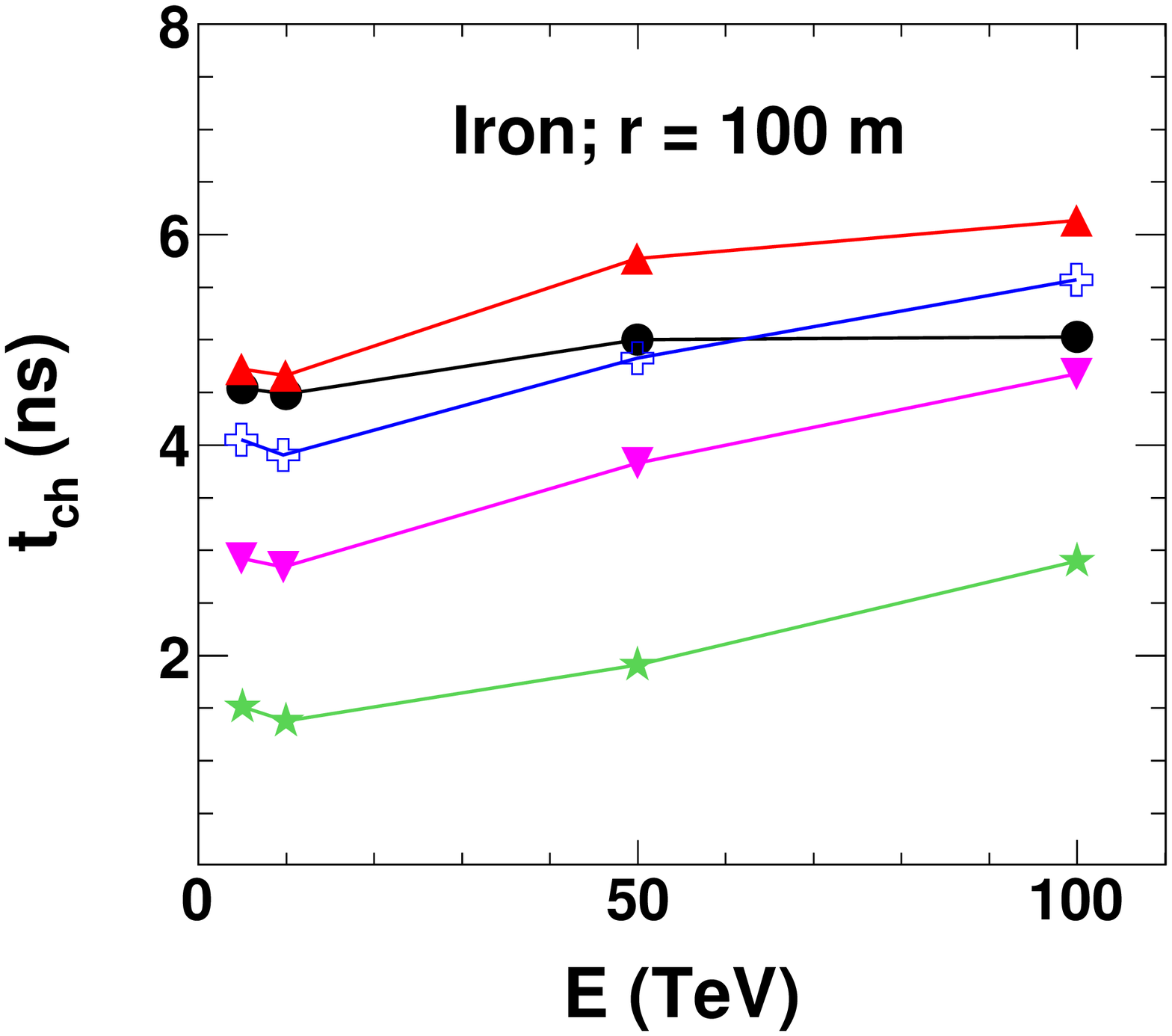}}

\caption{Variation of $t_{ch}$ as a function of energy of primary at 100 m
distance from the shower core of $\gamma$, proton and iron
primaries.}
\label{fig11}
\end{figure*}

\subsubsection{Primary particle and energy dependence}
In Fig.\ref{fig9} the average arrival time of Cherenkov photons ($t_{ch}$) 
with respect to the distance from the
shower core is plotted for the three primaries of same energy (in each plot)
inclined at
10$^{\circ}$ and 40$^{\circ}$. It is seen that almost in all the cases 
$t_{ch}$ is 
higher over all core distances for the $\gamma$-ray primary than that for the 
iron primary, but comparable with that for the proton primary. Hence in general 
the thickness of Cherenkov photons' light-pool 
initiated by a $\gamma$-ray primary is wider on average than 
that from the iron primary and is comparable with that for the proton primary
of same energy. $\gamma$-ray and proton
primaries show very similar pattern, but the iron primary initiated photons
distinctly have flatter average arrival time. 

At a given energy and zenith angle the shower maximum of the iron primary is 
produced at a 
significantly higher altitude in comparison to $\gamma$-ray and proton 
primaries. But its value is comparable for the $\gamma$-ray and proton 
primaries (see the table \ref{tab2}). Moreover, the nature of production of 
shower is different 
for all these three primaries as stated earlier.   
So, at a given energy and zenith angle, the $\gamma$-ray initiated shower 
takes distinctly longer average time (because of wider time spread) than the 
iron initiated shower and 
slightly different average time than the shower of proton primary to reach 
the observation 
level. This is due to the reasons explained earlier related with the shower 
maximum and the nature of shower production.

In Fig.\ref{fig11} we have shown the variation of $t_{ch}$ at 100 m
distance from the shower core with respect to energy of the $\gamma$-ray,
proton and iron primaries incident at various zenith angles. As stated earlier
$t_{ch}$ increases with increasing energy of all primary particles due to the
decreasing altitude of the position of shower maximum with increasing energy. 
This increase of $t_{ch}$ is non-linear for the $\gamma$-ray (highly) and 
proton initiated showers, but almost linear for the iron initiated showers. 
The pattern of increase of $t_{ch}$ for the vertically incident shower of all 
primaries is almost similar with a slight variation due to the nature of shower 
production for the primary type.  


\subsubsection{Photon arrival time fluctuations}
We have calculated the ratio of r.m.s. to mean of the photon arrival time 
($\sigma_{pm}$) for all three primaries 
having different energy and zenith angle combination, like we did for 
the density fluctuations. These are plotted in the Fig.\ref{fig6} to see 
the fluctuations in the $t_{ch}$ distributions. It is clearly visible that, 
for all showers irrespective of the primary particle, the energy and zenith 
angle, the fluctuation is large near the shower core 
but it decreases with increasing 
distance from the core. On average the proton primary shows maximum fluctuation 
(highest at 250 GeV) and the iron shows the least. With increasing zenith angle
the fluctuations 
increases further near the shower core
with same decreasing pattern with increasing distance. 
Furthermore, for different primaries $\sigma_{pm}$
behaves differently. For example, for both $\gamma$-ray and proton primaries 
its value decreases with increasing energy, almost for all core distances and
and for all zenith angles with slight deviations. But exactly such a behaviour 
can not be seen for the iron primary.        

\begin{figure*}[hbt]
\centerline
\centerline{
\includegraphics[width=5.6cm, height=4.55cm]{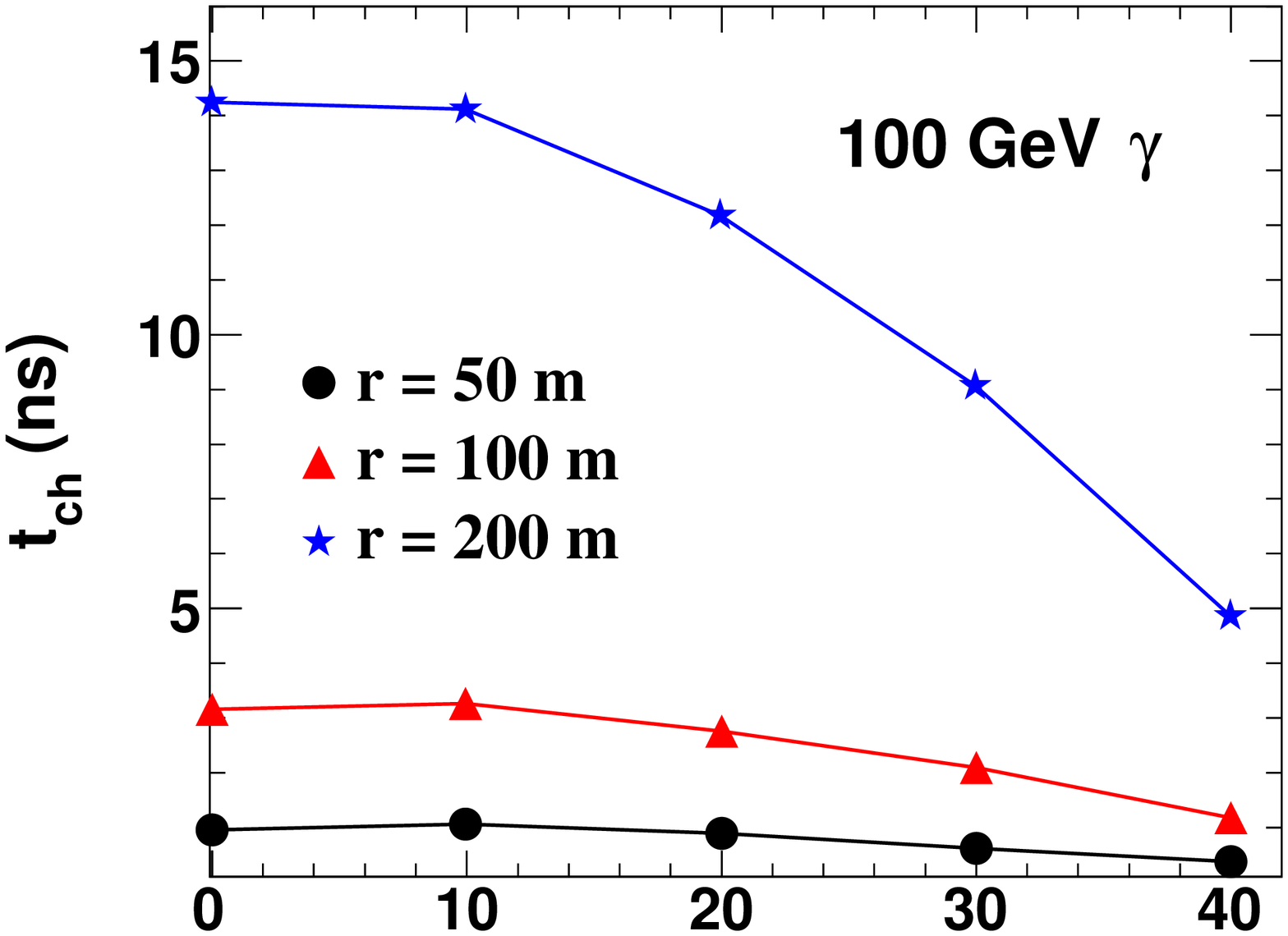}\hspace{-1.5mm}
\includegraphics[width=5.6cm, height=4.55cm]{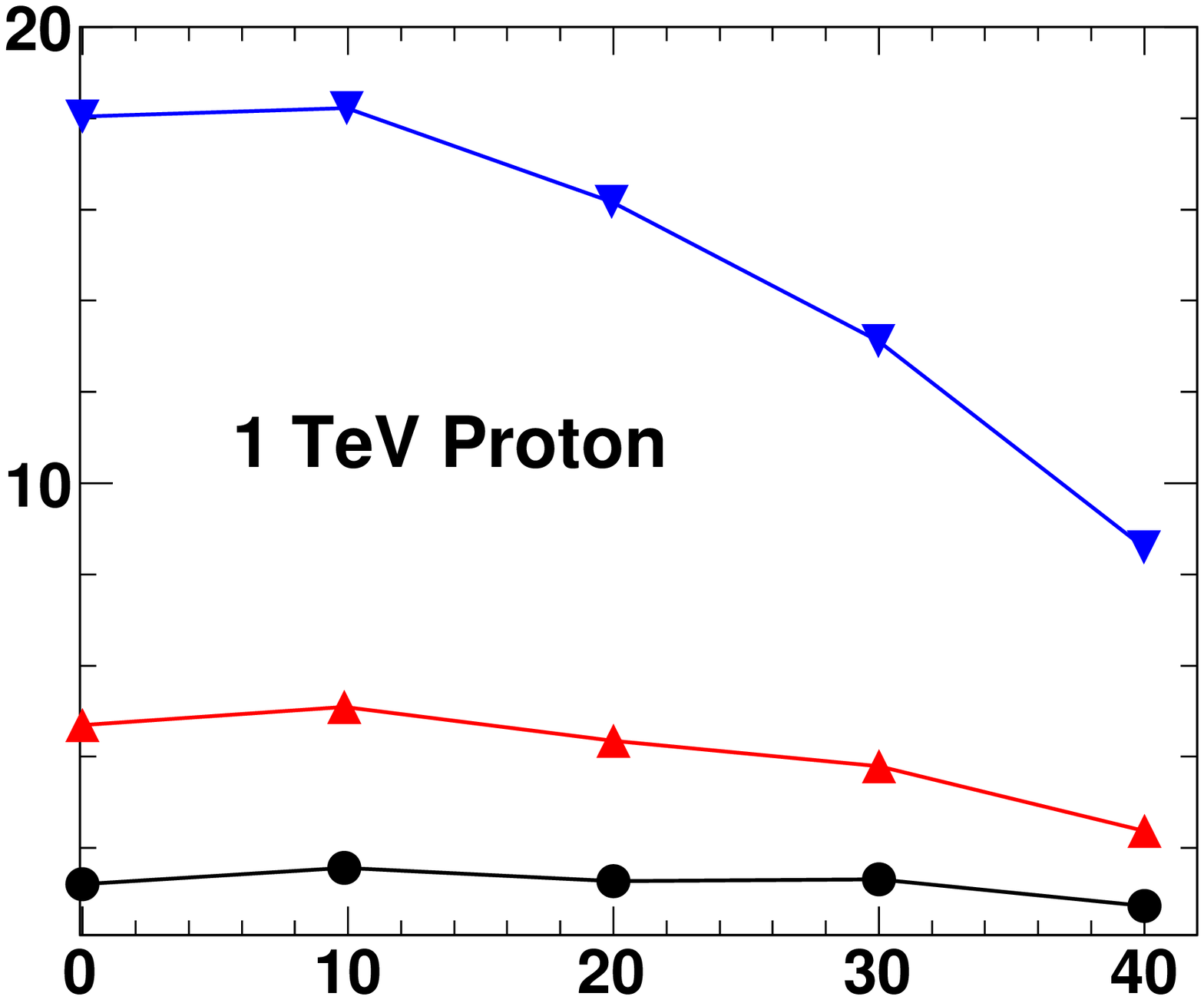}\hspace{-1.5mm}
\includegraphics[width=5.6cm, height=4.55cm]{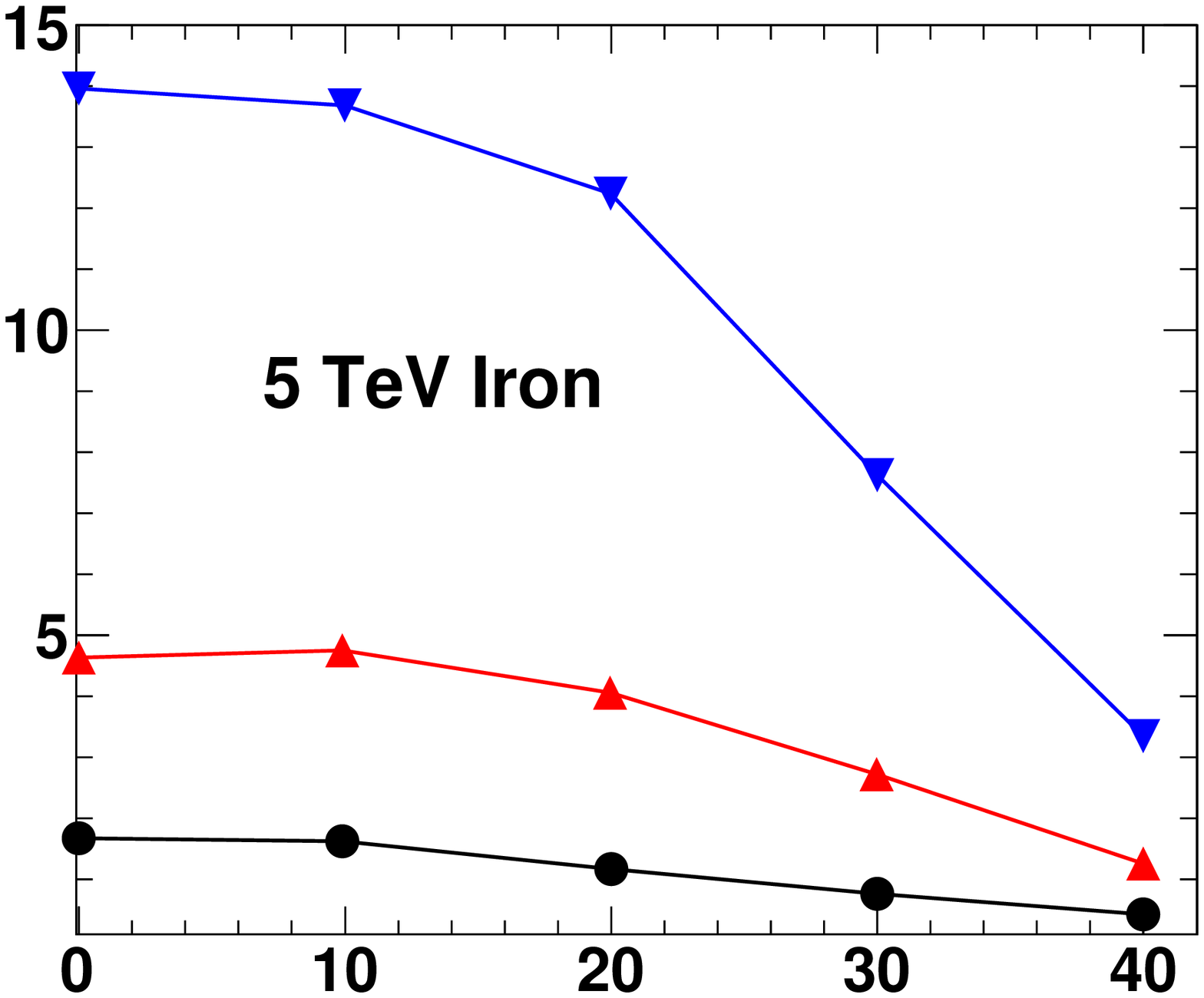}
}
\vspace{-2mm}
\centerline{
\includegraphics[width=5.6cm, height=4.55cm]{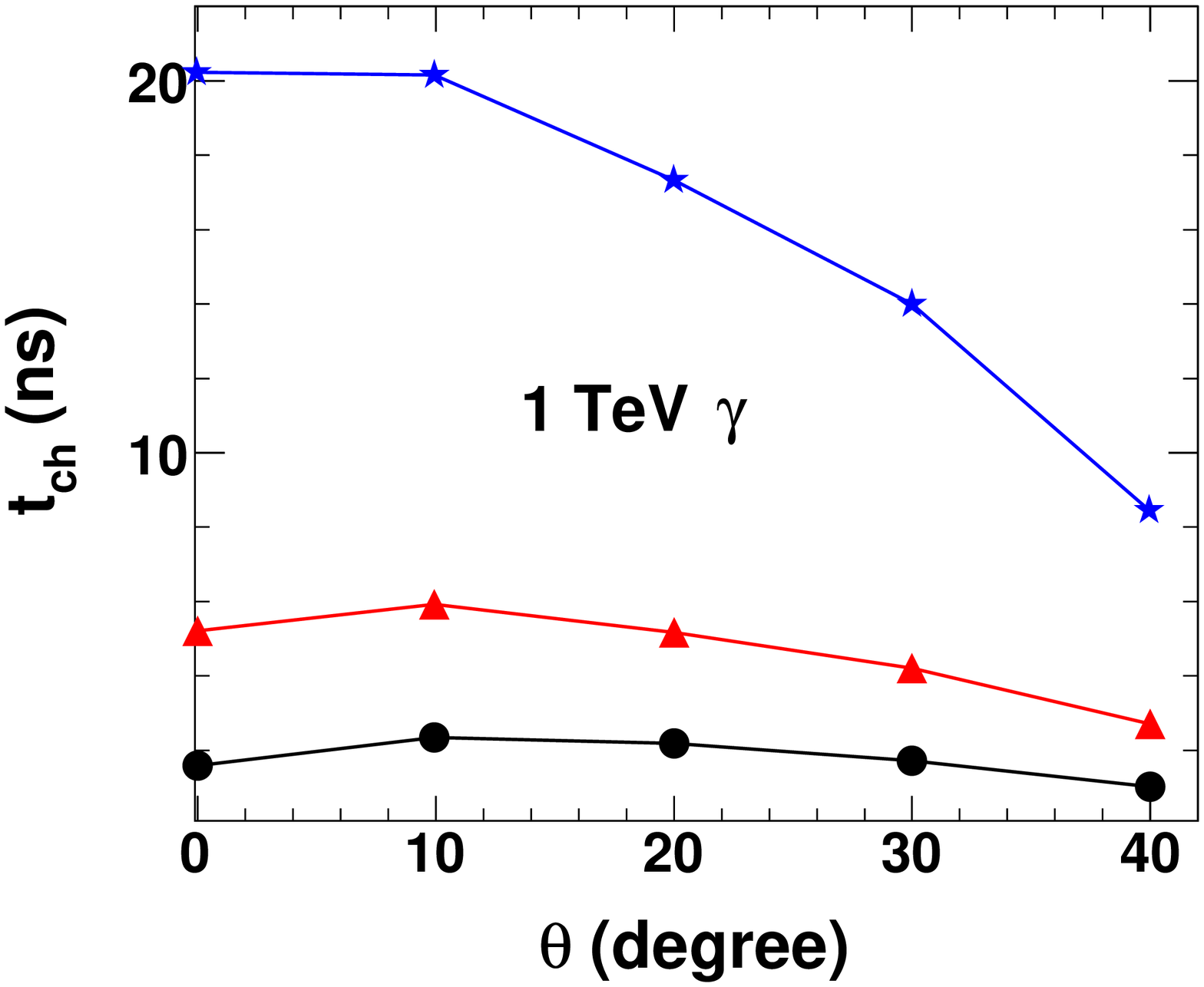}\hspace{-1.5mm}
\includegraphics[width=5.6cm, height=4.55cm]{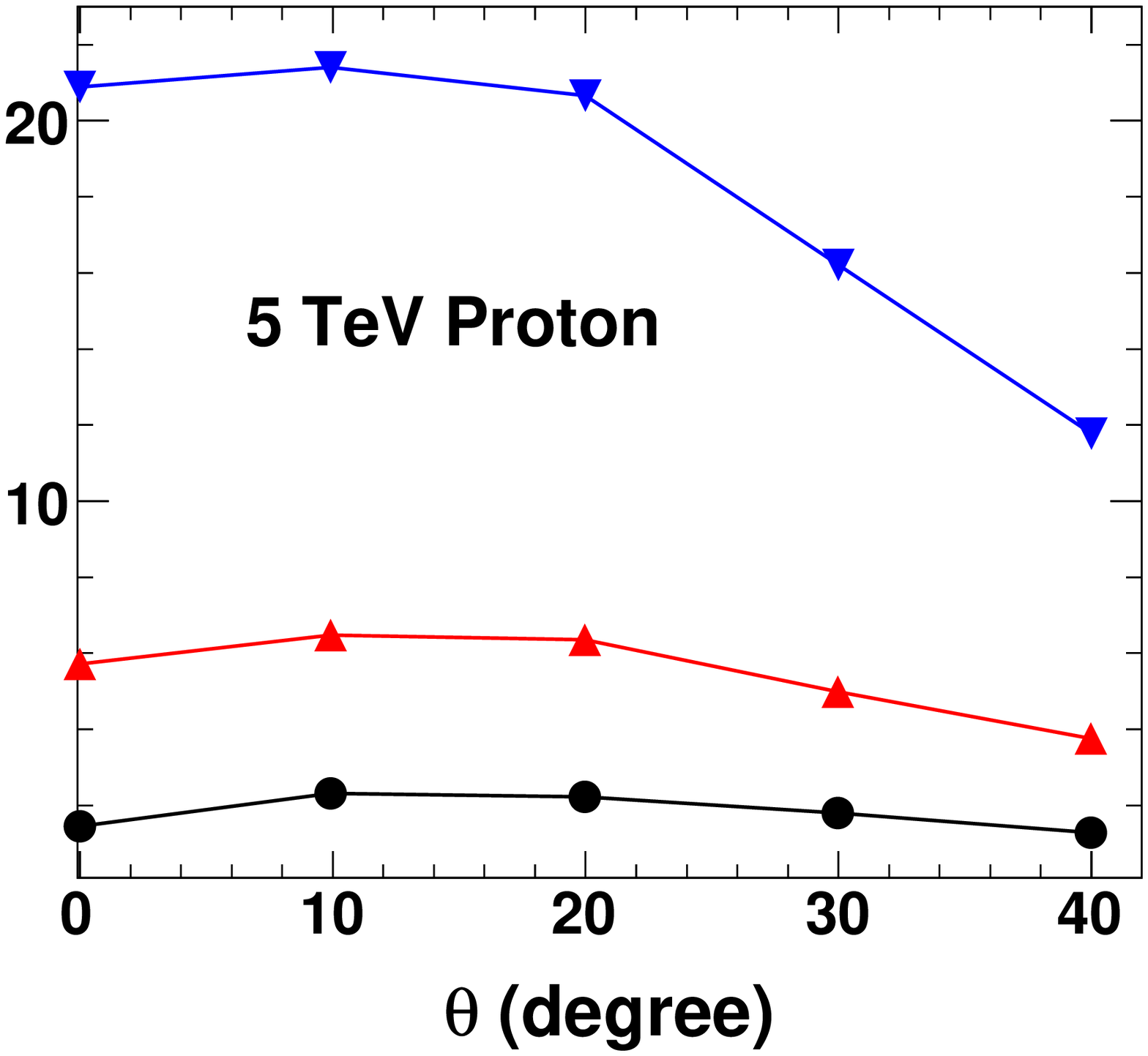} \hspace{-1.5mm}
\includegraphics[width=5.6cm, height=4.55cm]{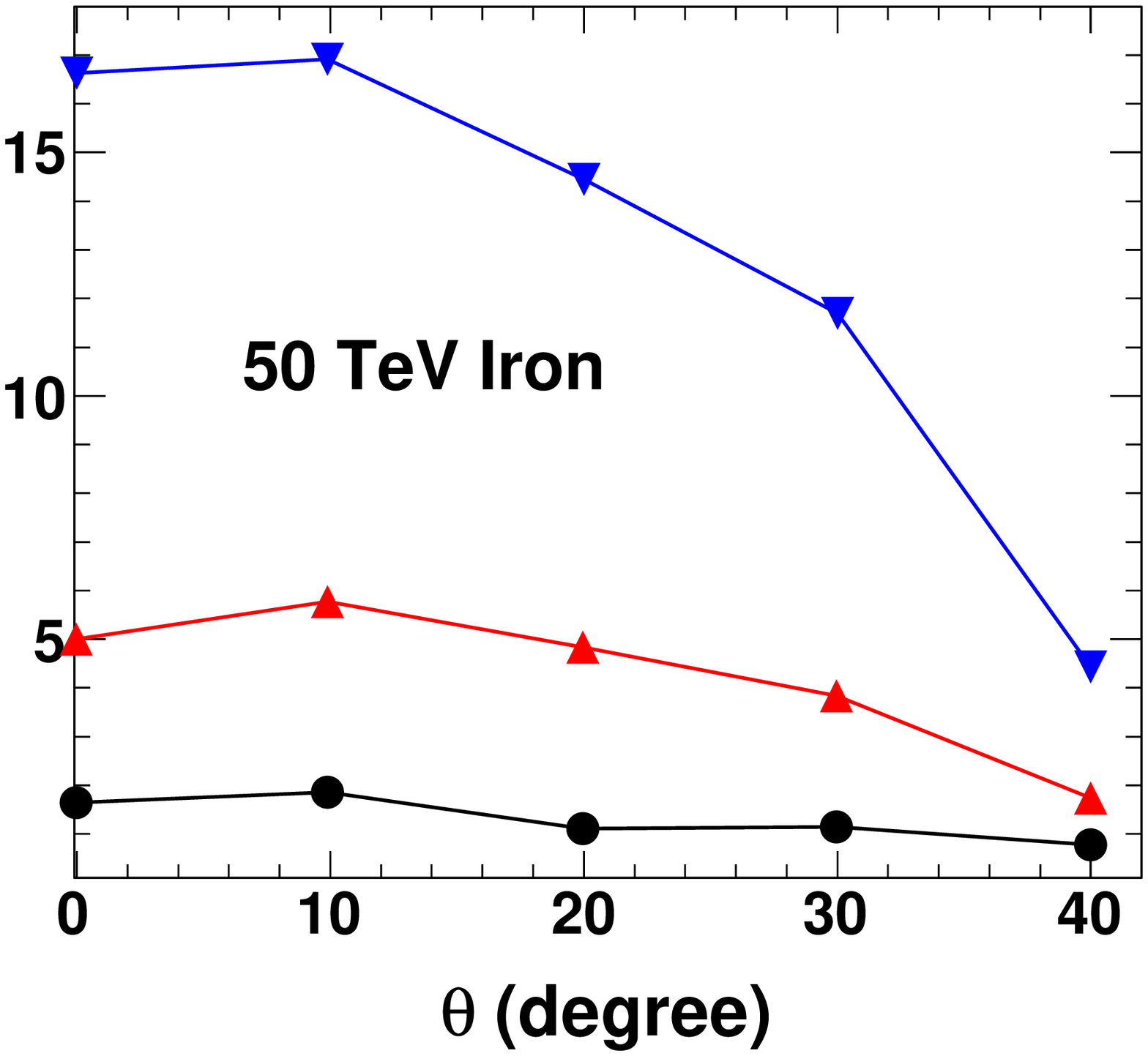}}

\caption{Variation of $t_{ch}$ with respect to zenith angle at 50 m,
100 m and 200 m from the core of the showers of $\gamma$, proton and iron
primaries.}
\label{fig10}
\end{figure*}

\subsubsection{Zenith angle dependence}
The variation of $t_{ch}$ with zenith angle for the $\gamma$-ray, 
proton and iron primaries are shown in Fig.\ref{fig10}. The figure shows only 
plots for two different energies of each primary at a core distance of 50 
m, 100 m and 200 m. It is observed that, there is an overall falling trend of 
$t_{ch}$ with respect to zenith angle for all primary particles, 
energies and at all core distances. This is due to the reason of position of
shower maximum in relation to zenith angle as discussed earlier. 
In the case of $\gamma$-ray primaries the trend is most smooth, whereas it is 
least smooth for the iron primaries. It is because the
air shower produced by a $\gamma$-ray primary is homogeneous in nature, on the
other hand air shower produced by an iron primary is most inhomogeneous.  
Moreover, with increasing distance from the core of shower, the rate of 
falling of the $t_{ch}$ increases with the zenith angle. This behaviour
is again due to the fact that with the increasing distance of a detector from 
the shower core, the distance to the shower maximum increases.

\begin{figure*}[hbt]
\centerline
\centerline{
\includegraphics[scale=0.28]{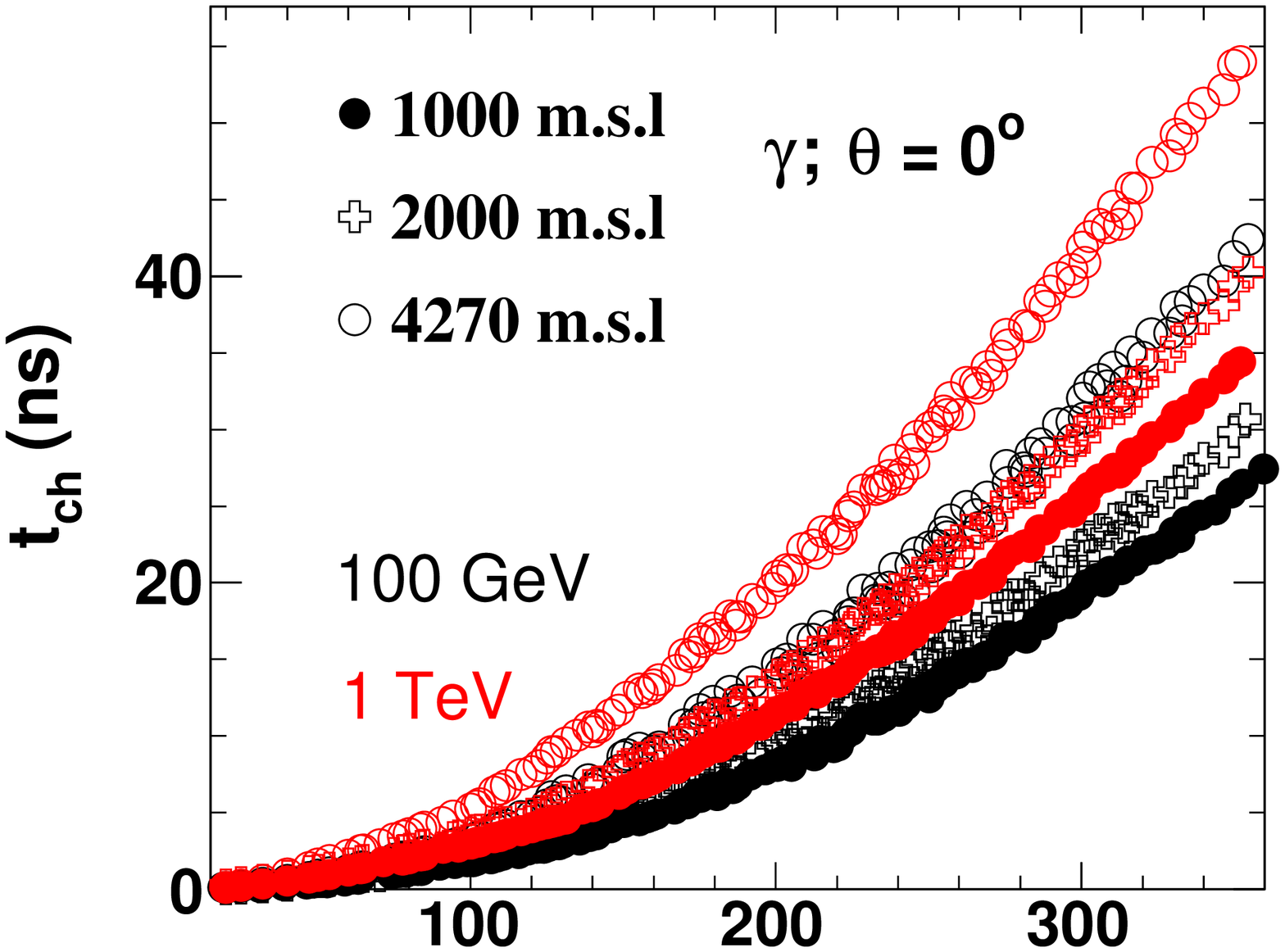}\hspace{-2mm}
\includegraphics[scale=0.28]{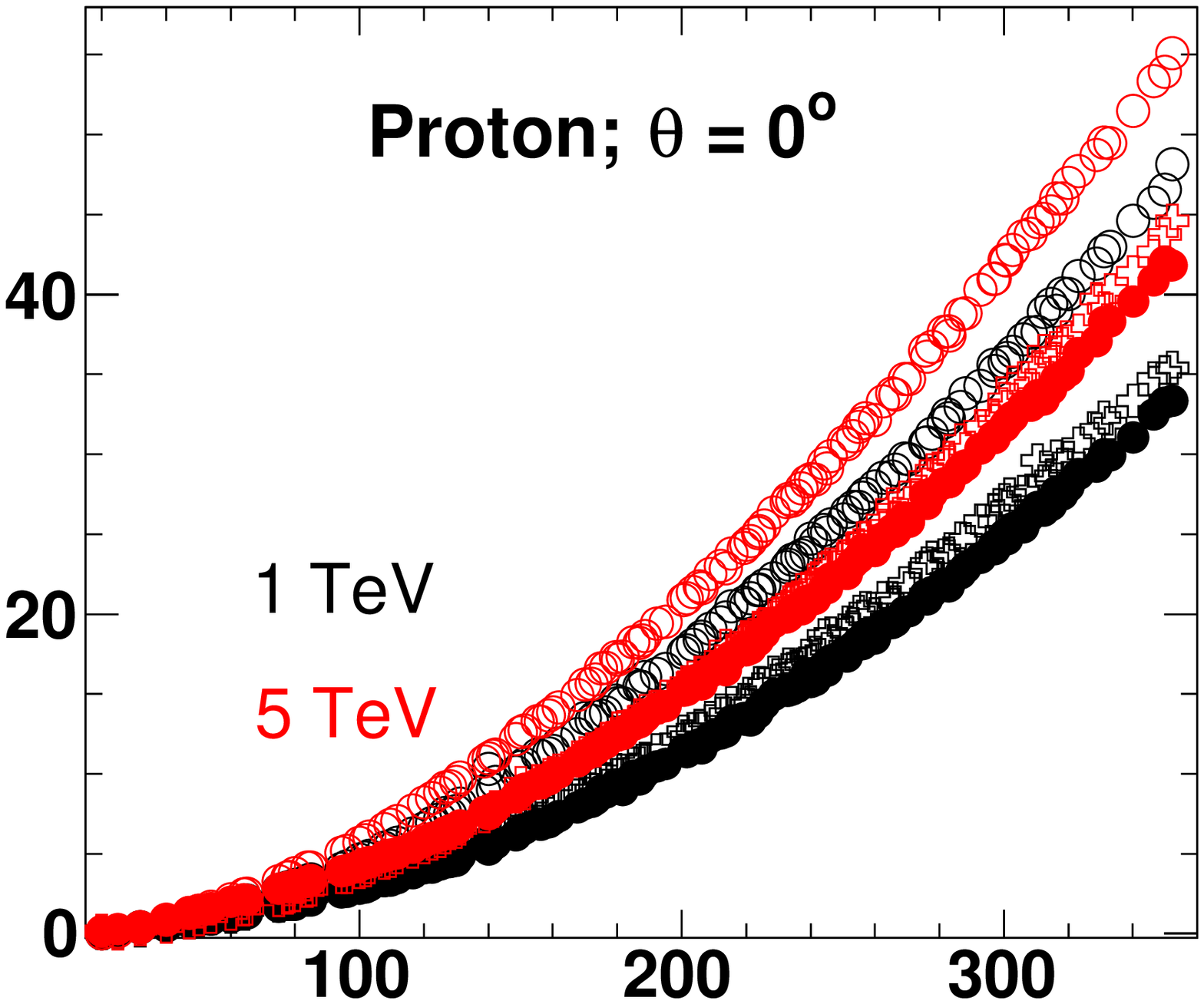} \hspace{-2mm}
\includegraphics[scale=0.28]{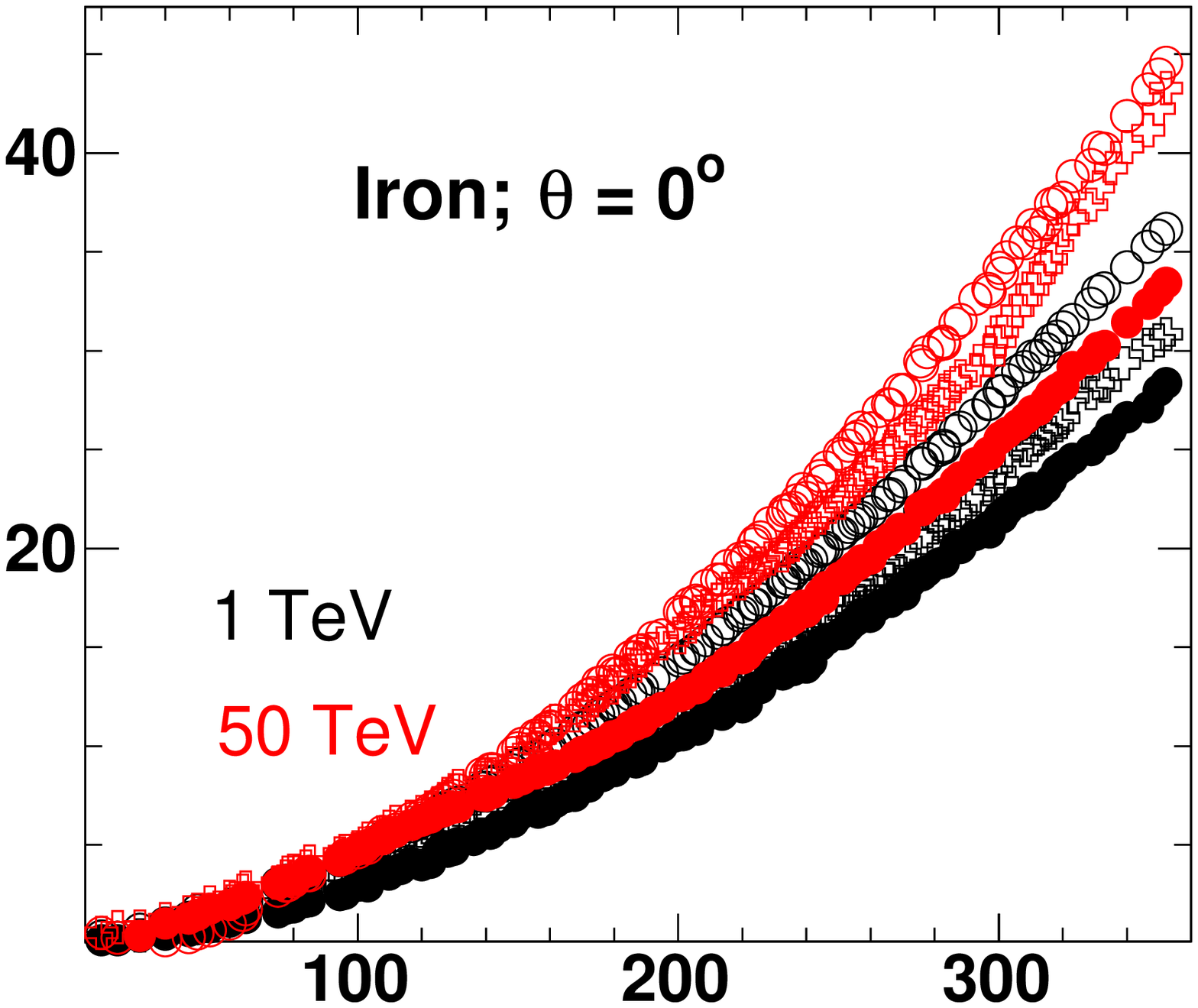}
}

\vspace{-2mm}
\centerline{\hspace{-1mm}
\includegraphics[scale=0.28]{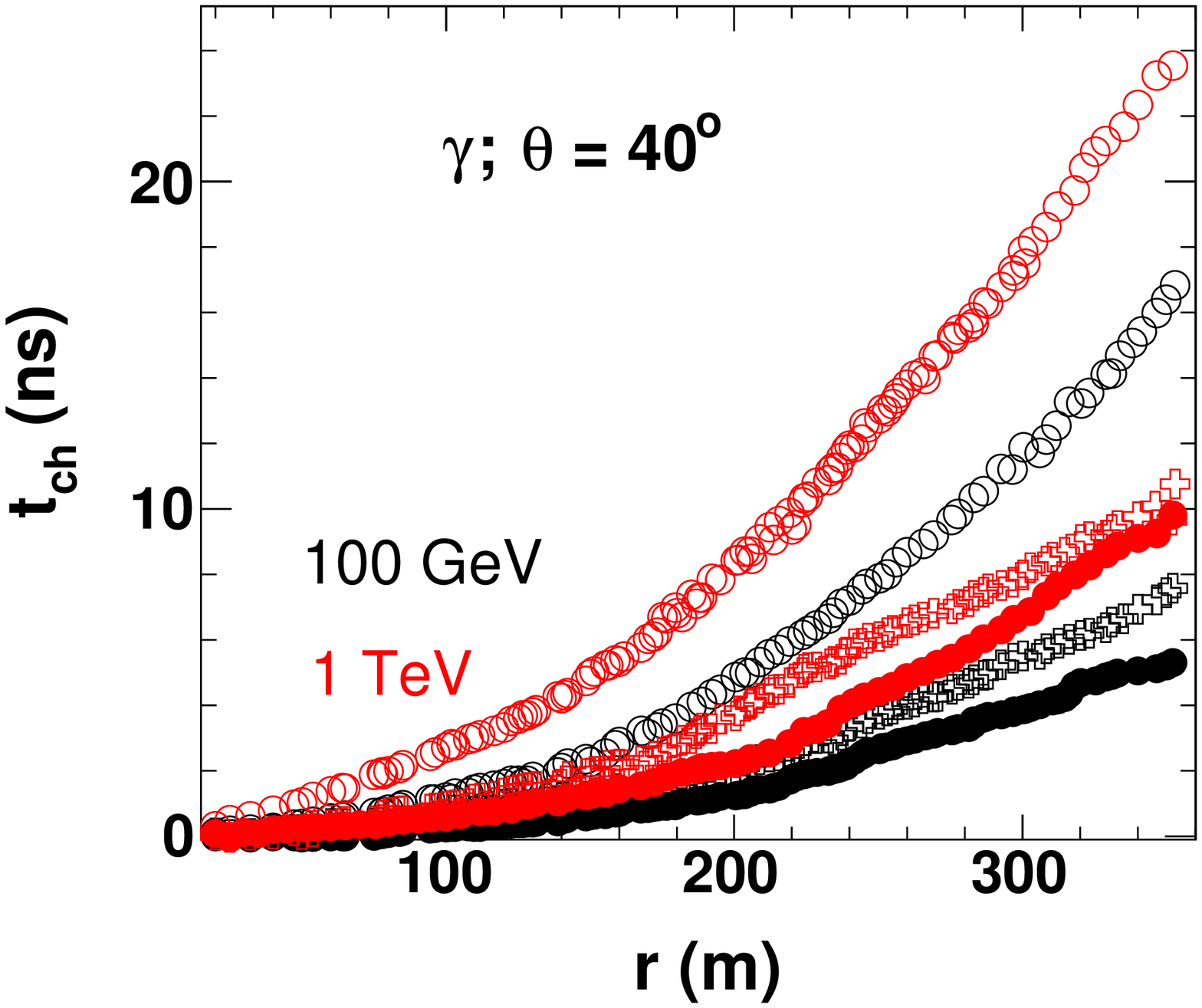} \hspace{-2mm}
\includegraphics[scale=0.28]{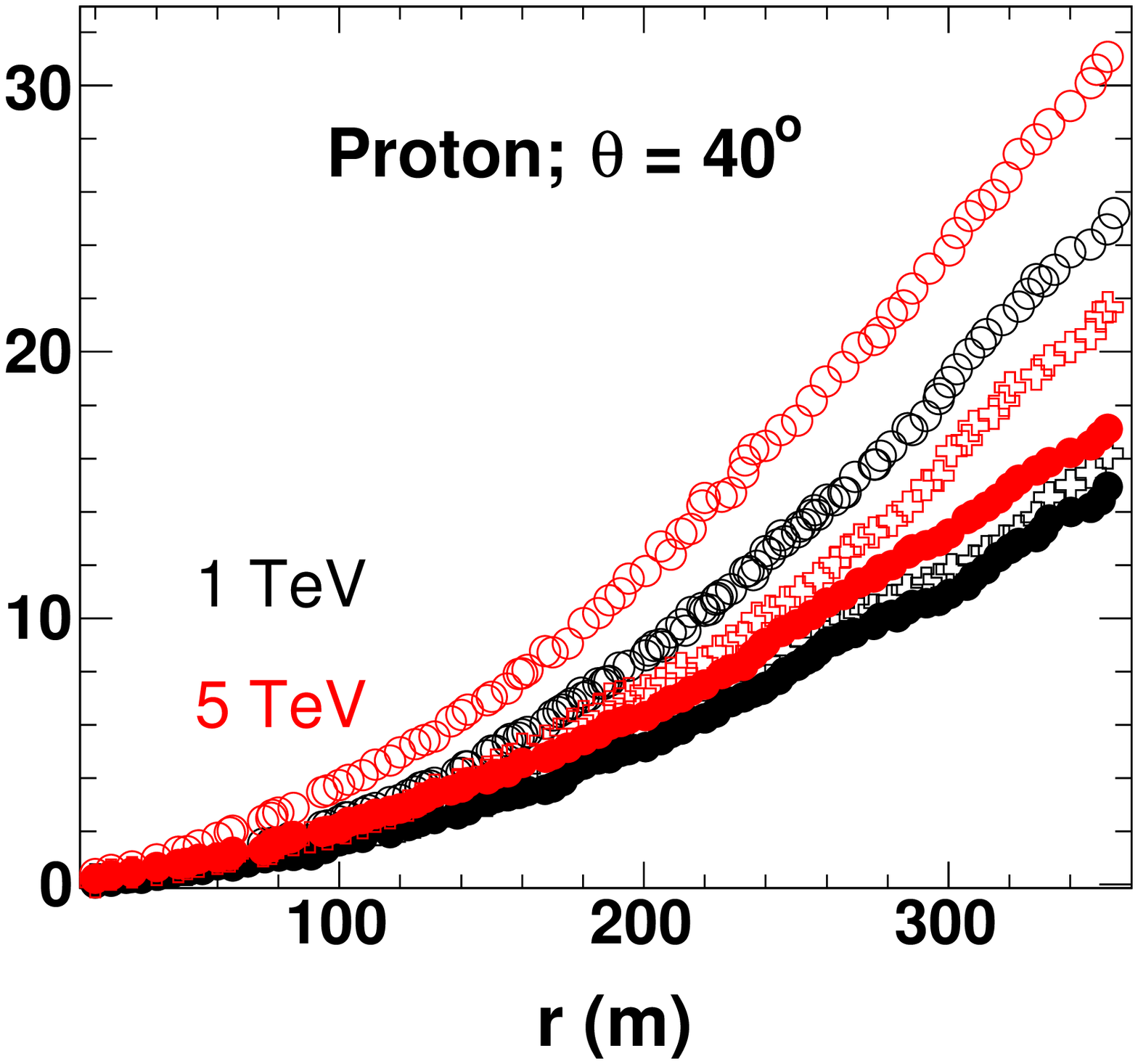}    \hspace{-2mm}
\includegraphics[scale=0.28]{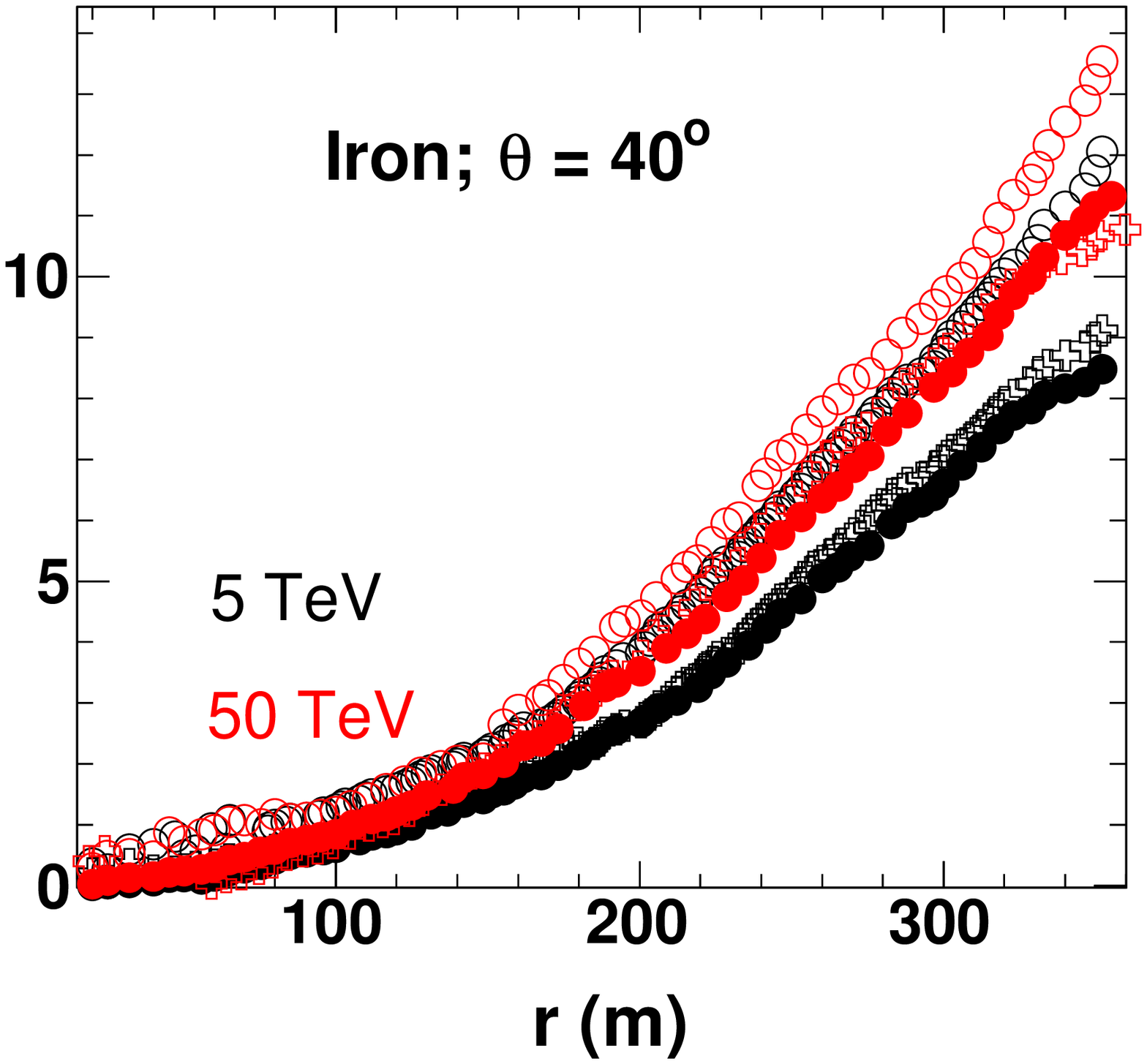}}

\caption{Variation of $t_{ch}$ with distance from shower core at three
different altitudes of observation level for
$\gamma$, proton and iron primary incident at 0$^{\circ}$ and 40$^{\circ}$
zenith angles. Low energy plots are indicated by the black colour, whereas the
high energy plots are represented by the red colour.}
\label{fig11a}
\end{figure*}

\subsubsection{Altitude dependence}
With increasing altitude of an observation level, the observation level comes 
closure to the shower maximum of a shower. Thus for the reason explained 
earlier, the Cherenkov light front becomes wider with 
increasing altitude of the observation level and becomes flatter with 
decreasing altitude (see the Fig.\ref{fig11a}).
This trend will continue if an observation level lies below the position of 
shower maximum. Similarly for the same reason 
the arrival time for a inclined shower is found to be shorter in comparison to 
a vertical shower at a given observation level. 
These behaviours of arrival time of Cherenkov photons are most prominent for
the $\gamma$-ray primary and are least prominent for the iron primary for the
reason already mentioned in the previous section. 
Thus the disk thickness 
\cite{Goswami1} of Cherenkon photons at high altitude observation level for a 
high energy vertically incident primary particle is wider in comparison to
that at low altitude observation level and for a low energy primary
particle incident with some zenith angle.         

\begin{figure*}[hbt]
\centerline
\centerline{
\includegraphics[scale=0.28]{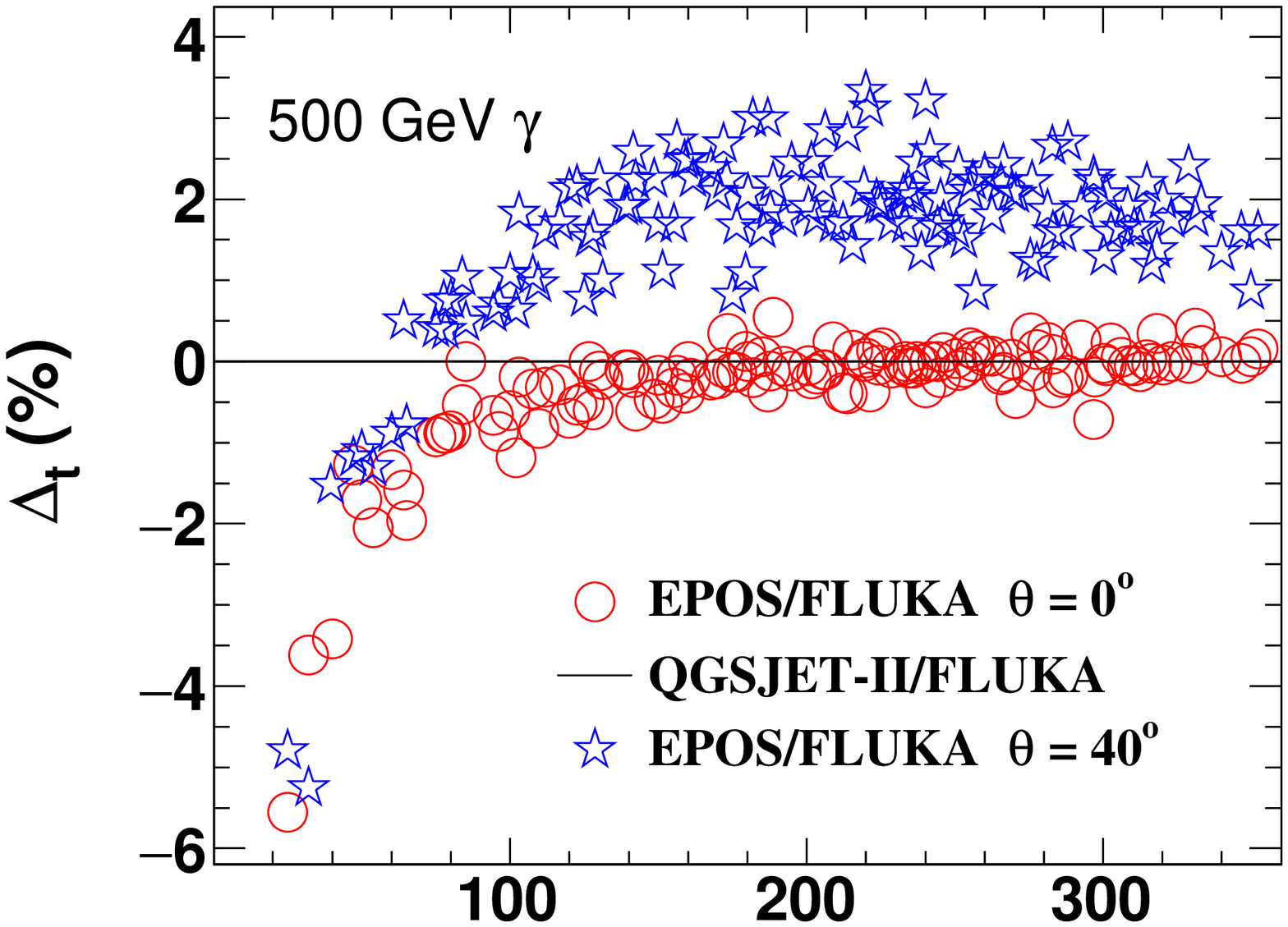} \hspace{-3.5mm}
\includegraphics[scale=0.28]{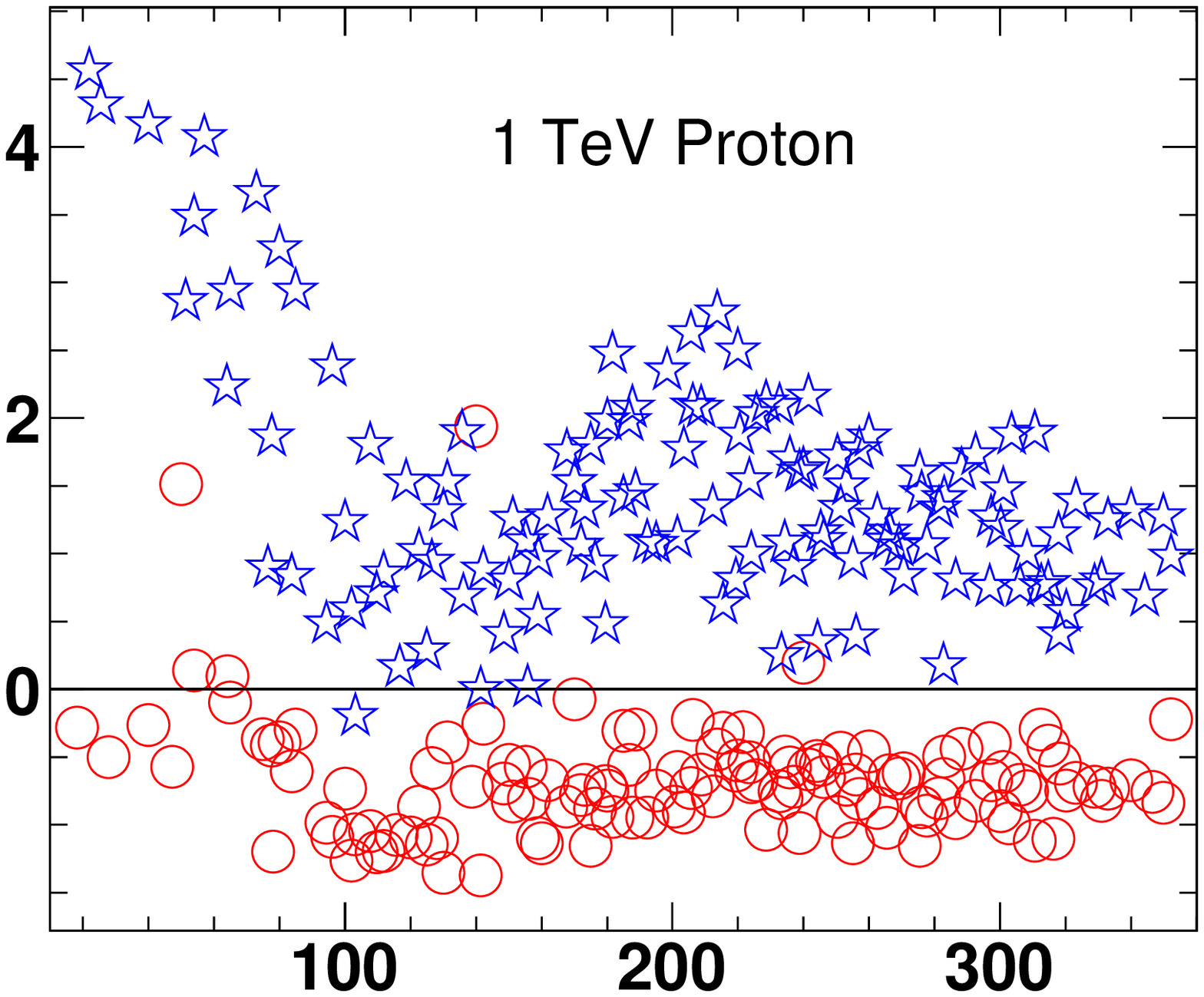} \hspace{-3.5mm}
\includegraphics[scale=0.28]{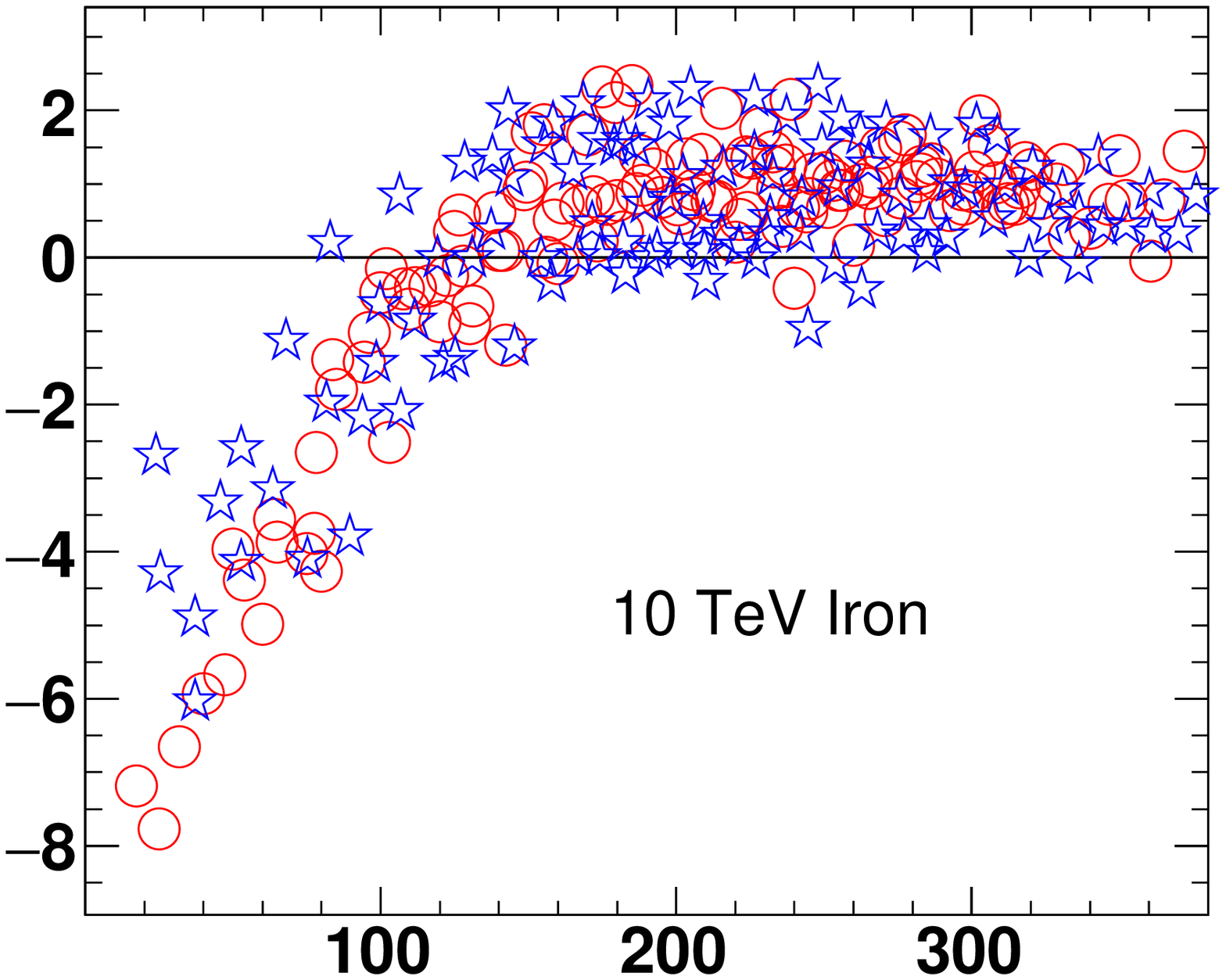}
}
\vspace{-5mm}
\centerline{\hspace{-0.5mm}
\includegraphics[scale=0.28]{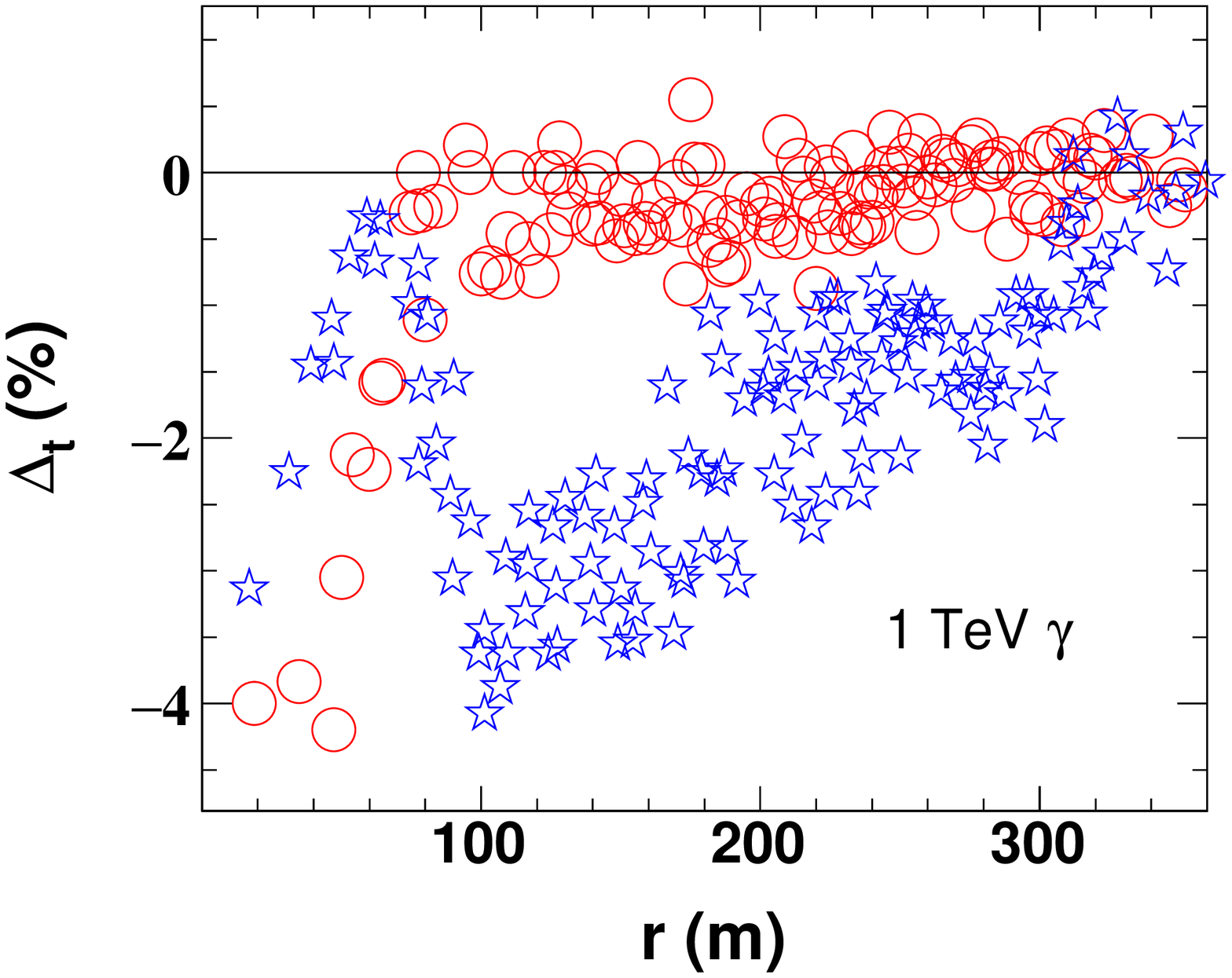} \hspace{-3.5mm}
\includegraphics[scale=0.28]{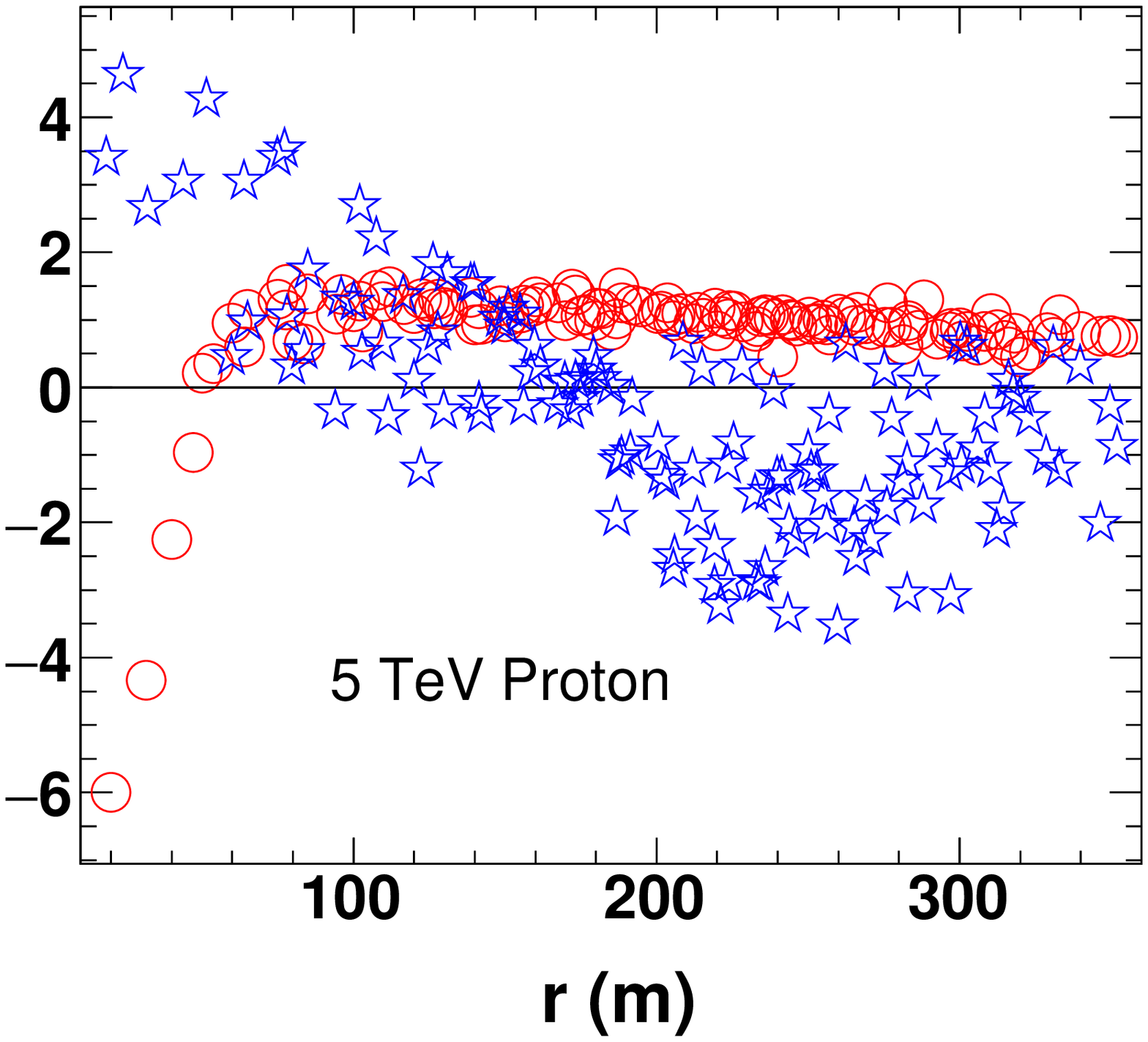} \hspace{-3.5mm}
\includegraphics[scale=0.28]{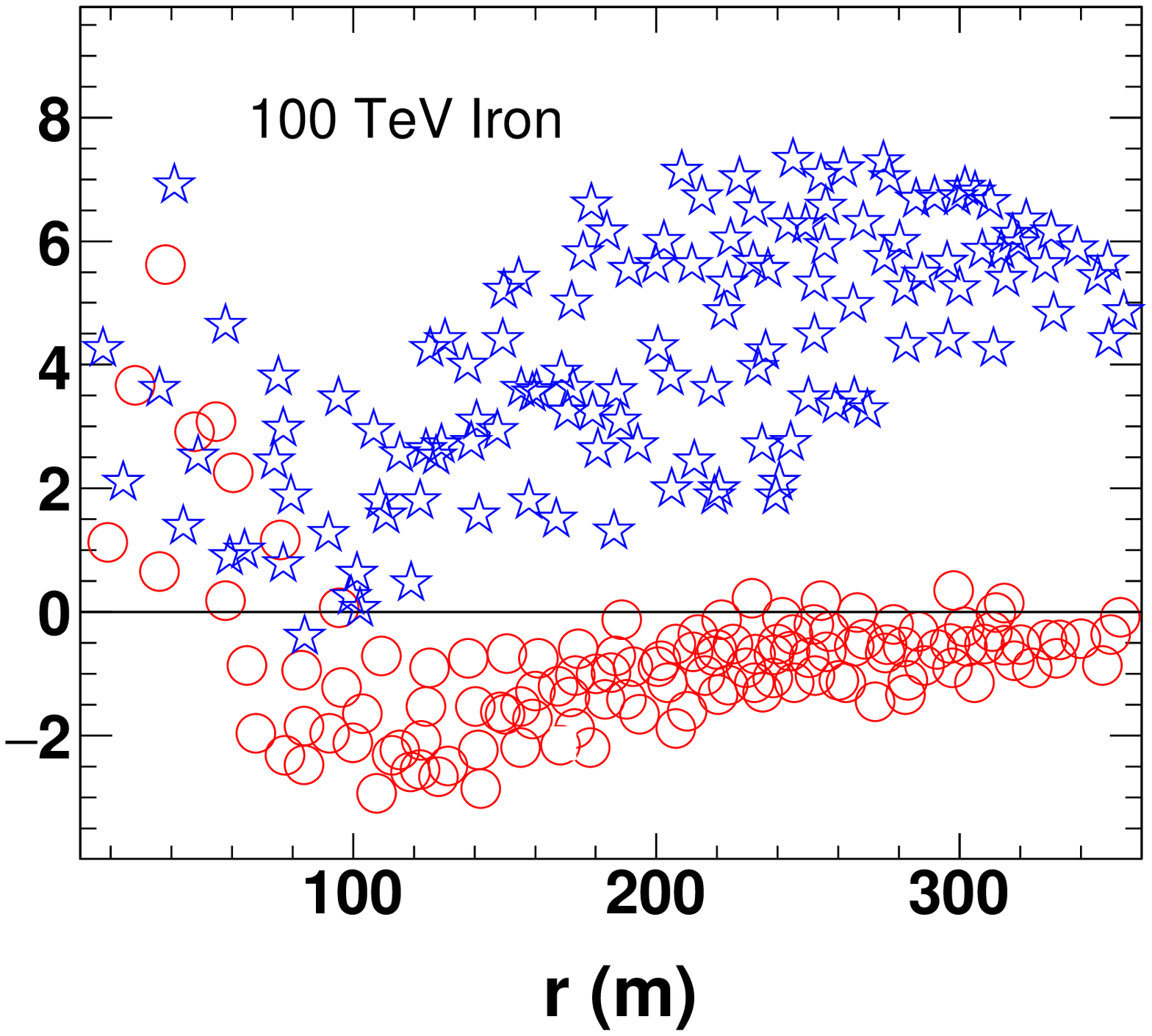}}

\vspace{5mm}
\caption{\% Relative deviation of Cherenkov photon's arrival times
($\Delta_t$s) with respect to core distance of showers of different
primaries for the QGSJETII and EPOS models of high energy hadronic
interactions. The QGSJETII-FLUKA model combination is indicated by the
horizontal solid lines in all plots, which is considered as the reference for
the calculation.}
\label{fig8}
\end{figure*}

\subsubsection{Hadronic interaction model sensitivity}
As in the case of Cherenkov photon density mentioned above, we have studied
also the relative deviation in the $t_{ch}$ considering the
model combination QGSJETII-FLUKA as the reference model. This is for a better
idea about the hadronic interaction model's sensitivity to the Cherenkov
photon's arrival time with respect to shower core for different
primary particles with different energies and zenith angles. 
Fig.\ref{fig8} shows the relative deviations in percentage of the arrival 
times ($\Delta_t$s) of the Cherenkov light fronts obtained from the EPOS-FLUKA 
and QGSJETII-FLUKA model combinations for various monoenergetic primary 
particles, incident vertically and at an angle 40$^{\circ}$. It is seen that, 
the deviations in the $t_{ch}$s due to two the model combinations are 
small (within $\pm$7\%) for all the vertically incident as well as inclined 
primaries. Moreover, for the vertical showers, the difference
between the two models is remarkable mainly for the core distances below 100 m,
beyond which the \% relative variation in $t_{ch}$ is quite small,
mostly less than $\pm$ 3\%. But for the inclined showers, this difference
lies within $\pm$ 4\% to $ 7\%$ (100 TeV iron) almost throughout the distance 
from the core.

\begin{figure*}[hbt]
\centerline
\centerline{
\includegraphics[scale=0.29]{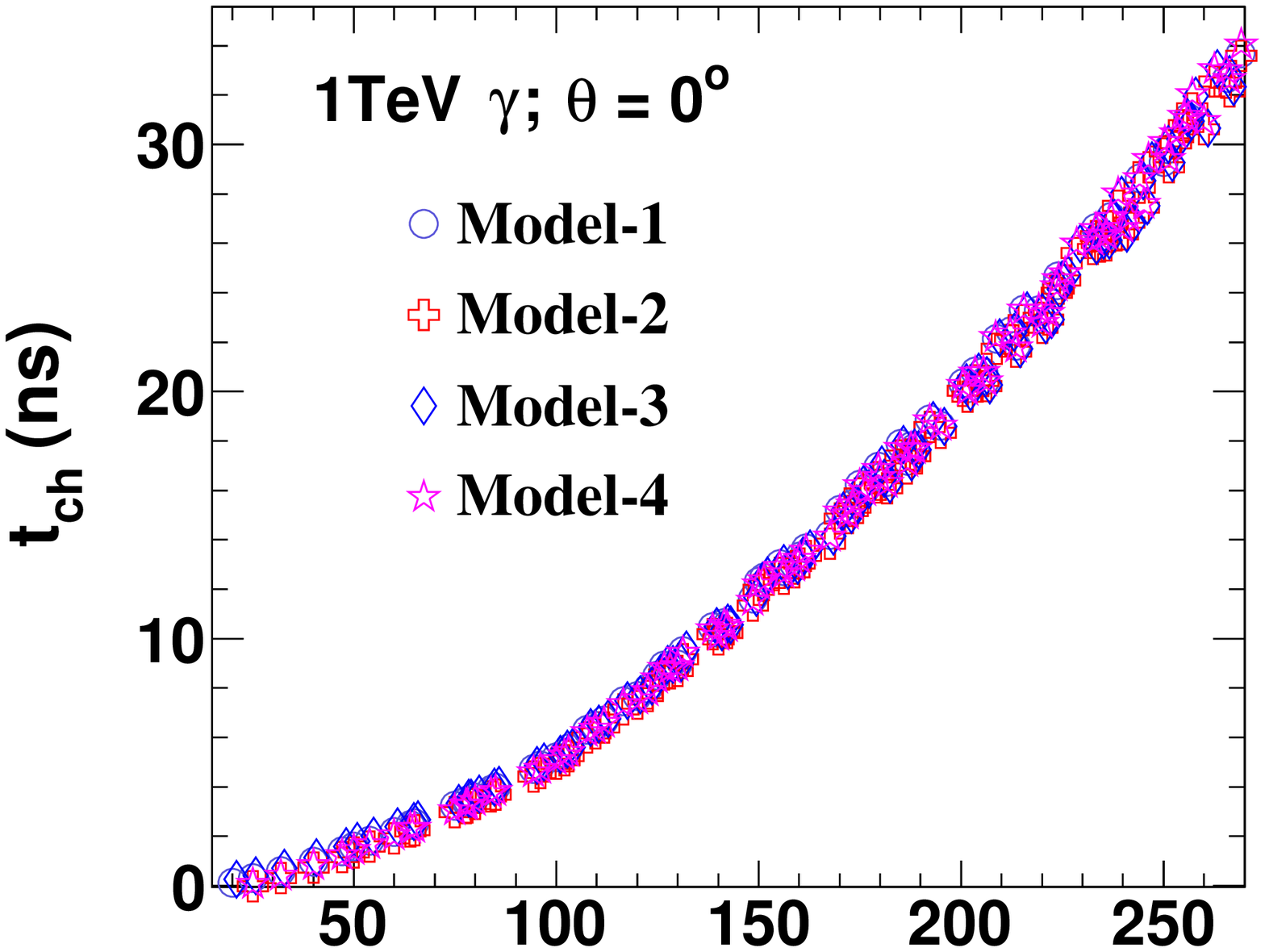}\hspace{-0.3cm}
\includegraphics[scale=0.29]{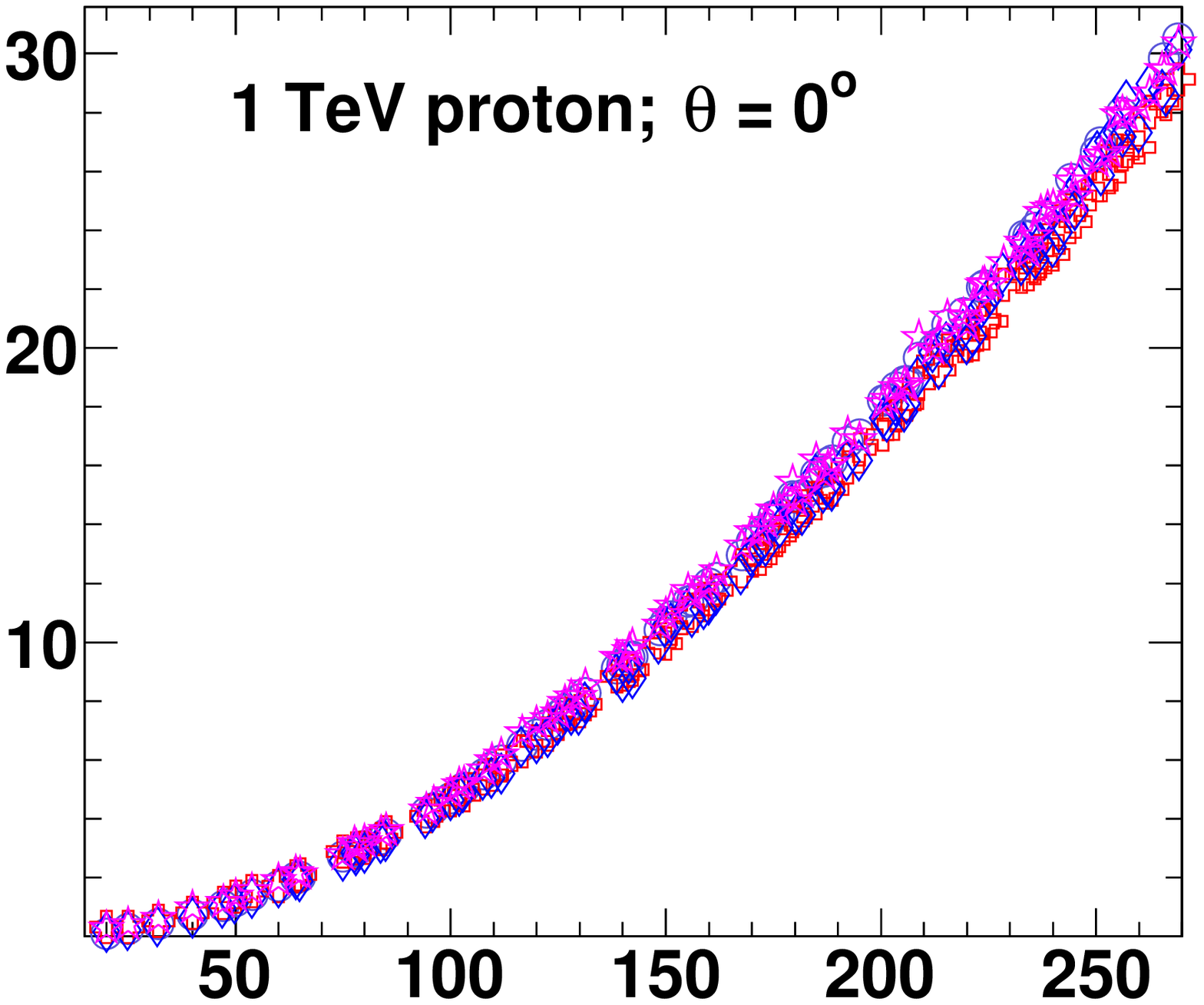} \hspace{-0.3cm}
\includegraphics[scale=0.29]{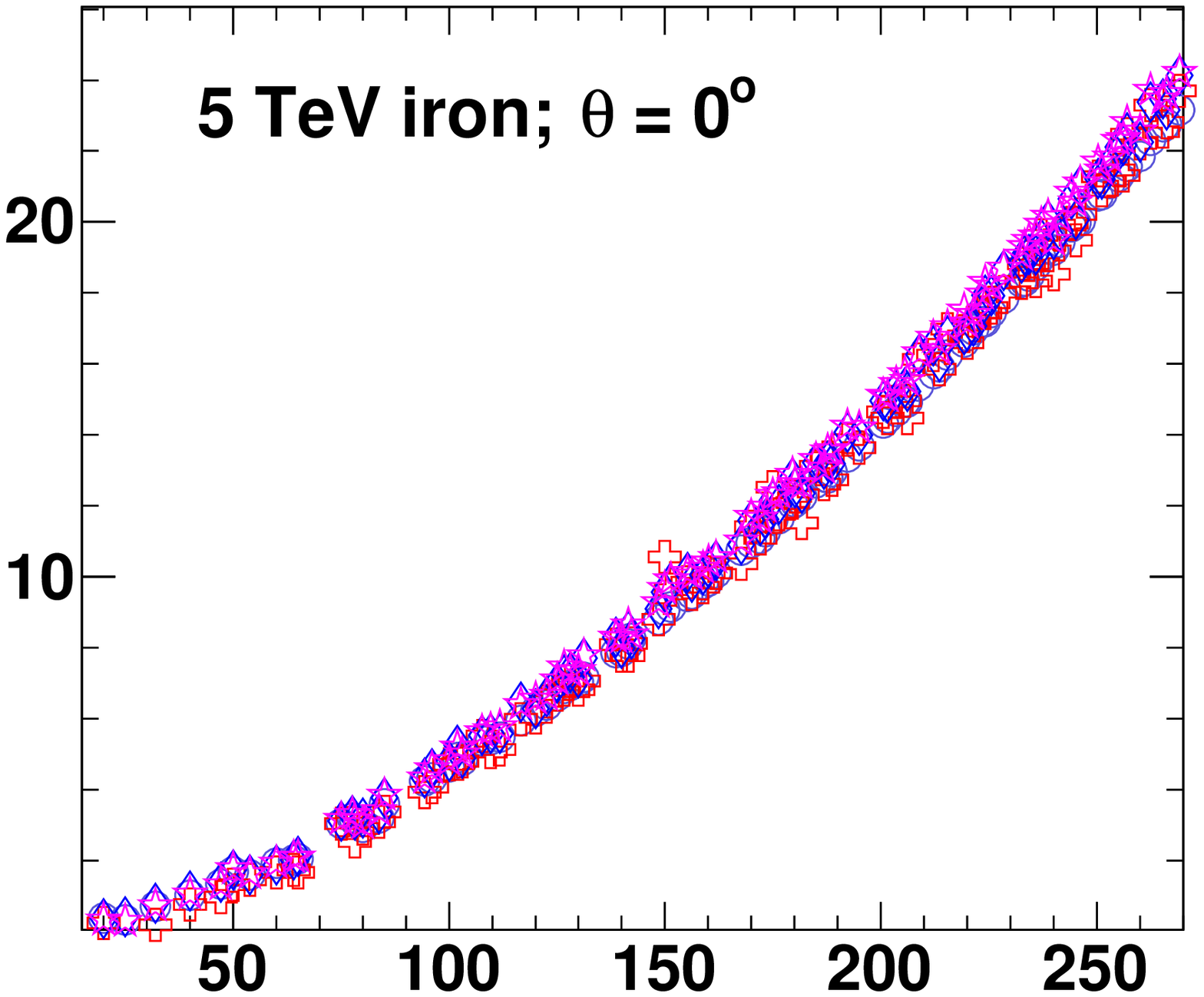}}

\hspace{-0.2cm}
\centerline{
\includegraphics[scale=0.29]{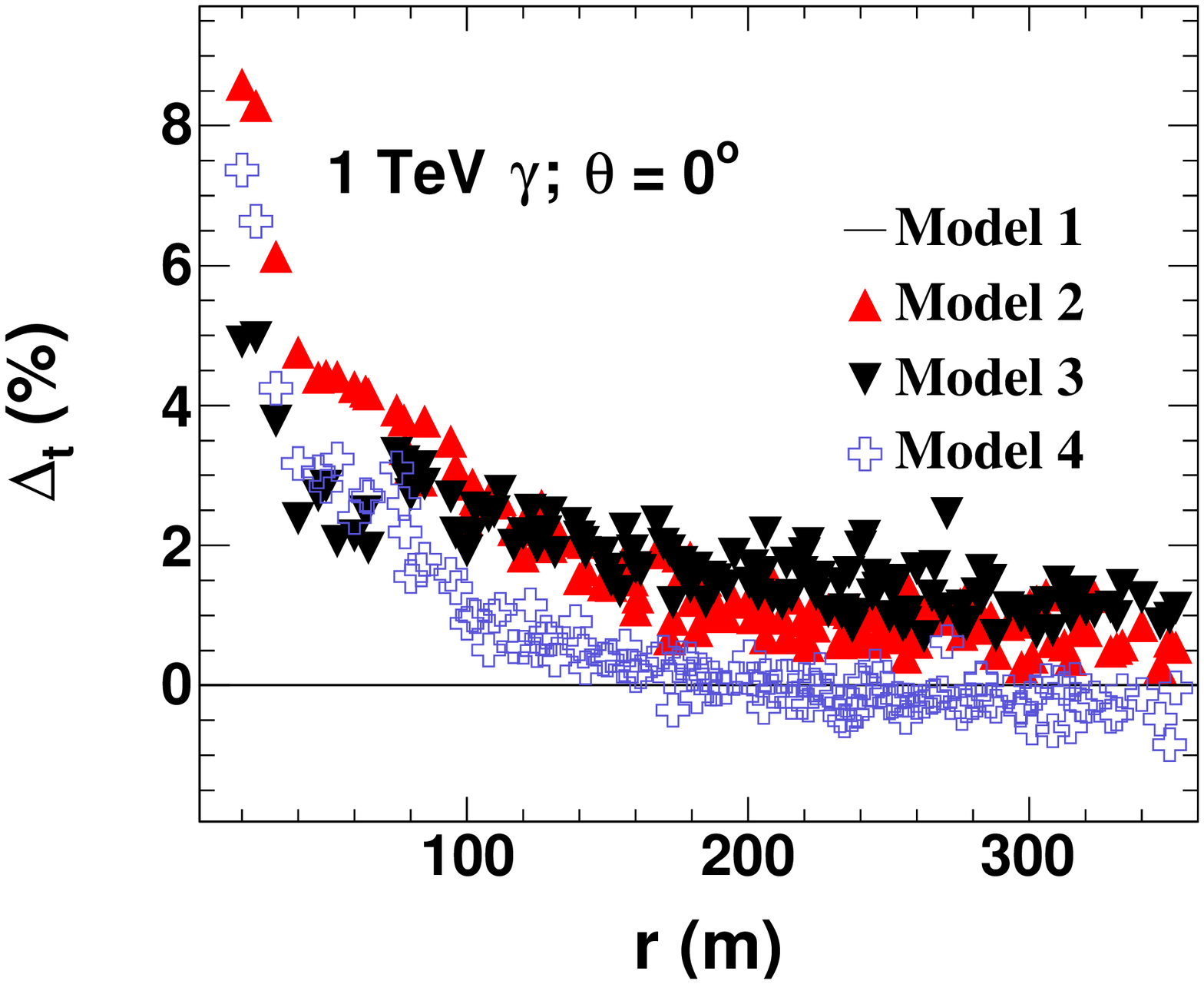} \hspace{-0.3cm}
\includegraphics[scale=0.29]{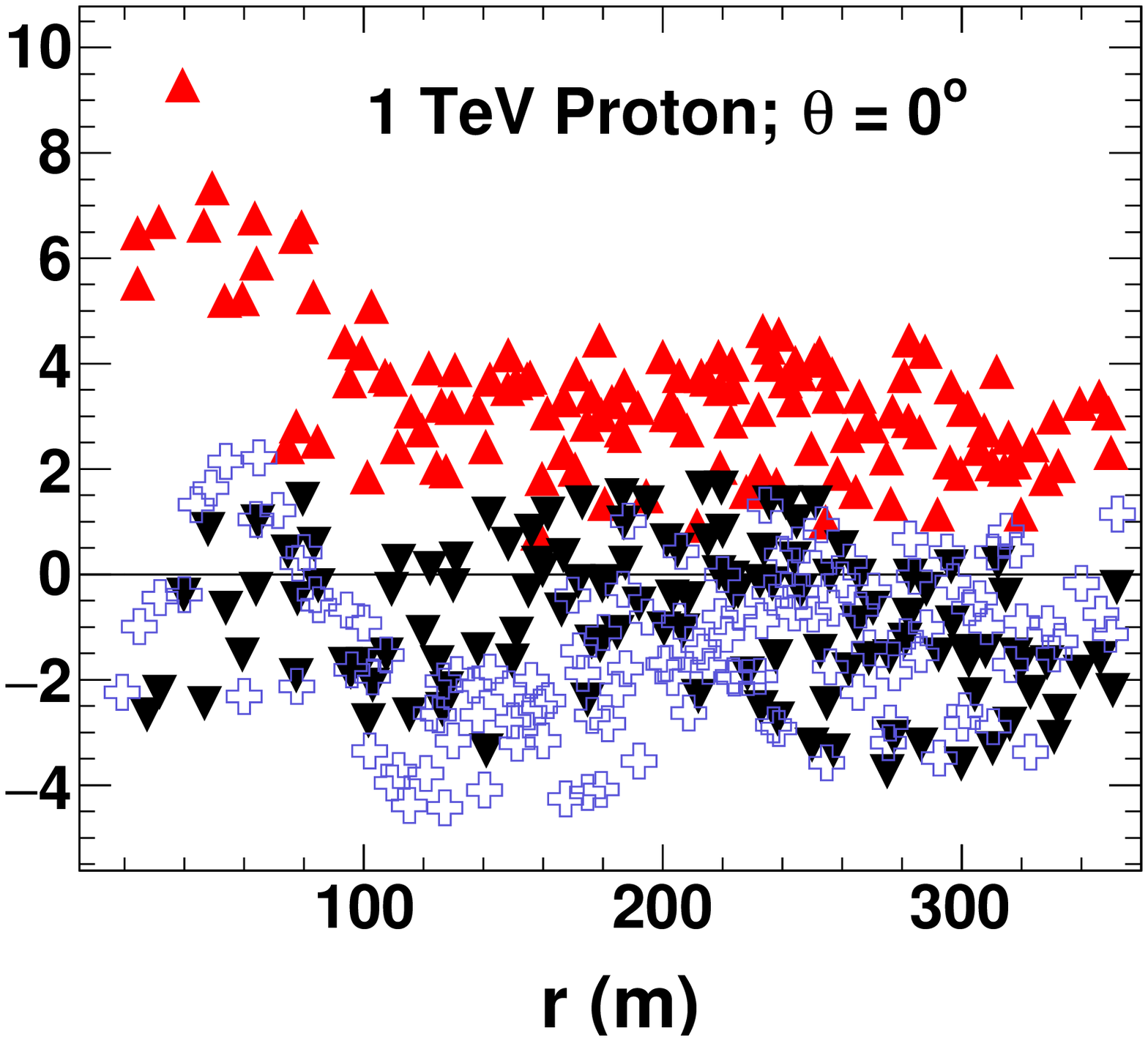} \hspace{-0.3cm}
\includegraphics[scale=0.29]{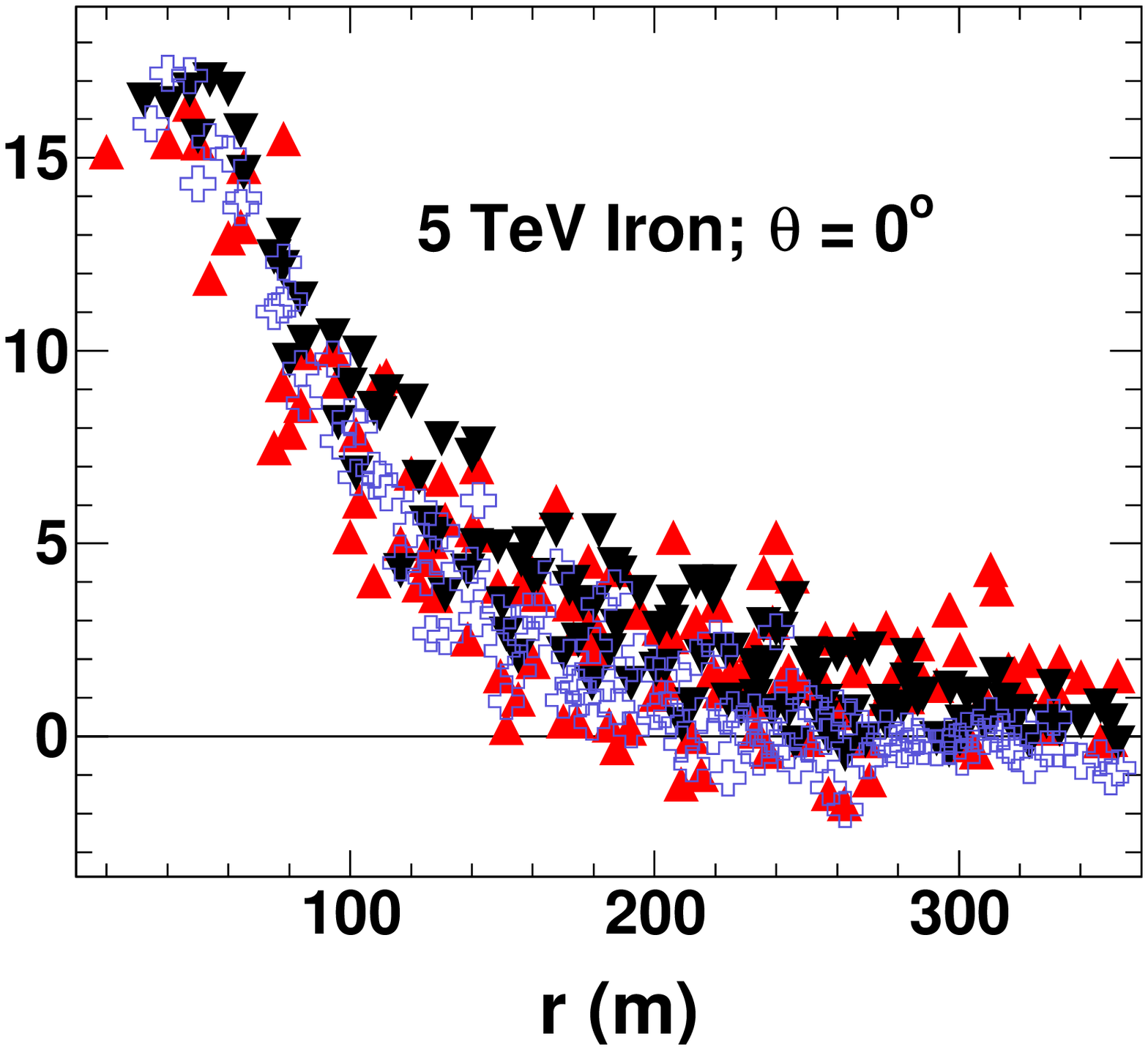}}

\caption{Variation of $t_{ch}$ with distance
from the shower core  of $\gamma$-ray, proton and iron primary incident at
0$^{\circ}$ zenith angle as obtained for four different atmospheric models
(top panels) and the corresponding \% relative differences ($\Delta_t$)
between different model predictions.
The Model-1, Model-2, Model-3 and Model-4 represent the U.S. standard
atmosphere as parameterized by Linsley, AT 115 Central European atmosphere for Jan.
15, 1993, Central European atmosphere for Feb. 23, 1993 and Central European
atmosphere for May 11, 1993 \cite{Heck} respectively.}
\label{fig11b}
\end{figure*}

\subsubsection{Influence of atmospheric model}
Similarly to the case of Cherenkov photon density, we have done the calculation
of arrival time of Cherenkov photons using the same atmospheric models to 
check the influence of atmospheric model consideration in our analysis as 
shown in the top panels of the Fig.\ref{fig11b}. It is seen that the difference
in arrival times as obtained from these models are 
$<$ 10\% for the $\gamma$-ray, $<$ 11\% for the proton and $<$ 19\% for the 
iron primaries (see the lower panels of Fig.\ref{fig11b}). 
Furthermore, it is observed that for the case of $\gamma$-ray primary, only 
very near the shower core ($<$ 30 m) the deviations go beyond 5\%, beyond 
which the deviations are below 3\% only. Hence the
effect of the atmospheric model on the calculation of arrival time of Cherenkov
photons can be considered negligible.   
  
\subsection{Angular distribution of Cherenkov photons}
In this section we study the distribution of Cherenkov photon's average 
angular position with respect to the axis of a shower of different primary
type, energy and zenith angle. The study of this
section is basically guided by our theoretical interest rather than 
experimental implementation point of view. However, this study may provide
us an idea to know about the average angular position of photons incident
on an imaging detector of IACTs located at a particular distance from the
core of a particular shower. On the theoretical aspect, such comparison with
experiment with an in-depth view may also help to improve further the high 
energy hadronic interaction models.           
\subsubsection{General characteristics}
\begin{figure*}[hbt]
\centerline
\centerline{
\includegraphics[scale=0.50]{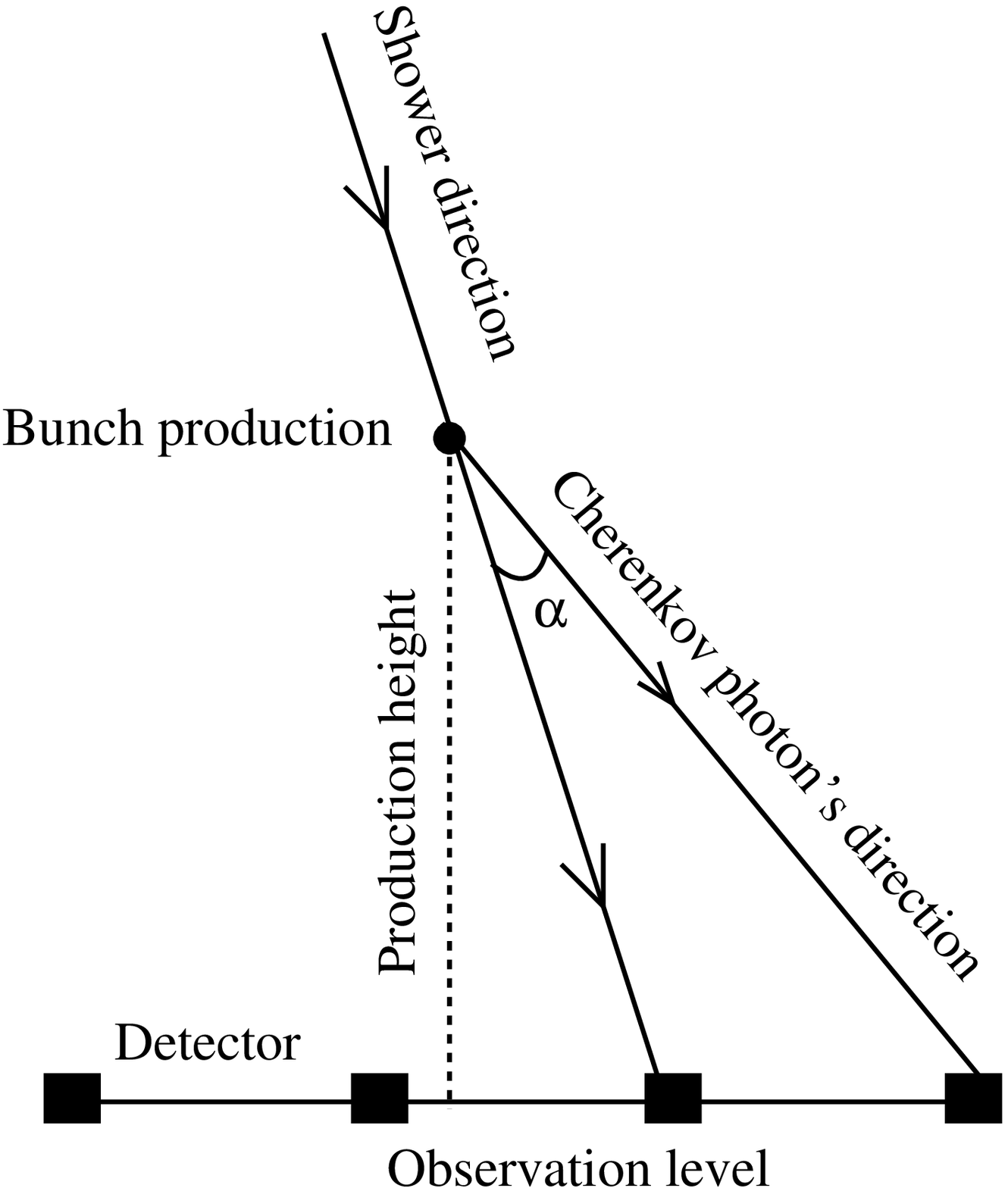}}
\caption{Schematic of Cherenkov photon's bunch production and the related 
angular position ($\alpha$) with respect to the shower axis.}
\label{fig12a}
\end{figure*}

\begin{figure*}[hbt]
\centerline
\centerline{
\includegraphics[scale=0.29]{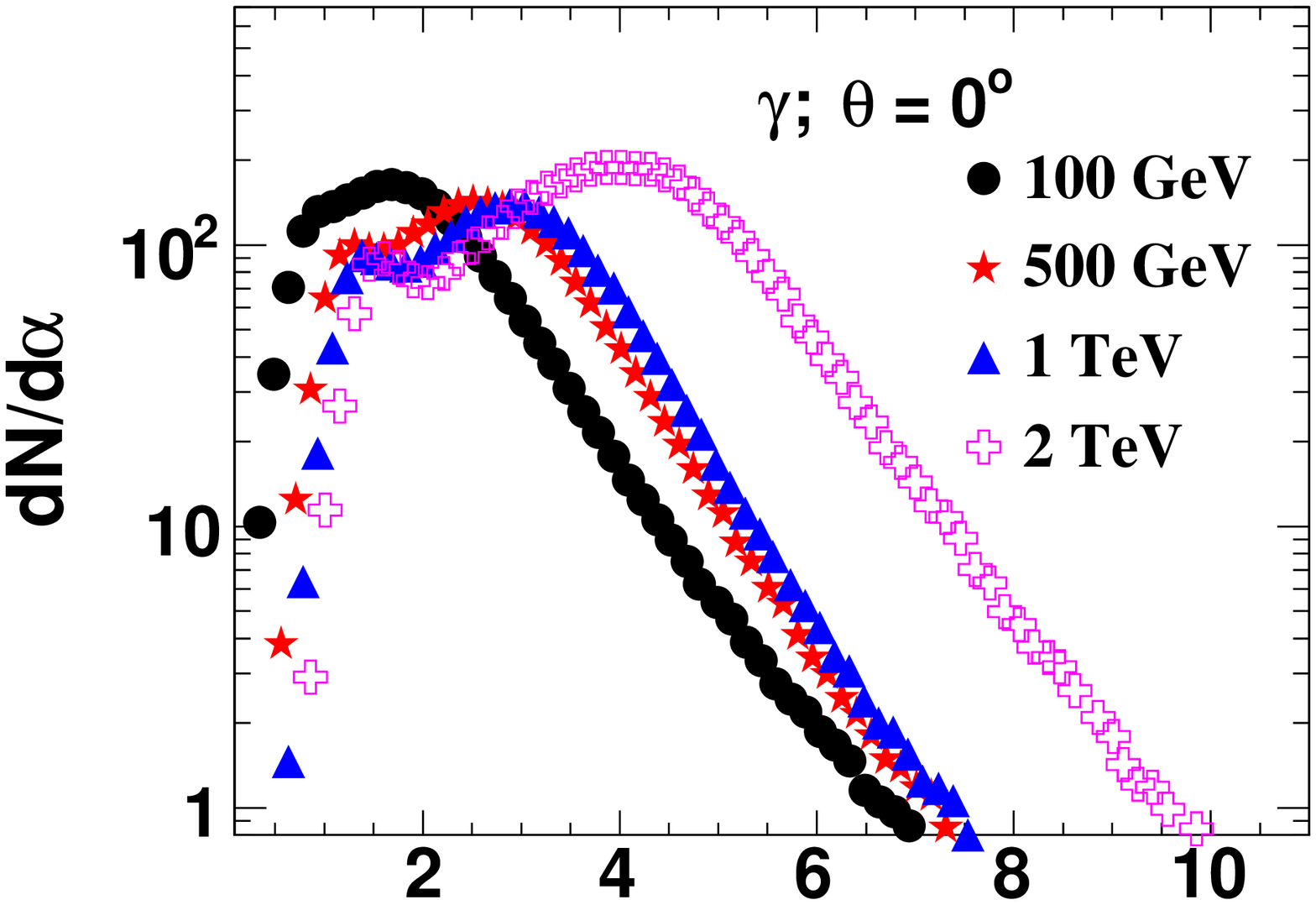} \hspace{-3mm}
\includegraphics[scale=0.29]{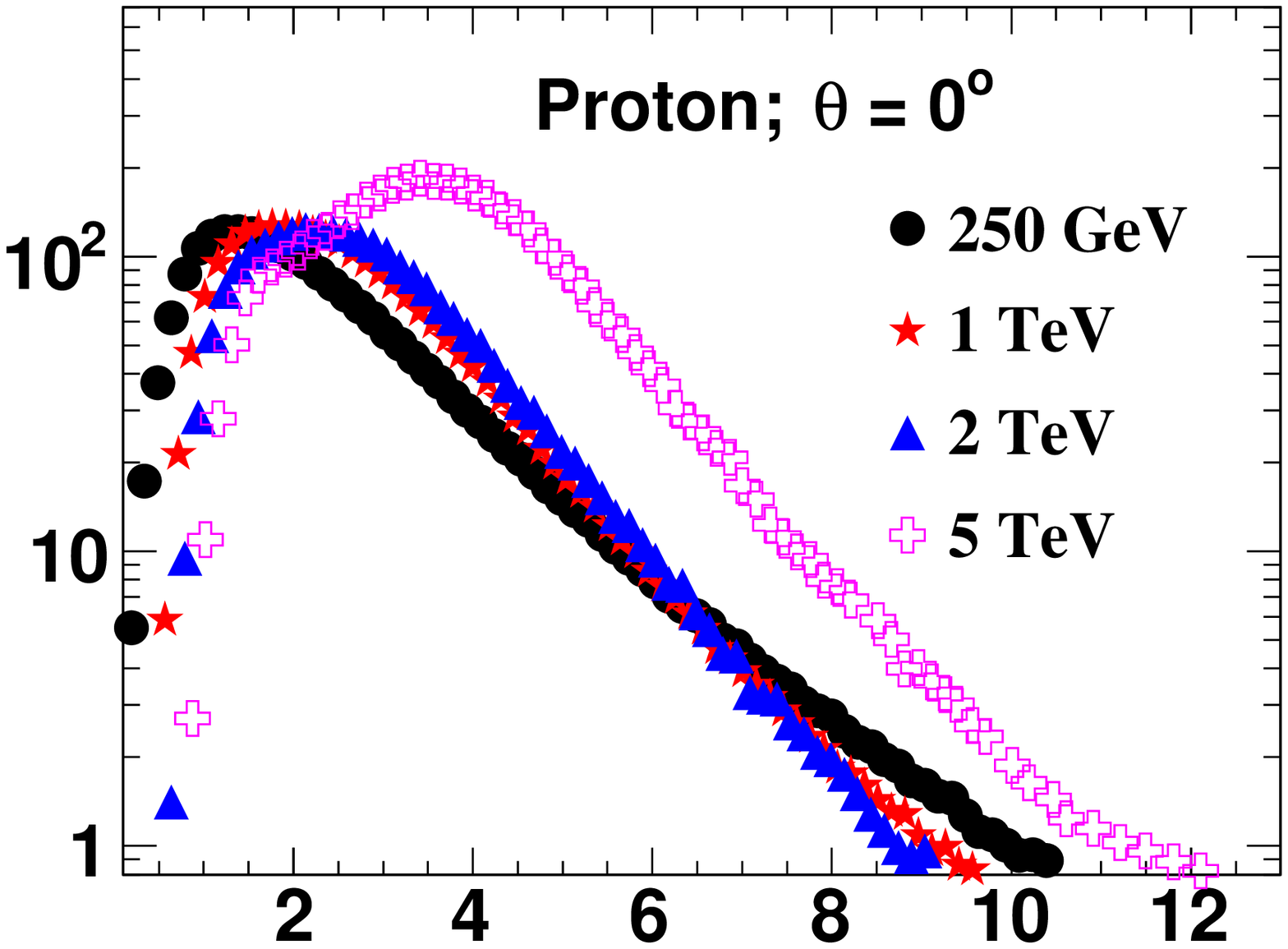} \hspace{-3mm}
\includegraphics[scale=0.29]{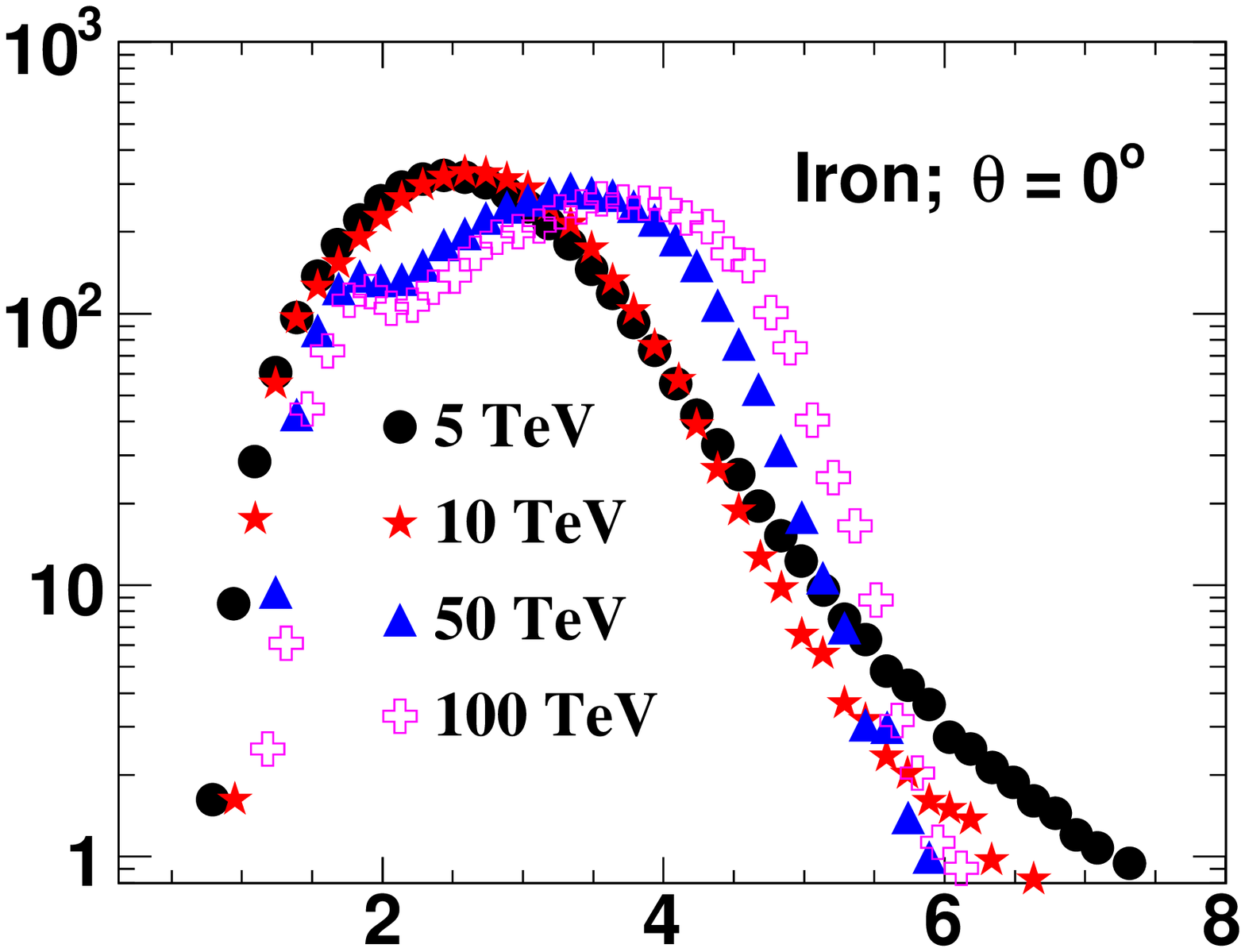}}
\vspace{-3mm}

\centerline{
\includegraphics[scale=0.29]{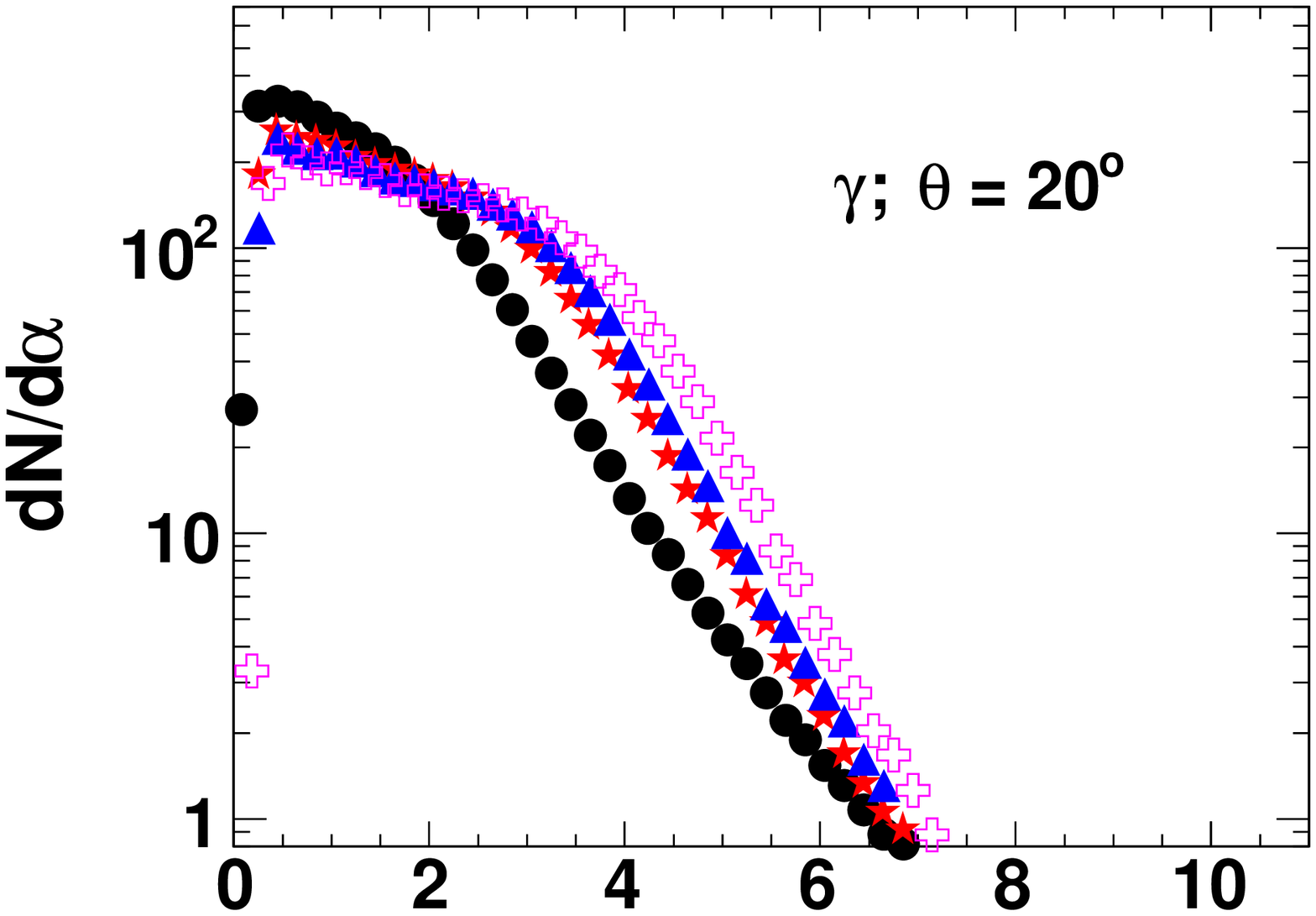} \hspace{-3mm}
\includegraphics[scale=0.29]{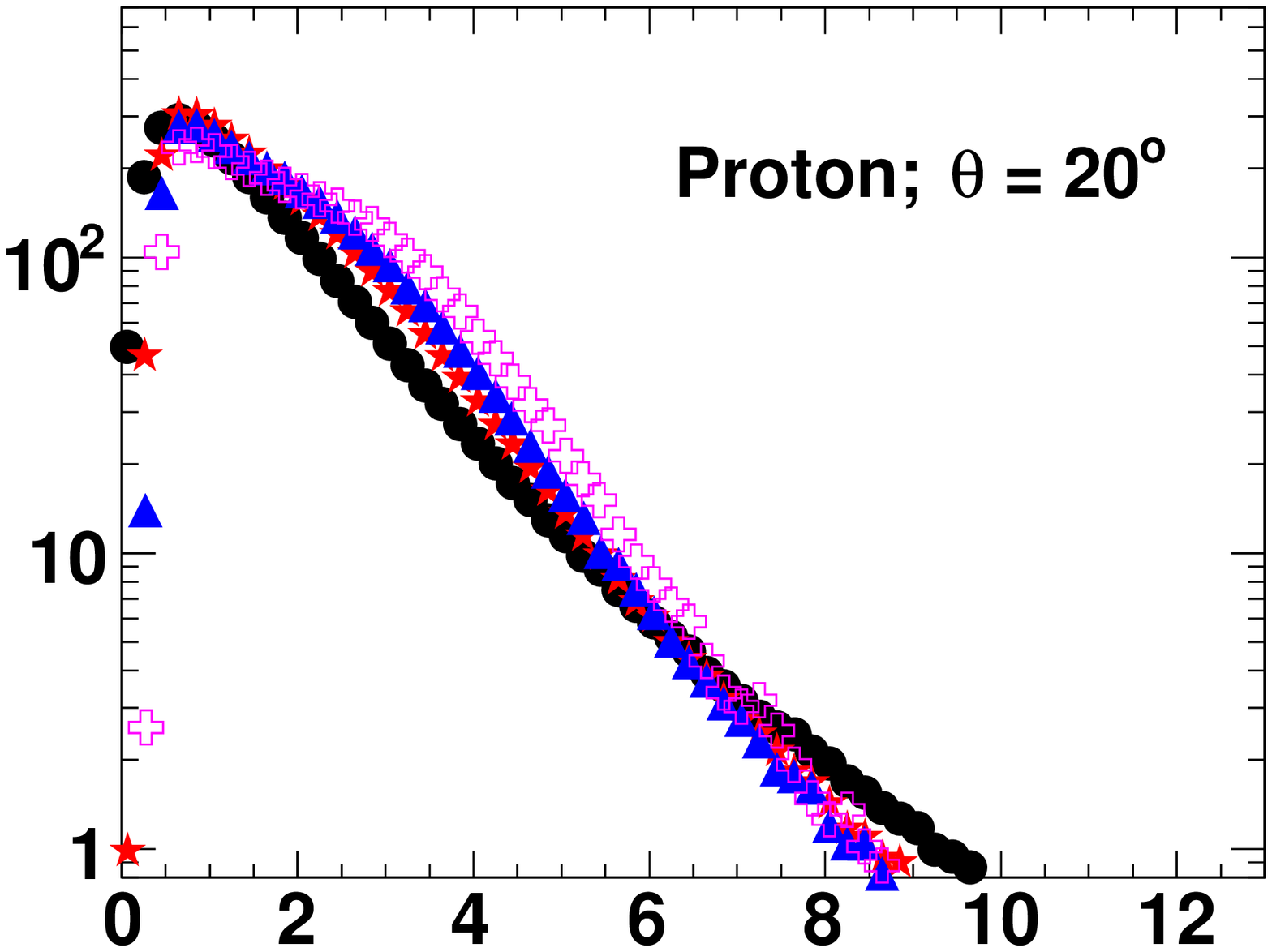} \hspace{-3mm}
\includegraphics[scale=0.29]{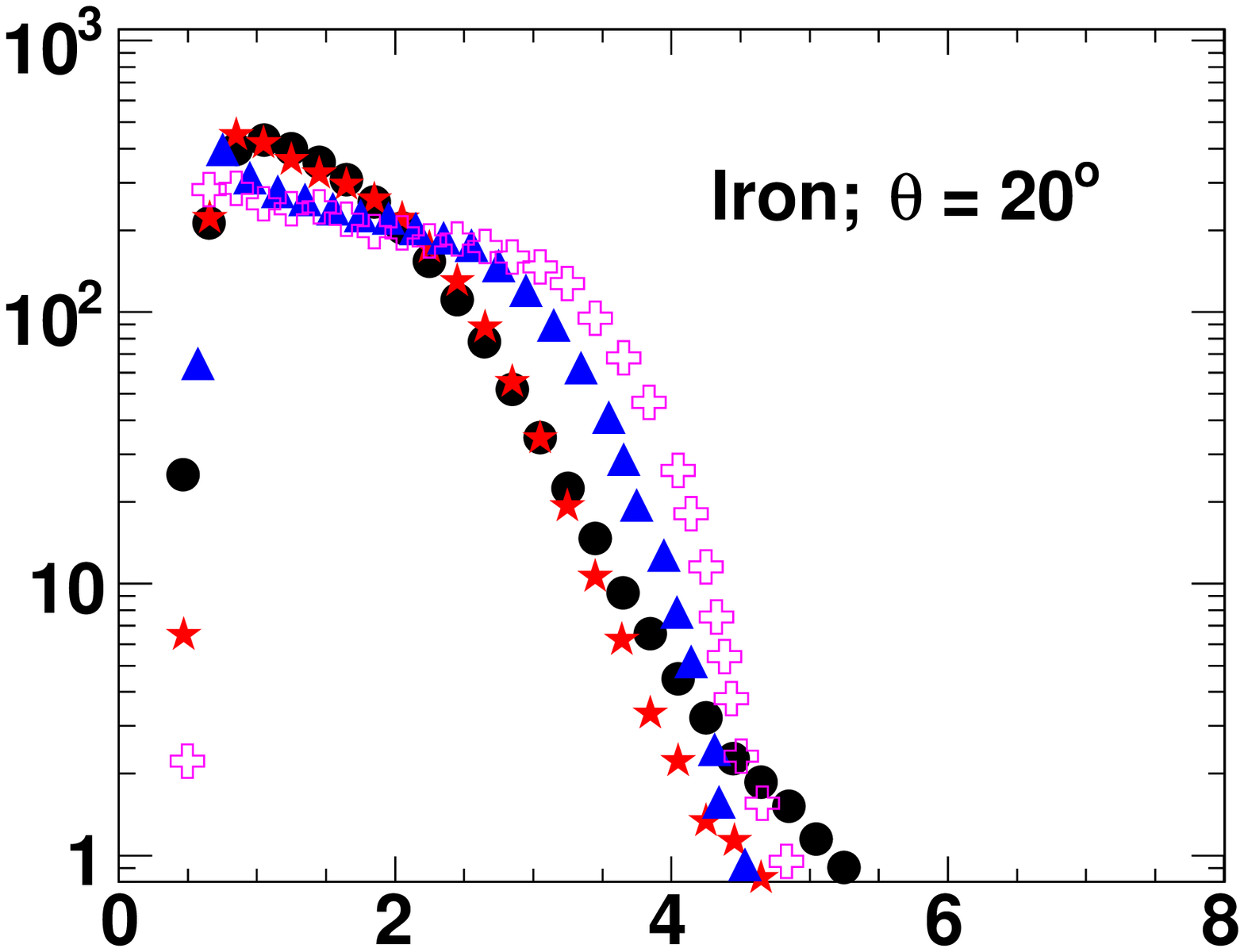}}
\vspace{-3mm}

\centerline{
\includegraphics[scale=0.29]{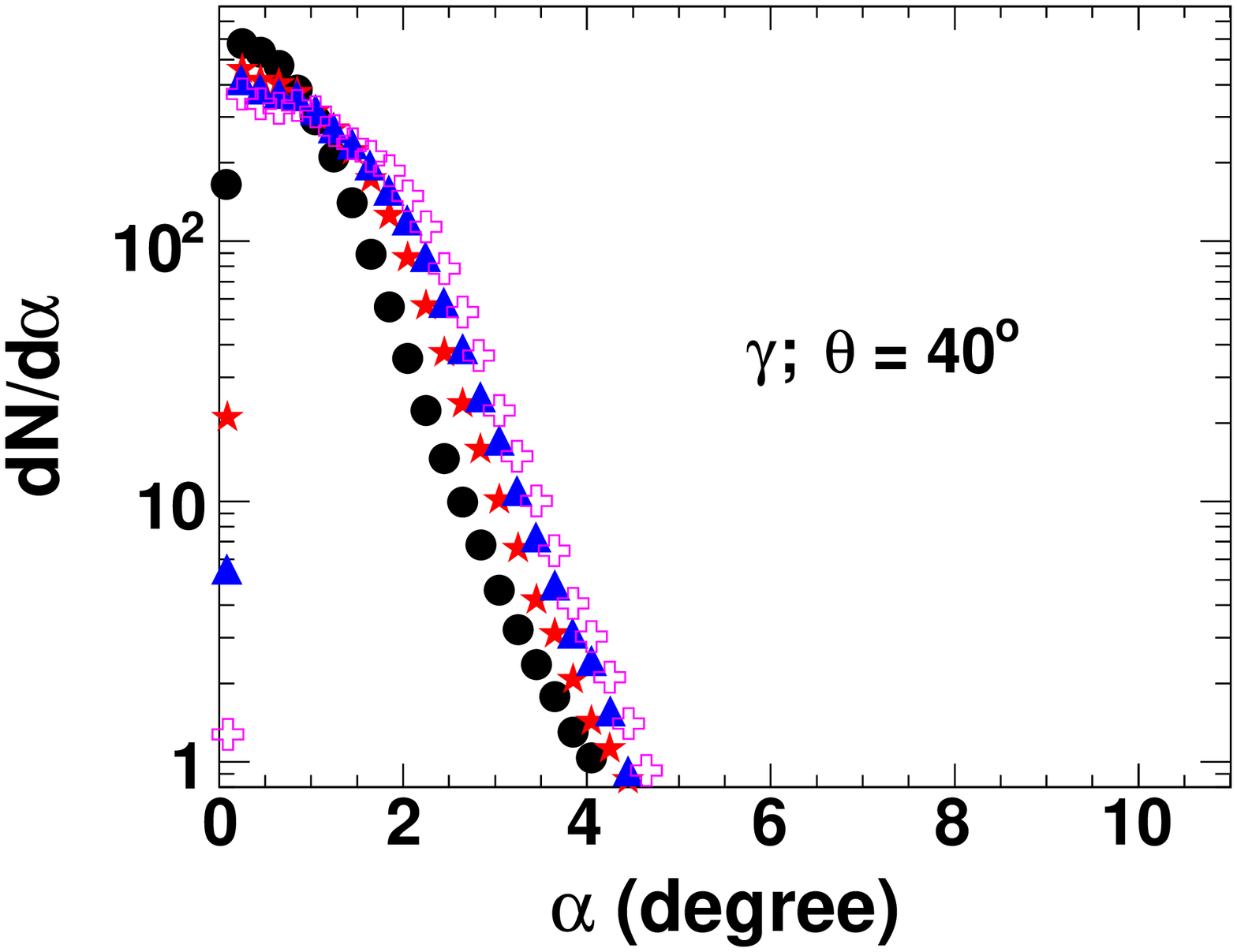}\hspace{-3mm} 
\includegraphics[scale=0.29]{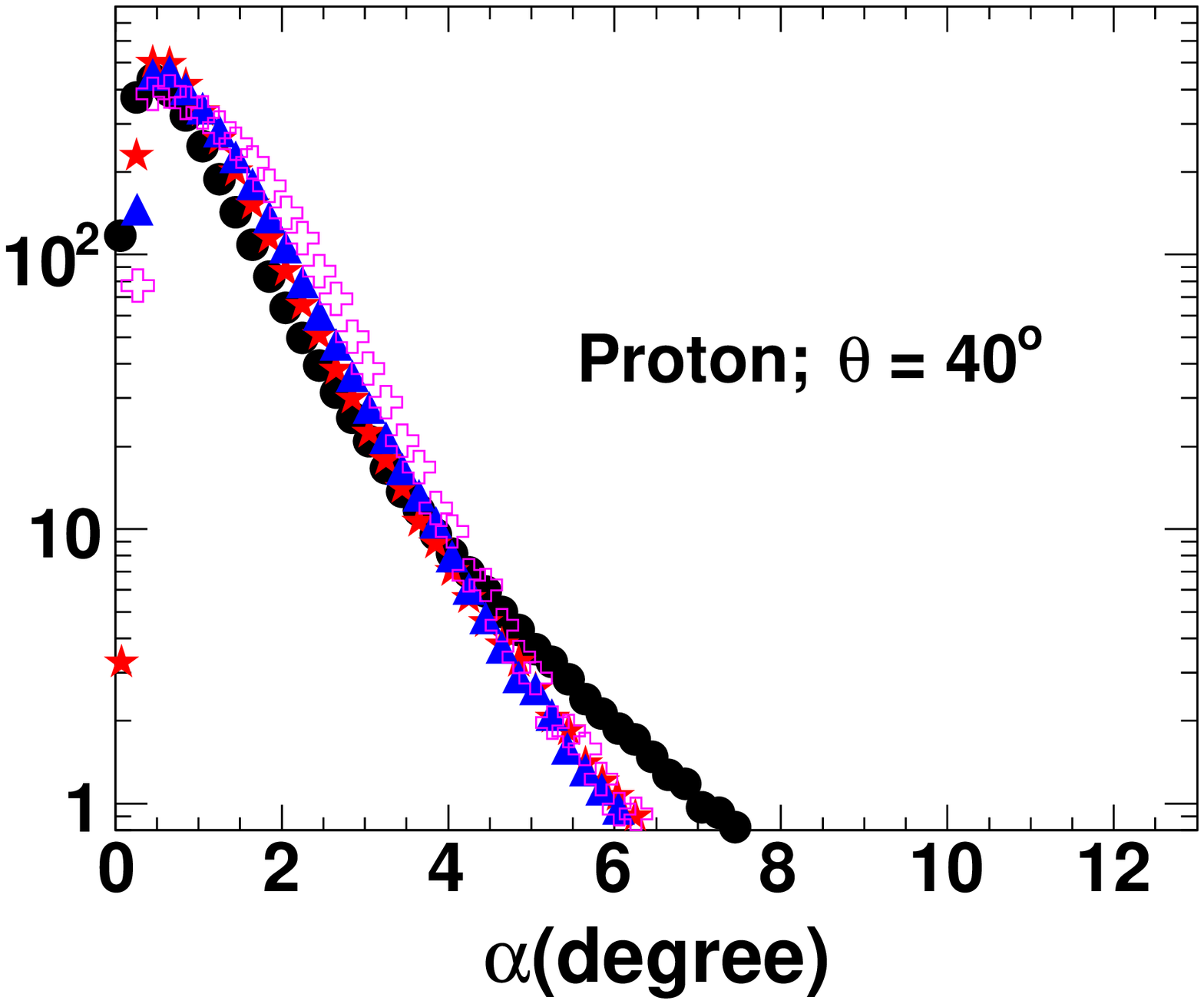} \hspace{-3mm}
\includegraphics[scale=0.29]{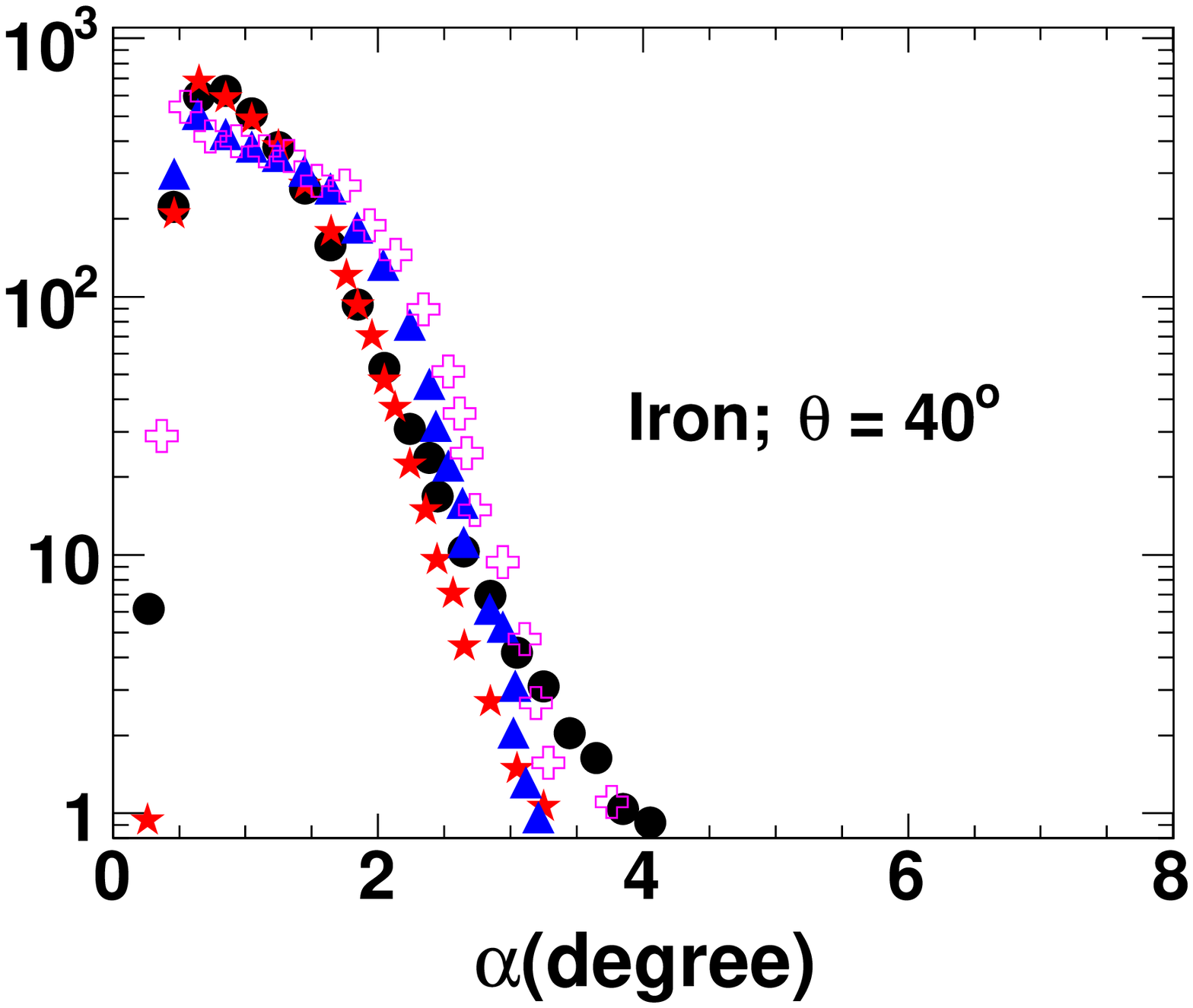}}

\caption{Cherenkov photon's angular distributions obtained from different
primaries with different energies corresponding to some fixed value of zenith
angle of primary particle.}
\label{fig12}
\end{figure*}
To obtain the angular distribution of Cherenkov photons, the number of photons 
produced per angular bin ($\frac{dN}{d\alpha}$ (degree$^{-1}$)) of
the Cherenkov photon's angular position ($\alpha$) with respect to the 
shower axis (see the Fig.\ref{fig12a}) over each detector are counted. Some of 
such distributions are shown in the Fig.\ref{fig12} for various 
combinations of energy and zenith angle for all three primaries. 
It is seen that the distribution becomes increasingly 
wider with increase in energy of the primary, specially for the vertically
incident particle. With increasing the angle of incidence this tendency
decreases gradually and at large incident angle ($> 20^\circ$) it is almost 
negligible.  
The first behaviour clearly suggests that, as energy of the vertically incident 
primary increases, the number of 
particles deflected at larger angles from the shower core also increases with
the proportionate increase in the shower size. Moreover, this tendency of the 
distribution is most prominent for the $\gamma$-ray shower and least prominent 
for the iron shower. This is because the size of the EM part of a shower (which
is deflected most and is responsible for the production of Cherenkov 
photons) produced by the iron primary is the lowest, whereas the shower 
produced by $\gamma$-ray primary is wholly EM in nature. In 
addition, it is seen that for the vertically incident showers of all primaries,
the value of the angle $\alpha$ at which maximum number of Cherenkov photons 
are emitted with respect shower axis 
shifts to higher value with the increasing energy of the primary particle.  
This shift is highest for the $\gamma$-ray primary and least for the iron
primary. 
As discussed in the earlier section, the increase of 
density of Cherenkov photons with energy is relatively more for the 
$\gamma$-ray primary than that for the proton and specially for the iron 
primary. 
So the average value of the angle at which maximum photons are emitted 
is shifted more towards its higher value for the $\gamma$-ray primary than 
that for proton and iron primaries. May be because of the 
complex nature of the hadron initiated shower, the largest value of $\alpha$ 
move towards its lower value with the increasing energy of 
primary particle, specially for the iron (see the top right panel of 
Fig.\ref{fig12}). Similar behaviour can be seen for all primaries incident at
larger zenith angle, noticeably for the proton primary in this case. There 
might be some contributions to these observed behaviours from the height of 
the shower maximum also, which are different for 
different primary at a given energy and zenith angle as stated earlier.            

It has also been observed that for a particular energy of the primary, the 
angular distribution profile becomes increasingly narrower as the angle of 
incidence of primary increases. The reason is as follows. With increasing zenith angle 
the slant depth of the shower maximum decreases (see the Table \ref{tab2}),
i.e. shower maximum is produced gradually at higher altitude with increasing 
zenith angle. Hence, the charged particles of a shower, which emit Cherenkov 
photons, has to cross gradually longer distance with 
increasing angle of incidence of the primary particle to reach the observation 
level. Consequently most of the low energetic as well as highly scattered
charged particles are absorbed by the atmosphere before they reach on the
observation level. Only sufficiently energetic particles with a smaller angular 
position $\alpha$ can reach on the observation level depending on the 
zenith angle of the primary particle, and hence the angular distribution 
profiles 
with increasing zenith angles of primary particles become gradually narrower. 

Further, the value of $\alpha$ at which maximum photons are emitted with 
respect to the axis of a shower decreases very slowly with increasing zenith 
angle. This is
also due to the increasing altitude of shower maximum with increasing zenith
angle as stated above. Because of the significantly different composition of
the iron initiated shower, as mentioned in the previous section, the angular 
distribution pattern of Cherenkov photons for the iron primary as a whole is
quite different that from $\gamma$-ray and proton primaries. The 
distinction between distribution patterns for $\gamma$-ray and proton primaries
is also noticeable at higher zenith angles.


\begin{figure*}[hbt]
\centerline
\centerline{
\includegraphics[width=5.6cm, height=4.7cm]{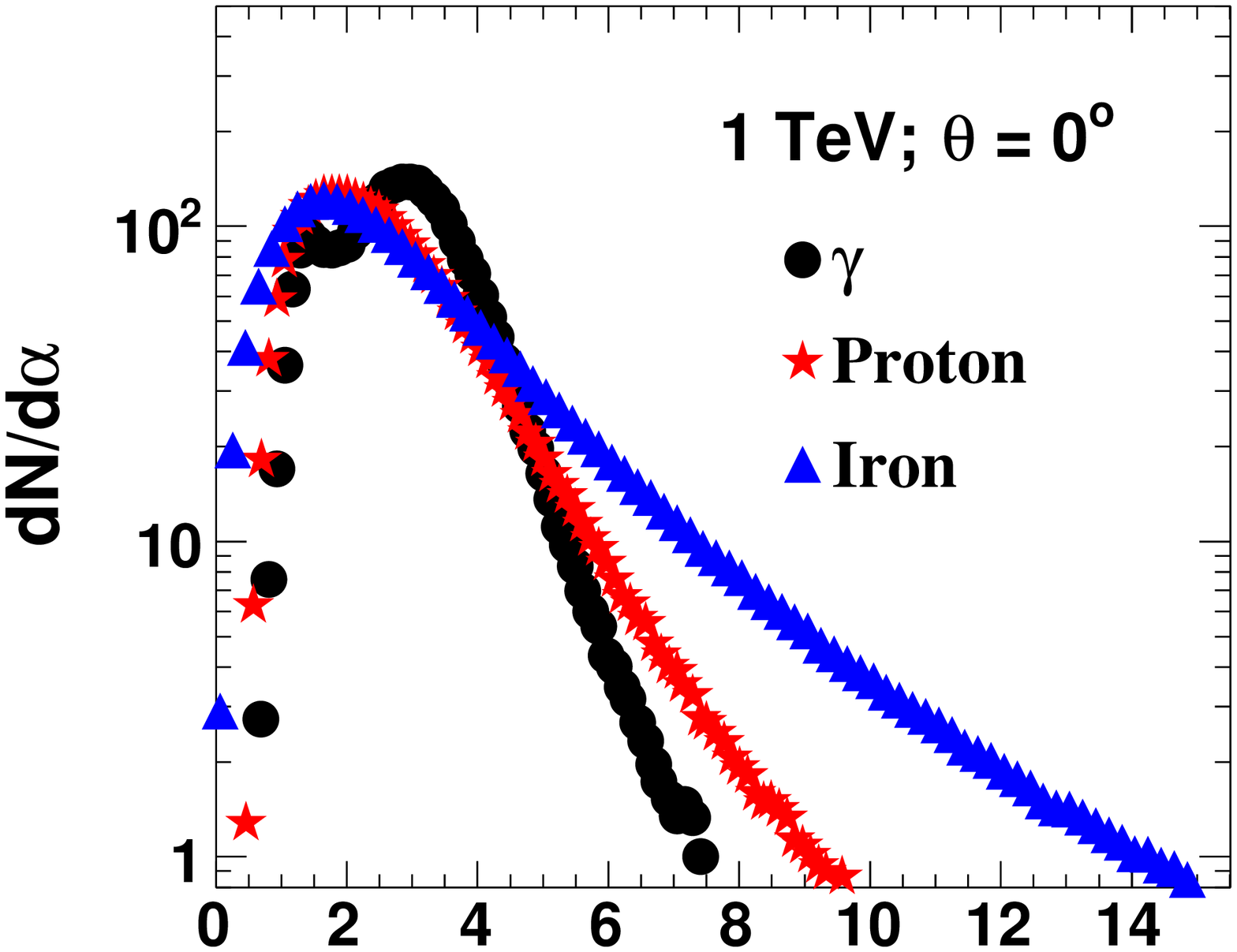} \hspace{-2mm}
\includegraphics[width=5.6cm, height=4.7cm]{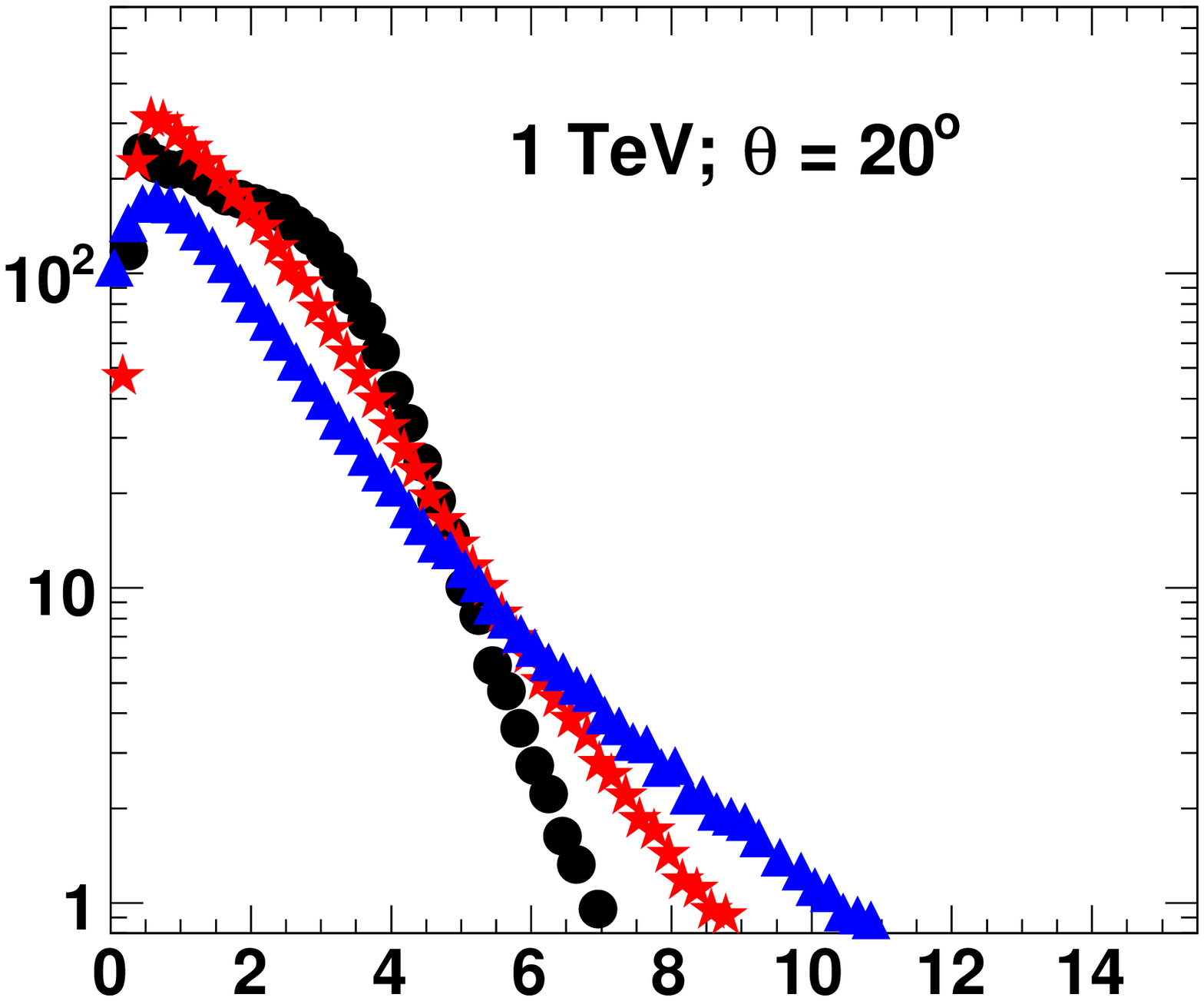} \hspace{-1mm}
\includegraphics[width=5.6cm, height=4.7cm]{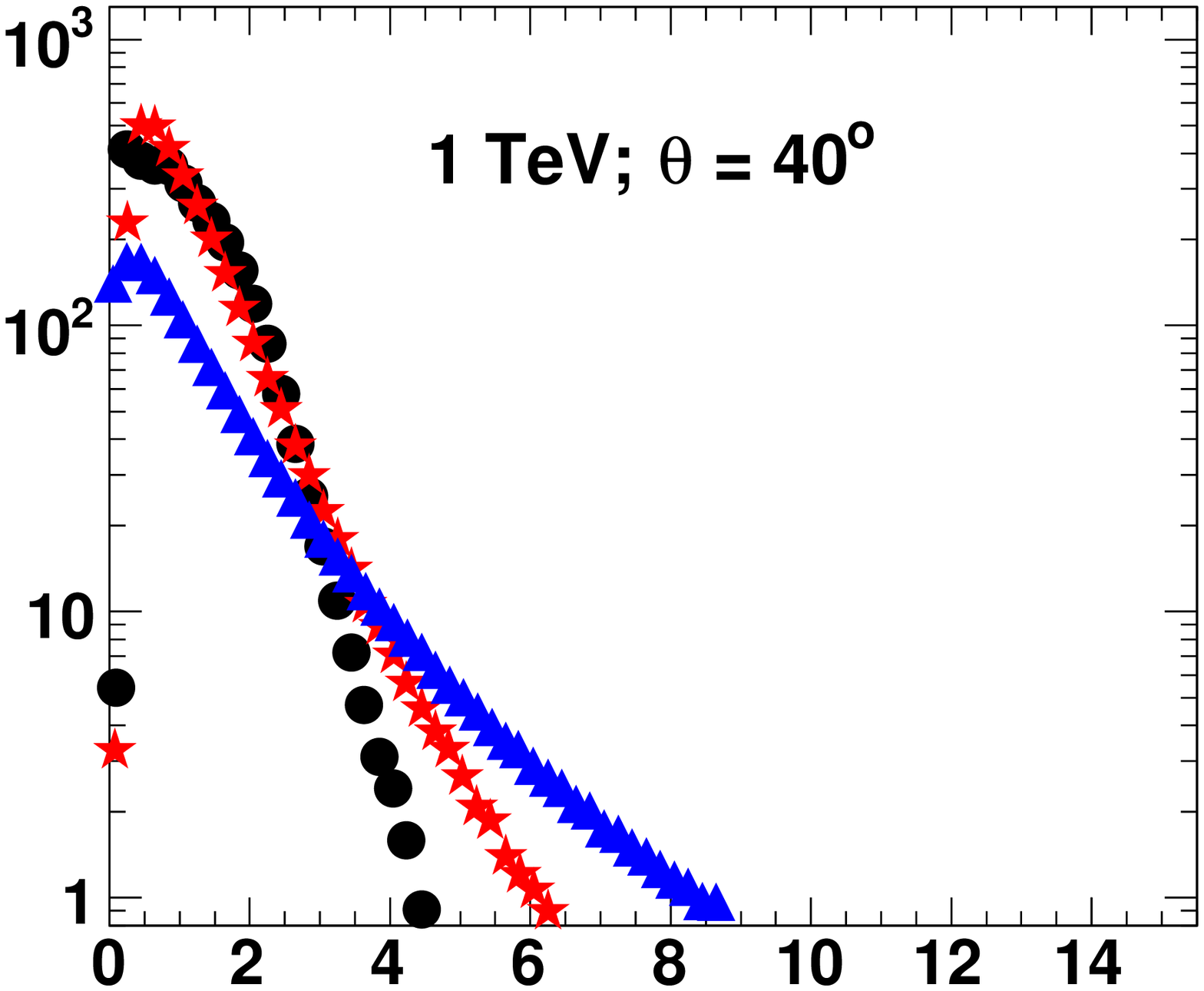}}

\vspace{-3mm}
\centerline{
\includegraphics[width=5.6cm, height=4.7cm]{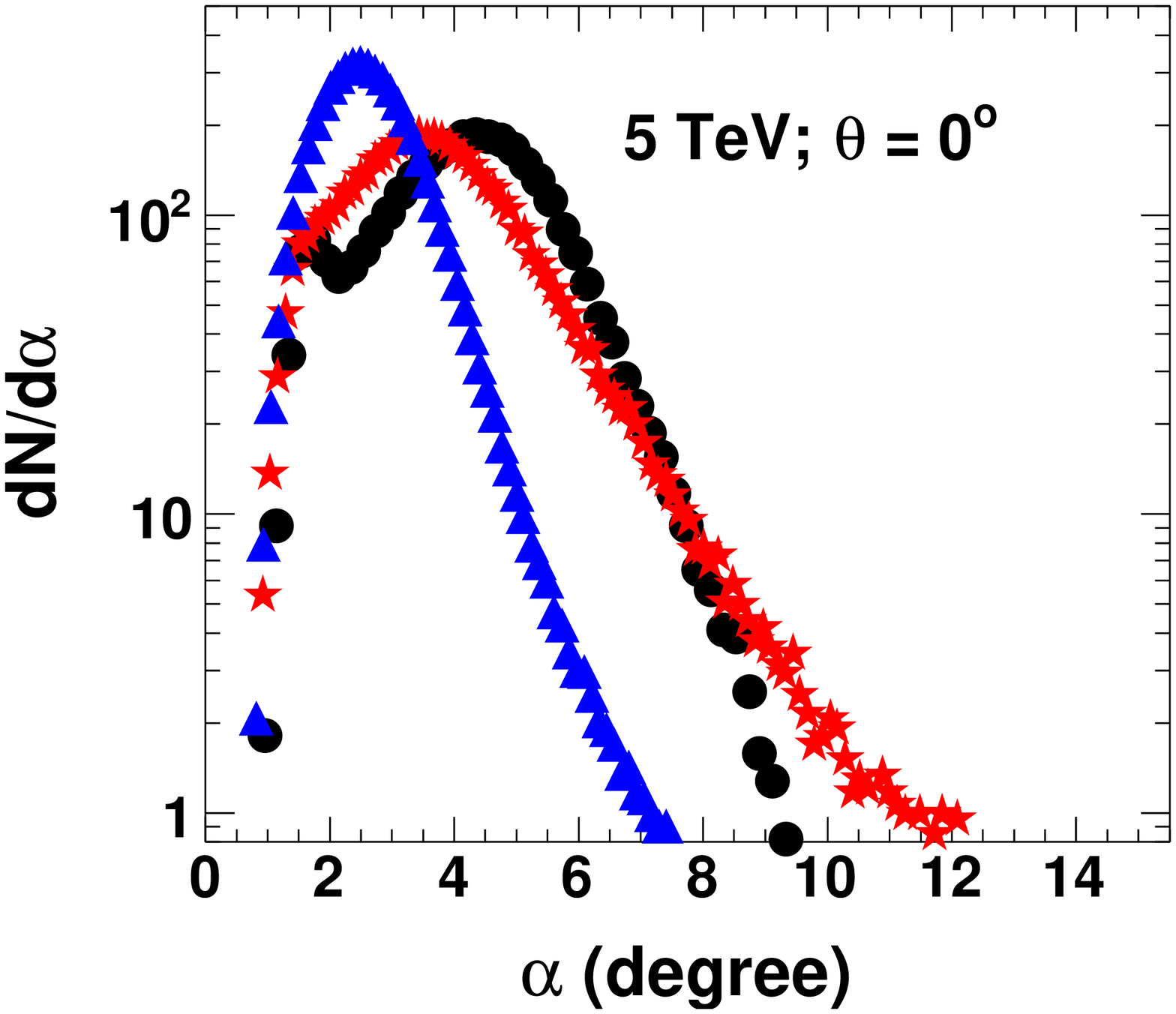}\hspace{-2mm}
\includegraphics[width=5.6cm, height=4.7cm]{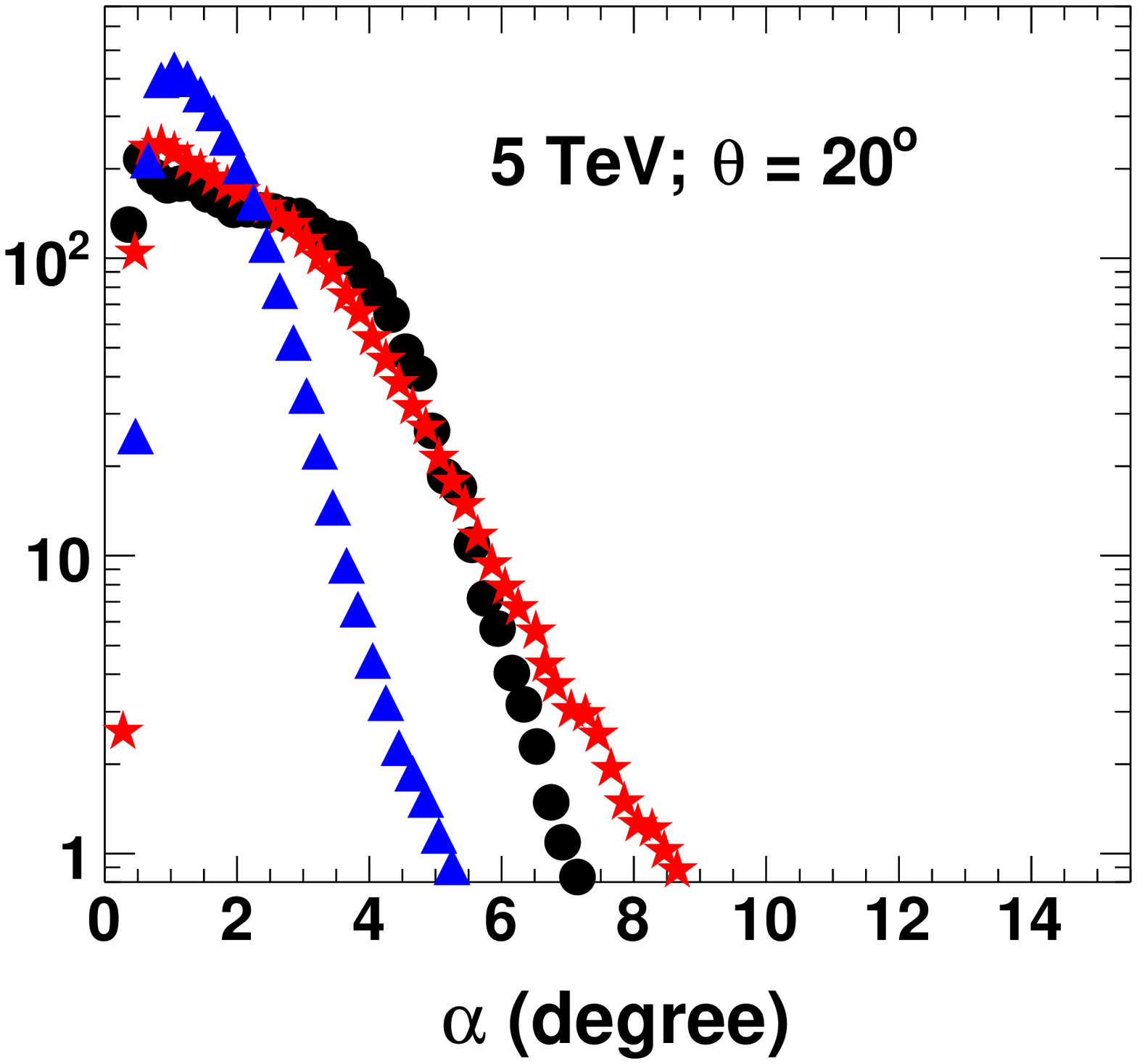} \hspace{-1mm}
\includegraphics[width=5.6cm, height=4.7cm]{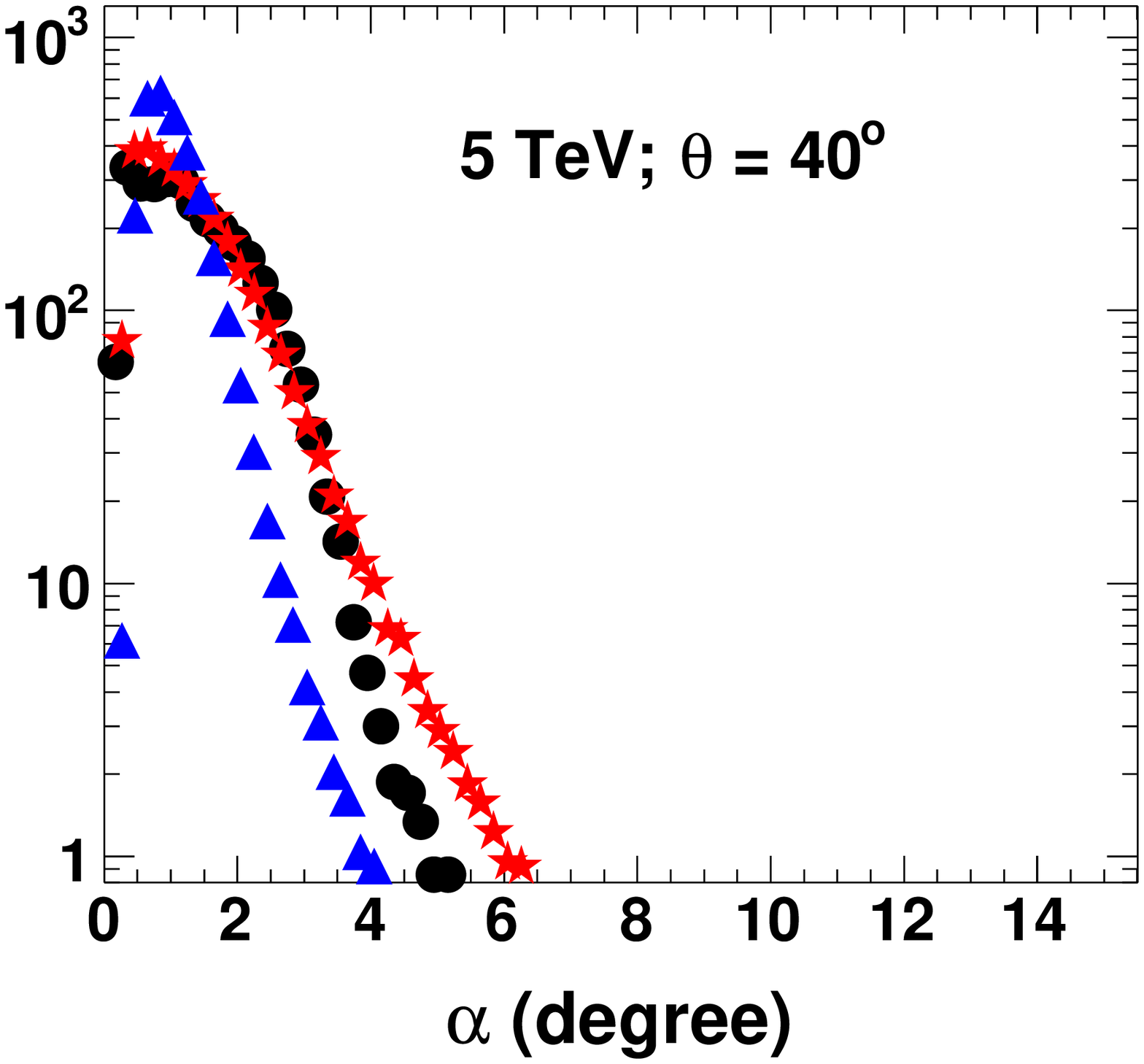}}

\caption{Cherenkov photon's angular distributions obtained from different 
primaries with two different energies and zenith angles.}
\label{fig14}
\end{figure*}

\begin{figure*}[hbt]
\centerline
\centerline{
\includegraphics[scale=0.28]{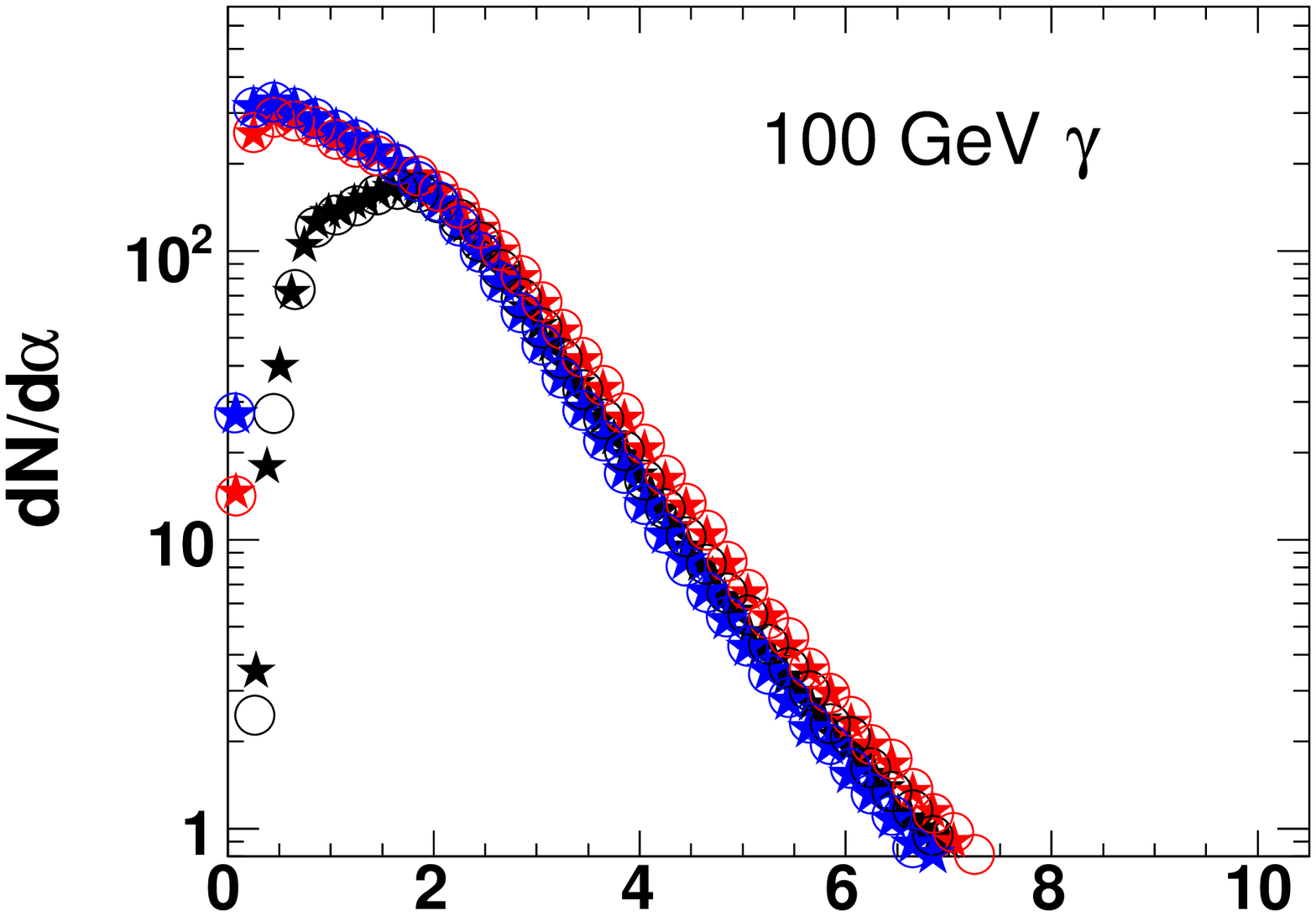} \hspace{-3mm}
\includegraphics[scale=0.28]{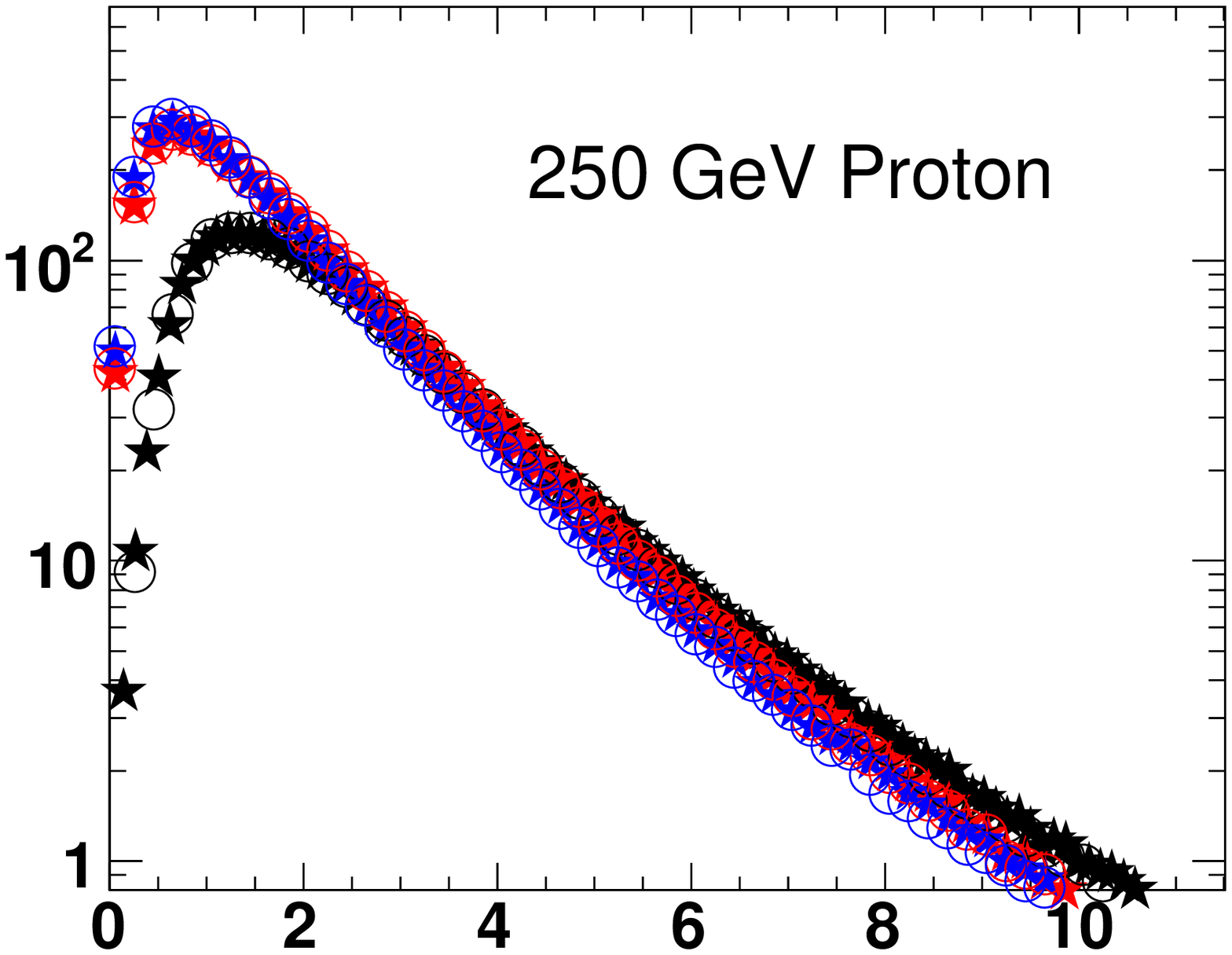} \hspace{-3mm}
\includegraphics[scale=0.28]{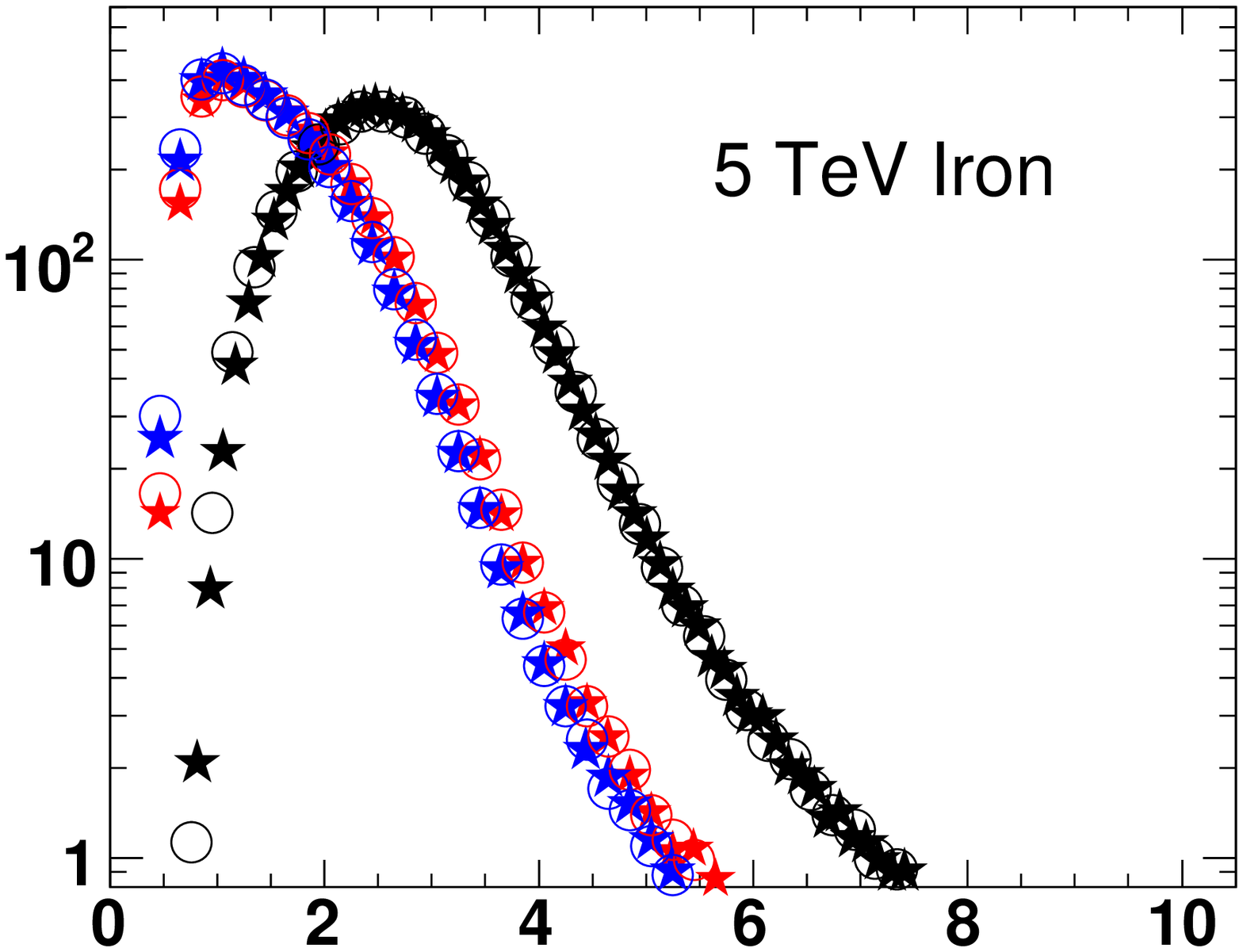}}
\vspace{-3mm}
\centerline{
\includegraphics[scale=0.28]{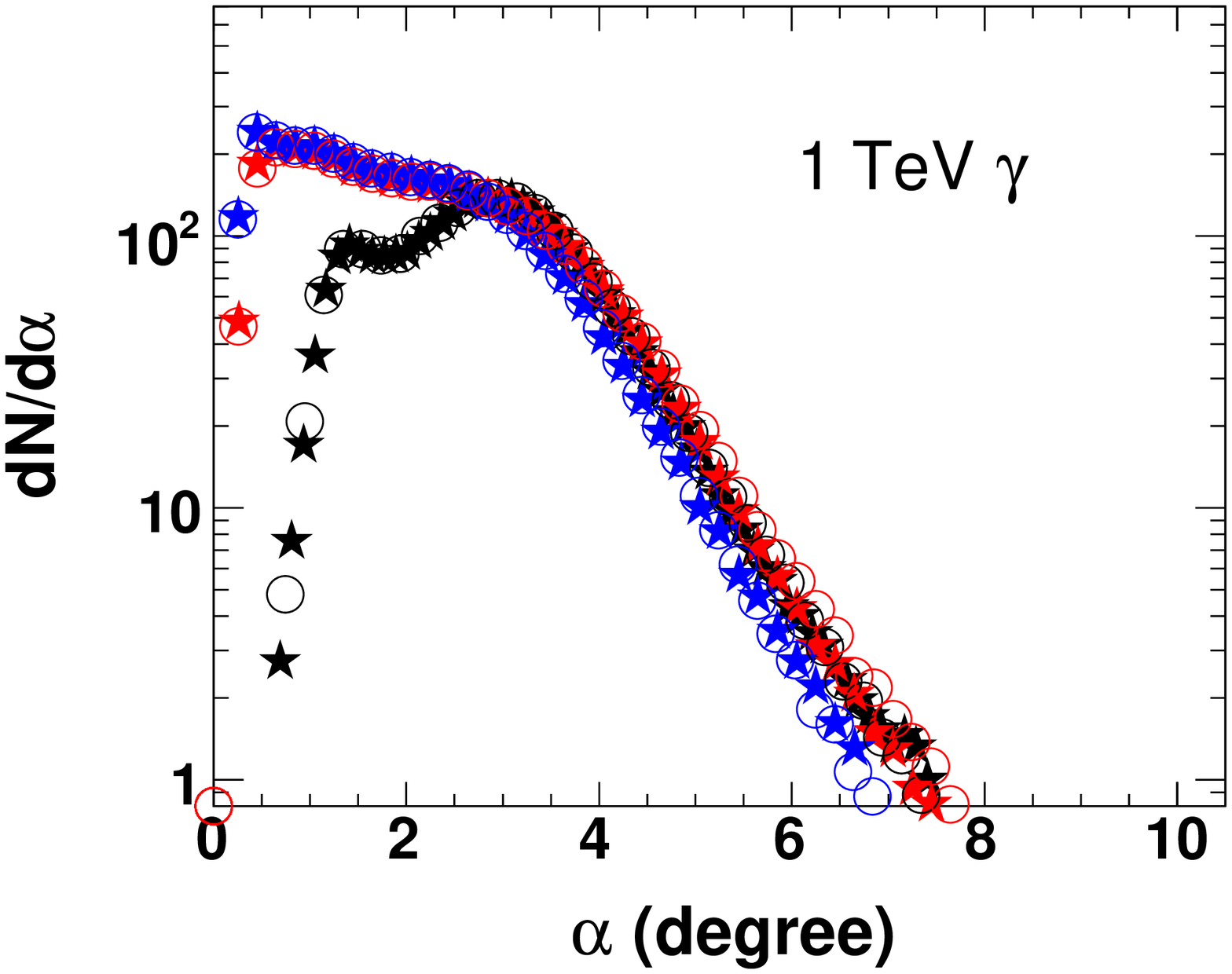} \hspace{-3mm}
\includegraphics[scale=0.28]{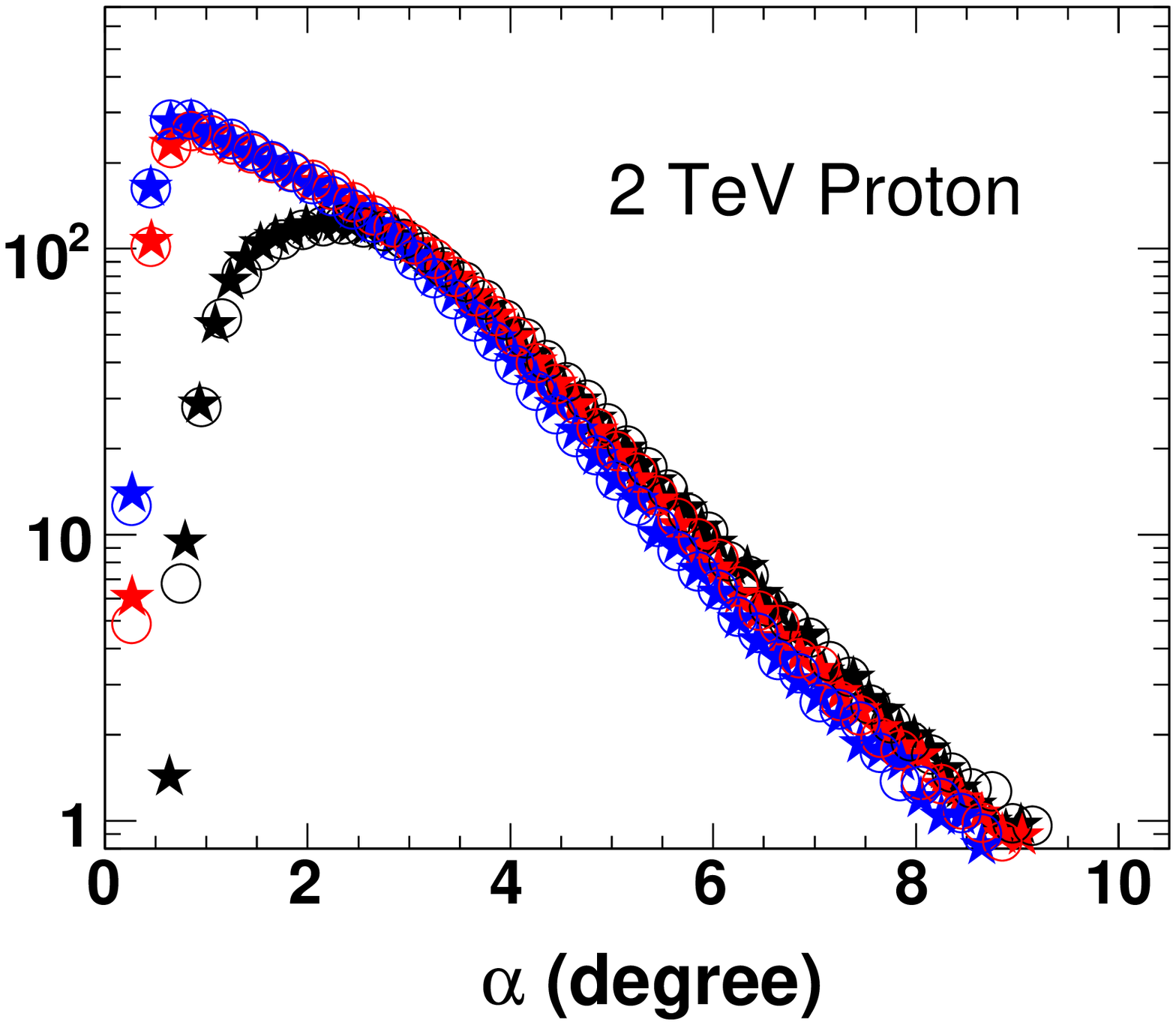} \hspace{-3mm}
\includegraphics[scale=0.28]{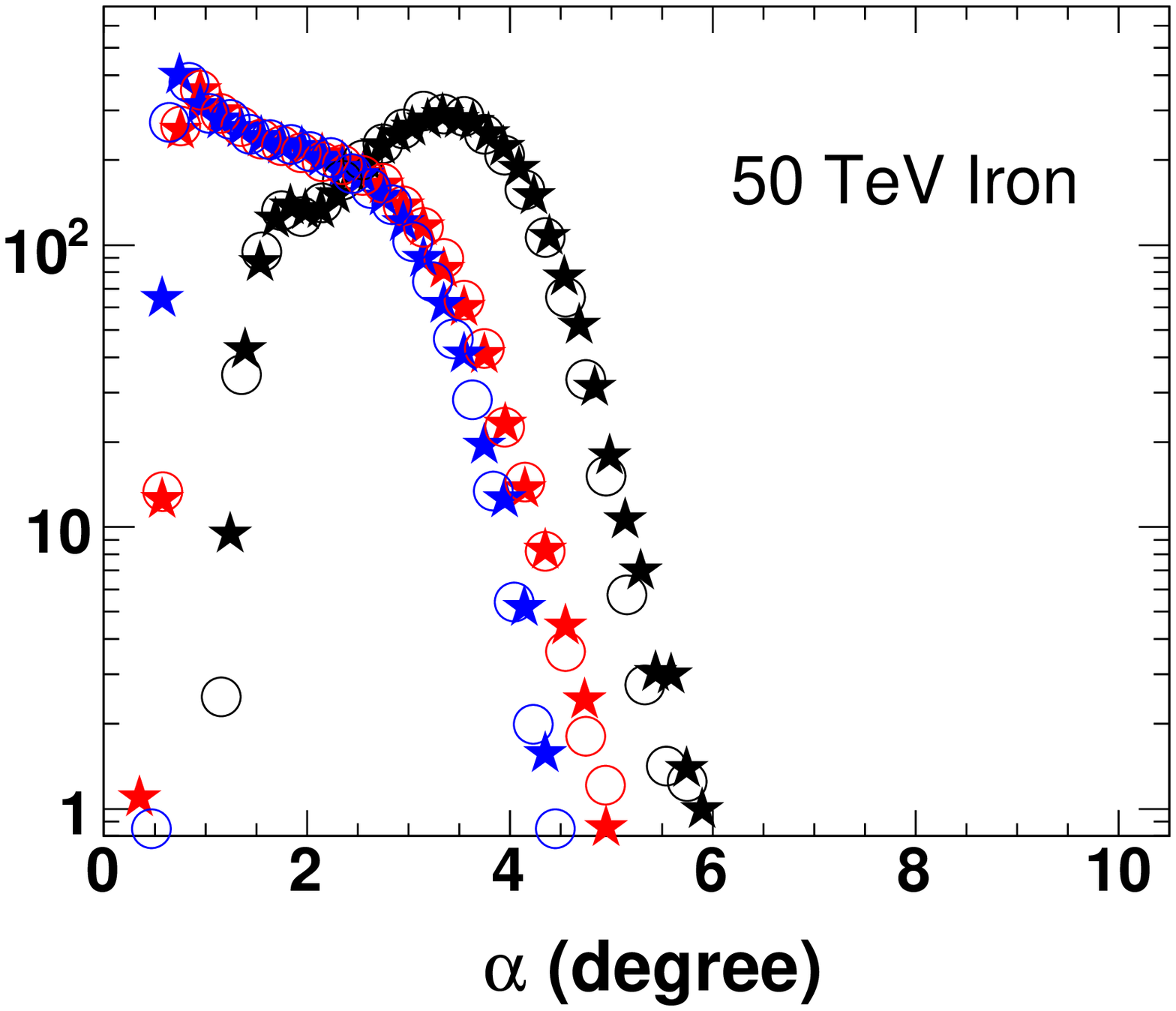}}

\caption{Cherenkov photon's angular distributions for different primaries that
are obtained by using the model combinations, viz., EPOS-FLUKA and
QGSJETII-FLUKA. In the plots {\large $\star$} and {\large $\circ$} indicate
the EPOS-FLUKA and QGSJETII-FLUKA respectively. Black colour represents
$0^{\circ}$, red colour $10^{\circ}$ and blue colour $20^{\circ}$ zenith
angle of the primary particle. }
\label{fig13}
\end{figure*}

\subsubsection{Primary particle, energy and interaction model sensitivity}
To see clearly the effect of the type of primary particle on the angular 
distribution
pattern, we have plotted the Cherenkov photon's angular distributions in the 
showers of $\gamma$-ray, proton and iron primaries for two different 
energies incident at two different zenith angles in the Fig.\ref{fig14}. It is 
seen that at low energy, Cherenkov photons 
initiated by vertically incident $\gamma$-ray are distributed in a small 
angular range in comparison to the corresponding proton and iron primaries, 
where iron has the widest distribution. That is, at low energy the vertically
incident iron shower is distributed in a wider range in comparison 
to proton and $\gamma$-ray primaries of same incident angle. But the maximum
number of Cherenkov photons are produced very near to shower core similar to
the proton. This situation has changed with 
increase in energy, where iron has the smallest angular range and the proton 
has the highest. Perhaps, this feature is attributed by the subsequent 
secondary interactions in the hadronic part of the shower of proton and iron
primaries.  
Moreover, the average angle for the maximum number of Cherenkov photons for 
the proton primary at higher energy is shifted to comparatively higher side 
than that for the iron primary of same energy. But this angle is almost same 
for all three primary types of same energy incident at zenith angles other 
than zero with additional features related with the zenith angle as discussed 
above. 

The Cherenkov photon's angular distributions that are obtained by using the
EPOS-FLUKA and QGSJETII-FLUKA model combinations are shown in the Fig.\ref{fig13}.
It is observed that, there is hardly any difference between the results
obtained from these two models for the $\gamma$-ray, proton and low energy
iron primaries. However, for the iron primary at higher energy there is a 
slight difference between these two models' predictions, basically towards 
the higher side of the value of $\alpha$.
Higher the energy of the primary, more high energy secondary interactions,
which results such differences in the distributions. It is also observed from
this figure that apart from the position of maximum number of photons for
the vertically incident shower, the distributions are almost similar for 
the same energy primaries of same type, incident at different zenith angles, 
specially for the case of $\gamma$-ray and proton primaries.       

\section{Summary and conclusion}
Taking into consideration of the importance of effective techniques for the 
gamma hadron separation in ACT, we have made an effort here to study the 
Cherenkov photon's
density, arrival time and angular distributions in EASs of vertically incident 
as well as inclined $\gamma$-ray, proton and iron primaries with different 
energies using the simulation package, the CORSIKA 6.990 \cite{Heck}. This is 
the sequel of our earlier work \cite{Hazarika} to generalize the study in 
extending energy and zenith angles of primary particles. 

Cherenkov photon's density increases 
with energy for all primaries and decreases almost exponentially with increase 
in distance from the shower core. This is an obvious experimental fact that
at a particular observation level it is easier to detect a high energy shower
than a low energy shower and for a proper estimation of energy of a shower,
the shower has to be well contained within the detector array. Also with 
increase 
in angle of inclination, the density decreases gradually near the shower
core, but remains almost constant far away from the core. This result
also supports the well known observational situation that, it is harder to detect
an inclined than a vertical shower of same energy. $\gamma$-rays have the 
highest Cherenkov photon yield followed by proton and iron for any combination 
of energy, angle and hadronic interaction model. Thus, the equivalent energy of the iron
primary must be highest followed by proton and $\gamma$-ray for a given 
Cherenkov photon yield for any cited combination.     

The average arrival time of Cherenkov photon is found to increase according to 
an exponential function (see equation (3)) with increase in distance from the 
shower core for all combinations of energy and zenith angle. With 
increase in energy, the general trend shows an overall increase in the arrival 
time for all primaries. However, with the 
increasing zenith angle the arrival time profile becomes flatter and  
hence there is a decrease in arrival time. At a particular energy and an angle
of incidence, the average arrival time is highest for the $\gamma$-ray primary and least
for the iron primary. All these information along with the features of density
distribution may be useful to disentangle the showers of $\gamma$-ray from the 
hadronic showers while analyzing the experimental EAS data, apart from usual 
determination of direction of a shower from the arrival time information.  

In general, the shower to shower fluctuation for density and arrival time of 
Cherenkov photons decreases with increasing energy of primary particle,
and is highest for proton primary and least for the $\gamma$-ray primary at
all zenith angle. While estimating the systematic uncertainties in the
data of a $\gamma$-ray experiment, this information will provide an important
input to be considered.        

As we have seen in our earlier work \cite{Hazarika}, the density of photons 
increases with increasing altitude of observation level for all primaries, but 
decreases with the increasing  zenith angle of all these particles. 
At 
highest zenith angle (40$^\circ$) the characteristic hump is also seen for the 
iron primary at lower altitude of the observation level. Similarly, the 
Cherenkov light front is flatter for the lower observation altitude as well as
for the larger zenith angle of all primary particles. Thus the recording 
time of an inclined shower for a $\gamma$-ray telescope array at a lower 
observation level is less than that for a vertical shower of same energy and
observed at higher observation level.

In the calculation of density and arrival time distributions of Cherenkov 
photons, on average four different atmospheric models (viz., U.S. standard
atmosphere as parameterized by Linsley, AT 115 Central European atmosphere for 
Jan. 15, 1993, Malarg\"ue winter atmosphere I after Keilhauer and U.S. standard atmosphere as parameterized by Keilhauer) available in the CORSIKA give
almost similar results. Thus this analysis shows that any one of them may be 
used within a reasonable limit of error.      

The angular distributions of Cherenkov photons have distinct features
in connection with the type of primary particle, its energy and  zenith angle.
For vertical showers of all primaries, the Cherenkov photon's angular position
with respect to shower axis at which
maximum photons are concentrated shifts to higher value with increasing energy
of the primary. This tendency is highest for the $\gamma$-ray primary and least 
for
the iron primary. With increasing zenith angle, the maximum photons are
found to be remained within very near to the shower axes for all energy 
primaries. Also distributions for all primaries and energies gradually become 
narrower with increasing zenith angle. At low energy the iron primary 
has largest angular distribution at all
zenith angles, whereas at higher energy it is the proton which has the 
largest distribution.


Moreover, the QGSJETII-FLUKA and EPOS-FLUKA model combinations produce almost 
similar results in the density, arrival time and angular distributions. So, any 
of these two high energy models can be used to analyze the experimental data of
$\gamma$-ray astronomy.
 
A clear understanding of Cherenkov photon's density, arrival
time and angular distributions for different primary particle with different 
energy and at different zenith angle, and also their possible 
interdependence will be greatly helpful to develop a more efficient technique 
of gamma-hadron separation in future. In this 
context a full paramerizations study on these sensitive parameters 
is very essential. We hope to report such work in future as a part of complete 
simulation study on atmospheric Cherenkov photons.   

\section*{Acknowledgments}
We are thankful to anonymous referees for their useful comments, which leads to
further improve the work. UGD is thankful to the Inter-University Centre for 
Astronomy and Astrophysics (IUCAA), Pune for hospitality during his visit as
a Visiting Associate.

\end{document}